\documentclass[acmsmall,natbib=false]{acmart}

\renewcommand\footnotetextcopyrightpermission[1]{} 

\AtBeginDocument{%
  \providecommand\BibTeX{{%
    Bib\TeX}}}

\usepackage{amsmath,amssymb}
\usepackage{algorithmic}
\usepackage{graphicx}
\usepackage{textcomp}
\usepackage{xcolor}
\usepackage{algorithmic}
\usepackage{algorithm}
\usepackage{array}
\usepackage{textcomp}
\usepackage{stfloats}
\usepackage{url}
\usepackage{verbatim}
\usepackage{graphicx}
\usepackage{float}

\usepackage{listings} 
\usepackage{gensymb}
\usepackage{graphicx}
\usepackage{mathtools}

\usepackage[inline]{enumitem}
\usepackage{amsmath}
\usepackage{amssymb}
\usepackage{optidef}
\usepackage{lineno,hyperref}
\usepackage{amsthm}
\usepackage{xcolor}
\usepackage{algorithm,algorithmic}
\usepackage{subcaption}
\usepackage{bbm}
\usepackage{eqparbox}
\usepackage{makecell}
\usepackage{booktabs}

\usepackage{etoolbox}
\usepackage{calc}
\usepackage{upgreek}
\usepackage{multirow}
\usepackage{adjustbox}
\usepackage{lscape}
\usepackage{pdfpages}
\usepackage{subcaption}
\usepackage{wrapfig}
\usepackage{pifont} 
\newcommand{\yesmark}{\ding{51}}  
\newcommand{\nomark}{\ding{55}}

\newcounter{innerfig}

\usepackage{tikz}
\usetikzlibrary{decorations.pathreplacing}
\usepackage{amsmath}

\newcommand{\stepnum}[1]{%
\tikz[baseline=(char.base)]{
\node[shape=circle, fill=black, text=white, inner sep=1pt] (char) {\small #1};}
}

\DeclareMathOperator*{\argmin}{arg\,min}
\def\BibTeX{{\rm B\kern-.05em{\sc i\kern-.025em b}\kern-.08em
    T\kern-.1667em\lower.7ex\hbox{E}\kern-.125emX}}

\RequirePackage[
  datamodel=acmdatamodel,
  style=numeric,
  sorting=none,
  ]{biblatex}
\newcommand{\suresh}[1] {{\color{black} #1}}
\newcommand{\method}{\textsc{SEED}}

\begin{document}
\textcolor{blue}{\large \textbf{Accepted to ACM Transactions on AI Security and Privacy (June 2026)}}
\\
\title{\suresh{\method}: \underline{S}emi-supervised Continual Malwar\underline{E}
Detection for Tackling Conc\underline{E}pt Drift on a Bu\underline{D}get}

\author{Suresh Kumar Amalapuram}
\email{apskumarkrc@gmail.com}
\orcid{0009-0007-6789-997X}
\affiliation{%
  \institution{Indian Institute of Technology Ropar}
  \country{India}
}
\author{Bikraj Shresta}
\orcid{0009-0006-1321-0156}
\email{bikrajshrestha2@gmail.com}
\affiliation{%
  \institution{Indian Institute of Technology Hyderabad}
  \country{India}
}

\author{Siva Ram Murthy Chebiyyam}
\orcid{0000-0002-1032-8857}
\email{murthy@cse.iith.ac.in}
\affiliation{%
  \institution{Indian Institute of Technology Hyderabad}
  \country{India}
}

\author{Bheemarjuna Reddy Tamma}
\orcid{0000-0002-4056-7963}
\email{tbr@cse.iith.ac.in}
\affiliation{%
  \institution{Indian Institute of Technology Hyderabad}
  \country{India}
}

\author{Sumohana Channappayya}
\orcid{0000-0002-5687-0887}
\email{sumohana@ee.iith.ac.in}
\affiliation{%
  \institution{Indian Institute of Technology Hyderabad}
  \country{India}
}
\begin{abstract}
 Machine learning based malware detectors become obsolete over a period of time due to \emph{concept drift} in the benign and malware apps. Recent method requires fully labeled data and combines hierarchical contrastive \suresh{loss} with active learning (HCL), \suresh{leveraging the semantic structure of malware representations to improve robustness against drift}. \suresh{However, obtaining labeled data remains an arduous task in the security domain.}  Alternately, if training is performed using partially labeled data, HCL suffers a performance degradation in detecting unseen malware (in terms of area under the time (AUT) metric \suresh{computed over precision recall area under the curve), particularly on datasets (such as BODMAS) where such semantic structures may not be present}. Specifically, on the BODMAS dataset the reduction is from 0.795 to 0.613 and on AndroZoo, it is from 0.879 to 0.780.
 
In this paper, we propose a semantic-structure-agnostic method for malware detection. The proposed method dubbed \method{} is robust under limited supervision, outperforming contrastive learning in the absence of strong semantic structure while remaining competitive when such structure exists. Specifically \method{} adopts a tailored binary cross-entropy, and integrates it with semi-supervised continual learning and active learning. The proposed method finds a suitable labeled sample for each unlabeled sample (from the seen tasks) by projecting it into the representation space  constructed from the basis vectors of the previously seen data using singular value decomposition and encourages the malware detector to learn the same representation for these pairwise samples. On unseen tasks (fully unlabeled data), we quantify the uncertainty of the new sample using cosine distance in representation space and select the most uncertain samples for the security analyst to label. Later, we train the unseen tasks using seen tasks procedure. We evaluated the \method{} using both Windows and Android malware datasets. Our experiments show that the proposed method, using 20\% labeled data on seen tasks, achieves an average improvement (in terms of AUT) of 40\% on the BODMAS dataset and 14\% on the AndroZoo dataset for detecting unseen malware across varying labeling budgets, compared to HCL$^*$ (the semi-supervised adaptation of HCL). Further, its performance is competitive with HCL$^*$ on APIGraph dataset. Eventually, we introduce a delayed buffer update strategy to mitigate the effect of label noise in unseen tasks. This reduces noise propagation during replay and improves learning stability. Our code is available at the following link~\footnote{\url{https://github.com/amalapuram/SEED}}.
\end{abstract}






\maketitle

\section{Introduction}

Recently, sophisticated machine learning (ML) algorithms surpassed human intelligence in many cognitive tasks ~\cite{krizhevsky2012imagenet,NIPS2017_3f5ee243}. Inspired by this success, many security researchers are adopting ML solutions for developing malware detectors. However, the progress of ML is slower in security applications as compared to other domains, due to the fixed \textit{closed world} assumption, in which training and testing data are drawn from the same probability distribution. On contrary, security applications are deployed in a hostile environment in which testing data distributions are different from training data (this phenomenon is known as \textit{concept drift}), causing gradual performance decay in detection rate of the ML classifier.

\noindent \textbf{Concept drift:} This may occur in both benign (often known as goodware) and malware executables. For example, thousands of software companies release new types of benign executables that are significantly different from those seen in train data~\cite{kasperskywhitepaper}. On the other hand, adversaries such as malware authors constantly write novel malware or extend the functionality of existing malware by code obfuscation techniques/tools~\cite{themida,codevirtualizer}  with an intent to evade detection by malware detectors. Concept drift in malware can also arise due to changes in the behavior of an application's maliciousness. For instance, earlier mobile ransomware aimed at stealing personal information has shifted to collecting users' expense consumption details~\cite{chow2023drift}. To mitigate the effects of concept drift, malware classifiers must continuously evolve, maintaining both the stability of learned malware knowledge and the plasticity to identify and adapt to novel malware. The plasticity property is known as \textit{open world learning} (OWL)~\cite{guo2022robust,cao2022openworld}. The classifier that fails to maintain stability property is said to exhibiting the \textit{catastrophic forgetting} ~\cite{wang2024comprehensive,lopez2017gradient} of the learned knowledge. We assume that the basic unit of training is a \textit{task} in this setting~\cite{chaudhry2018efficient,aljundi2019gradient,chrysakis2020online}. In malware detection, a task contains a mix of benign and malware samples collected over a period of time (say a month). Catastrophic forgetting occurs when learning a new task hurts the performance of previously learned tasks~\cite{lopez2017gradient}.

\begin{figure*}[!t]
\centering
\begin{subfigure}{0.49\textwidth}
\centering
\includegraphics[scale=0.2]{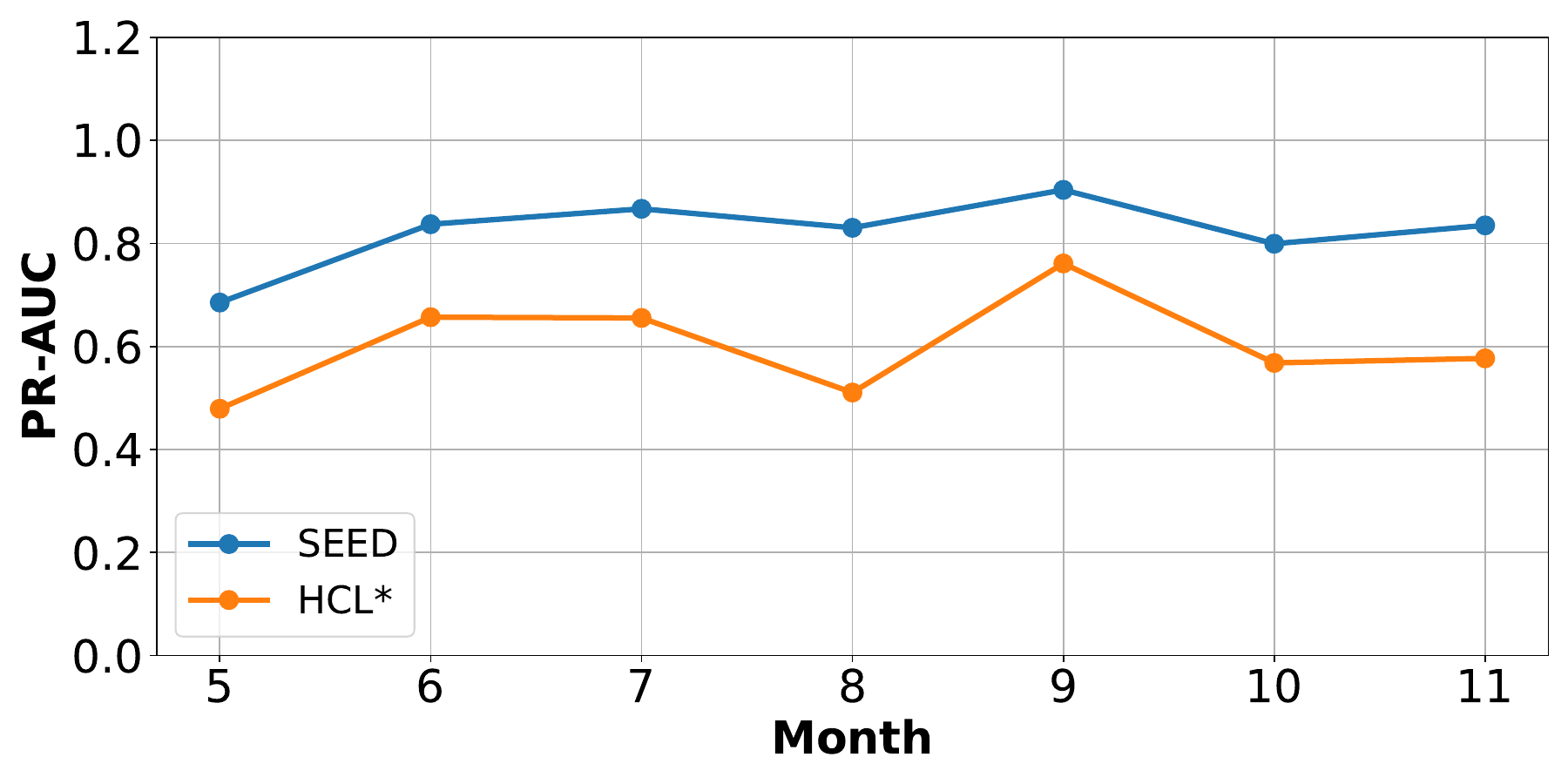}
\caption{PR-AUC on unseen malware on BODMAS dataset.}
\label{fig:bodmas_motivation}
\end{subfigure}
\begin{subfigure}{0.49\textwidth}
\centering
\includegraphics[scale=0.2]{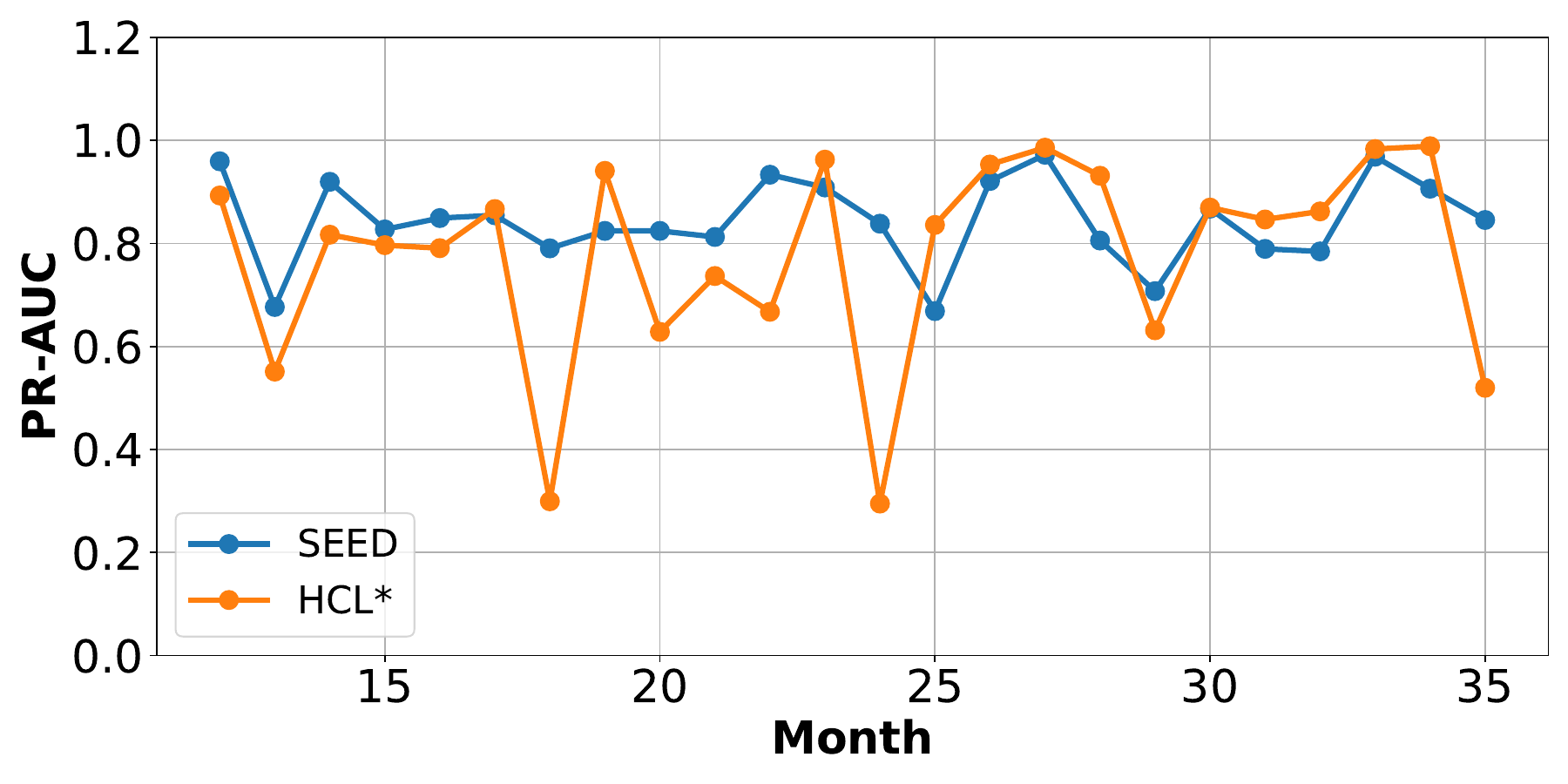}
\caption{PR-AUC on unseen malware on AndroZoo dataset.}
\label{fig:androzoo_motivation}
\end{subfigure}


\begin{subfigure}{0.99\textwidth}
\centering
\caption{AUT over unseen malware on BODMAS and AndroZoo datasets}
\footnotesize
\begin{tabular}{lccc}

 & \multicolumn{3}{c}{AUT over unseen malware} \\
\cmidrule(lr){2-4}
Dataset & HCL~\cite{291253} & HCL$^*$~\cite{291253} & \method{} (proposed) \\
\midrule
BODMAS~\cite{bodmas} 
& 0.795 
& 0.613 ($\textcolor{red}{\downarrow}$) 
& 0.810 ($\textcolor{green}{\uparrow}$) \\

AndroZoo~\cite{291253}
& 0.879 
& 0.780 ($\textcolor{red}{\downarrow}$)  
& 0.855 ($\textcolor{green}{\uparrow}$) \\
\bottomrule
\end{tabular}

\end{subfigure}

\Description{(a) PR-AUC performance on unseen BODMAS tasks. 
(b) PR-AUC performance on unseen AndroZoo tasks. 
(c) AUT comparison under partially labeled training.}
\caption{\textcolor{black}{Comparison of SEED with HCL$^*$ (HCL trained on partially labeled data) for unseen malware detection on the BODMAS and AndroZoo datasets. The BODMAS dataset spans twelve months; SEED and HCL$^*$ are trained on the first five months and evaluated on the remaining seven months. Similarly, AndroZoo spans three years (36 months); the malware detectors are trained on the first twelve months and evaluated on the remaining months. The top two figures (Fig.~\ref{fig:bodmas_motivation} and Fig.~\ref{fig:androzoo_motivation}) illustrate the performance decay of the respective methods using the precision–recall area under the curve (PR-AUC) metric. The x-axis (labeled as month) denotes the testing month for each dataset. The table in the bottom row summarizes performance degradation in detecting unseen malware across all evaluation months using a single aggregate metric, the Area Under the Time curve (AUT), computed over the PR-AUC trajectory. SEED, based on semi-supervised continual learning with active learning, improves AUT by 32\% on BODMAS and 9\% on AndroZoo under partially labeled settings.}}

\label{fig:unseen_malware_motivation}
\end{figure*}

\noindent \textbf{Mitigating the catastrophic forgetting:} Forgetting can occur when the balance between stability and plasticity is disrupted. This situation can primarily occur when learning novel malware (or benign) features interferes with the existing knowledge. \textcolor{black}{Although individual malware samples may become obsolete within a short period, catastrophic forgetting remains a practical concern in learning-based malware detection, as malware families, feature patterns often persist, evolve, and reappear over time~\cite{arp2014drebin,pendlebury2019tesseract,yang2021cade,wu2023grim}.
Malware families evolve incrementally through polymorphism and code reuse, where new variants retain structural and behavioral similarities with historical samples~\cite{arp2014drebin,yang2021cade}. Forgetting prior representations can therefore impair detection performance. Moreover, drift in the benign samples have been shown to degrade long-term robustness of malware detectors~\cite{pendlebury2019tesseract,wu2023grim}, reinforcing the need for stability–plasticity balance.} Therefore, to mitigate the effects of catastrophic forgetting, our work uses the continual learning (CL) paradigm for sequential learning. The reason for selecting CL framework is two fold: to maintain the learned knowledge for a longer period the CL algorithms require limited access to the past data and CL algorithms are flexible to integrate novel knowledge. Despite its merits, the direct application of CL for malware detection is under explored~\cite{rahman2022limitations}. Furthermore, the existing works in CL heavily focus on supervised settings and assume a closed-world setting. 

\noindent \textbf{Labeling issues for security:} CL algorithms require access to labeled data from the evolving distribution to remain effective in the OWL setting. Unlike other domains (computer vision), annotations in cybersecurity domain are influenced by many other factors such as labeling dynamics (flips), label unification, labeling budget, lack of correctness verification in label generation methods, among others. 
\textcolor{black}{Labeling budget is the amount of money spent on human analyst to manually obtain the labels. A typical human analyst can review about 80 samples per day~\cite{miller2016reviewer}.}
\textcolor{black}{Consequently, a low budget limits the number of labeled samples, which are crucial for adapting to concept drift in OWL. Further, the correctness of the auto labeling process in case of cybersecurity domain is questionable}~\cite{liu2022error}. 
\textcolor{black}{Even when assuming access to relatively clean analyst-verified labels, constructing and maintaining a fully labeled dataset over time is prohibitively expensive and operationally unrealistic in open-world malware detection. The instability of VirusTotal (VT)-based labeling further increases the cost of obtaining consistent supervision, reinforcing the impracticality of relying on large-scale fully labeled continual training. Therefore, instead of explicitly modeling label noise, this work focuses on adapting to concept drift under limited but reliable labeled supervision, reducing dependency on exhaustive labeling while preserving robustness in evolving threat landscapes. Thus, we focus on the tackling labeling budget issue in concept drift adaptation under limited labeled samples}.

\par Existing state of the art (SOTA) approaches~\cite{291253,yang2021cade} for handling concept drift in malware detection utilize contrastive representation learning. These methods work by minimizing intra-class distances while maximizing inter-class distances. The effectiveness of such representations heavily relies on the availability of fully labeled train data to construct positive and negative sample pairs. 


Further, our research also finds that using partial labeled data on contrastive methods result in suboptimal performance (refer to Figure~\ref{fig:unseen_malware_motivation}). Motivated by our findings and labeling issues we propose a novel semi-supervised continual learning and active learning, dubbed \method{}, that effectively leverages partially labeled data without using contrastive loss and a delayed buffer update strategy to mitigate the effects of noisy labels. 



\noindent \textbf{Two-stage learning strategy in \method{}:}  \textcolor{black}{Semi-supervised learning (SSL)~\cite{zhu2022introduction} leverages the unlabeled data as an additional source of knowledge to improve the performance of malware detectors. The traditional SSL setting assumes that classifier can accurately generate labels for unlabeled data after training on a partially labeled dataset, due to the closed world assumption (labeled and unlabeled data are drawn from the same distribution). However, this assumption may not hold for malware detection as the unlabeled data contains drifted and  novel samples that are different from partially labeled data.  To solve this, in this paper, we design training strategies to handle two scenarios when the partial labeled data is available (\textit{seen} tasks) and unavailable (\textit{unseen} tasks). Concretely, seen tasks contain partially labeled and unlabeled data, whereas unseen tasks contain only unlabeled data.}

\noindent \textbf{Training with seen tasks:} 
\textcolor{black}{Given a task $`t$' with labeled exemplars ${D}_l^t$ and unlabeled set $D_u^t$, our goal is to find a suitable labeled sample for each sample $\textbf{x} \in D_u^t$. This is achieved by projecting samples into the latent space encoded by the classifier, using the partially labeled data.}  
\textcolor{black}{However, the effectiveness of this discrimination is limited by the absence of fully labeled data. Further, in the CL setting, the previously learned discriminative features can be hindered by interference from the learning of new tasks. So, to improve this we use orthogonal projections along with a novel buffer memory management method to minimize interference.}

\noindent \textbf{Training with unseen tasks:} \textcolor{black}{Our second contribution is a novel training procedure for a task involving solely unlabeled data.} 
\textcolor{black}{Given a unseen task $`t^\prime$' with unlabeled data $D_{u}^{t^{\prime}}$, our goal is to transform the unlabeled data ($D_{u}^{t^{\prime}}$) to a partially labeled data $\{D_{l}^{t^{\prime}},D_{u^*}^{t^{\prime}}\}$, where $D_{u^*}^{t^{\prime}} \subset D_{u}^{t^{\prime}}$. Later, we train $D_{u^*}^{t^{\prime}}$ using training process of seen task.}

 \noindent \textbf{Delayed Buffer Update Strategy for Noisy Label Mitigation:} \textcolor{black}{To mitigate the adverse effects of noisy labels, we propose a simple delayed buffer update strategy that avoids immediately storing potentially unreliable samples in the replay buffer. Instead, samples are incorporated after a fixed delay, allowing labels to stabilize over time and thereby reducing noise propagation during replay-based learning.}

\textcolor{black}{Through extensive experiments on both Windows and Android malware datasets , our method \method{}, compared to HCL$^*$, achieves an average (with different labeling budgets) improvement of 40\% on the BODMAS dataset and a 14\% improvement on the AndroZoo dataset in detecting unseen malware using the AUT metric. Furthermore, the proposed method demonstrates competitive performance on the APIGraph dataset. We also conduct comprehensive ablation studies and sensitivity analyses. Collectively, these experiments demonstrate that \method{} is robust under limited supervision, outperforming contrastive learning when strong semantic structure is absent while remaining competitive when such structure is present.}

Accordingly, this paper makes the following contributions to the field of malware detection.
 \begin{itemize}
     \item We shed light on the difficulty of maintaining higher detection performance of the contrastive learning-based malware detector under limited supervision (Figure~\ref{fig:unseen_malware_motivation}).
     \item We propose a novel method, dubbed SEED, based on semi-supervised continual learning and active learning. SEED quantifies the uncertainty of an unknown sample by projecting it into a representation space constructed from the basis vectors using the available partially labeled data (Sections~\ref{sec:seen_train} and ~\ref{sec:unseen_train}).
     \item We introduce a simple delayed buffer update strategy (Section~\ref{sec:delayed_buffer_update}) to mitigate noisy label effects and evaluate its robustness under varying label noise levels (10\%–80\%).
     \item \suresh{With extensive evaluation (Section ~\ref{sec:results}), we demonstrate that \method{} achieves better performance on datasets where semantic structure is absent and remains competitive when such structure is present. To encourage reproducible research and promote open science, we publicly release our code.}
 \end{itemize}


 \section{Related Work}
\label{app:related-work}
\par In this section, we describe the related work, covering the concept drift in malware detection, various approaches including contrastive learning and other incremental learning methods to address concept drift, and existing strategies for leveraging unlabeled data. In each of these, we clearly highlight how the proposed approach differs from the existing works.

\textbf{Concept drift in malware detection:} Concept drift can arise in benign apps due to its natural evolution (e.g., introducing new functionality), whereas in malware this drift can be attributed to the novel malware variants introduced or modifying the existing malware using code obfuscation techniques. Traditionally, handling concept drift in security applications is a two-step process~\cite{yang2021cade,jordaney2017transcend,hananomaly,barbero2022transcending,arp2014drebin,yang2024recda}; drift detection and drift adaptation. One of the earliest works that attempts to find the drift in the samples is \texttt{TRANSCENDENT}~\cite{jordaney2017transcend,barbero2022transcending}. This method rejects the drifted samples using the non-conformity measure (NCM) based on algorithm confidence and credibility. Confidence is the likelihood of the given test object belonging to a particular class and credibility quantifies how relevant the training set is to the prediction.

\noindent \textbf{Contrastive learning for malware detection:} Another set of recent works focuses on supervised contrastive loss to deal with concept drift for malware detection. Previously, the unsupervised version of this loss function has widely used for contemporary learning tasks (image classification, object detection) in the computer vision domain. It works by constructing positive and negative pairs for a given anchor sample using the label preserving data augmentation techniques. However, such augmentation techniques may not be effective under concept drift~\cite{wu2023grim} for malware analysis. \texttt{CADE} uses contrastive auto encoder with a distance (median absolute deviation) based function to identify the drifted sample as an out of distribution (OOD) sample. The OOD samples labeled by a security analyst are incorporated into training for adapting the classifier to the drift. Inspired by \texttt{CADE}, recently~\cite{291253} uses hierarchical contrastive loss along with cross entropy loss (pseudo-label confidence) to identify the uncertain test samples. These test samples are labeled by the analyst and incorporated in the training set. Another recent work~\cite{thirumuruganathan2024detecting} uses two techniques to detect sampling bias; one is based on domain discrimination (using logistic regression) and the other method is non-parametric (\textit{k}-NN based bias detection combined with contrastive loss). Authors of ~\cite{thirumuruganathan2024detecting} also proposed two methods to adapt to concept drift; the first approach is based on the contrastive loss function combined with generating pseudo labels for unseen test samples, whereas the other approach uses cyclic consistency loss.

\noindent \textbf{Continuous/Incremental learning in malware detection}: The authors of~\cite{rahman2022limitations} investigated the suitability of CL for malware classification using various CL techniques spanning regularization, generative replay, and exemplar replay family of approaches over three malware datasets. Their findings suggest that partial replay of historical data will improve malware classification performance. 

Another set of works relies on the active learning scheme for identifying hard samples for labeling by an oracle (security analyst). For instance, ~\cite{291253} uses hierarchal contrastive loss value for identifying the most uncertain samples. \texttt{BODMAS}~\cite{bodmas} compares the various active learning methods based sample selection schemes. 

There are other works that do not rely on a human oracle, but rather on a trained classifier providing labels for unseen test samples. INSOMNIA~\cite{andresini2021insomnia} is one of the earliest to auto generate labels for the unseen network traffic using a nearest centroid neighborhood classifier. Droidevolver~\cite{xu2019droidevolver} also generates pseudo labels for the new malware variant avoiding the manual labeling overhead. However, self-labeling strategies are prone to contaminate the learning through self poisoning~\cite{kan2021investigating}. 

One way to mitigate this label contamination issue is to solve the malware detection problem in an unsupervised fashion. OWAD~\cite{hananomaly} formulates drift detection and adoption as an anomaly detection (AD) problem. This work has shown that adapting to distribution shift in benign applications is sufficient to handle various malware variants. Recently, \cite{escudero2023application} formulates the malware detection problem as AD to mitigate the effects of class imbalance and distribution shifts. However, methods formulated as AD may fail to distinguish whenever the novel malware variants bear similarity to the drifted benign samples. 
Orthogonal to these existing works, our proposed method focuses on semi-supervised continual learning and handles concept drift using partially labeled data.

\noindent \textbf{Leveraging unlabeled data in malware detection:} Here, we discuss two popular ways of leveraging unlabeled data to improve performance of the malware detector. The first method is via the self supervision with carefully designed pretext tasks. One such pretext task is to use consistency regularization loss that encourages the classifier to have the same distribution for the feature vector and its augmented version. These augmented operations involve masking the values of feature vectors and filling the masked values with using some empirical distribution~\cite{yoon2020vime}. The pretext task involves recovering original sample from corrupted sample by estimating mask vector and feature vector.
Recently, ~\cite{wu2023grim} used this feature mask estimation for malware classification tasks and has shown that such a transformation is ineffective under concept drift. Further, such transformations may not preserve the original label~\cite{apruzzese2024when,thirumuruganathan2024detecting} as compared to domains such as computer vision. 

Another way of leveraging the unlabeled data is to use \textit{pseudo labeling} in a teacher-student training paradigm. The working principle behind this strategy is knowledge distillation. The teacher model is expected to have more wisdom in generating the pseudo labels for the unlabeled data. The student model considers these labels as ground truth and reduces the entropy of its prediction using a supervised classification loss function (such as cross entropy). However, the generated pseudo labels may contain false positives, resulting in contaminating the learning of the student model. To mitigate this, often high confidence pseudo labels above a certain predefined threshold ($\tau$) are considered for learning. In the CL setting, the classifier trained until the last task is used as a teacher model for the immediate next task. Meanwhile, the student model can be an entirely new classifier initialized with random weights or the teacher model continued to be trained on the new task~\cite{lechat2021pseudo,wang2021ordisco,suresh}.  

\par In contrast to the existing methods, our approach finds the most suitable labeled example corresponding to each unlabeled example and encourages the classifier to learn similar representations in the latent space.


\section{Preliminaries}
This section provides the foundational definitions essential for a comprehensive understanding of the work, including the definition of a task in the context of malware detection, how tasks are created from datasets, and a description of the problem of malware detection within the framework of SSCL.
\par \textbf{Notion of a `task':} A task consists of a subset of training examples. The process of creating a task is sensitive to the application domain. For instance, in image classification on CIFAR-10~\cite{krizhevsky2009learning} dataset, a task contains one (or more) class(es) of images. For security applications,  we find that two characteristics are meaningful while creating a task.
\begin{itemize}
    \item \textit{Class Imbalance} (CI): It is the ratio of the number of benign to malware samples. In real-world scenarios, malware datasets exhibit higher CI with most of the samples being benign~\cite{291253,li2022imbalanced,zhang2020enhancing,androzoo,bodmas,2018arXiv180404637A}. 
    \item \textit{Granularity of data:} It describes the temporal granularity of the task. Specifically, it describes the time span of the data used in task creation. For instance, whenever the granularity of the data is one month and a malware dataset is spread over one year, it will have a total of $12$ tasks.
\end{itemize}

\noindent \textbf{Datasets:} The training dataset is a sequence of ${T}$ temporally ordered tasks, where each task  `$t$' $ \in \{1,2,\dots, {T}\}$. 
Our work assumes a more general setting in which all seen tasks will appear prior to the encountering the first unseen task. Intuitively, this means unseen tasks represent unlabeled data samples collected after training the classifier with seen tasks. 


\par Let us assume that initial $\{1,2,\dots, t\}$ tasks are seen tasks. For simplicity, the data of each task is denoted by $D^t$. These tasks contain only partially labeled samples and unlabeled samples, represented as $D^t = \{D^{t}_{l}, D^{t}_{u}\}$.
Here, $D^{t}_{l} = \{(\mathbf{x}_{1},\mathbf{y}_{1}),(\mathbf{x}_{2},\mathbf{y}_{2}),\dots,(\mathbf{x}_{n},\mathbf{y}_{n})\}$ where $\mathbf{x} \in \mathbf{R}^{d}$ is a feature vector extracted from a benign/malware application and $\mathbf{y} \in \{0,1\}$ is the label associated with the feature vector with $0$ for benign and $1$ for malware. $D^{t}_{u} = \{\mathbf{x}_{n+1},\dots, \mathbf{x}_{n+N}\}$ such that $n \ll N$. The remaining $\{t+1,t+2,\dots, {T}\}$ are the unseen tasks, where each unseen task ($t^{\prime}$) contains only unlabeled data. Specifically, $D^{t^{\prime}} = \{ D^{t^{\prime}}_{u}\}$ and $D^{t^{\prime}}_{u} = \{\mathbf{x}_{1}, \mathbf{x}_{2}, \dots\}$. 


\noindent \textbf{Problem description:} Our goal is to continuously train a binary classifier $f(\mathbf{x};\theta)$ aiming to reduce the labeling budget while adapting to the concept drift. Towards this, we keep a human analyst in the loop to manually label a minimal number of training samples in the SSCL setting. On a technical note, our work follows the fundamental non-stationary assumption of CL~\cite{lopez2017gradient}, where the joint probability distribution $P(.)$ of each task is unique and distinct from others \textit{i.e.,} $P(i) \neq P(j)$ for $i \neq j$, where $i,j \in \{1,2,\dots, {T}\}$. This will implicitly disqualify the need for additional methods detecting the presence of concept drift, unlike SOTA methods~\cite{yang2021cade,thirumuruganathan2024detecting}. This work operates in a domain incremental learning setting~\cite{amalapuram2023augmented,rahman2022limitations}, where each task has fixed label space (\texttt{0} or \texttt{1}) but varying data space (owing to concept drift in benign/malware apps). \textcolor{black}{In other words, our work addresses domain-incremental concept drift, encompassing both changes in sample frequency (prior shift) and changes in feature distributions within existing classes (concept evolution), while maintaining a fixed label space (benign vs malware). The framework is designed to adapt to evolving feature representations over time} 

\noindent \textbf{SSCL for malware detection:} The training dataset is a collection of ${T}$ tasks. In our CL training setting, we denote a malware detector parameterized by $\theta$ as $f{(\mathbf{x};\theta)}$. The training is a two step process; training with seen tasks and unseen tasks. A seen task contains labeled data and the corpus of unlabeled data to improve the detection performance. Training with label data is a straight forward approach, where an objective function is used to train the detector and errors in predictions are corrected by adjusting the classifier parameters via back propagating the gradients computed over error vectors. The supervised objective function for the labeled data of the task `$t$' is as follows.  

\begin{equation}
  \begin{gathered}  
  \label{eq:sup_loss}
  \hspace{-14mm}
 {L}_{sup}(t)= {L} (D^{t}_{l}) =\frac{1}{|D^{t}_{l}|} \sum_{\substack{(\mathbf{x}_{l}^{t},\mathbf{y}_{l}^{t})\sim D^{t}_{l}}} \ell_{c}(f(\mathbf{x}_{l}^{t}) ,\mathbf{y}_{l}^{t})
    \end{gathered}   
  \end{equation}

\noindent where, $\ell_{c}(.)$ is a per sample loss function such as cross entropy. For each unlabeled sample in seen task, we identify its most relevant labeled sample and enforce representational similarity between them in the latent space. In the second step, for unseen tasks comprising entirely unlabeled data, labels for a small set of informative samples are obtained through human annotation and subsequently used to train the $f{(\mathbf{x};\theta)}$  following the procedure established for seen tasks.

\section{Threat Model and Assumptions}
\textcolor{black}{
In this section we describe our assumptions about the attacker's knowledge, the environment, ML-based defense mechanism, the labeling procedure and drift in the apps.}
\par \textit{Attacker's goals and knowledge}: \textcolor{black}{The attacker aims to steal sensitive information such as passwords and financial transactions details in android mobile phones, and encrypting the files of the victim to demand ransom payment for access restoration. The adversary does not know the datasets or algorithms used in the ML-based defense mechanism deployed.}
\par \par \textit{ML-based malware detector:} \textcolor{black}{The ML-based malware detector uses static features of applications to classify them as benign (goodware) or malware. However, such detectors can become obsolete due to drift in benign and malicious application characteristics over time. To mitigate performance degradation caused by this drift, the malware detector must be periodically updated, while access to historical training data is limited to only a partial subset of the original data.}
\par \textit{Labeling and labeling budget}: \textcolor{black}{We assume that the available partially labeled data from seen tasks. The proposed method considers the budget (money spent) for labeling efforts required by a human analyst.}
\par \textit{Environment:} \textcolor{black}{This work focuses on general malware detection tasks under concept drift, so the proposed methods are applicable to Windows portable executable (PE) and android based malware.}
\par \textit{Drift in benign and malware apps:} \textcolor{black}{We assume that benign and malware apps evolve over time. This evolution in benign apps can be attributed to additional functionalities or new features added by the developers of the apps. The drift in malware apps can be due to the introduction of novel malware variants to evade detection.}

\section{Methodology}
\label{sec:methodology}
\par In this section, we first describe the training process of the seen, unseen tasks, and the buffer memory organization policies.

\subsection{Training with Seen Tasks}
\label{sec:seen_train}

Training with seen tasks constitutes samples retrieved from the current task, including both labeled and unlabeled data, along with replay samples from the buffer memory. Analogous to semi-supervised learning, two types of losses are computed over each batch. The loss over labeled samples is computed using the cross-entropy objective. Our model consists of two sub-networks: an encoder and a classifier, where the encoder produces latent representations of the input samples.

\begin{figure}[!t]
    \centering
    \includegraphics[scale=0.15]{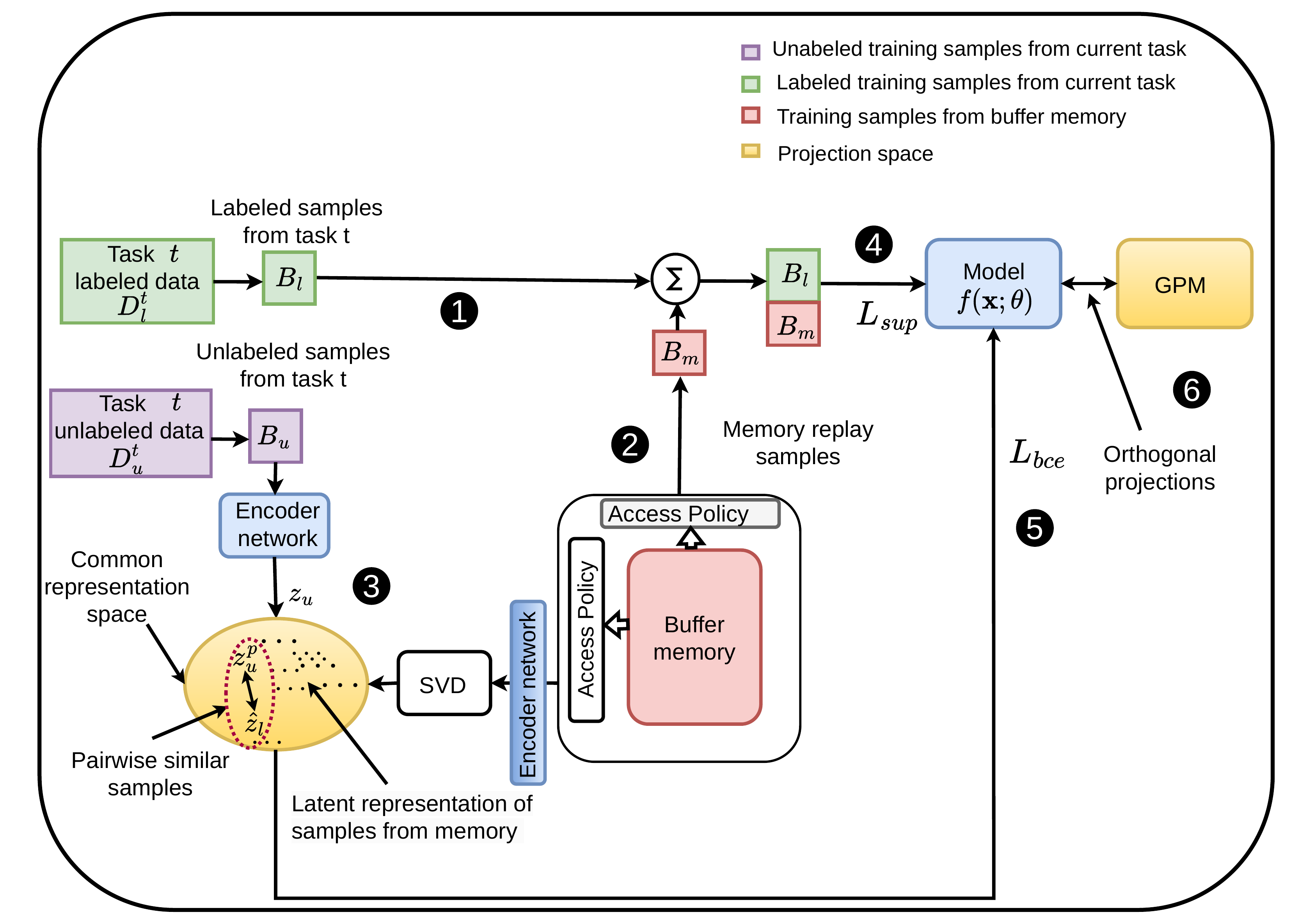}

    \Description{Architecture of the proposed semi-supervised continual malware detection framework showing representation learning, uncertainty estimation, and active labeling.}
    \caption{\suresh{Graphical illustration of the end-to-end training process of the proposed SSCL method for \emph{seen tasks}}.}   
    \label{fig:seen_task}
\end{figure}

For unlabeled data, instead of relying on pseudo-labeling or augmentation strategies, our method identifies a suitable labeled exemplar for each sample. Specifically, the latent representation of an unlabeled sample is projected into a representation space constructed from labeled exemplars of previous tasks. Cosine distances are then computed between the projected sample and stored exemplars, and a soft thresholding mechanism is used to select the most appropriate match. The model is subsequently trained to align the representations of the unlabeled sample with its selected labeled exemplar. Figure~\ref{fig:seen_task} provides a graphical illustration of this process. In the following, we describe each component in detail.
\subsubsection{\textcolor{black}{Training samples from current-task and buffer memory (steps \protect\stepnum{1} and \protect\stepnum{2})}}
\noindent The data of each seen task `$t$' is represented as $D^t = \{D^{t}_{l}, D^{t}_{u}\}$. During training, each batch of samples is composed of three sources: buffer memory, labeled and unlabeled data from the current task: ($D_m, D^{t}_{l}, D^{t}_{u}$). From these, we sample labeled instances $B_l$ from $D^{t}_{l}$ and unlabeled instances $B_u$ from $D^{t}_{u}$ for training, these constitute the primary inputs from the current task. The buffer memory contains the partial labeled data of the past seen tasks and $B_m$ samples from $D_m$ are retrieved from the buffer memory. The complete details of the buffer memory organization policies (storage, retrieval) are described in Section~\ref{subsec:buffer_management}

\subsubsection{\textcolor{black}{Representation space construction and suitable exemplar selection (step \protect\stepnum{3})}}
\noindent \textbf{Representation space of all past tasks:} We find the basis vectors of the labeled exemplars of all the past tasks. We begin by noting that while a key assumption of the CL paradigm is to avoid access to past task exemplars, our method follows the memory replay-based CL approach, which allows limited access to past task samples. These samples are a small, labeled subset of previous task data stored in a buffer memory  (refer to Section~\ref{subsec:buffer_management} for more details).

\par Initially, we obtain the latent representation ($Z^m)$ of the `\textit{n}' past task examples ($X^m)$ from memory using the encoder sub-network.
\begin{equation}
\label{eq:encoderep}
    \begin{gathered}
    Z^{m} = enc(X^{m}), 
    \end{gathered}
\end{equation}
where $enc(.)$ represents the encoder subnetwork, $X^m \in R^{n \times d_{1}}$, $Z^m \in R^{n \times d_{2}}$. 

\noindent \textbf{Motivation for using singular value decomposition:} Although, the dimensionality of the encoded representation is smaller ( $d_2 << d_1$), we can further reduce its dimensionality by considering the basis vectors that span the encoded representation. By focusing on the basis vectors, we effectively capture the most salient features of the encoded data, reducing storage costs associated with storing all the encoded representations (Eq~\ref{eq:encoderep}) particularly for longer sequence of tasks. The basis vectors of the latent representation are derived using singular value decomposition (SVD). SVD decomposes $Z^m$ into the product of three matrices: $Z^{m} = U \Sigma V^{T}$, where $U \in R^{n \times n}$, $\Sigma \in R^{n \times d_{2}}$, $V \in R^{d_2 \times d_2}$, $V^T$ is the transpose of $V$, and $\Sigma$ contains the singular values sorted along its diagonal. 
The column vectors of $V$, which span the matrix $Z^m$, serve as the basis vectors. For convenience, we denote these basis vectors as $V^\prime$.
\begin{equation}
\label{eq:basisvector}
    \begin{gathered}
    V^\prime = \text{span} \{\mathbf{v}_1,\mathbf{v}_2,\cdots,\mathbf{v}_k \},
    \end{gathered}
\end{equation}

\noindent where $k$ is the number of basis vectors. The number of basis vectors $(k)$ is determined by analyzing the singular values obtained from the decomposition. The squared singular values are normalized to compute their cumulative ratio. The basis vectors are then selected based on a threshold (95\%) applied onto the cumulative ratios of the singular values to ensure that the chosen vectors retain the most important information from the original data.


We also empirically validate the effectiveness of using SVD approach through ablation study, showing that detection performance benefit from using the orthonormal basis constructed from SVD over the original feature space.
\begin{table}[!tbh]
  \caption{Ablation study demonstrating the sensitivity of the proposed method using SVD in constructing the representation space. The best values are marked in \textbf{bold}.}
  \label{tab:ablation_study_svd_sensitivity}
  \centering
 

   
    
  \begin{tabular}{llllllll}
  \multicolumn{4}{c}{unseen-AUT (A)}\\
      \cmidrule(lr){2-4} 
   SVD & BODMAS  &  AndroZoo &  API Graph\\
   \midrule
   
     \nomark & 0.752 $\pm$ 0.061 & 0.821 $\pm$ 0.045 & 0.919 $\pm$ 0.003\\
      \yesmark & \textbf{0.810 $\pm$ 0.027} & \textbf{0.855 $\pm$ 0.018} & \textbf{0.930 $\pm$ 0.002}\\  
    
    \bottomrule    
  \end{tabular}


   \end{table}

The ablation study results are presented in Table~\ref{tab:ablation_study_svd_sensitivity}, and we make the following observations. First, the inclusion of SVD consistently improves detection performance on unseen tasks across all datasets, as the orthonormal basis vectors capture more salient features than the original encoded features. Second, the use of SVD reduces variance in malware detection. Finally, we empirically observe that its impact is more pronounced on datasets with higher dimensionality. Specifically, compared to AndroZoo (2,381 dimensions) and APIGraph (1,150 dimensions), the absence of SVD has a greater negative impact on the BODMAS dataset, which has a dimensionality (16,978 dimensions) \suresh{and does not inherently preserve structural semantic representations (unlike APIGraph).}


\noindent \textbf{Finding the suitable labeled exemplar:} After finding the basis vectors, the latent vector of each unlabeled sample ($\mathbf{z}_u$) is projected into representation space. This projection step can be interpreted as finding a point ($\mathbf{z}_u^p$) within space $V^\prime$ that is closest to $\mathbf{z}_u$.
\begin{equation}
\label{eq:approxsample}
    \begin{gathered}
    \mathbf{z}_u^p = \arg\min_{\mathbf{z} \in V^\prime} \|\mathbf{z}_u - \mathbf{z}\|_2
    \end{gathered}
\end{equation}
Now, the distances between $\mathbf{z}_u^p$ and $Z^m$ are computed to find the suitable labeled example for $\mathbf{x}_u$. \textcolor{black}{Specifically, we find cosine distance is particularly well-suited to our setting because the projected representations reside in a continuous, $\ell_2$-normalized latent space where relational information is encoded through angular alignment rather than magnitude. By measuring angular dissimilarity, cosine distance remains scale-invariant and robust to variations in feature magnitude introduced by projection or drift. In high-dimensional embedding spaces, angular metrics are generally more stable and discriminative than magnitude-based distances. In contrast, set-based similarity measures (e.g., Jaccard) disregard geometric structure and are not designed to capture relationships in dense continuous representations, making them less appropriate for our framework.}

\noindent \textbf{Adaptive thresholding strategy:} To identify the most suitable labeled example we use an adaptive (soft) thresholding technique that has an initial threshold value ($\tau_{init}$) and a maximum threshold value ($\tau_{max}$). Starting with $\tau_{init}$, we increase the threshold by a specified step size until it reaches a defined $\tau_{max}$. For each $\mathbf{z}_u^p$, we find a list of samples in $Z^m$ that have cosine distance values below the current threshold. We group these list of samples using their labels and select a target group with most number of samples. From the target group, a sample with small cosine distance is selected. If such a sample ($\hat{\mathbf{x}_l}$) is found, the search operation is halted to improve efficiency. Otherwise, we reject the unlabeled sample $\mathbf{x}_u$. Similarly, we also find a suitable example for each labeled sample as well. This step intuitively clusters the samples with similar latent structure. However, choosing a fixed $\tau_{max}$ may result in the selection of samples that are very distant potentially leading to false positives. To address this issue, we adaptively choose the maximum threshold based on the labeled ratio ($r$). Intuitively, a higher $r$ indicates the availability of more labeled data in memory, allowing the maximum threshold to shrink quickly; otherwise, it decreases more gradually. Mathematically, the dynamic threshold ($\tau^{\prime}_{max}$) is defined as follows:
\begin{equation}
\label{eq:tau}
    \begin{gathered}    
\tau^{\prime}_{max} = \tau_{max}* \exp(-r*\beta),
    \end{gathered}
\end{equation}

\noindent where `$\beta$' is the temperature factor. Let the latent representation (encoder output) of $\hat{\mathbf{x}_l}$ be $\hat{\mathbf{z}_l}$ which is an input to the classifier subnetwork $f_c (.)$. 

\subsubsection{\textcolor{black}{Pairwise similarity learning for unlabeled data (steps \protect\stepnum{4} and \protect\stepnum{5})}}
Now, our goal is to bring $\hat{\mathbf{x}_l}$ and $\mathbf{x}_u$ (or $\mathbf{x}_l)$ closer together in the latent space. To achieve this (without relying on contrastive loss), we modify the binary cross entropy (BCE) loss. We define the similarity score ($ss$) as follows:

\begin{equation}
\label{eq:simscore}
    \begin{gathered}    
ss = \langle\sigma( f_c(\hat{\mathbf{z}_l})), \sigma(f_c(\mathbf{z}^\prime))\rangle,
    \end{gathered}
\end{equation}
\noindent where $\sigma(.)$ is the softmax function that translates the latent space vectors into probability distributions, $\langle.,.\rangle$ is the dot product, and $\mathbf{z}^\prime$ is the latent vector of $\mathbf{x}_u$ or $\mathbf{x}_l$. The BCE loss is used to predict whether two instances are similar (\( y = 1 \)) or dissimilar (\( y = 0 \)), which is given by the following equation.
\begin{equation}
\label{eq:originalbce}
    \begin{gathered}    
\text{BCE Loss} = - \left[ y \log(p) + (1 - y) \log(1 - p) \right],
    \end{gathered}
\end{equation}

\noindent where $p$ is the predicted similarity score. For the task of positive pair learning (only considering $y=1$), the loss simplifies to the following:
\begin{equation}
\label{eq:originalbcepp}
    \begin{gathered}    
\text{BCE Loss} = - \log(p),
    \end{gathered}
\end{equation}

Intuitively, our goal of bringing samples with lower cosine distance becomes positive pair learning, so $p=ss$. Therefore, the modified BCE loss for a given task `$t$' is defined by
\begin{equation}
\label{eq:bce}
    \begin{gathered}    
L_{bce}(t)= \frac{1}{|D^t|} \sum_{\substack{\hat{\mathbf{z}}_l \in enc(X^m)\\ \mathbf{z}^\prime \in enc(\mathbf{x}^\prime), \mathbf{x}^\prime \sim D^t}} -\log\langle\sigma( f_c(\hat{\mathbf{z}_l})), \sigma(f_c(\mathbf{z}^\prime))\rangle.
    \end{gathered}
\end{equation}
The overall objective of the seen tasks is a combination of cross-entropy loss over labeled exemplars and BCE over all the exemplars of the task `$t$', and it is given by:

\begin{equation}
\label{eq:overallloss}
    \begin{gathered}    
            L(t) = L_{sup}(t)+ L_{bce}(t).
    \end{gathered}
\end{equation}

\par The gradients computed for Eq.~\ref{eq:overallloss} may be noisy due to the smaller intra-class variance in labeled data compared to unlabeled data\cite{cao2022openworld}. This discrepancy leads to faster convergence for labeled data (especially the majority class i.e., benign) compared to unlabeled data and subsequent unseen tasks. Class imbalance further negatively affects the minority class (malware) learning due to skewed gradients~\cite{francazi2023theoretical,he2024gradient}. Additionally, interference from past tasks, along with the aforementioned issues, jointly hinders the learned representation of malware samples.

\subsubsection{\textcolor{black}{Improved representation through orthogonal projections (Step \protect\stepnum{6})}} Inspired by the prior works to improve the performance of the representations by minimizing the interference from past tasks~\cite{farajtabar2020orthogonal,saha2020gradient,zhao2023rethinking}, the gradients of the current task are projected in the direction orthogonal to the past tasks gradients. Specifically, our method uses gradient projection memory (GPM)~\cite{saha2020gradient} that contains the basis vectors that span the gradient directions of the past tasks. 
\begin{equation}
  \begin{gathered}      
  \text{span} \{\nabla_{\theta}{L}_1, \nabla_{\theta}{L}_2, ..,\nabla_{\theta}{L}_k,.. \nabla_{\theta}{L}_{t-1}\} = {M}_{gpm}
    \end{gathered}
    \label{eq:gpm}    
  \end{equation}
\noindent where $\nabla_{\theta}{L}_k$ is the gradient of $k^{th}$ task. Here, we store the gradient directions of malware class samples in GPM. Thus, all the previously learned definitions of malware are preserved in GPM.  The orthogonal gradient direction (to GPM) will also minimize the effects of the drifted malware sample on previously learned definitions. Eventually, the overall training objective function of the seen task is given by the following equation.
\begin{equation}
  \begin{gathered} 
  \theta^* = \argmin_{\theta}\hspace{0.1cm} {L}(t)  \\
    \hspace{-0.2cm}\text{subject to} \hspace{0.2cm} \nabla_{\theta}{L}_{t} \perp M_{gpm} 
    \end{gathered}
    \label{eq:obj_fun_new3}
  \end{equation}
\par Note that, although GPM is an existing framework incorporated into our pipeline, our contribution lies in strategically utilizing it to enhance malware detection performance by leveraging limited labeled data. Additionally, we utilized the partial labeled data to its optimal capacity in three ways: to construct the representation space for GPM, for sample replay during training, and to identify the most suitable sample for each unlabeled sample in the unseen task training process. The way we use GPM is also a critical factor in improving malware detection performance. Experimentally, we demonstrated the importance of this exploration through an ablation study, as presented in Table~\ref{tab:ablation_study}. The pseudo code for training the seen task is presented in Algorithm~\ref{alg:seentask}. 
\begin{algorithm}[t]
\footnotesize
   \caption{Training with Seen Tasks}
   \label{alg:seentask}
\begin{algorithmic}[1]
   \STATE {\bfseries Input:} 
   the sequence of seen tasks $\{1,2,\cdots,{T}-1,{T}\}$, data of task `t' ${D}^{t} = \{{D}^{t}_{l},{D}^{t}_{u}\}$, batch size $b$, model $f_{\theta}$, buffer memory ${M}$ (initially empty), pair set $P$ (initially empty), gradient projection memory $M_{gpm}$ (initially empty)

    \STATE {\bfseries Output:} $f_{\theta}$ model trained on all ${T}$ seen tasks
    \FOR{each task '$t$'}
    \STATE copy $D^t_l$ to $M$
    \WHILE {${D}^{t}$ is non-empty}
        \STATE sample labeled data of size $b_l \sim {D}^{t}_{l}$, call it $B_l$
         \STATE sample labeled data from M of size $b_m \sim {M}$, call it $B_m$
         \STATE sample unlabeled data of size $b_u \sim {D}^{t}_{u}$, call it $B_u$
         \FOR{each labeled sample $\mathbf{x}_l$ in $B_l$}
         \STATE find another labeled sample $\mathbf{x}_l^{\prime}$ with same label as $\mathbf{x}_l$
         \STATE add $\mathbf{x}^{\prime}_{l}$ to $P$
         \ENDFOR
         \FOR{each unlabeled sample $\mathbf{x}_u$ in $B_u$}
         \STATE find a suitable labeled sample $\mathbf{x}_u^{\prime}$ using Eqs.~\ref{eq:encoderep},~\ref{eq:basisvector}, and ~\ref{eq:approxsample}
         \STATE add $\mathbf{x}_u^{\prime}$ to $P$
         \ENDFOR
         \STATE compute $L_{sup}$ on $B_l$ and $B_m$ using the Eq~\ref{eq:sup_loss}
         \STATE compute $L_{bce}$ on $B_l$,$B_m$,$B_u$ and $P$ using the Eq.~\ref{eq:bce}
         \STATE compute total loss $L$ using Eq.~\ref{eq:overallloss}
        \STATE compute gradient $\nabla_{\theta}{L}$ of classification loss 
        \IF{`t' $>$ 1}
        \STATE project $\nabla_{\theta}{L}$ orthogonal to $M_{gpm}$
        \ENDIF
        \ENDWHILE
       \STATE using $D^t_l$, compute basis vectors of gradients and add them to $M_{gpm}$ 
    \ENDFOR
    
    \RETURN $f_{\theta}$  
\end{algorithmic}
\end{algorithm}

\subsection{Training with Unseen Tasks}
\label{sec:unseen_train}
\par The training mechanism here operates on the fully unlabeled exemplars in an open world learning (OWL) setting. 
Given an unlabeled task $t^\prime$, our goal is to transform it from $D^{t^\prime}_u$ to partially labeled task $\{D_l^{t^\prime},D_{u^*}^{t^\prime}\}$, where $D_{u^*}^{t^\prime} \subset D_{u}^{t^\prime}$.  To achieve this, unlabeled exemplars are ranked and a few of them are chosen for labeling (by the security analyst) under the allowable budget constraints. A SOTA approach~\cite{291253} uses the hierarchical contrastive loss values to rank the samples, chooses the samples based on higher loss values for labeling. Later, it augments the existing fully labeled dataset with new exemplars to \textit{retrain} the classifier and repeats this for every new task. However, this requires fully labeled data. 
In stark contrast, our method works with partial labeled data. 
In the following, we describe our method.

\noindent \textbf{Distance based metric in latent space:} The BCE loss defined in Eq.~\ref{eq:bce} encourages samples with similar structures (i.e., lower cosine distances) to align closely in the latent space. Building on this intuition, our method utilizes the latent space to estimate the uncertainty of unlabeled exemplars. Specifically, for each unlabeled sample, we compute its distances to two groups of labeled samples (benign and malware) stored in the buffer memory. Initially, we partition $M$ into two groups based on their class labels: $group0$ (class $0$) and $group1$ (class $1$). Let $n_0$ and $n_1$ denote the number of samples in $group0$ and $group1$, respectively. For each unlabeled sample, we calculate its cosine distances to all exemplars in $group0$ and $group1$, resulting in $n_0$ distance values for $group0$ and $n_1$ values for $group1$. To simplify the comparison, we take the mean of these distance values for each group. These mean values are then used as unified measures to evaluate the similarity between the unlabeled exemplar and each group.

\noindent \textbf{Ranking the samples:} Once the mean cosine distance is computed for each unlabeled sample from $group0$ and $group1$, the next step is to rank these samples for selection by the analyst for labeling. A common strategy is to select samples that are farthest from the groups for labeling. However, during validation on the seen tasks (where ground truth labels are available), we observe a peculiar behavior with this approach. Specifically, when using the farthest distance criterion, most samples selected for labeling belong to the benign class (label $0$). As a result, the classifier becomes increasingly biased, leading to degraded performance on unseen tasks. A plausible explanation for this behavior is the combined effect of class imbalance and drift in the data. While drift is observed in both benign and malware classes, the class imbalance results in a higher number of benign samples compared to malware. Consequently, selecting the farthest samples disproportionately targets benign samples. In contrast, choosing samples with the closest distances helps mitigate this issue and rejuvenates the classifier’s performance on unseen tasks.

\noindent \textbf{Transition from unseen to seen task:} After ranking the samples, we select the appropriate ones for labeling based on the ranking mechanism and the monthly labeling budget (i.e., the number of samples to be labeled). Specifically,  half of the monthly labeling budget samples are selected based on their ranking relative to $group0$, and the remaining half are selected from $group1$. The intuition behind this selection strategy is rooted in the motivation of our work—to adapt to distribution shifts in both benign and malware exemplars. By ensuring a balanced selection, we account for potential shifts in both groups. Once the selected samples are labeled, the unseen task now has a few labeled examples. At this point, we can apply the training procedure for seen tasks, as described in Algorithm~\ref{alg:seentask}, to continue the training process.

\subsection{Delayed Buffer Update for Noisy Label Mitigation in Unseen Tasks}
\label{sec:delayed_buffer_update}
\par \textcolor{black}{To mitigate the adverse impact of noisy labels, we introduce a simple yet effective delayed buffer update heuristic. Based on our empirical observations, storing noisy labeled samples in the replay buffer significantly degrades performance, as replay-based training repeatedly revisits buffer samples, thereby amplifying the effect of label noise across tasks. To address this, we avoid immediately storing newly observed samples in the buffer when their labels are likely to be unreliable. Instead, we delay their inclusion and incorporate them only after a certain time period, under the assumption that their labels become more accurate.}

\textcolor{black}{Building on this intuition, we determine the delay duration based on insights from prior studies~\cite{wang2023re,251586,botacin2025towards}, which show that labels obtained from multi-engine platforms (e.g., VirusTotal) are inherently dynamic and may stabilize over weeks to several months. Motivated by these findings, we consider multiple delay intervals (1 month, 3 months, and 11 months) to capture different stages of label refinement. The shorter delay (1 month) accounts for early corrections in rapidly evolving threat landscapes, while longer delays (3 and 11 months) allow sufficient time for labels to converge toward more reliable ground truth. In our setting, samples from task `\method $+\Delta$ months' are thus incorporated into the replay buffer at a future time step and retrain the detector with them, where $\Delta$ corresponds to the chosen delay interval. This design enables the buffer to preferentially retain higher-quality labels, thereby reducing noise propagation during replay.}

\subsection{Reorganizing Buffer Memory}
\label{subsec:buffer_management}

\par
\begin{figure}[htb!]
    \centering
    \includegraphics[scale=0.3]{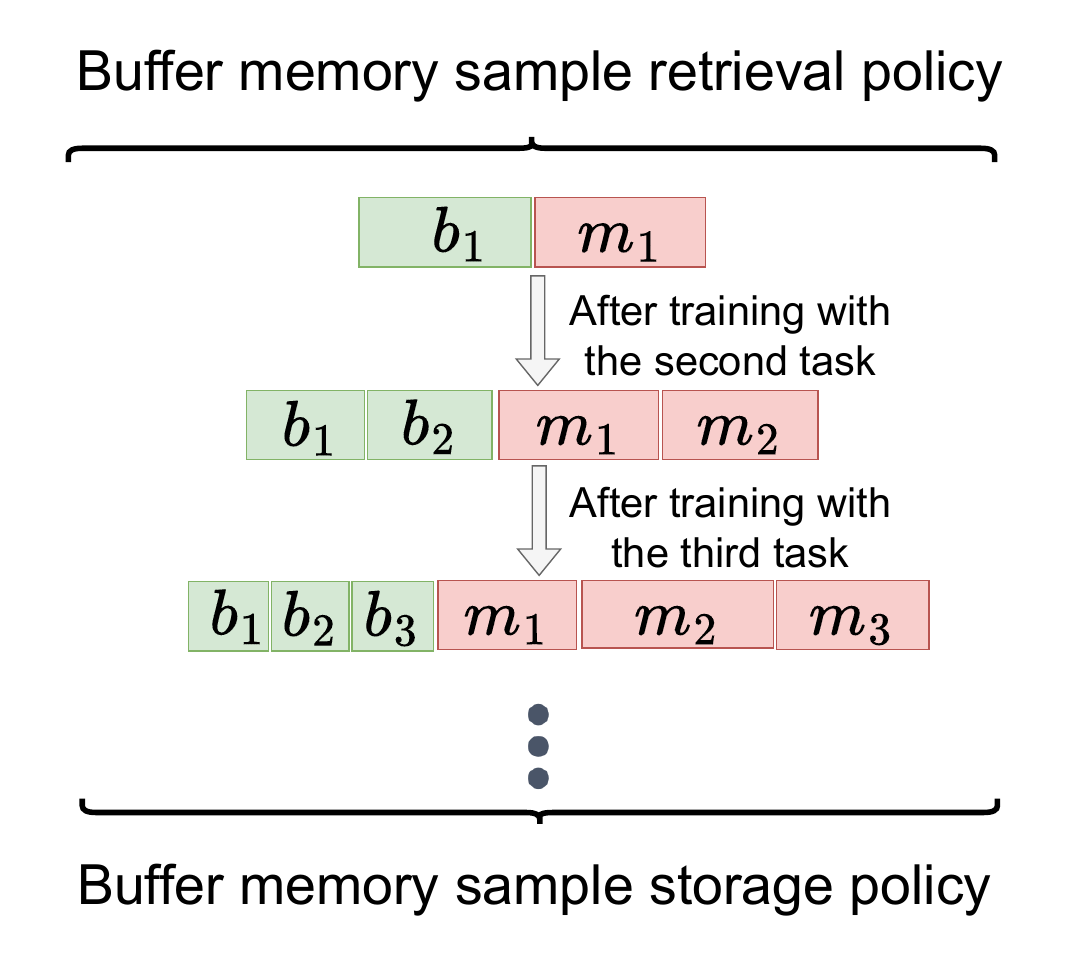}
    \Description{about the figure}
    \caption{Illustration of buffer memory sample storage and retrieval policy. At the beginning, memory is partitioned into two chunks and initialized with train data ($b_{1}-benign$ and $m_{1}-malware$ from the first task) whose size equals the number of labeled benign and malware samples. Prior to training with the second task, buffer memory ($M$) is augmented with two additional chunks containing the labeled data of this task ($b_2$ and $m_2$). This memory augmentation process is repeated for all the next tasks. During the sample selection, the sample retrieval policy views the entire memory as two partitions, irrespective of the number of smaller chunks.}   
    \label{fig:memory_org}
\end{figure}

 Buffer memory ($M$) is an additional memory used to save the subset of the past tasks labeled training samples. 
Besides data replay usage, prior works~\cite{chaudhry2018efficient} also used memory to revise the gradient directions. Thus, the organization of memory is driven by its usage. The buffer memory organization policy defines how to store and retrieve samples in memory. The visual mechanism of this approach is illustrated in Figure~\ref{fig:memory_org}.

\par Initially, the memory is populated with the labeled data from the first task. For this task, we store benign and malware samples in two separate chunks of different sizes. When a new task arrives, two additional chunks are allocated, bringing the total number of chunks to four. It is important to note that, upon the arrival of each new task, existing memory is allocated two new chunks. The sizes of these chunks are proportional to the number of labeled benign and malware training samples in that task. This ensures that all future tasks are always guaranteed to have their labeled exemplars stored in $M$.

 
 \par During the training process, a subset of samples is \textit{retrieved} for replay along with the current task training exemplars. The replay samples will help revise the knowledge of the past tasks. 
 In each retrieval operation, the number of benign and malware samples to be selected is to be regularized to handle the effect of class imbalance on the classifier.
\par Besides organization policies, the detection performance may also depend on the size of the memory (the maximum number of samples allowed to be stored). The size is constrained by many factors, such as storage cost incurred per sample, memory constraints of the deployed environment, size of each sample, and data privacy issues. As, our work operates in a constrained setting (open world SSCL), we relax the assumption on the availability of finite size of ${M}$ until the training of the seen tasks. After that, from the unseen tasks $M$ grows slowly and constantly as the availability of the number of labeled samples is regularized by the labeling budget constraint.


 \section{Experiments and Analysis}
  \label{sec:results}
  In this section, we describe the experimental setup, including details of the dataset, preprocessing, task splitting for training, baseline selections, evaluation metrics, and protocol. We also discuss comparisons between the proposed method and baseline methods, the sensitivity of the proposed method to different labeling budgets and false positives, ablation studies, and, finally, the limitations.
  \subsection{Experimental Setup}
  \textbf{Datasets:} We use three Malware datasets in our experiments. Specifically, APIGraph~\cite{zhang2020enhancing,291253} and AndroZoo~\cite{291253,androzoo} are Android malware based datasets containing evolved malware, and BODMAS~\cite{bodmas} is a windows PE based malware dataset. APIGraph contains Android apps spread over seven years, from 2012 to 2018, and nearly 90\% of them are benign Android apps. AndroZoo data spans over three years from 2019 to 2021, and the ratio of benign to malicious apps is 9:1 for each month of the data. The BODMAS dataset has a PE based malware samples collected from August 2019 to September 2020 and benign apps collected from January 2007 to September 2020 from the real world security company Blue Hexagon. Thus, it is one of most up-to-date datasets for validation, containing nearly 57\% of benign apps with dynamically evolving malware. The details of the datasets are presented in Table~\ref{datasets_span}.
  \begin{table}[!htb]
    \centering
    \caption{Details of the various Android and Windows PE malware benchmark datasets used in our experiments. These details include the time span and the number of benign and malicious apps per dataset. Note that \textit{k} is for thousand.}
      
    \begin{tabular}{lllll}
    
    \cmidrule(lr){1-5}
         Dataset& Timespan & \#Benign& \#Malware & \#Total  \\
     \midrule   
     BODMAS & 1.1 year & 77142 & 57,293 & 134.4 \textit{k}\\
     AndroZoo & 3 years & 89853 & 10200& 100 \textit{k}\\
         APIGraph & 7 years & 289511 & 30804 & 320.3 \textit{k}\\

         \bottomrule
    \end{tabular}
    
    \label{datasets_span}
\end{table}

\par \textbf{Feature extraction and preprocessing:} For APIGraph and AndroZoo datasets, we used DREBIN features for training that contain eight sets of features related to permission access and different API calls of the Android apps. The preprocessed DREBIN features are available in~\cite{291253} and are used for our experimentation. The number of features of the APIGraph and AndroZoo datasets are 1159 and 16978, respectively. 

The raw features of windows PE malware dataset (BODMAS) contain parsed features, format agnostic histograms, and count of strings. These raw features are  translated into model features using the \href{https://lief.re/}{LIEF} project; strings and exported names, etc are captured using feature hashing trick. Thus, each object is converted into a vector of $2381$ dimensions.

\noindent \textbf{Tasks creation}: For the APIGraph and AndroZoo datasets, tasks are created at the granularity of a \textit{month}. This results in $84$ tasks for APIGraph and $36$ tasks for AndroZoo, each with a class imbalance ratio of 9:1. Similarly, for BODMAS, twelve tasks are created, each with a class imbalance ratio of 10:7.


\noindent \textbf{Tasks split for training:} For BODMAS, we consider the first five tasks as seen tasks and next seven tasks as unseen tasks. For AndroZoo, the first twelve tasks are used as seen tasks, and next twenty-four tasks as unseen tasks. Similarly, for APIGraph, thirty-five tasks are used as seen tasks, with the remaining thirty-five tasks as unseen tasks. In fully supervised training, each task is split into 70\% for training, 5\% for validation, and 25\% for testing. In the semi-supervised setting, the labeled ratio determines the amount of data used as labeled examples, while the remaining data is treated as unlabeled. However, the validation (5\%) and testing (25\%) percentages remain the same as in the fully supervised setting.

  \noindent \textbf{Baselines selection:} 
 We consider classical supervised methods from different families of CL as baselines. These methods include elastic weight consolidation (EWC~\cite{kirkpatrick2017overcoming}) from regularization based CL methods, average-gradient episodic memory (A-GEM~\cite{chaudhry2018efficient}) from projection based methods, maximal interfered retrieval (MIR~\cite{aljundi2019online}) and class-balanced reservoir sampling (CBRS~\cite{chrysakis2020online}) from memory replay based methods. These methods are trained under full supervision.

   \par We also consider two recent approaches CADE~\cite{yang2021cade} and HCL~\cite{291253} that use contrastive learning loss functions to handle concept drift in malware classification tasks. These methods are initially not designed for continual learning setting. We took the implementations available in the code repository of ~\cite{291253} and tailored them for our CL setting. These methods require fully labeled exemplars to operate in seen tasks.
  
\noindent \textbf{Evaluation metrics}: 
In our experiments, we use the precision-recall area under the curve (PR-AUC) as the base evaluation metric and area under the time (AUT)~\cite{pendlebury2019tesseract} as derived metric. Unlike the F1-score, which is sensitive to threshold selection, PR-AUC does not rely on any threshold. Further, PR-AUC reflects both \emph{false positive} and \emph{false negative} rates through precision and recall, making it suitable for imbalanced tasks~\cite{quiring2022and}. To evaluate the effectiveness of the proposed method against continuous concept drift adaptation in both closed world and open world setting, we measure malware detector's performance decay over time. Towards this, we compute AUT using base metric PR-AUC, as shown in Eq.~\ref{eq:aut}.
\begin{equation} \label{eq:aut}
\begin{split}
AUT(f,N) = \frac{1}{N-1}\sum_{k=1}^{N-1}\frac{[f(x_{k+1})+f(x_{k})]}{2}
\end{split}
\end{equation}
where $f(x_{k})$ is the point estimate computed using the performance metric $f$ at the time period $k$, $N$ is the number of time slots used to test performance degradation, and $1/(N-1)$ is the normalization constant.

\noindent \textbf{Computing AUT over PR-AUC value:} We used a strategy similar to that in~\cite{dragoi2022anoshift, amalapuram2023augmented} to compute the PR-AUC for the benign and malware class. For the malware class (positive class), we use the predicted probabilities ($lr\_probs$) and set the $pos\_label$ (of the $precision\_recall\_curve$ function from $sklearn.metrics$) to $1$ in order to compute the precision and recall values. We then calculate the AUC from these precision and recall values using the $auc$ function. Similarly, for the benign class (negative class), we compute the Precision-Recall curve using $1 - lr\_probs$ and set $pos\_label=0$ to evaluate the AUC for the negative class. Once the PR-AUC values for each class in each task are computed, we calculate the AUT values on the PR-AUC values for the benign and malware classes across seen, unseen, and overall tasks.

\noindent \textbf{Evaluation protocol:} The goal of our work is to evaluate performance on unseen malware with a limited monthly labeling budget, aligning with the evaluation strategy commonly used in related work~\cite{291253}. 
However, our approach is based on a CL framework, whose primary evaluation focuses on measuring the effectiveness of preserving the previously learned distributions of the benign and malware samples. To bridge these differences and maximize the insights gained from both evaluation approaches, we measure three key quantities:
\begin{itemize}
    \item \textit{Closed world evaluation:} We evaluate the detection performance of the proposed approach on known benign and malware samples encountered in seen tasks, which have partially labeled data. Notationally, we use \emph{seen-AUT (B)} for benign samples and \emph{seen-AUT (A)} for malware samples.
    \item \textit{Open-world evaluation:} We evaluate the detection performance of the proposed approach on unknown benign and malware samples encountered in unseen tasks, for which no labels are available (unlabeled). Notationally, we use \emph{unseen-AUT (B)} for benign samples and \emph{unseen-AUT (A)} for malware samples.
    \item \textit{Overall evaluation:} This combines the joint evaluation of both seen and unseen tasks. For overall tasks, \emph{overall-AUT (B)} is used for benign samples and \emph{overall-AUT (A)} for malware samples.
\end{itemize}  

\begin{table}[!tbh]
 \captionsetup{type=table}  \caption{Comparing the performance results of the proposed method with baselines on BODMAS, AndroZoo, and APIGraph datasets. We report AUT values for benign and malware classes on seen and unseen tasks and also overall tasks. \textcolor{black}{The proposed method (\method{}) and HCL$^*$ use 20\% labeled data (\suresh{selected randomly}) on seen tasks across all the experiments on all the datasets}. Each experiment is repeated for three times and the mean values along standard deviations are reported and best values are marked in \textbf{bold}. If the best values are not found among CADE, HCL$^*$, and our method, then the best values among these are marked in \textcolor{blue}{blue}.}
  \label{tab:main_results}
  \centering
  \begin{adjustbox}{width=1\textwidth}
  \begin{tabular}{lllllll}
    & \multicolumn{6}{c}{BODMAS} \\
    \cmidrule(lr){1-2} \cmidrule(lr){2-7}
   Baseline Methods & seen-AUT (B) & seen-AUT (A) & unseen-AUT (B) & unseen-AUT (A) & overall-AUT (B) & overall-AUT (A)\\
   \midrule
  
   EWC~\cite{kirkpatrick2017overcoming} & 0.911 $\pm$ 0.010 & 0.528 $\pm$ 0.014 & 0.935 $\pm$ 0.013 & 0.674 $\pm$ 0.021  & 0.925 $\pm$ 0.011 & 0.616 $\pm$ 0.018 \\ 
 AGEM~\cite{chaudhry2018efficient} & 0.910 $\pm$ 0.008 & 0.527 $\pm$ 0.009 & 0.933 $\pm$ 0.011 & 0.670 $\pm$ 0.011 & 0.924 $\pm$ 0.009 & 0.612 $\pm$ 0.009 \\ 
 CBRS~\cite{chrysakis2020online} & 0.898 $\pm$ 0.002 & 0.621 $\pm$ 0.037 & 0.908 $\pm$ 0.009 & 0.680 $\pm$ 0.034 & 0.900 $\pm$ 0.005 & 0.648 $\pm$ 0.003 \\
 MIR~\cite{aljundi2019online} & 0.898 $\pm$ 0.005 & 0.599 $\pm$ 0.056 & 0.904 $\pm$ 0.004 & 0.668 $\pm$ 0.022 & 0.899 $\pm$ 0.004 & 0.635 $\pm$ 0.022 \\
  \midrule
  CADE~\cite{yang2021cade} & 0.845 $\pm$ 0.000 & 0.654 $\pm$ 0.000 & 0.845 $\pm$ 0.000 & 0.654 $\pm$ 0.000 & 0.845 $\pm$ 0.000  & 0.654 $\pm$ 0.000  \\
  HCL~\cite{291253} & 0.924 $\pm$ 0.007 &  0.699 $\pm$ 0.044  & 0.949 $\pm$ 0.006  & 0.795 $\pm$ 0.010 &  0.936  $\pm$  0.003 &  0.746  $\pm$ 0.017  \\
  HCL$^*$~\cite{291253} & 0.891 $\pm$ 0.018 ($\textcolor{red}{\downarrow}$) & 0.580  $\pm$ 0.050($\textcolor{red}{\downarrow}$) &  0.888  $\pm$ 0.032  ($\textcolor{red}{\downarrow}$) & 0.613 $\pm$ 0.045($\textcolor{red}{\downarrow}$) & 0.885 $\pm$ 0.028 ($\textcolor{red}{\downarrow}$) & 0.589 $\pm$ 0.037  ($\textcolor{red}{\downarrow}$) \\
  
  \midrule
 \method{} & \textbf{0.930} $\pm$ \textbf{0.002} & \textbf{0.701} $\pm$ \textbf{0.017} & \textbf{0.956} $\pm$ \textbf{0.002} & \textbf{0.810} $\pm$ \textbf{0.027} & \textbf{0.942} $\pm$ \textbf{0.000} & \textbf{0.755} $\pm$ \textbf{0.018} \\  
    
  \end{tabular}   
\end{adjustbox}
\begin{adjustbox}{width=1\textwidth}
  \begin{tabular}{lllllll}
    \toprule   
    & \multicolumn{6}{c}{AndroZoo} \\
    \cmidrule(lr){1-2} \cmidrule(lr){2-7}
   Baseline Methods & seen-AUT (B) & seen-AUT (A) & unseen-AUT (B) & unseen-AUT (A) & overall-AUT (B) & overall-AUT (A)\\
   \midrule
  
   EWC & 0.997 $\pm$ 0.000 & 0.868 $\pm$ 0.018 & 0.998 $\pm$ 0.000 & 0.930 $\pm$ 0.003 & 0.998 $\pm$ 0.000 & 0.910 $\pm$ 0.008 \\ 
 AGEM & \textbf{0.998 $\pm$ 0.000} & 0.901 $\pm$ 0.005 & \textbf{0.998 $\pm$ 0.000} & \textbf{0.938 $\pm$ 0.003} & \textbf{0.998 $\pm$ 0.000} & \textbf{0.926 $\pm$ 0.002} \\ 
 CBRS & 0.998 $\pm$ 0.000 & \textbf{0.925 $\pm$ 0.007} & 0.995 $\pm$ 0.001 & 0.781 $\pm$ 0.032 & 0.996 $\pm$ 0.000 & 0.831 $\pm$ 0.019 \\
 MIR & 0.996 $\pm$ 0.000 & 0.849 $\pm$ 0.026 & 0.996 $\pm$ 0.000 & 0.905 $\pm$ 0.005 & 0.996 $\pm$ 0.000 & 0.887 $\pm$ 0.008 \\
  \midrule
  CADE & 0.956 $\pm$  0.038 & 0.664  $\pm$ 0.088 &  0.971 $\pm$ 0.013 & 0.667 $\pm$  0.085  &  0.966 $\pm$ 0.016 & 0.667 $\pm$ 0.086  \\
  HCL & 0.997 $\pm$ 0.000 & 0.877 $\pm$ 0.058 & 0.995 $\pm$ 0.000 & 0.879 $\pm$ 0.010 & 0.996  $\pm$ 0.000 & 0.880 $\pm$  0.017 \\
  HCL$^*$ & \textcolor{blue}{0.996  $\pm$ 0.000}($\textcolor{red}{\downarrow}$) & 0.812 $\pm$ 0.052 ($\textcolor{red}{\downarrow}$) & \textcolor{blue}{0.994 $\pm$ 0.000}($\textcolor{red}{\downarrow}$) & 0.780  $\pm$ 0.004 ($\textcolor{red}{\downarrow}$) & \textcolor{blue}{0.995  $\pm$ 0.000}($\textcolor{red}{\downarrow}$) &  0.791 $\pm$ 0.021 ($\textcolor{red}{\downarrow}$) \\
  \midrule
 \method{} & 0.994 $\pm$ 0.002 & \textcolor{blue}{0.878 $\pm$ 0.002} & 0.993 $\pm$ 0.000 & \textcolor{blue}{0.855 $\pm$ 0.018} & 0.993 $\pm$ 0.001 & \textcolor{blue}{0.863 $\pm$ 0.011} \\
    
    \bottomrule
  \end{tabular}  
  \end{adjustbox}
  \begin{adjustbox}{width=1\textwidth}
  \begin{tabular}{lllllll}
    & \multicolumn{6}{c}{APIGraph} \\
    \cmidrule(lr){1-2} \cmidrule(lr){2-7}
   Baseline Methods & seen-AUT (B) & seen-AUT (A) & unseen-AUT (B) & unseen-AUT (A) & overall-AUT (B) & overall-AUT (A)\\
   \midrule
  
   EWC & 0.999 $\pm$ 0.000 & 0.922 $\pm$ 0.002 & 0.999 $\pm$ 0.000 & 0.960 $\pm$ 0.001 & 0.999 $\pm$ 0.000 & 0.944 $\pm$ 0.000 \\ 
 AGEM & \textbf{0.999 $\pm$ 0.000} & \textcolor{blue}{0.939 $\pm$ 0.003} &  \textbf{0.999 $\pm$ 0.000} & \textbf{0.965 $\pm$ 0.000} & \textbf{0.999 $\pm$ 0.000} & \textbf{0.954 $\pm$ 0.001} \\ 
 CBRS & 0.999 $\pm$ 0.000 & 0.923 $\pm$ 0.010 & 0.999 $\pm$ 0.000 & 0.931 $\pm$ 0.006 & 0.999 $\pm$ 0.000 & 0.928 $\pm$ 0.006 \\
 MIR & 0.999 $\pm$ 0.000 & 0.916 $\pm$ 0.006 & 0.999 $\pm$ 0.000 & 0.959 $\pm$ 0.001 & 0.999 $\pm$ 0.000 & 0.941 $\pm$ 0.001 \\
  \midrule
  CADE & 0.962   $\pm$ 0.043 &   0.874  $\pm$ 0.022 &  0.971 $\pm$  0.021 & 0.894 $\pm$ 0.005 & 0.968 $\pm$ 0.029  &  0.886 $\pm$  0.008 \\
  HCL & 0.999  $\pm$ 0.000 & 0.955 $\pm$ 0.003 & 0.998 $\pm$ 0.000 & 0.958 $\pm$ 0.004 & 0.999  $\pm$ 0.000 & 0.957  $\pm$ 0.004 \\
  HCL$^*$ & 0.998 $\pm$ 0.000 & 0.925  $\pm$ 0.002($\textcolor{red}{\downarrow}$) & \textcolor{blue}{0.998 $\pm$ 0.000} & \textcolor{blue}{0.948  $\pm$ 0.003}($\textcolor{red}{\downarrow}$) & 0.998 $\pm$ 0.000 & 0.938 $\pm$ 0.002($\textcolor{red}{\downarrow}$)  \\
  \midrule
 \method{} & \textcolor{blue}{0.999 $\pm$ 0.000} & \textbf{0.953 $\pm$ 0.009} & \textcolor{blue}{0.998 $\pm$ 0.000} & 0.930 $\pm$ 0.002 & \textcolor{blue}{0.999 $\pm$ 0.000} & \textcolor{blue}{0.940 $\pm$ 0.004} \\ 
    
    \bottomrule
  \end{tabular}  
  
  \end{adjustbox} 
\end{table}

\subsection{Comparing with the Baseline Methods}
\par The performance of the proposed method is compared with that of classical CL baselines and two recent methods CADE~\cite{yang2021cade}, HCL~\cite{291253} for malware detection and the results are presented in Table~\ref{tab:main_results}. First, we discuss the difference between the training strategies of the different methods and later present the performance comparison between different methods. 

\par \textbf{Training CL baselines.} Classical CL baseline methods operate under a fully supervised training mechanism across all tasks, both seen and unseen. In other words, full labeled data is used for training CL baselines when trained on seen and unseen tasks. The rationale behind this is rooted in the primary evaluation focus of the CL setting: maintaining the good performance on the (mitigate the catastrophic forgetting on) learned tasks.

\par \textbf{Training contrastive learning based baselines.} Contrastive learning baselines, such as CADE and HCL, utilize a fully supervised approach for seen tasks. For unseen tasks, within the constraints of a labeling budget, analysts label selected samples along with their family label information, which is then used for training in a CL framework. 
\par \textbf{Training the proposed approach.} In stark contrast to these approaches, our method employs a semi-supervised strategy for seen tasks, where only partial label information is available. This approach necessitates generating labels for the unlabeled data within the seen tasks. For unseen tasks (fully unlabeled data), analysts generate labels for selected samples. These labeled samples from unseen tasks are then utilized to train the model following the same procedure applied to seen tasks.

\par \textbf{Results.} The performance results of the baseline methods CADE, HCL and our proposed method are presented in Table~\ref{tab:main_results}. Here, a fixed labeling budget of $100$ samples per month is set. The results on CL baselines (EWC, AGEM, CBRS, and AGEM) serve  
as the most competitive values, as they operate under full supervision. However, they may not reliably be considered as upper-bound AUT values. For instance, on the AndroZoo and APIGraph datasets, the AGEM method achieves the best AUT values. In contrast, the AGEM's AUT values on the BODMAS dataset are lower compared to the proposed approach and the CADE. 

Our method on all the datasets uses a fixed labeled data ratio of 20\% (\suresh{selected randomly}) on seen tasks in all the experiments. The proposed method consistently outperforms CADE on all benchmark datasets (on evaluation metrics), HCL on BODMAS dataset and on par with HCL on AndroZoo dataset. On APIGraph, the difference between the proposed method and HCL (unseen-AUT (A)) is 0.028, thus it becomes the competitive baseline on unseen task. However, our method is competitive to HCL on seen tasks AUT and unseen-AUT (B) on APIGraph dataset. To understand the effectiveness of HCL in semi-supervised setting, we trained it with 20\% labeled data (similar to our method) on seen tasks, called it HCL$^*$. The performance of HCL$^*$ is lower compared to HCL on BODMAS and AndroZoo (marked this situation with $\textcolor{red}{\downarrow}$ in Table~\ref{tab:main_results}). We notice, our method outperforms HCL$^*$ on BODMAS and AndroZoo. The HCL$^*$ is competitive with our method on the APIGraph dataset. 
Specifically, this is due to the curation process of the APIGraph dataset and contrastive loss function of the HCL. We briefly describe this in the following.

\begin{figure*}[!b]

\centering 
\subfloat[\small{BODMAS}]{\label{bodmas} \includegraphics[scale=0.15]{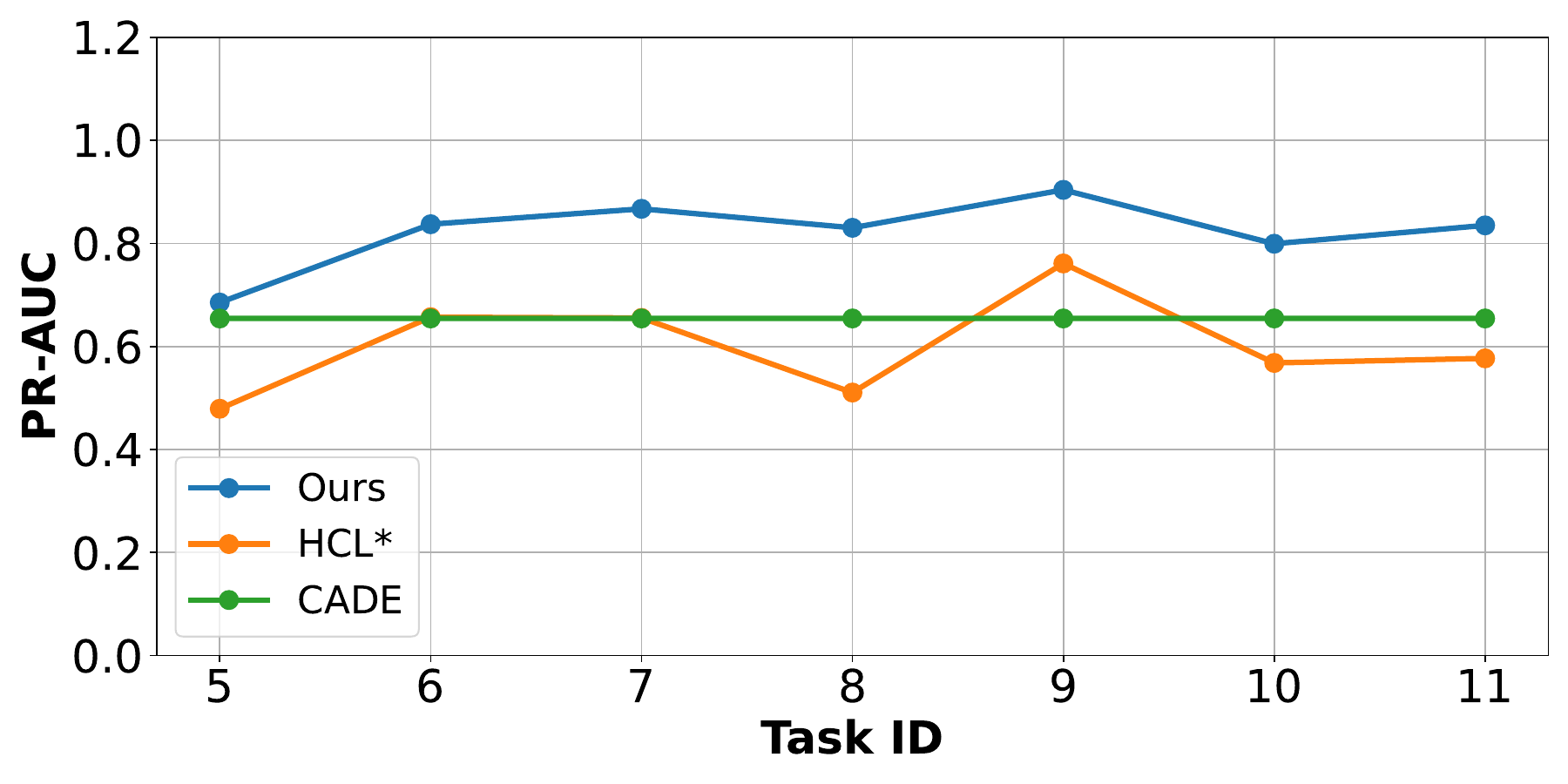}}
\subfloat[AndroZoo]{\label{androzoo} \includegraphics[scale=0.15]{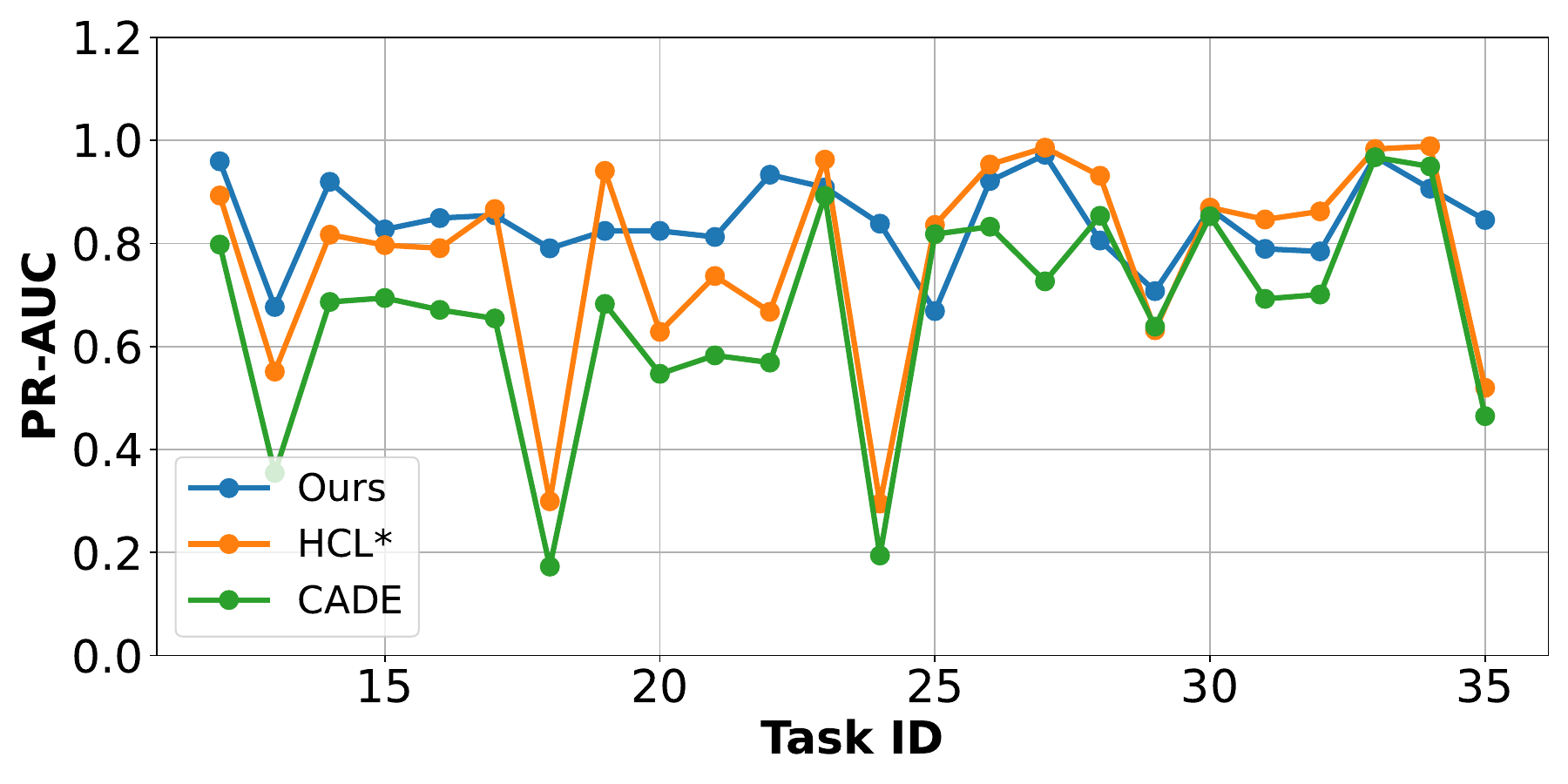}}
\subfloat[APIGraph]{\label{apigraph} \includegraphics[scale=0.15]{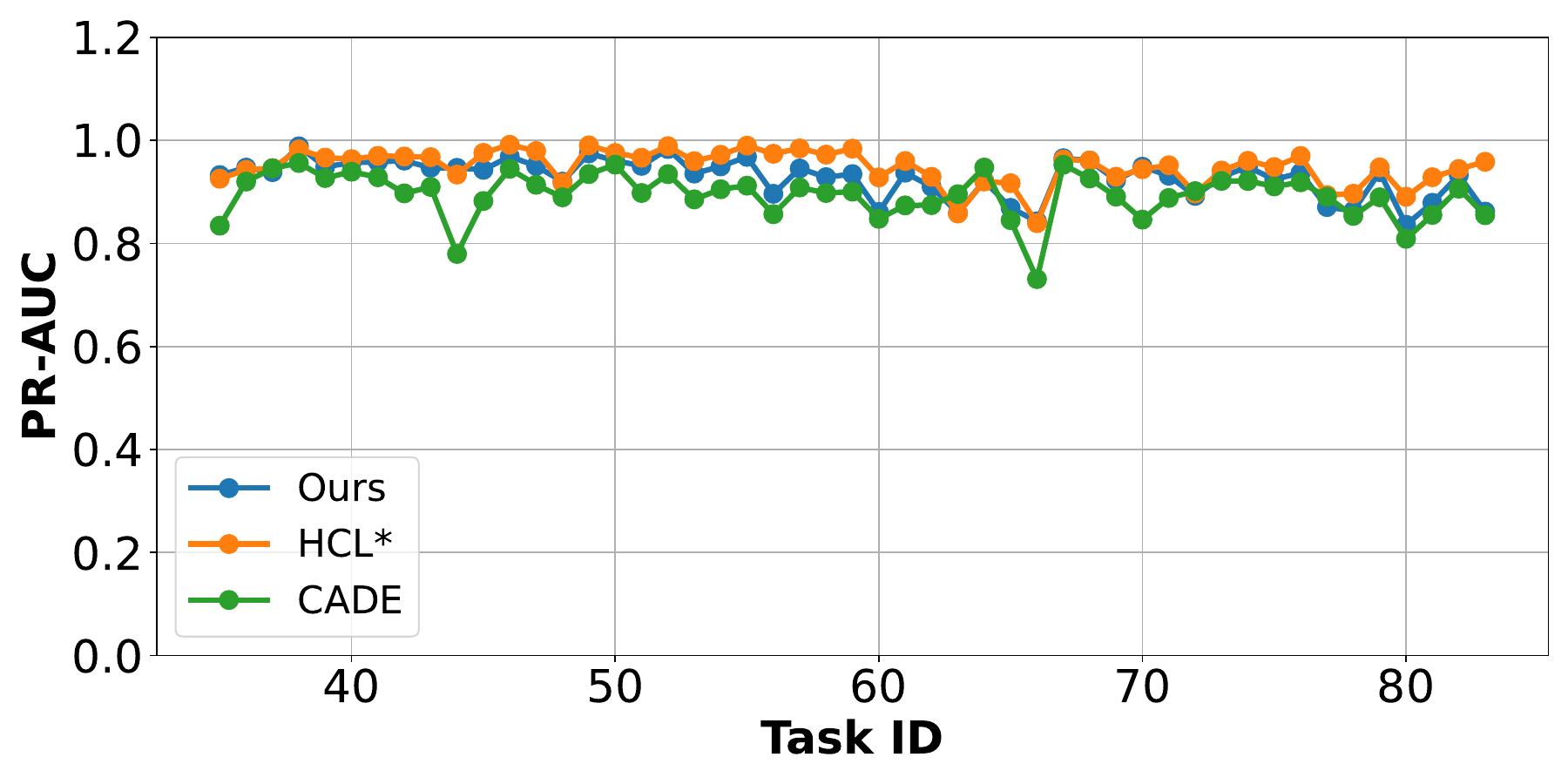}}%
 \Description{about the figure}
\caption{Comparing the proposed method (Ours/\method{} with labeling budget of 100 samples per month) with the CADE and  HCL$^*$ using the PR-AUC metric on the detection capability of unseen tasks malware. Each experiment is repeated for three different seed values and the mean PR-AUC values are used for demonstration. Our method consistently outperforms CADE and HCL$^*$ on BODMAS and AndroZoo on PR-AUC metric, stays competitive on APIGraph dataset.}
\label{fig:unseen-tasks-malware-pr-auc_cade_hcl}
\end{figure*}

\par The data curation process of APIGraph dataset inherently focuses on improving the performance of existing malware classifiers for tackling concept drift by bringing the common semantic structure among malware families using the API calls. 
Thus this dataset inherently maintains the semantic structure between the evolved malware samples. Further, contrastive learning loss used in HCL captures the semantic structure for constructing the positive and negative pair examples. Thus, HCL is competitive on the APIGraph dataset. On contrary, on other datasets which may not have such semantic structure, HCL$^*$ performance is degraded. In other words,  HCL$^*$ performance on unseen-AUT (A) is dropped from 0.795 to 0.613 on BODMAS dataset (lower than that of CADE) and on AndroZoo dataset it is dropped from 0.879 to 0.780. Thus, HCL$^*$ fails to stay competitive on other datasets, which do not necessarily have any semantic structure. It is also interesting to note that, on APIGraph difference in the performance between HCL$^*$ and our method on unseen-AUT (A) metric is $0.02$ which is lower than the average performance difference loss between HCL and HCL$^*$ ($0.1405$) compared to our method on other datasets.

We further analyzed this situation in greater depth over a time horizon using the PR-AUC metric. This experiment also serves to validate the reliability of the single-value metric AUT used across our experiments, as the same AUT score can sometimes arise from different detection behaviors (e.g., consistent moderate performance vs a decline from very good to poor). To this end, we compare the efficacy of the proposed approach with CADE and HCL$^*$ in detecting unseen-task malware across all datasets, as presented in Figure~\ref{fig:unseen-tasks-malware-pr-auc_cade_hcl}. Our approach consistently outperforms CADE and HCL$^*$ on the BODMAS and AndroZoo datasets, and remains competitive on APIGraph. On the other hand, our hypothesis—that the presence of API semantic structure in APIGraph benefits HCL—is robustly validated using the PR-AUC metric. In other words, while CADE, another contrastive learning-based method, performs poorly compared to our approach on other datasets, it becomes competitive on APIGraph. Furthermore, HCL’s use of multi-level (hierarchical) contrastive loss likely contributes to its strong performance, even with partially labeled data (HCL$^*$), on APIGraph. Thus, unlike existing methods, our approach is more versatile in handling situations, even in the absence of semantic structure. Further, curating a dataset like APIGraph requires a lot of effort to parse the available API documentations to find such common semantic structure. In many real-world settings, creating such structured dataset using millions of samples from an anti-virus company at a scale is challenging in terms of cost and effort.  

\begin{figure}[!ht]
    \centering 
\subfloat[\small{BODMAS}]{\label{bodmas3} \includegraphics[scale=0.15]{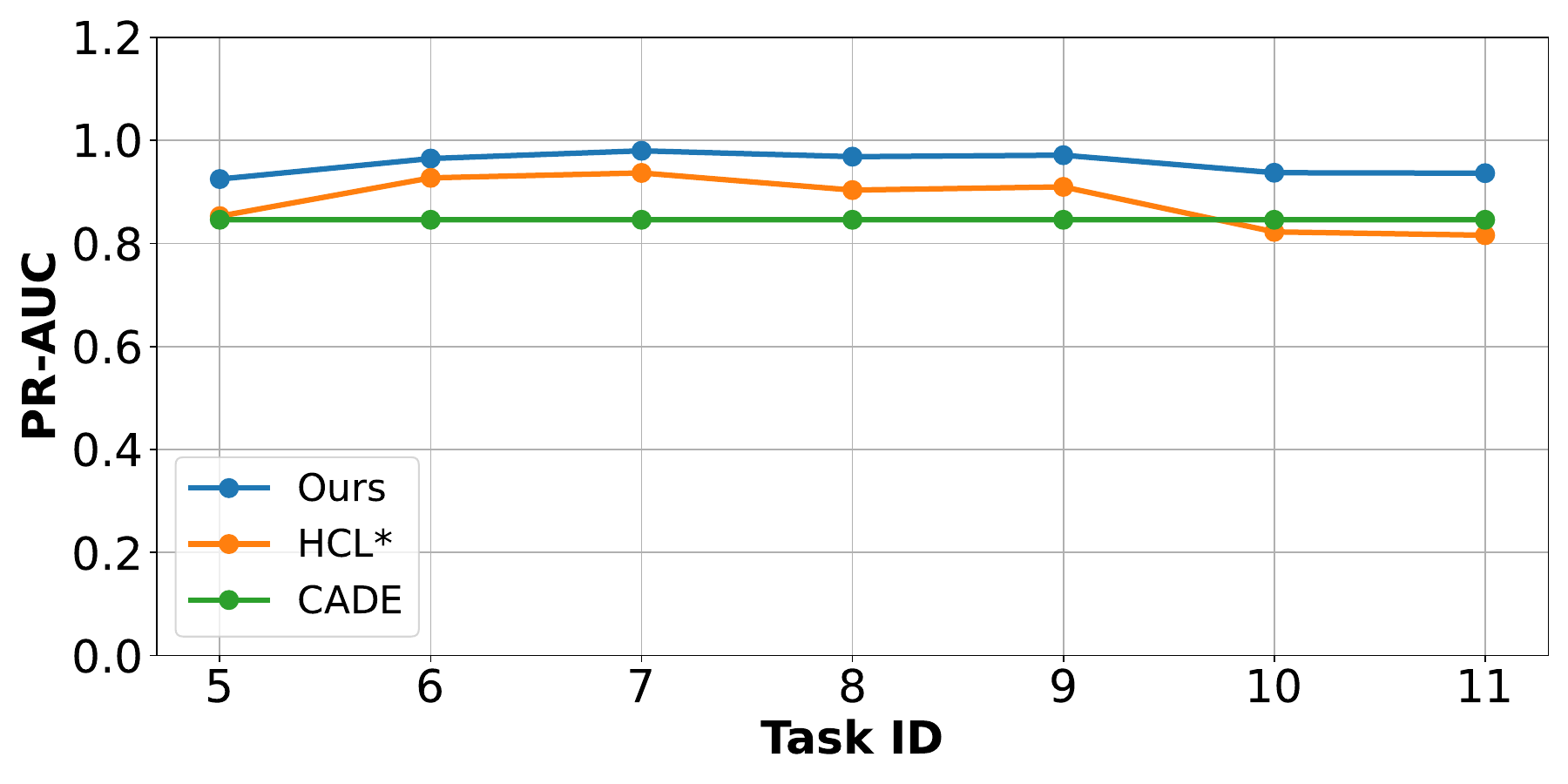}}
\subfloat[AndroZoo]{\label{androzoo3} \includegraphics[scale=0.15]{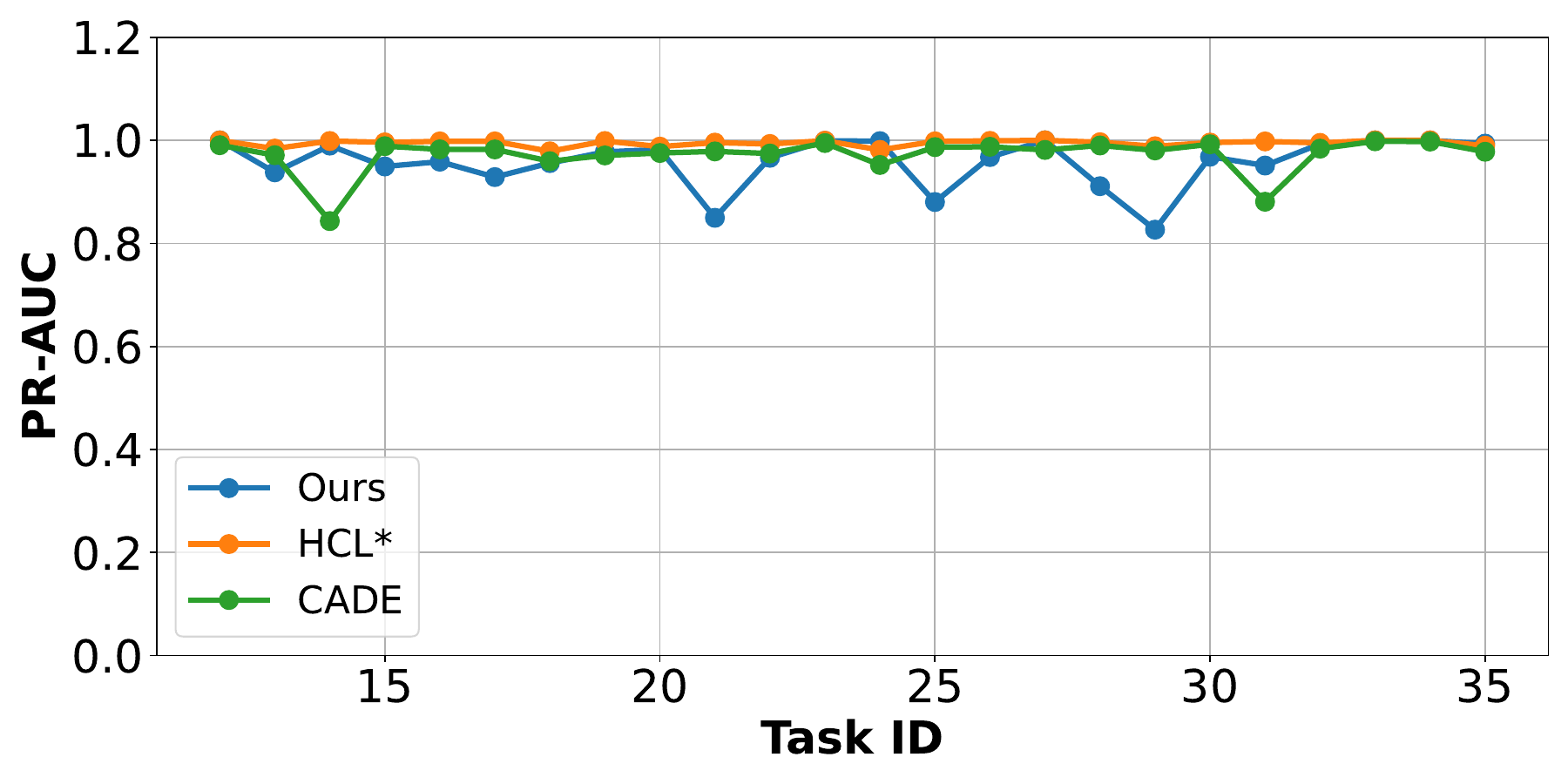}}
\subfloat[APIGraph]{\label{apigraph3} \includegraphics[scale=0.15]{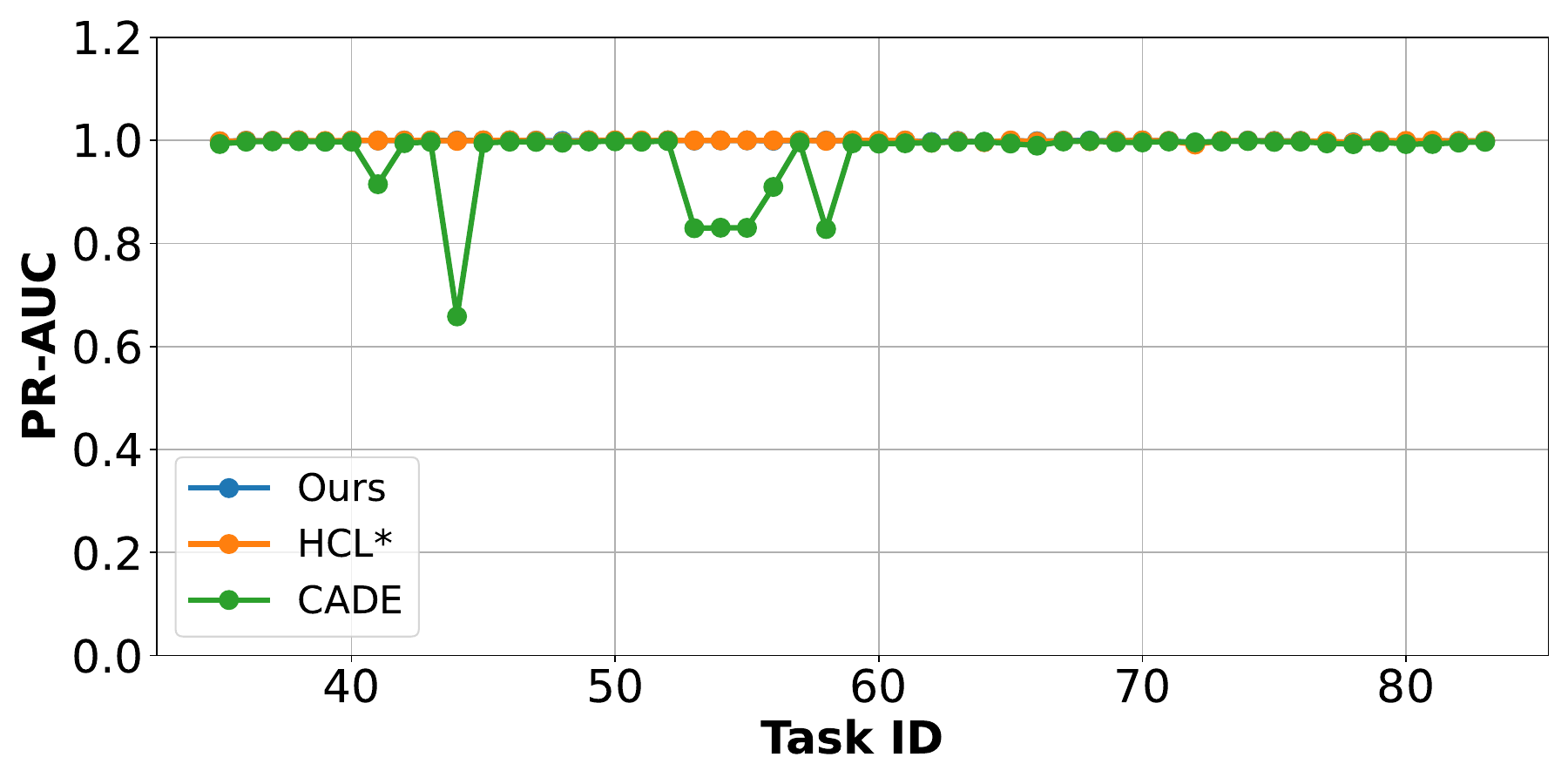}}%
 \Description{about the figure}
 
\caption{Comparing the proposed method (Ours/\method{} with labeling budget 100) with the CADE and  HCL$^*$ using the PR-AUC metric on the detection capability of unseen tasks benign samples. Each experiment is repeated for three different seed values and the mean PR-AUC values are used for demonstration. Our method stays competitive/outperforms CADE and HCL$^*$ on BODMAS, AndroZoo and APIGraph on PR-AUC metric, except on couple of tasks in AndroZoo dataset. }
\label{fig:unseen-tasks-benign-pr-auc}
\end{figure}
Our method consistently outperforms (refer to Figure~\ref{fig:unseen-tasks-benign-pr-auc}) CADE and HCL$^*$ in detecting benign samples from unseen tasks (open-world setting) on the BODMAS and APIGraph datasets. On the AndroZoo dataset, it remains competitive with the baselines, except on six tasks. However, in the context of a longer task sequence (thirty-six tasks), such low performance on only two tasks has minimal impact on the overall detection capability.

\begin{figure}[ht]

 \centering 
\subfloat[\small{BODMAS}]{\label{bodmas2} \includegraphics[scale=0.15]{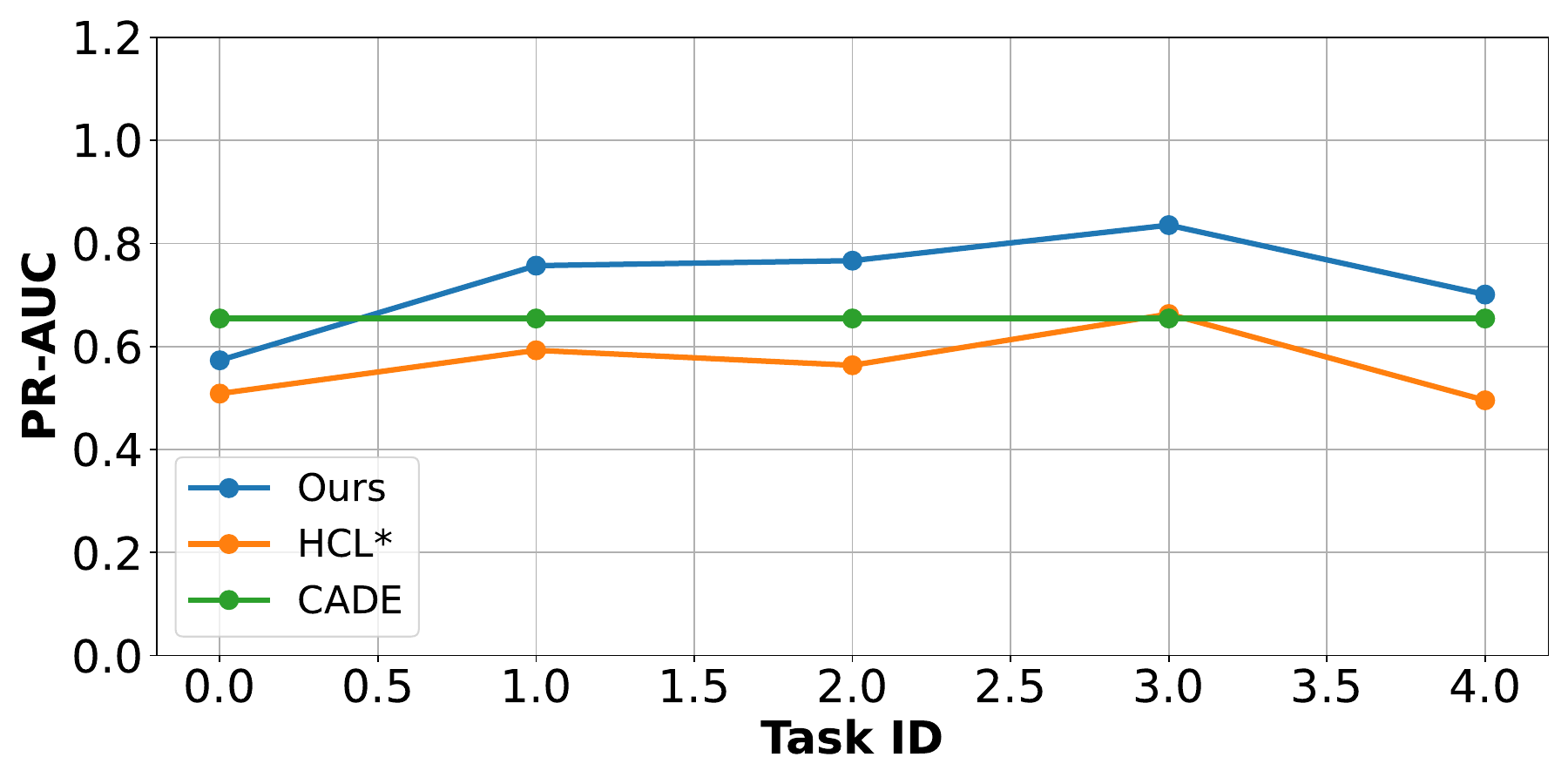}}
\subfloat[AndroZoo]{\label{androzoo2} \includegraphics[scale=0.15]{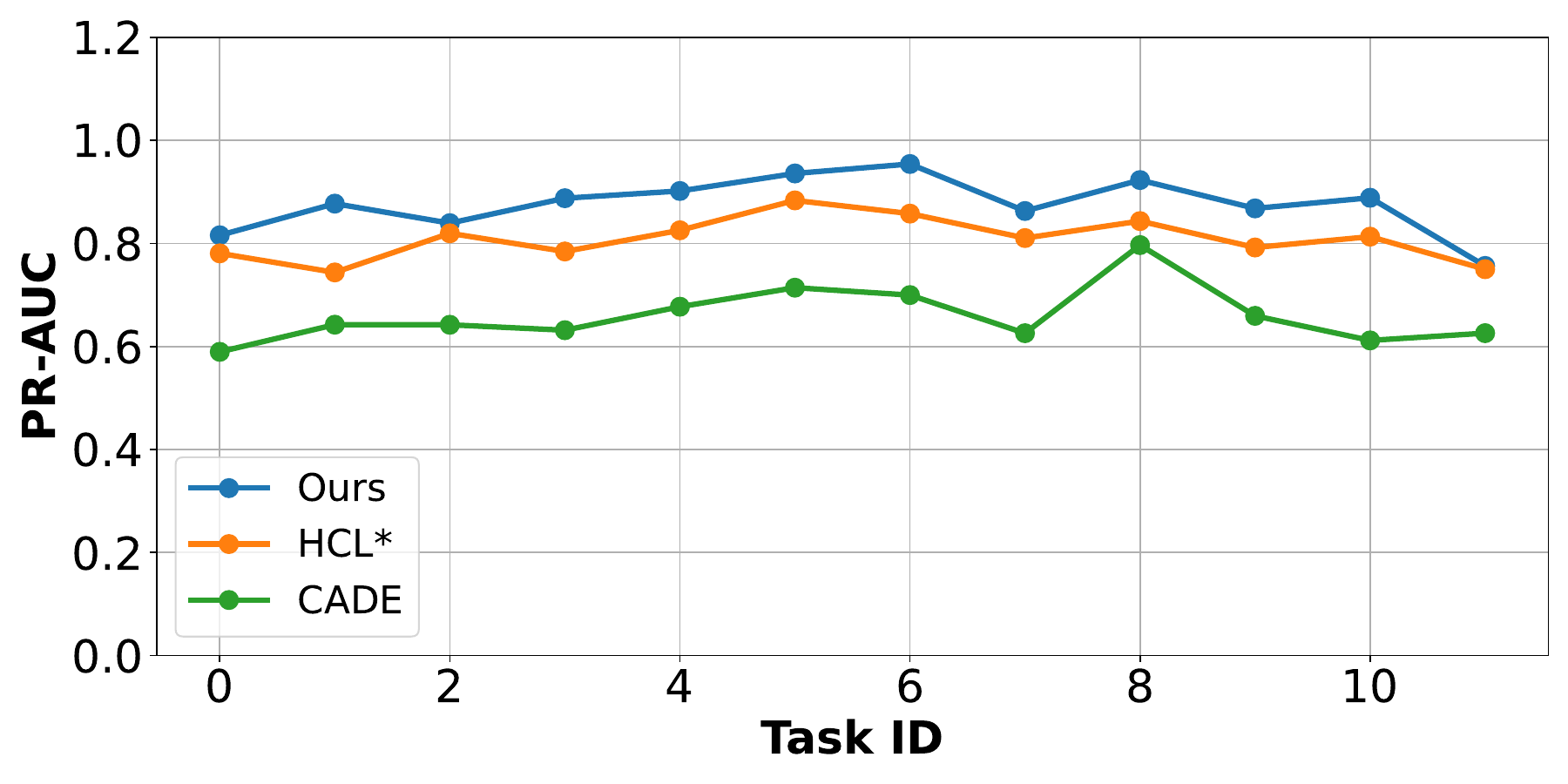}}
\subfloat[APIGraph]{\label{apigraph2} \includegraphics[scale=0.15]{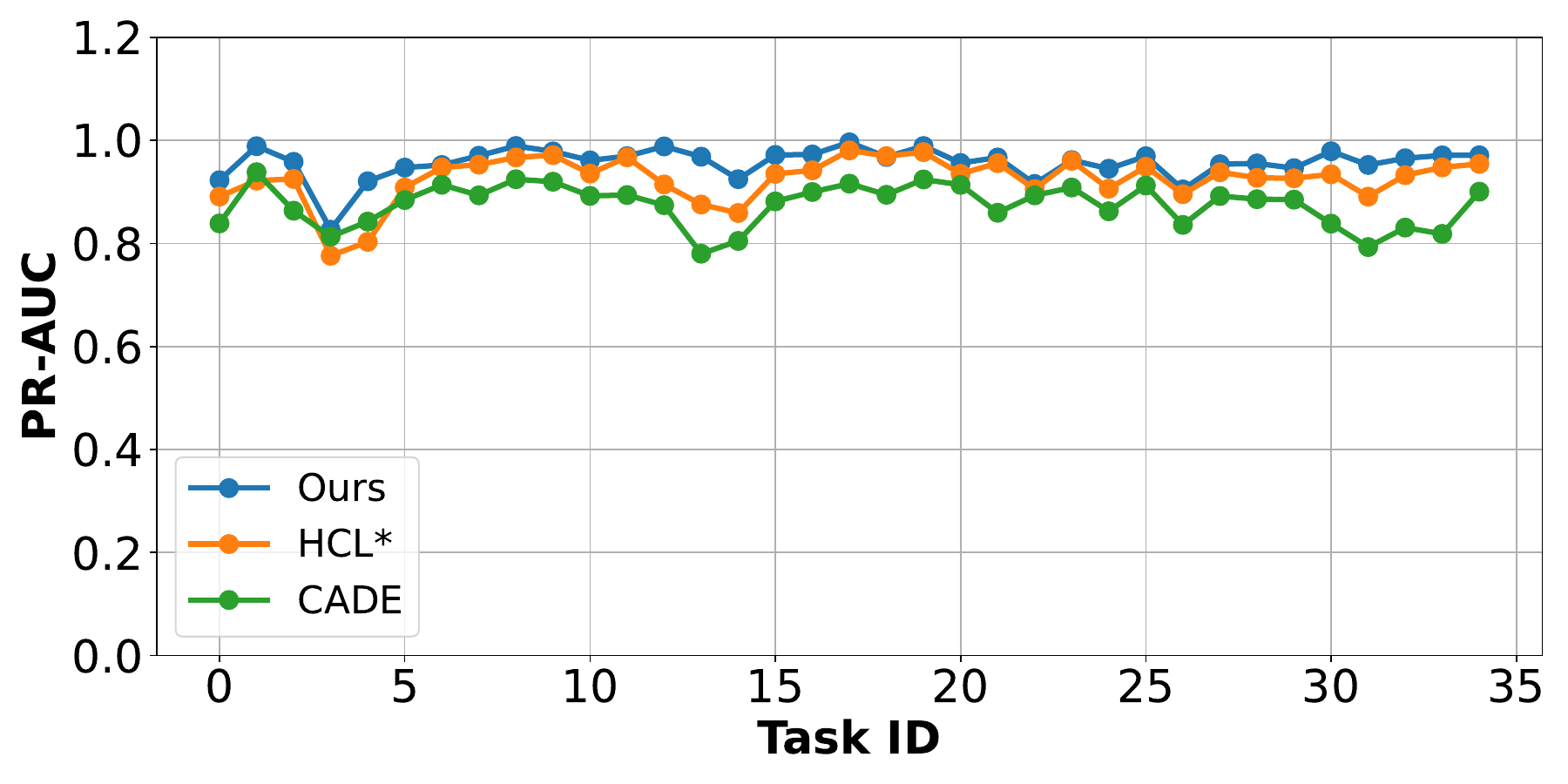}}%
 \Description{about the figure}
\caption{Comparing the proposed method (Ours with labeling budget 100) with the CADE and  HCL$^*$ using the PR-AUC metric on the detection capability of seen tasks malware. Each experiment is repeated for three different seed values and the mean PR-AUC values are used for demonstration. Our method consistently outperforms CADE and HCL$^*$ on BODMAS and AndroZoo on PR-AUC metric, stays competitive on APIGraph dataset. }
\label{fig:seen-tasks-malware-pr-auc}

\end{figure}
Our method consistently outperforms (refer to Figure~\ref{fig:seen-tasks-malware-pr-auc}) CADE and HCL$^*$ in detecting malware samples from seen tasks (closed-world setting) across all datasets. On the BODMAS and AndroZoo datasets, it outperforms the baselines by a significant margin. On the APIGraph dataset, the margin is smaller—this, as discussed earlier, can be attributed to the contrastive loss in CADE and HCL$^*$ effectively leveraging the API semantic structure present in the dataset.

\begin{figure}
     \centering 
\subfloat[\small{BODMAS}]{\label{bodmas1} \includegraphics[scale=0.15]{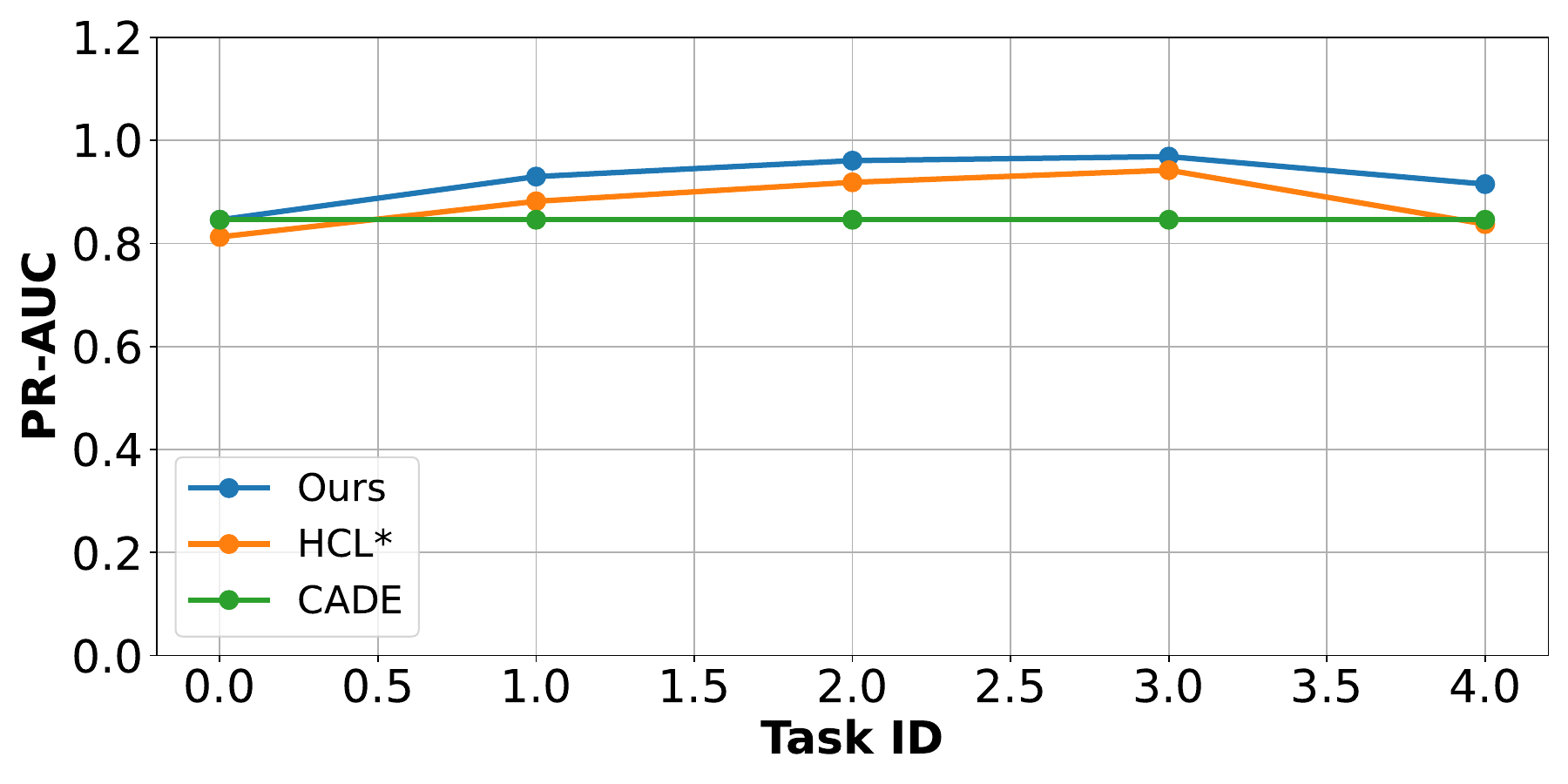}}
\subfloat[AndroZoo]{\label{androzoo1} \includegraphics[scale=0.15]{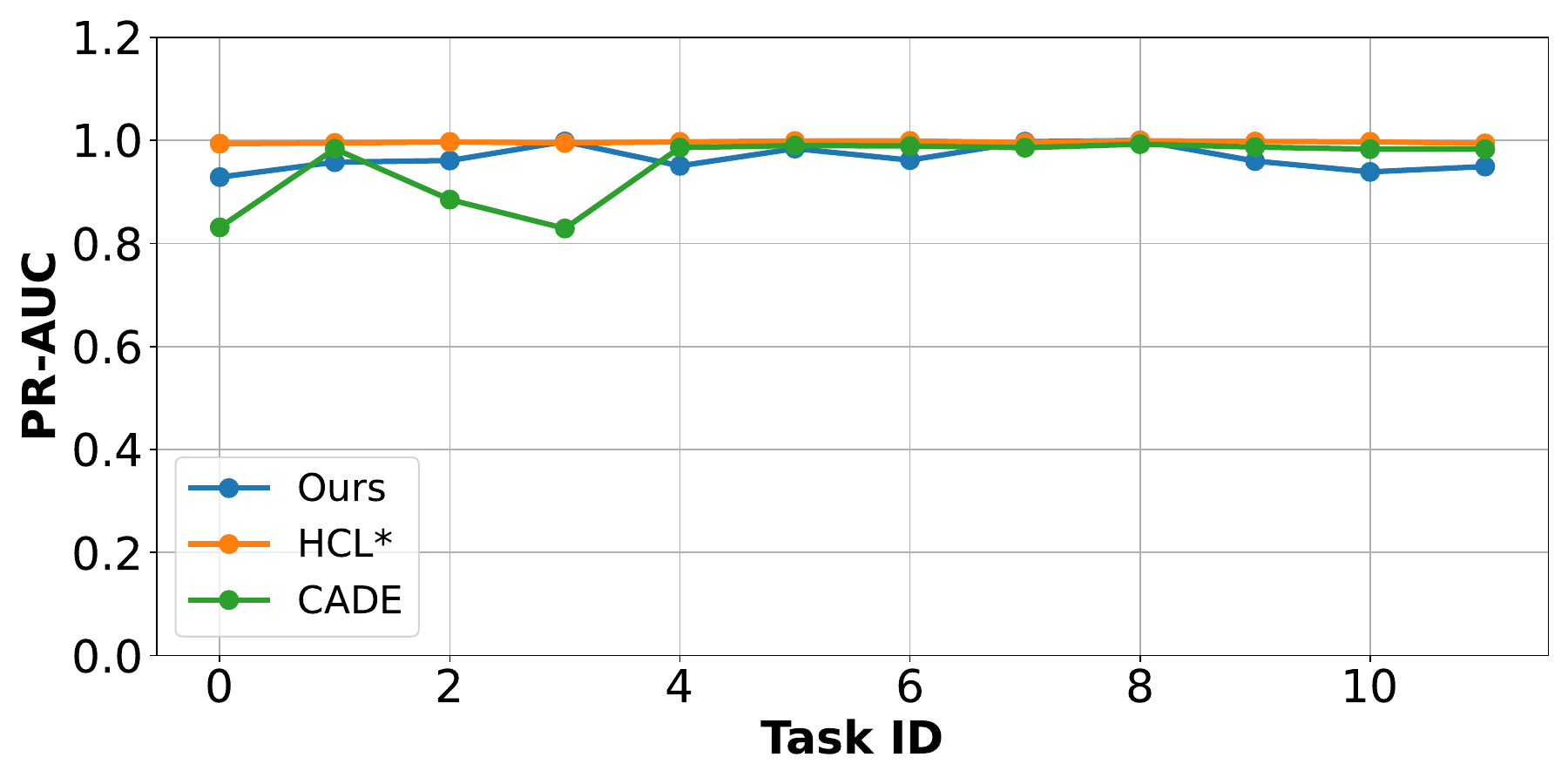}}
\subfloat[APIGraph]{\label{apigraph1} \includegraphics[scale=0.15]{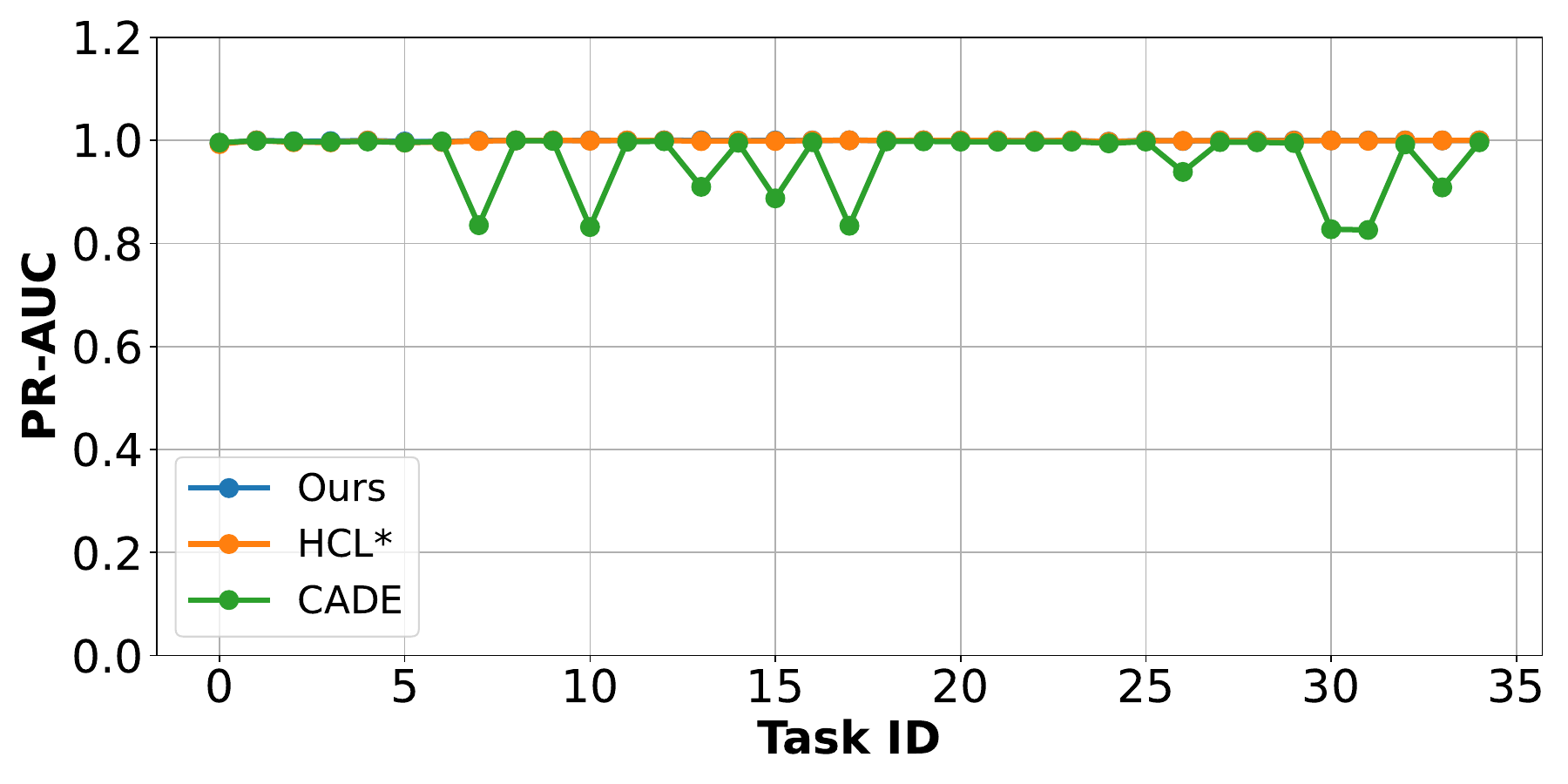}}%
 \Description{about the figure}
\caption{Comparing the proposed method (Ours/\method{} with labeling budget 100) with the CADE and  HCL$^*$ using the PR-AUC metric on the detection capability of seen tasks benign samples. Each experiment is repeated for three different seed values and the mean PR-AUC values are used for demonstration. Our method consistently outperforms CADE and HCL$^*$ on BODMAS and APIGraph on PR-AUC metric, stays competitive on AndroZoo dataset. }
\label{fig:seen-tasks-benign-pr-auc}
\end{figure}
Our method consistently outperforms (refer to Figure~\ref{fig:seen-tasks-benign-pr-auc}) CADE and HCL$^*$ in detecting benign samples from seen tasks (closed-world setting) on the BODMAS and AndroZoo dataset. On the APIGraph dataset, it outperforms CADE and performs on par with HCL$^*$, achieving the maximum possible PR-AUC value (PR-AUC = 1).

Ultimately, we observed that our findings remain consistent across all experiments with varying labeling budgets on all datasets, when comparing the proposed method with CADE and HCL$^*$.

\subsection{Statistical Analysis of Detection Performance}
\textcolor{black}{Across the datasets, we observe that the proposed \method{} demonstrates strong performance on unseen attack detection under limited labeled data, with statistical testing further supporting these observations. For significance analysis, we perform paired t-tests on seed-wise unseen-AUT values, where each seed forms a paired observation across methods. On BODMAS, which lacks strong inherent semantic structure, \method{} significantly outperforms CADE (\emph{p} = 0.015) and shows improvements over HCL$^*$ with marginal significance (\emph{p} = 0.054), highlighting its effectiveness when semantic signals are weak. On AndroZoo, \method{} achieves statistically significant gains over HCL* (\emph{p} = 0.044) and also outperforms CADE across all seeds; however, the difference is not statistically significant (\emph{p} = 0.127), likely due to the limited number of seeds and variability in the magnitude of improvement. On APIGraph, which exhibits richer semantic structure, \method{} remains competitive, significantly outperforming CADE (\emph{p} = 0.008) while performing comparably to HCL$^*$ (\emph{p} = 0.011). Overall, these results indicate that \method{} performs robustly across varying degrees of semantic structure, achieving strong results in its absence under limited supervision while maintaining competitive performance when such structure is present.}

\subsection{Sensitivity to Varying the Labeling Budgets}
\par In this section, we study the effectiveness of the proposed method under various labeling budgets. Specifically, we compare our method with CADE, HCL, and HCL$^*$ on the labeling budgets of 50, 150, 200, and 250. The results are presented in Tables~\ref{tab:varying_label_budget_bodmas}, \ref{tab:varying_label_budget_androzoo} and \ref{tab:varying_label_budget_apigraph}, while the results on the budget of 100 are present in Table~\ref{tab:main_results}. 
We made the following observations from these results. 
\begin{table}[ht]
\captionsetup{type=table}
  \caption{Performance comparison of the proposed method with baselines on BODMAS dataset by varying the monthly labeling budget (MB). We report AUT values for benign and malware classes on seen and unseen tasks, and also on the overall tasks. HCL$^*$ represents the case where experiments are conducted with 20\% labeled data on seen tasks. The proposed method (\method{}) uses 20\% labeled data on seen tasks across all the experiments on all the datasets. Each experiment is repeated for three times and the mean value along with standard deviation are reported. Best values are marked in \textbf{bold} and second best values are marked in \textcolor{blue}{blue}.}
  \label{tab:varying_label_budget_bodmas}
  \centering
  
  \begin{adjustbox}{width=1\textwidth}

\begin{tabular}{llllllll}
    & \multicolumn{6}{c}{BODMAS} \\
    \cmidrule(lr){1-8}
   MB & Method & seen-AUT (B) & seen-AUT (A) & unseen-AUT (B) & unseen-AUT (A) & overall-AUT (B) & overall-AUT (A)\\
   \midrule
  
     & CADE  & \textcolor{blue}{0.845 $\pm$ 0.000} & \textcolor{blue}{0.654 $\pm$ 0.000} & \textcolor{blue}{0.845 $\pm$ 0.000} & \textcolor{blue}{0.654 $\pm$ 0.000} & \textcolor{blue}{0.845 $\pm$ 0.000} & \textcolor{blue}{0.654 $\pm$ 0.000} \\ 
   50&  HCL & 0.920 $\pm$ 0.006 & 0.636 $\pm$ 0.055  & 0.955  $\pm$ 0.001 & 0.784$\pm$ 0.034 & 0.940 $\pm$ 0.001 & 0.719 $\pm$ 0.04 \\
   & HCL$^*$ & 0.864 $\pm$ 0.006 ($\textcolor{red}{\downarrow}$)& 0.530 $\pm$ 0.043 ($\textcolor{red}{\downarrow}$)&  0.847   $\pm$0.006($\textcolor{red}{\downarrow}$)   & 0.532 $\pm$  0.021 ($\textcolor{red}{\downarrow}$) & 0.849 $\pm$ 0.006($\textcolor{red}{\downarrow}$)  & 0.518 $\pm$ 0.029 ($\textcolor{red}{\downarrow}$)\\
   & \method{} & \textbf{0.928 $\pm$ 0.011} & \textbf{0.736 $\pm$ 0.013} & \textbf{0.940 $\pm$ 0.018} & \textbf{0.793 $\pm$ 0.038} & \textbf{0.933 $\pm$ 0.014} & \textbf{0.764 $\pm$ 0.023} \\   
    
  \midrule
  
 & CADE  & \textcolor{blue}{0.845 $\pm$ 0.000} & \textcolor{blue}{0.654 $\pm$ 0.000} & 0.845 $\pm$ 0.000 & \textcolor{blue}{0.654 $\pm$ 0.000} & 0.845 $\pm$ 0.000 & \textcolor{blue}{0.654 $\pm$ 0.000} \\ 
150 & HCL & 0.922 $\pm$  0.005 & 0.684  $\pm$ 0.011 & 0.953 $\pm$ 0.014 & 0.804 $\pm$  0.030  & 0.938 $\pm$ 0.011 & 0.745 $\pm$ 0.018  \\
 & HCL$^*$ &  0.861  $\pm$0.005 ($\textcolor{red}{\downarrow}$)& 0.513 $\pm$ 0.037 ($\textcolor{red}{\downarrow}$)& \textcolor{blue}{0.842 $\pm$ 0.010}($\textcolor{red}{\downarrow}$) & 0.521 $\pm$ 0.030 ($\textcolor{red}{\downarrow}$) & \textcolor{blue}{0.844 $\pm$ 0.008}($\textcolor{red}{\downarrow}$) & 0.505  $\pm$ 0.031 ($\textcolor{red}{\downarrow}$)\\
& \method{} & \textbf{0.917 $\pm$ 0.002} & \textbf{0.719 $\pm$ 0.006} & \textbf{0.934 $\pm$ 0.012} & \textbf{0.782 $\pm$ 0.038} & \textbf{0.925 $\pm$ 0.006} & \textbf{0.747 $\pm$ 0.020} \\
 \midrule

  & CADE & 0.844  $\pm$ 0.003 & 0.523 $\pm$ 0.014 & 0.828 $\pm$ 0.041 &  0.521 $\pm$ 0.037  & 0.833 $\pm$ 0.027 & 0.520 $\pm$ 0.034 \\ 
 200 & HCL & 0.923 $\pm$ 0.008 & 0.693  $\pm$ 0.029 & 0.951 $\pm$ 0.016 & 0.802 $\pm$ 0.033 & 0.938  $\pm$ 0.013 & 0.748 $\pm$ 0.027 \\ 
& HCL$^*$ & \textcolor{blue}{0.865  $\pm$ 0.013 } ($\textcolor{red}{\downarrow}$)& \textcolor{blue}{0.526 $\pm$ 0.061}($\textcolor{red}{\downarrow}$)& \textcolor{blue}{0.855 $\pm$ 0.016}($\textcolor{red}{\downarrow}$) & \textcolor{blue}{0.552   $\pm$ 0.044} ($\textcolor{red}{\downarrow}$) & \textcolor{blue}{0.854 $\pm$   0.015}($\textcolor{red}{\downarrow}$) & \textcolor{blue}{0.530 $\pm$   0.051} ($\textcolor{red}{\downarrow}$)\\
 & \method{} & \textbf{0.926 $\pm$ 0.010} & \textbf{0.664 $\pm$ 0.074} & 0.\textbf{953 $\pm$ 0.006} & 0\textbf{.777 $\pm$ 0.057} & \textbf{0.940 $\pm$ 0.007} & \textbf{0.727 $\pm$ 0.060} \\
  \midrule

 & CADE  & 0.845 $\pm$ 0.000 & \textcolor{blue}{0.654 $\pm$ 0.000} & 0.845 $\pm$ 0.000 & \textcolor{blue}{0.654 $\pm$ 0.000} & 0.845 $\pm$ 0.000 & \textcolor{blue}{0.654 $\pm$ 0.000} \\ 
250 & HCL & 0.924  $\pm$ 0.010 & 0.635  $\pm$ 0.088  & 0.955 $\pm$ 0.003 & 0.772 $\pm$ 0.027 & 0.941 $\pm$ 0.003 &  0.711 $\pm$ 0.047  \\ 
 &HCL$^*$ & \textcolor{blue}{0.881 $\pm$ 0.027} ($\textcolor{red}{\downarrow}$)& 0.529  $\pm$ 0.025 ($\textcolor{red}{\downarrow}$)& \textcolor{blue}{0.872$\pm$ 0.046}($\textcolor{red}{\downarrow}$) &  0.571 $\pm$ 0.071($\textcolor{red}{\downarrow}$) & \textcolor{blue}{ 0.872    $\pm$  0.040  }($\textcolor{red}{\downarrow}$) &   0.544 $\pm$ 0.051 ($\textcolor{red}{\downarrow}$)\\
  &  \method{} & \textbf{0.931 $\pm$ 0.003} & \textbf{0.709 $\pm$ 0.017} & \textbf{0.957 $\pm$ 0.001} & \textbf{0.810 $\pm$ 0.009} & \textbf{0.944 $\pm$ 0.001} & \textbf{0.761 $\pm$ 0.011} \\
    \bottomrule
  \end{tabular} 
\end{adjustbox}

\end{table}
\par First, the proposed method consistently outperforms CADE on all the benchmarks on all the labeling budgets. Our hypothesis on reductions in performance (on all evaluation metrics) of HCL with partial labeled data (HCL$^*$) remains valid on the BODMAS and AndroZoo benchmarks with varying labeling budgets. On APIGraph dataset, with partial labeled data, the performance of seen tasks malware (seen-AUT (A)) is consistently reduced across all the labeling budgets. HCL$
^*$ consistency on all the labeling budgets on APIGraph confirms our finding that semantic structure of the dataset combined with bias of contrastive loss that inherently leverages this structure can be a powerful mechanism to work even under the partial labeled data.

\begin{table}[ht]
\captionsetup{type=table}
  \caption{Performance comparison of the proposed method with baselines on AndroZoo dataset by varying the monthly labeling budget (MB). We report AUT values for benign and malware classes on seen and unseen tasks, and also on the overall tasks. HCL$^*$ represents the case where experiments are conducted with 20\% labeled data on seen tasks. The proposed method (\method{}) uses 20\% labeled data on seen tasks across all the experiments on all the datasets. Each experiment is repeated for three times and the mean value along with standard deviation are reported. Best values are marked in \textbf{bold} and second best values are marked in \textcolor{blue}{blue}.}
  \label{tab:varying_label_budget_androzoo}
  \centering
  
\begin{adjustbox}{width=1\textwidth}

\begin{adjustbox}{width=1\textwidth}
  \begin{tabular}{llllllll}
    & \multicolumn{6}{c}{Andro Zoo} \\
    \cmidrule(lr){1-8} 
  MB&  Method  & seen-AUT (B) & seen-AUT (A) & unseen-AUT (B) & unseen-AUT (A) & overall-AUT (B) & overall-AUT (A)\\
   \midrule

  & CADE  &  0.981 $\pm$ 0.003  & 0.676 $\pm$ 0.054 & 0.974 $\pm$ 0.01 & 0.593  $\pm$  0.059  & 0.976 $\pm$ 0.007 & 0.619 $\pm$  0.043  \\ 
  50 &  HCL & 0.997 $\pm$ 0.000 & 0.881 $\pm$ 0.024  & 0.994 $\pm$ 0.001 &  0.807  $\pm$ 0.017 & 0.995 $\pm$ 0.000 & 0.833 $\pm$  0.019 \\ 
  & HCL$^*$ & \textbf{0.996 $\pm$ 0.000}($\textcolor{red}{\downarrow}$) & \textcolor{blue}{0.850 $\pm$  0.054}($\textcolor{red}{\downarrow}$)& \textcolor{blue}{0.978 $\pm$ 0.009}($\textcolor{red}{\downarrow}$) & \textcolor{blue}{0.536 $\pm$ 0.086} ($\textcolor{red}{\downarrow}$) & \textcolor{blue}{0.985 $\pm$ 0.006}($\textcolor{red}{\downarrow}$) & \textcolor{blue}{0.644$\pm$ 0.039} ($\textcolor{red}{\downarrow}$)\\
& \method{} & \textcolor{blue}{0.995 $\pm$ 0.001} & \textbf{0.875 $\pm$ 0.016} & \textbf{0.994 $\pm$ 0.000} & \textbf{0.839 $\pm$ 0.010} & \textbf{0.994 $\pm$ 0.000} & \textbf{0.851 $\pm$ 0.004} \\ 
  \midrule
  
& CADE  & 0.976 $\pm$ 0.008 & 0.542  $\pm$ 0.263 & 0.951 $\pm$ 0.033 & 0.614 $\pm$ 0.134  & 0.959 $\pm$ 0.018 & 0.590 $\pm$ 0.179  \\ 
 150 &HCL & 0.998 $\pm$ 0.000 & 0.907 $\pm$ 0.024 & 0.996 $\pm$ 0.000 &  0.872  $\pm$ 0.025  & 0.997 $\pm$ 0.000 & 0.885 $\pm$ 0.023 \\ 
 &HCL$^*$ & \textbf{0.997 $\pm$ 0.000}($\textcolor{red}{\downarrow}$) & \textcolor{blue}{0.813 $\pm$ 0.025}($\textcolor{red}{\downarrow}$)& \textbf{0.994 $\pm$ 0.001}($\textcolor{red}{\downarrow}$) & \textcolor{blue}{ 0.821  $\pm$ 0.023 } ($\textcolor{red}{\downarrow}$) & \textbf{0.995 $\pm$ 0.001}($\textcolor{red}{\downarrow}$) & \textcolor{blue}{0.819   $\pm$ 0.024 } ($\textcolor{red}{\downarrow}$)\\
&\method{} & \textcolor{blue}{0.992 $\pm$ 0.001} & \textbf{0.860 $\pm$ 0.010} & \textcolor{blue}{0.994 $\pm$ 0.000} & \textbf{0.855 $\pm$ 0.034} & \textcolor{blue}{0.993 $\pm$ 0.000} & \textbf{0.857 $\pm$ 0.026} \\  
 \midrule
 
 &CADE  & 0.987 $\pm$  0.000 & 0.671 $\pm$ 0.035 & 0.987 $\pm$ 0.002 & 0.664 $\pm$ 0.108& 0.987  $\pm$  0.001& 0.668 $\pm$ 0.064  \\ 
200 &HCL & 0.998 $\pm$ 0.000 & 0.915 $\pm$ 0.012 & 0.997  $\pm$ 0.000 &  0.882 $\pm$ 0.031 & 0.997 $\pm$ 0.000 & 0.894 $\pm$ 0.018  \\ 
& HCL$^*$ & \textbf{0.996 $\pm$ 0.000}($\textcolor{red}{\downarrow}$) & \textcolor{blue}{0.812  $\pm$ 0.052}($\textcolor{red}{\downarrow}$)& \textbf{0.994 $\pm$ 0.000}($\textcolor{red}{\downarrow}$) & \textcolor{blue}{0.780 $\pm$ 0.004} ($\textcolor{red}{\downarrow}$) & \textbf{0.995   $\pm$ 0.000}($\textcolor{red}{\downarrow}$) & \textcolor{blue}{0.791  $\pm$  0.021} ($\textcolor{red}{\downarrow}$)\\

 &\method{} & \textcolor{blue}{0.992 $\pm$ 0.001} & \textbf{0.869 $\pm$ 0.014} & \textcolor{blue}{0.994 $\pm$ 0.000} & \textbf{0.879 $\pm$ 0.002} & \textcolor{blue}{0.993 $\pm$ 0.000} & \textbf{0.876 $\pm$ 0.002} \\ 
 \midrule

 &CADE & 0.957 $\pm$ 0.037 & 0.672 $\pm$ 0.082 & 0.977 $\pm$ 0.009  &  0.715  $\pm$ 0.092  & 0.971 $\pm$ 0.018 & 0.704 $\pm$ 0.047  \\ 
 250&HCL & 0.998 $\pm$ 0.000 & 0.914 $\pm$ 0.003  & 0.997 $\pm$ 0.000 & 0.907  $\pm$ 0.004 & 0.997 $\pm$ 0.000 & 0.910 $\pm$ 0.003  \\ 
 &HCL$^*$ & \textbf{0.996  $\pm$ 0.000}($\textcolor{red}{\downarrow}$) & \textcolor{blue}{ 0.829 $\pm$  0.026 }($\textcolor{red}{\downarrow}$)& \textbf{0.996  $\pm$ 0.000}($\textcolor{red}{\downarrow}$) & \textcolor{blue}{0.847 $\pm$ 0.012} ($\textcolor{red}{\downarrow}$) & \textbf{0.996 $\pm$ 0.000}($\textcolor{red}{\downarrow}$) & \textcolor{blue}{0.843 $\pm$ 0.009  } ($\textcolor{red}{\downarrow}$)\\
   &\method{} & \textcolor{blue}{0.993 $\pm$ 0.001} & \textbf{0.880 $\pm$ 0.004} & \textcolor{blue}{0.994 $\pm$ 0.001} & \textbf{0.878 $\pm$ 0.018} & \textcolor{blue}{0.994 $\pm$ 0.001} & \textbf{0.879 $\pm$ 0.012} \\  
    \bottomrule
  \end{tabular}  
  \end{adjustbox}
  \end{adjustbox}
  
\end{table}
\par On BODMAS dataset (refer to Table~\ref{tab:varying_label_budget_bodmas}), CADE outperforms HCL$^*$ in terms of unseen tasks AUT~(A) metric with labeling budgets of 50, 150, and 250. With reduced labeled data, HCL$^*$ experiences performance degradation when compared to HCL on unseen AUT~(A), with an average reduction of 0.246. This is the highest reduction observed across all other evaluation metrics. Our method continues to achieve higher malware detection on seen tasks (seen-AUT~(A)) compared to HCL on all labeling budgets and unseen tasks (unseen-AUT~(A)) on HCL on labeling budgets of 50, 100, and 250. The average performance gains achieved using our method compared to HCL$^*$ on unseen AUT~(A) metric is $0.236$ on partial labeled data. 

\par On AndroZoo dataset (refer to Table~\ref{tab:varying_label_budget_androzoo}), our proposed method outperforms HCL$^*$ and CADE on unseen-AUT (A) values on all the labeling budgets. The average performance gain our method achieved compared to HCL$^*$ on unseen AUT (A) metric is $0.108$.

\par On APIGraph dataset (refer to Table~\ref{tab:varying_label_budget_apigraph}), our method outperforms HCL$^*$ on seen-AUT (A) and seen-AUT (B) while it is competitive on unseen-AUT (B), overall-AUT (B), and overall-AUT (A). with increasing labeling budget, the performance difference between HCL$^*$ and our method on unseen AUT (A) continues to reduce. Specifically, the difference is 0.031 for the labeling budget of 50, 0.018 for 100, 0.018 for 150, 0.024 for 200, and 0.019 for 250.  

Overall, our method consistently outperforms HCL$^*$ across datasets, particularly on unseen AUT~(A) tasks under limited labeling budgets. The gains are most pronounced on BODMAS, AndroZoo, and competitive on APIGraph. 

\begin{table*}
\captionsetup{type=table}
  \caption{{Performance comparison of the proposed method with baselines on APIGraph datasets by varying the monthly labeling budget (MB). We report AUT values for benign and malware classes on seen and unseen tasks, and also on the overall tasks. HCL$^*$ represents the case where experiments are conducted with 20\% labeled data on seen tasks. The proposed method (\method{}) uses 20\% labeled data on seen tasks across all the experiments on all the datasets. Each experiment is repeated for three times and the mean value along with standard deviation are reported. Best values are marked in \textbf{bold} and second best values are marked in} \textcolor{blue}{blue}.}
  \label{tab:varying_label_budget_apigraph}
  \centering

  \begin{adjustbox}{width=1\textwidth}
\begin{tabular}{llllllll}
    & \multicolumn{6}{c}{APIGraph} \\
    \cmidrule(lr){1-8} 
  MB & Method & seen-AUT (B) & seen-AUT (A) & unseen-AUT (B) & unseen-AUT (A) & overall-AUT (B) & overall-AUT (A)\\
   \midrule
  
    &CADE  & 0.933 $\pm$ 0.062 & 0.881$\pm$ 0.024   & 0.987  $\pm$ 0.006  & 0.86 $\pm$ 0.023 & 0.965 $\pm$ 0.027 & 0.868 $\pm$ 0.004 \\ 
    50&HCL & 0.999 $\pm$ 0.000 & 0.962 $\pm$ 0.003  & 0.998 $\pm$ 0.000 &   0.950  $\pm$ 0.000 & 0.998 $\pm$ 0.000 & 0.955  $\pm$ 0.001 \\ 
    &HCL$^*$ & \textcolor{blue}{0.998 $\pm$ 0.000}($\textcolor{red}{\downarrow}$) & \textcolor{blue}{0.929 $\pm$ 0.007}($\textcolor{red}{\downarrow}$)& \textbf{0.998 $\pm$ 0.000} & \textbf{0.946 $\pm$ 0.001} ($\textcolor{red}{\downarrow}$) & \textbf{0.998 $\pm$ 0.000} &  \textbf{0.939$\pm$ 0.002} ($\textcolor{red}{\downarrow}$)\\
    &Ours & \textbf{0.999 $\pm$ 0.000} & \textbf{0.957 $\pm$ 0.003} & \textcolor{blue}{0.998 $\pm$ 0.000} & \textcolor{blue}{0.915 $\pm$ 0.002} & 0.998 $\pm$ 0.000 & \textcolor{blue}{0.932 $\pm$ 0.000} \\
  \midrule
   &CADE  & 0.887  $\pm$ 0.081 & 0.888   $\pm$   0.016 &  0.931 $\pm$ 0.058  & 0.898 $\pm$ 0.011    &   0.911  $\pm$ 0.068 &  0.893 $\pm$ 0.011 \\ 
   150&HCL & 0.999 $\pm$ 0.000 & 0.961 $\pm$  0.001 & 0.999  $\pm$ 0.000 & 0.961 $\pm$   0.003  & 0.999 $\pm$ 0.000 & 0.961 $ \pm$ 0.001  \\
   &HCL$^*$ & \textcolor{blue}{0.998 $\pm$ 0.000}($\textcolor{red}{\downarrow}$) & \textcolor{blue}{0.930  $\pm$   0.01  }($\textcolor{red}{\downarrow}$)& \textbf{0.999 $\pm$ 0.000} & \textbf{0.952 $\pm$ 0.003 } ($\textcolor{red}{\downarrow}$) & \textbf{0.998 $\pm$ 0.000}($\textcolor{red}{\downarrow}$) & \textbf{0.943 $\pm$ 0.006} ($\textcolor{red}{\downarrow}$)\\
&Ours & \textbf{0.999 $\pm$ 0.000} & \textbf{0.959 $\pm$ 0.003} & 0.998 $\pm$ 0.000 & \textcolor{blue}{0.934 $\pm$ 0.000} & 0.999 $\pm$ 0.000 & \textbf{0.945 $\pm$ 0.001} \\

   \midrule
  & CADE &  0.864 $\pm$ 0.068  &  0.881 $\pm$ 0.023  & 0.907 $\pm$  0.047  & 0.892 $\pm$  0.006 & 0.888 $\pm$ 0.052 & 0.887 $\pm$ 0.012\\  
 200 & HCL & 0.999 $\pm$ 0.000 & 0.950 $\pm$ 0.01  &  0.999 $\pm$ 0.000 & 0.960  $\pm$ 0.005 &  0.999 $\pm$ 0.000 & 0.956 $\pm$ 0.006 \\ 
 & HCL$^*$ & \textcolor{blue}{0.998 $\pm$ 0.000}($\textcolor{red}{\downarrow}$) & \textcolor{blue}{0.928 $\pm$ 0.003}($\textcolor{red}{\downarrow}$) & \textbf{0.999 $\pm$ 0.000} & \textbf{0.962 $\pm$ 0.001}($\textcolor{green}{\uparrow}$) & \textbf{0.999 $\pm$ 0.000} & \textbf{0.948 $\pm$ 0.001}($\textcolor{red}{\downarrow}$) \\
 & Ours & \textbf{0.999 $\pm$ 0.000} & \textbf{0.961 $\pm$ 0.002} & \textcolor{blue}{0.998 $\pm$ 0.000} & \textcolor{blue}{0.938 $\pm$ 0.001} & \textcolor{blue}{0.999 $\pm$ 0.000} & \textcolor{blue}{0.948 $\pm$ 0.000} \\
 
 \midrule
 & CADE & 0.981  $\pm$  0.009 &  0.897 $\pm$ 0.031 & 0.959   $\pm$ 0.029  &   0.909 $\pm$ 0.021 & 0.968 $\pm$ 0.018 & 0.904 $\pm$  0.025 \\ 
   250 & HCL & 0.999 $\pm$ 0.000 & 0.948 $\pm$ 0.006 & 0.999 $\pm$ 0.000 & 0.965  $\pm$ 0.006 & 0.999 $\pm$ 0.000 & 0.958 $\pm$ 0.005 \\ 
    &HCL$^*$ & \textcolor{blue}{0.998 $\pm$ 0.000}($\textcolor{red}{\downarrow}$) & \textcolor{blue}{0.920 $\pm$ 0.007}  $\textcolor{red}{(\downarrow)}$ & \textbf{0.999 $\pm$ 0.000} & \textbf{0.963 $\pm$  0.002}($\textcolor{red}{\downarrow}$) & \textbf{0.999 $\pm$ 0.000} & \textcolor{blue}{0.946 $\pm$ 0.003}$\textcolor{red}{(\downarrow)}$ \\

    & Ours & \textbf{0.999 $\pm$ 0.000} & \textbf{0.961 $\pm$ 0.000} & \textcolor{blue}{0.999 $\pm$ 0.000} & \textcolor{blue}{0.944 $\pm$ 0.001} & \textcolor{blue}{0.999 $\pm$ 0.000} & \textbf{0.951 $\pm$ 0.000} \\

    \bottomrule
  \end{tabular} 
  
  \end{adjustbox}


\end{table*}

\subsection{sensitivity to Varying Labeled Data Ratios}
\textcolor{black}{We compare the performance of \method{} on unseen malware detection with varying label ratios across different labeling budgets. Across all datasets, increasing the label ratio consistently improves unseen-AUT (A).}

\paragraph{BODMAS} \textcolor{black}{For BODMAS, increasing the label ratio leads to clear improvements in unseen-AUT (A) across all budgets (refer to Table~\ref{tab:varying_label_bdget_label_ratio_bodmas}). At a budget of 50, the score increases from 0.718 (5\%) to 0.793 (20\%) and further to 0.809 (40\%). A similar pattern is observed at a budget of 100, where performance improves from 0.754 to 0.810 and 0.821. Even at higher budgets (e.g., 250), the gains persist, rising from 0.764 to 0.810 and 0.822. Overall, this corresponds to relative improvements in the range of approximately 6\%--12\%, indicating a strong dependence on labeled data for generalization.}
\begin{table*}
\captionsetup{type=table}
\caption{\textcolor{black}{Detection performance of \method{} on the BODMAS dataset under varying labeled data ratios and labeling budgets. Each experiment is repeated for three times and the mean value along with standard deviation are reported.}}
\label{tab:varying_label_bdget_label_ratio_bodmas}
\centering  

\begin{adjustbox}{width=1\textwidth}
\begin{tabular}{llllllll}
 & \multicolumn{6}{c}{BODMAS} \\
\cmidrule(lr){1-8}
Label budget & Label ratio & seen-AUT (B) & seen-AUT (A) & unseen-AUT (B) & unseen-AUT (A) & overall-AUT (B) & overall-AUT (A)\\
\midrule

&5\%  & $0.916 \pm 0.009$ & $0.637 \pm 0.079$ & $0.926 \pm 0.033$ & $0.718 \pm 0.023$ & $0.921 \pm 0.018$ & $0.684 \pm 0.023$ \\ 
50&10\% & $0.924 \pm 0.005$ & $0.724 \pm 0.009$ & $0.933 \pm 0.005$ & $0.760 \pm 0.011$ & $0.926 \pm 0.004$ & $0.734 \pm 0.009$ \\ 
&20\% & $0.928 \pm 0.011$ & $0.736 \pm 0.013$ & $0.940 \pm 0.018$ & $0.793 \pm 0.038$ & $0.933 \pm 0.014$ & $0.764 \pm 0.023$ \\
&30\% & $0.927 \pm 0.011$ & $0.718 \pm 0.037$ & $0.942 \pm 0.006$ & $0.776 \pm 0.022$ & $0.933 \pm 0.008$ & $0.743 \pm 0.027$ \\
&40\% & $0.930 \pm 0.005$ & $0.745 \pm 0.001$ & $0.947 \pm 0.009$ & $0.809 \pm 0.025$ & $0.938 \pm 0.007$ & $0.774 \pm 0.013$ \\

\midrule

&5\%  & $0.920 \pm 0.002$ & $0.672 \pm 0.009$ & $0.929 \pm 0.031$ & $0.754 \pm 0.023$ & $0.925 \pm 0.017$ & $0.717 \pm 0.009$ \\ 
100&10\% & $0.911 \pm 0.020$ & $0.674 \pm 0.065$ & $0.921 \pm 0.022$ & $0.719 \pm 0.071$ & $0.913 \pm 0.021$ & $0.689 \pm 0.070$ \\
 &20\% & $0.930 \pm 0.002$ & $0.701 \pm 0.017$ & $0.956 \pm 0.002$ & $0.810 \pm 0.027$ & $0.942 \pm 0.000$ & $0.755 \pm 0.018$ \\ 
&30\% & $0.929 \pm 0.009$ & $0.703 \pm 0.032$ & $0.946 \pm 0.011$ & $0.779 \pm 0.010$ & $0.934 \pm 0.007$ & $0.739 \pm 0.011$ \\
&40\% & $0.935 \pm 0.008$ & $0.721 \pm 0.016$ & $0.959 \pm 0.007$ & $0.821 \pm 0.009$ & $0.947 \pm 0.007$ & $0.772 \pm 0.010$ \\

\midrule

&5\% & $0.916 \pm 0.004$ & $0.634 \pm 0.084$ & $0.936 \pm 0.016$ & $0.732 \pm 0.003$ & $0.927 \pm 0.008$ & $0.690 \pm 0.031$ \\  
150&10\% & $0.927 \pm 0.004$ & $0.715 \pm 0.012$ & $0.948 \pm 0.010$ & $0.788 \pm 0.024$ & $0.937 \pm 0.005$ & $0.750 \pm 0.012$ \\ 
&20\% & $0.917 \pm 0.002$ & $0.719 \pm 0.006$ & $0.934 \pm 0.012$ & $0.782 \pm 0.038$ & $0.925 \pm 0.006$ & $0.747 \pm 0.020$ \\
&30\% & $0.924 \pm 0.003$ & $0.714 \pm 0.010$ & $0.945 \pm 0.003$ & $0.795 \pm 0.019$ & $0.934 \pm 0.003$ & $0.752 \pm 0.019$ \\
&40\% & $0.930 \pm 0.007$ & $0.716 \pm 0.058$ & $0.942 \pm 0.013$ & $0.782 \pm 0.027$ & $0.936 \pm 0.005$ & $0.750 \pm 0.029$ \\

\midrule

&5\% & $0.920 \pm 0.007$ & $0.634 \pm 0.083$ & $0.943 \pm 0.019$ & $0.738 \pm 0.013$ & $0.932 \pm 0.011$ & $0.693 \pm 0.035$ \\ 
200&10\% & $0.923 \pm 0.003$ & $0.701 \pm 0.028$ & $0.942 \pm 0.016$ & $0.773 \pm 0.026$ & $0.932 \pm 0.011$ & $0.735 \pm 0.005$ \\ 
&20\% & $0.926 \pm 0.010$ & $0.664 \pm 0.074$ & $0.953 \pm 0.006$ & $0.777 \pm 0.057$ & $0.940 \pm 0.007$ & $0.727 \pm 0.060$ \\
&30\% & $0.922 \pm 0.014$ & $0.665 \pm 0.029$ & $0.950 \pm 0.007$ & $0.773 \pm 0.014$ & $0.936 \pm 0.010$ & $0.719 \pm 0.022$ \\
&40\% & $0.923 \pm 0.005$ & $0.671 \pm 0.079$ & $0.938 \pm 0.015$ & $0.766 \pm 0.017$ & $0.930 \pm 0.010$ & $0.720 \pm 0.034$ \\

\midrule

&5\% & $0.922 \pm 0.010$ & $0.662 \pm 0.093$ & $0.945 \pm 0.014$ & $0.764 \pm 0.054$ & $0.934 \pm 0.008$ & $0.718 \pm 0.065$ \\ 
250&10\% & $0.924 \pm 0.005$ & $0.672 \pm 0.060$ & $0.944 \pm 0.011$ & $0.756 \pm 0.042$ & $0.935 \pm 0.009$ & $0.717 \pm 0.044$ \\ 
&20\% & $0.931 \pm 0.003$ & $0.709 \pm 0.017$ & $0.957 \pm 0.001$ & $0.810 \pm 0.009$ & $0.944 \pm 0.001$ & $0.761 \pm 0.011$ \\
&30\% & $0.936 \pm 0.002$ & $0.736 \pm 0.008$ & $0.951 \pm 0.008$ & $0.806 \pm 0.011$ & $0.942 \pm 0.005$ & $0.766 \pm 0.004$ \\
&40\% & $0.935 \pm 0.002$ & $0.734 \pm 0.020$ & $0.955 \pm 0.013$ & $0.822 \pm 0.024$ & $0.944 \pm 0.007$ & $0.777 \pm 0.016$ \\

\bottomrule
\end{tabular}
\end{adjustbox}
\end{table*}

\paragraph{AndroZoo} \textcolor{black}{On AndroZoo, the effect of increasing the label ratio is more subtle (refer to Table~\ref{tab:varying_label_bdget_label_ratio_androzoo}). At a budget of 50, performance changes marginally from 0.843 (5\%) to 0.846 (40\%), and at 100, from 0.855 to 0.859. Larger budgets show slightly more noticeable gains, such as an increase from 0.860 to 0.879 at 200, although fluctuations appear at higher ratios (e.g., 250). In general, the improvements remain modest, typically around 0.5\%--2.5\%, suggesting diminishing returns with additional labeled data.}

\begin{table*}
\captionsetup{type=table}
\caption{\textcolor{black}{Detection performance of \method{} on the AndroZoo dataset under varying labeled data ratios and labeling budgets. Each experiment is repeated for three times and the mean value along with standard deviation are reported.}}
\label{tab:varying_label_bdget_label_ratio_androzoo}
\centering  

\begin{adjustbox}{width=1\textwidth}
\begin{tabular}{llllllll}
 & \multicolumn{6}{c}{AndroZoo} \\
\cmidrule(lr){1-8}
Label budget & Label ratio & seen-AUT (B) & seen-AUT (A) & unseen-AUT (B) & unseen-AUT (A) & overall-AUT (B) & overall-AUT (A)\\
\midrule

&5\%  & $0.995 \pm 0.001$ & $0.885 \pm 0.013$ & $0.994 \pm 0.001$ & $0.843 \pm 0.015$ & $0.994 \pm 0.001$ & $0.857 \pm 0.014$ \\ 
50&10\% & $0.996 \pm 0.001$ & $0.872 \pm 0.028$ & $0.994 \pm 0.000$ & $0.824 \pm 0.009$ & $0.995 \pm 0.000$ & $0.841 \pm 0.012$ \\ 
&20\%  & $0.995 \pm 0.001$ & $0.875 \pm 0.016$ & $0.994 \pm 0.001$ & $0.839 \pm 0.010$ & $0.994 \pm 0.001$ & $0.851 \pm 0.004$ \\ 
&30\% & $0.995 \pm 0.000$ & $0.869 \pm 0.025$ & $0.993 \pm 0.000$ & $0.844 \pm 0.011$ & $0.994 \pm 0.000$ & $0.852 \pm 0.014$ \\
&40\% & $0.995 \pm 0.001$ & $0.886 \pm 0.009$ & $0.994 \pm 0.001$ & $0.846 \pm 0.008$ & $0.994 \pm 0.001$ & $0.859 \pm 009$ \\

\midrule

&5\%  & $0.995 \pm 0.001$ & $0.863 \pm 0.014$ & $0.995 \pm 0.000$ & $0.855 \pm 0.014$ & $0.995 \pm 0.000$ & $0.858 \pm 0.013$ \\ 
100&10\% & $0.996 \pm 0.001$ & $0.893 \pm 0.010$ & $0.996 \pm 0.000$ & $0.859 \pm 0.004$ & $0.996 \pm 0.000$ & $0.871 \pm 0.002$ \\ 
&20\%  & $0.994 \pm 0.002$ & $0.878 \pm 0.002$ & $0.993 \pm 0.000$ & $0.855 \pm 0.018$ & $0.993 \pm 0.001$ & $0.863 \pm 0.011$ \\ 
&30\% & $0.995 \pm 0.000$ & $0.850 \pm 0.018$ & $0.994 \pm 0.000$ & $0.846 \pm 0.026$ & $0.994 \pm 0.000$ & $0.848 \pm 0.023$ \\
&40\% & $0.864 \pm 0.186$ & $0.868 \pm 0.012$ & $0.966 \pm 0.040$ & $0.832 \pm 0.021$ & $0.933 \pm 0.088$ & $0.844 \pm 0.017$ \\

\midrule

&5\% & $0.996 \pm 0.001$ & $0.865 \pm 0.049$ & $0.996 \pm 0.001$ & $0.866 \pm 0.020$ & $0.996 \pm 0.001$ & $0.866 \pm 0.020$ \\  
150&10\% & $0.995 \pm 0.001$ & $0.871 \pm 0.030$ & $0.995 \pm 0.001$ & $0.857 \pm 0.004$ & $0.995 \pm 0.001$ & $0.862 \pm 0.010$ \\ 
&20\%  & $0.992 \pm 0.001$ & $0.860 \pm 0.010$ & $0.994 \pm 0.000$ & $0.855 \pm 0.034$ & $0.993 \pm 0.000$ & $0.857 \pm 0.026$ \\ 
&30\% & $0.996 \pm 0.001$ & $0.857 \pm 0.012$ & $0.995 \pm 0.000$ & $0.862 \pm 0.016$ & $0.995 \pm 0.000$ & $0.861 \pm 0.015$ \\
&40\% & $0.995 \pm 0.001$ & $0.847 \pm 0.018$ & $0.994 \pm 0.000$ & $0.855 \pm 0.029$ & $0.994 \pm 0.000$ & $0.853 \pm 0.013$ \\

\midrule

&5\% & $0.996 \pm 0.000$ & $0.855 \pm 0.018$ & $0.994 \pm 0.001$ & $0.860 \pm 0.039$ & $0.995 \pm 0.001$ & $0.869 \pm 0.031$ \\ 
200&10\% & $0.996 \pm 0.001$ & $0.896 \pm 0.006$ & $0.996 \pm 0.000$ & $0.851 \pm 0.022$ & $0.996 \pm 0.000$ & $0.866 \pm 0.017$ \\ 
&20\%  & $0.992 \pm 0.001$ & $0.869 \pm 0.014$ & $0.994 \pm 0.001$ & $0.879 \pm 0.002$ & $0.993 \pm 0.001$ & $0.876 \pm 0.005$ \\ 
&30\% & $0.995 \pm 0.002$ & $0.883 \pm 0.015$ & $0.994 \pm 0.001$ & $0.861 \pm 0.033$ & $0.995 \pm 0.001$ & $0.869 \pm 0.027$ \\
&40\% & $0.994 \pm 0.000$ & $0.859 \pm 0.026$ & $0.993 \pm 0.001$ & $0.840 \pm 0.003$ & $0.993 \pm 0.000$ & $0.847 \pm 0.024$ \\

\midrule

&5\% & $0.996 \pm 0.001$ & $0.877 \pm 0.011$ & $0.996 \pm 0.000$ & $0.882 \pm 0.015$ & $0.996 \pm 0.000$ & $0.881 \pm 0.006$ \\ 
250&10\% & $0.995 \pm 0.001$ & $0.891 \pm 0.010$ & $0.996 \pm 0.000$ & $0.875 \pm 0.009$ & $0.996 \pm 0.000$ & $0.880 \pm 0.009$ \\ 
&20\%  & $0.993 \pm 0.001$ & $0.880 \pm 0.004$ & $0.994 \pm 0.001$ & $0.878 \pm 0.018$ & $0.994 \pm 0.001$ & $0.879 \pm 0.012$ \\ 
&30\% & $0.995 \pm 0.001$ & $0.866 \pm 0.019$ & $0.994 \pm 0.000$ & $0.863 \pm 0.011$ & $0.993 \pm 0.002$ & $0.864 \pm 0.008$ \\
&40\% & $0.893 \pm 0.144$ & $0.874 \pm 0.008$ & $0.965 \pm 0.039$ & $0.870 \pm 0.003$ & $0.941 \pm 0.074$ & $0.872 \pm 0.022$ \\

\bottomrule
\end{tabular}
\end{adjustbox}
\end{table*}

\paragraph{APIGraph} \textcolor{black}{In contrast, APIGraph exhibits steady and consistent improvements with higher label ratios (refer to Table~\ref{tab:varying_label_bdget_label_ratio_api_graph}). At a budget of 50, unseen-AUT (A) increases from 0.888 (5\%) to 0.920 (40\%), and at 100, from 0.910 to 0.931. This upward trend continues at larger budgets, with values improving from 0.923 to 0.941 (150), 0.926 to 0.942 (200), and 0.932 to 0.948 (250). These gains correspond to relative improvements of approximately 1.5\%--3.5\%, reflecting better generalization as more labeled data becomes available.}

\begin{table*}
\captionsetup{type=table}
\caption{\textcolor{black}{Detection performance of \method{} on the APIGraph dataset under varying labeled data ratios and labeling budgets. Each experiment is repeated for three times and the mean value along with standard deviation are reported.}}
\label{tab:varying_label_bdget_label_ratio_api_graph}
\centering  

\begin{adjustbox}{width=1\textwidth}
\begin{tabular}{llllllll}
 & \multicolumn{6}{c}{APIGraph} \\
\cmidrule(lr){1-8}
Label budget & Label ratio & seen-AUT (B) & seen-AUT (A) & unseen-AUT (B) & unseen-AUT (A) & overall-AUT (B) & overall-AUT (A)\\
\midrule

&5\%  & $0.998 \pm 0.000$ & $0.939 \pm 0.001$ & $0.997 \pm 0.000$ & $0.888 \pm 0.015$ & $0.998 \pm 0.000$ & $0.910 \pm 0.009$ \\ 
50&10\% & $0.999 \pm 0.000$ & $0.946 \pm 0.007$ & $0.998 \pm 0.000$ & $0.898 \pm 0.009$ & $0.998 \pm 0.000$ & $0.918 \pm 0.007$ \\ 
&20\% & $0.999 \pm 0.000$ & $0.957 \pm 0.003$ & $0.998 \pm 0.000$ & $0.915 \pm 0.002$ & $0.998 \pm 0.000$ & $0.932 \pm 0.000$ \\
&30\% & $0.999 \pm 0.000$ & $0.956 \pm 0.003$ & $0.998 \pm 0.000$ & $0.912 \pm 0.009$ & $0.998 \pm 0.000$ & $0.931 \pm 0.005$ \\
&40\% & $0.999 \pm 0.000$ & $0.961 \pm 0.001$ & $0.998 \pm 0.000$ & $0.920 \pm 0.008$ & $0.999 \pm 0.000$ & $0.937 \pm 0.005$ \\

\midrule

&5\%  & $0.998 \pm 0.000$ & $0.940 \pm 0.001$ & $0.998 \pm 0.000$ & $0.910 \pm 0.007$ & $0.998 \pm 0.000$ & $0.923 \pm 0.004$ \\ 
100&10\% & $0.999 \pm 0.000$ & $0.945 \pm 0.000$ & $0.998 \pm 0.000$ & $0.917 \pm 0.002$ & $0.998 \pm 0.000$ & $0.929 \pm 0.001$ \\
&20\% & $0.999 \pm 0.000$ & $0.953 \pm 0.009$ & $0.998 \pm 0.000$ & $0.930 \pm 0.004$ & $0.998 \pm 0.000$ & $0.940 \pm 0.004$ \\
&30\% & $0.999 \pm 0.000$ & $0.956 \pm 0.001$ & $0.998 \pm 0.000$ & $0.924 \pm 0.004$ & $0.998 \pm 0.000$ & $0.937 \pm 0.003$ \\
&40\% & $0.999 \pm 0.000$ & $0.958 \pm 0.002$ & $0.998 \pm 0.000$ & $0.931 \pm 0.002$ & $0.999 \pm 0.000$ & $0.942 \pm 0.001$ \\

\midrule

&5\% & $0.998 \pm 0.000$ & $0.942 \pm 0.001$ & $0.998 \pm 0.000$ & $0.923 \pm 0.007$ & $0.998 \pm 0.000$ & $0.931 \pm 0.004$ \\  
150&10\% & $0.999 \pm 0.000$ & $0.948 \pm 0.005$ & $0.998 \pm 0.000$ & $0.921 \pm 0.003$ & $0.998 \pm 0.000$ & $0.933 \pm 0.000$ \\ 
&20\% & $0.999 \pm 0.000$ & $0.959 \pm 0.003$ & $0.998 \pm 0.000$ & $0.934 \pm 0.000$ & $0.999 \pm 0.000$ & $0.945 \pm 0.001$ \\
&30\% & $0.999 \pm 0.000$ & $0.959 \pm 0.003$ & $0.998 \pm 0.000$ & $0.933 \pm 0.005$ & $0.999 \pm 0.000$ & $0.944 \pm 0.002$ \\
&40\% & $0.999 \pm 0.000$ & $0.959 \pm 0.001$ & $0.999 \pm 0.000$ & $0.941 \pm 0.001$ & $0.999 \pm 0.000$ & $0.948 \pm 0.001$ \\

\midrule

&5\% & $0.999 \pm 0.000$ & $0.945 \pm 0.002$ & $0.998 \pm 0.000$ & $0.926 \pm 0.005$ & $0.998 \pm 0.000$ & $0.934 \pm 0.003$ \\ 
200&10\% & $0.999 \pm 0.000$ & $0.953 \pm 0.001$ & $0.998 \pm 0.000$ & $0.929 \pm 0.001$ & $0.999 \pm 0.000$ & $0.939 \pm 0.001$ \\ 
&20\% & $0.999 \pm 0.000$ & $0.961 \pm 0.002$ & $0.998 \pm 0.000$ & $0.938 \pm 0.001$ & $0.999 \pm 0.000$ & $0.948 \pm 0.000$ \\
&30\% & $0.999 \pm 0.000$ & $0.957 \pm 0.001$ & $0.998 \pm 0.000$ & $0.939 \pm 0.006$ & $0.999 \pm 0.000$ & $0.946 \pm 0.003$ \\
&40\% & $0.999 \pm 0.000$ & $0.959 \pm 0.001$ & $0.999 \pm 0.000$ & $0.942 \pm 0.002$ & $0.999 \pm 0.000$ & $0.942 \pm 0.000$ \\

\midrule

&5\% & $0.998 \pm 0.000$ & $0.947 \pm 0.002$ & $0.998 \pm 0.000$ & $0.932 \pm 0.005$ & $0.998 \pm 0.000$ & $0.938 \pm 0.003$ \\ 
250&10\% & $0.999 \pm 0.000$ & $0.950 \pm 0.001$ & $0.998 \pm 0.000$ & $0.937 \pm 0.001$ & $0.998 \pm 0.000$ & $0.942 \pm 0.001$ \\ 
&20\% & $0.999 \pm 0.000$ & $0.961 \pm 0.000$ & $0.998 \pm 0.000$ & $0.944 \pm 0.001$ & $0.999 \pm 0.000$ & $0.951 \pm 0.000$ \\
&30\% & $0.999 \pm 0.000$ & $0.959 \pm 0.002$ & $0.998 \pm 0.000$ & $0.942 \pm 0.004$ & $0.999 \pm 0.000$ & $0.949 \pm 0.002$ \\
&40\% & $0.999 \pm 0.000$ & $0.960 \pm 0.002$ & $0.999 \pm 0.000$ & $0.948 \pm 0.000$ & $0.999 \pm 0.000$ & $0.953 \pm 0.001$ \\

\bottomrule
\end{tabular}
\end{adjustbox}
\end{table*}

\subsection{Handling False Positives}
\begin{table}[!t]
\footnotesize
  \caption{Ablation study demonstrating effect of various maximum threshold $\tau_{max}$ values on the unseen-AUT (A) of the proposed method, which serves as a proxy for understanding the effect of false positives samples incurred during the finding of the suitable example corresponding to each unlabeled example. The best values are marked in \textbf{bold}.}
  \label{tab:vary_tau}
 
  
   \begin{tabular}{llllll}
  \multicolumn{6}{c}{\large$\tau_{max}$}\\
      \cmidrule(lr){3-6} 
   Dataset & Metrics & 0.01 & 0.03  &  0.05 & 0.09 \\ 
   \midrule
   
     & Unseen-AUT (A) & \textbf{0.810 $\pm$ 0.027} & 0.798 $\pm$ 0.053 & 0.790 $\pm$ 0.037 & 0.780 $\pm$ 0.023 \\ 
    
  BODMAS  & Avg. \#Rej in seen& 1543 $\pm$ 34 & 557 $\pm$ 190 & 435 $\pm$ 55 & 121 $\pm$ 25 \\ 
    
   & Avg. \#Rej in unseen & 2511 $\pm$ 593 & 852 $\pm$ 433 & 521 $\pm$ 165 & 126 $\pm$ 42 \\ 
    \midrule
   & Unseen-AUT (A)  & 0.821 $\pm$ 0.014 & 0.811 $\pm$ 0.012 & \textbf{0.855 $\pm$ 0.018} & 0.844 $\pm$ 0.010 \\ 

AndroZoo & Avg. \#Rej in seen  & 1426 $\pm$ 22 & 849 $\pm$ 110 & 578 $\pm$ 69 & 184 $\pm$ 33 \\ 

  & Avg. \#Rej in unseen  & 1146 $\pm$ 60 & 682 $\pm$ 111 & 498 $\pm$ 95 & 125 $\pm$ 31 \\ 
   \midrule
      & Unseen-AUT (A) & 0.921 $\pm$ 0.002 & \textbf{0.930 $\pm$ 0.002} & 0.927 $\pm$ 0.007 & 0.919 $\pm$ 0.010 \\ 
      API Graph  & Avg. \#Rej in seen& 2058 $\pm$ 228 & 1934 $\pm$ 292 & 1629 $\pm$ 212 & 1303 $\pm$ 52 \\ 
    
   & Avg. \#Rej in unseen & 2626 $\pm$ 6 & 2508 $\pm$ 138 & 2221 $\pm$ 301 & 1650 $\pm$ 80 \\ 

    \bottomrule    
  \end{tabular}
   \end{table}
In this section, we study how the performance of the proposed method is impacted by the threshold $\tau_{max}$ on the false positive samples, which are used to find the most suitable labeled example for each unlabeled example. $\tau_{max}$ indicates the maximum cosine distance between the unlabeled sample and labeled samples to find the most suitable labeled example. Towards this, we compare the performance in terms of unseen AUT (A), average number of samples rejected during training on seen (Avg.~\#Rej in seen) and unseen tasks (Avg.~\#Rej in unseen) by varying $\tau_{max}$ values. The rejected samples are the unlabeled samples that are rejected whenever the suitable labeled example is not found. This rejection process is controlled by the dynamic threshold $\tau_{max}$ that decides number of labeled samples to be considered while finding the most suitable example. Intuitively, larger the threshold, more unrelated labeled samples (false positives) may participate in suitable sample selection process. Lower the threshold, related samples may not participate. Thus, we decided to study its impact and these results are presented in Table~\ref{tab:vary_tau}. We made the following observations from these results.


\par The best $\tau_{max}$ value, across all the datasets, in the range of $0.01$ and $0.05$ indicates the selected suitable labeled example is always lying near the unlabeled sample in the latent space. As $\tau_{max}$ approaches zero, finding the suitable labeled samples becomes finding the most similar sample. As a result, more suitable samples are rejected. This increased rejection rate with lowering $\tau_{max}$ is consistently observed on all the datasets, with an exception that rate of rejection is slower on APIGraph. Increasing $\tau_{max}$ brings the most distant samples, thus increases the number of \textit{false positive} labeled samples for consideration. Due to this, we observe a drop on unseen-AUT (A) on BODMAS dataset from $0.810$ to $0.798$, from $0.855$ to $0.844$ on AndroZoo dataset, and from $0.930$ to $0.927$ on APIGraph dataset. Similarly, the reduction in number of rejected samples is also consistently observed with increasing threshold on all the datasets.

\textbf{Practitioner note on choosing the value of $\tau_{max}$:}
\textcolor{black}{Based on the experimental results, we observe that a suitable range for $\tau_{max}$ (which serves as a proxy for cosine distance) lies between 0.01 and 0.05. This corresponds to selecting labeled samples whose cosine similarity with an unlabeled novel sample lies approximately in the range of 95\% to 99\% in the latent representation space. Intuitively, this ensures that each unlabeled sample is associated with a highly similar labeled instance. In practice, we recommend tuning the cosine similarity threshold within a slightly broader range of 90\% to 99\% as a safer choice.}


\begin{table*}
\captionsetup{type=table}
\caption{\textcolor{black}{Detection performance of \method{} under varying label noise ratios and different labeled data ratios on the APIGraph dataset. Each experiment is repeated three times, and the mean along with standard deviation is reported.}}
\label{tab:varying_label_noise_label_ratio_api_graph_}
\centering  

\begin{adjustbox}{width=1\textwidth,height=9.5cm}
\begin{tabular}{llllllll}
 & \multicolumn{6}{c}{APIGraph} \\
\cmidrule(lr){1-8}
 Label ratio & Label noise ratio & seen-AUT (B) & seen-AUT (A) & unseen-AUT (B) & unseen-AUT (A) & overall-AUT (B) & overall-AUT (A)\\
\midrule

&10\%+\method{}  & $0.998 \pm 0.000$ & $0.928 \pm 0.002$ & $0.995 \pm 0.001$ & $0.843 \pm 0.009$ & $0.996 \pm 0.001$ & $0.879 \pm 0.006$ \\ 
&10\%+\method{}+1 & $0.998 \pm 0.000$ & $0.945 \pm 0.001$ & $0.997 \pm 0.000$ & $0.902 \pm 0.014 \, $ & $0.998 \pm 0.000$ & $0.920 \pm 0.008$ \\
&10\%+\method{}+3 & $0.999 \pm 0.000$ & $0.944 \pm 0.001$ & $0.998 \pm 0.000$ & $0.899 \pm 0.018 \, $ & $0.998 \pm 0.000$ & $0.918 \pm 0.010$ \\
&10\%+\method{}+11 & $0.998 \pm 0.000$ & $0.942 \pm 0.004$ & $0.997 \pm 0.000$ & $0.887 \pm 0.017$ & $0.998 \pm 0.000$ & $0.910 \pm 0.010$ \\

\cmidrule(lr){2-8}

5\%&20\%+\method{} & $0.997 \pm 0.000$ & $0.895 \pm 0.006$ & $0.991 \pm 0.002$ & $0.683 \pm 0.024$ & $0.994 \pm 0.001$ & $0.772 \pm 0.015$ \\ 
&20\%+\method{}+1  & $0.998 \pm 0.000$ & $0.947 \pm 0.002$ & $0.997 \pm 0.000$ & $0.895 \pm 0.012 \, $ & $0.998 \pm 0.000$ & $0.917 \pm 0.007$ \\
&20\%+\method{}+3  & $0.999 \pm 0.000$ & $0.950 \pm 0.001$ & $0.998 \pm 0.000$ & $0.897 \pm 0.019 \, $ & $0.998 \pm 0.000$ & $0.919 \pm 0.011$ \\
&20\%+\method{}+11 & $0.999 \pm 0.000$ & $0.945 \pm 0.002$ & $0.997 \pm 0.000$ & $0.889 \pm 0.018$ & $0.998 \pm 0.000$ & $0.913 \pm 0.010$ \\

\cmidrule(lr){2-8}

&30\%+\method{} & $0.997 \pm 0.000$ & $0.877 \pm 0.005$ & $0.989 \pm 0.002$ & $0.591 \pm 0.020$ & $0.992 \pm 0.001$ & $0.710 \pm 0.012$ \\
&30\%+\method{}+1 & $0.998 \pm 0.000$ & $0.946 \pm 0.003$ & $0.997 \pm 0.000$ & $0.896 \pm 0.013 \, $ & $0.998 \pm 0.000$ & $0.917 \pm 0.008$ \\
&30\%+\method{}+3 & $0.998 \pm 0.000$ & $0.948 \pm 0.000$ & $0.997 \pm 0.000$ & $0.894 \pm 0.016 \, $ & $0.998 \pm 0.000$ & $0.917 \pm 0.009$ \\
&30\%+\method{}+11 & $0.998 \pm 0.000$ & $0.947 \pm 0.003$ & $0.997 \pm 0.000$ & $0.876 \pm 0.021$ & $0.997 \pm 0.000$ & $0.906 \pm 0.012$ \\

\cmidrule(lr){2-8}

&40\%+\method{} & $0.997 \pm 0.000$ & $0.846 \pm 0.005$ & $0.982 \pm 0.004$ & $0.359 \pm 0.031$ & $0.988 \pm 0.002$ & $0.562 \pm 0.018$ \\
&40\%+\method{}+1 & $0.999 \pm 0.000$ & $0.948 \pm 0.002$ & $0.997 \pm 0.000$ & $0.893 \pm 0.015 \, $ & $0.998 \pm 0.000$ & $0.916 \pm 0.001$ \\
&40\%+\method{}+3 & $0.998 \pm 0.000$ & $0.947 \pm 0.002$ & $0.997 \pm 0.000$ & $0.885 \pm 0.015 \, $ & $0.998 \pm 0.000$ & $0.911 \pm 0.009$ \\
&40\%+\method{}+11 & $0.998 \pm 0.000$ & $0.945 \pm 0.001$ & $0.997 \pm 0.000$ & $0.875 \pm 0.018$ & $0.998 \pm 0.000$ & $0.904 \pm 0.011$ \\

\cmidrule(lr){2-8}

&50\%+\method{} & $0.996 \pm 0.000$ & $0.795 \pm 0.017$ & $0.975 \pm 0.010$ & $0.248 \pm 0.055$ & $0.984 \pm 0.006$ & $0.476 \pm 0.039$ \\
&50\%+\method{}+1 & $0.998 \pm 0.000$ & $0.947 \pm 0.003$ & $0.997 \pm 0.000$ & $0.884 \pm 0.013 \, $ & $0.998 \pm 0.000$ & $0.911 \pm 0.006$ \\
&50\%+\method{}+3 & $0.998 \pm 0.000$ & $0.948 \pm 0.003$ & $0.997 \pm 0.000$ & $0.890 \pm 0.014 \, $ & $0.998 \pm 0.000$ & $0.914 \pm 0.009$ \\
&50\%+\method{}+11 & $0.998 \pm 0.000$ & $0.948 \pm 0.001$ & $0.997 \pm 0.000$ & $0.872 \pm 0.020$ & $0.998 \pm 0.000$ & $0.904 \pm 0.011$ \\

\cmidrule(lr){2-8}

&60\%+\method{} & $0.995 \pm 0.000$ & $0.777 \pm 0.021$ & $0.963 \pm 0.014$ & $0.153 \pm 0.012$ & $0.977 \pm 0.008$ & $0.412 \pm 0.012$ \\
&60\%+\method{}+1 & $0.998 \pm 0.000$ & $0.947 \pm 0.003$ & $0.997 \pm 0.000$ & $0.888 \pm 0.018 \, $ & $0.998 \pm 0.000$ & $0.913 \pm 0.011$ \\
&60\%+\method{}+3 & $0.998 \pm 0.000$ & $0.949 \pm 0.004$ & $0.997 \pm 0.000$ & $0.876 \pm 0.010 \, $ & $0.997 \pm 0.000$ & $0.907 \pm 0.007$ \\
&60\%+\method{}+11 & $0.998 \pm 0.000$ & $0.945 \pm 0.002$ & $0.996 \pm 0.000$ & $0.855 \pm 0.031$ & $0.997 \pm 0.000$ & $0.893 \pm 0.019$ \\

\cmidrule(lr){2-8}

&70\%+\method{} & $0.995 \pm 0.000$ & $0.753 \pm 0.016$ & $0.959 \pm 0.007$ & $0.106 \pm 0.004$ & $0.974 \pm 0.004$ & $0.375 \pm 0.009$ \\
&70\%+\method{}+1 & $0.998 \pm 0.000$ & $0.946 \pm 0.001$ & $0.997 \pm 0.000$ & $0.888 \pm 0.012 \, $ & $0.998 \pm 0.000$ & $0.913 \pm 0.007$ \\
&70\%+\method{}+3 & $0.998 \pm 0.000$ & $0.945 \pm 0.003$ & $0.996 \pm 0.001$ & $0.867 \pm 0.025 \, $ & $0.997 \pm 0.000$ & $0.900 \pm 0.014$ \\
&70\%+\method{}+11 & $0.998 \pm 0.000$ & $0.946 \pm 0.001$ & $0.996 \pm 0.001$ & $0.859 \pm 0.027$ & $0.997 \pm 0.000$ & $0.896 \pm 0.016$ \\

\cmidrule(lr){2-8}

&80\%+\method{} & $0.995 \pm 0.000$ & $0.749 \pm 0.012$ & $0.947 \pm 0.013$ & $0.076 \pm 0.011$ & $0.967 \pm 0.008$ & $0.354 \pm 0.010$ \\
&80\%+\method{}+1 & $0.998 \pm 0.000$ & $0.946 \pm 0.000$ & $0.996 \pm 0.001$ & $0.877 \pm 0.017 \, $ & $0.997 \pm 0.000$ & $0.906 \pm 0.009$ \\
&80\%+\method{}+3 & $0.998 \pm 0.000$ & $0.947 \pm 0.004$ & $0.997 \pm 0.000$ & $0.878 \pm 0.021 \, $ & $0.997 \pm 0.000$ & $0.907 \pm 0.012$ \\
&80\%+\method{}+11 & $0.998 \pm 0.000$ & $0.948 \pm 0.005$ & $0.996 \pm 0.000$ & $0.852 \pm 0.024$ & $0.997 \pm 0.000$ & $0.893 \pm 0.016$ \\

\midrule

&10\%+\method{}  & $0.998 \pm 0.000$ & $0.942 \pm 0.008$ & $0.996 \pm 0.000$ & $0.845 \pm 0.004$ & $0.997 \pm 0.000$ & $0.885 \pm 0.004$ \\ 
&10\%+\method{}+1 & $0.999 \pm 0.000$ & $0.955 \pm 0.002$ & $0.998 \pm 0.000$ & $0.904 \pm 0.001 \, $ & $0.998 \pm 0.000$ & $0.925 \pm 0.000$ \\
&10\%+\method{}+3 & $0.999 \pm 0.000$ & $0.955 \pm 0.000$ & $0.998 \pm 0.000$ & $0.906 \pm 0.011 \, $ & $0.998 \pm 0.000$ & $0.927 \pm 0.006$ \\
&10\%+\method{}+11 & $0.999 \pm 0.000$ & $0.959 \pm 0.002$ & $0.998 \pm 0.000$ & $0.904 \pm 0.010$ & $0.998 \pm 0.000$ & $0.927 \pm 0.005$ \\

\cmidrule(lr){2-8}

10\%&20\%+\method{} & $0.998 \pm 0.000$ & $0.917 \pm 0.007$ & $0.994 \pm 0.001$ & $0.720 \pm 0.029$ & $0.995 \pm 0.001$ & $0.802 \pm 0.019$ \\ 
&20\%+\method{}+1  & $0.999 \pm 0.000$ & $0.956 \pm 0.002$ & $0.998 \pm 0.000$ & $0.899 \pm 0.005 \, $ & $0.998 \pm 0.000$ & $0.923 \pm 0.002$ \\
&20\%+\method{}+3  & $0.999 \pm 0.000$ & $0.957 \pm 0.002$ & $0.998 \pm 0.000$ & $0.900 \pm 0.010 \, $ & $0.998 \pm 0.000$ & $0.924 \pm 0.005$ \\
&20\%+\method{}+11 & $0.998 \pm 0.000$ & $0.957 \pm 0.001$ & $0.997 \pm 0.000$ & $0.890 \pm 0.012$ & $0.998 \pm 0.000$ & $0.918 \pm 0.006$ \\

\cmidrule(lr){2-8}

&30\%+\method{} & $0.997 \pm 0.000$ & $0.890 \pm 0.015$ & $0.989 \pm 0.000$ & $0.587 \pm 0.020$ & $0.992 \pm 0.000$ & $0.7131 \pm 0.012$ \\
&30\%+\method{}+1 & $0.998 \pm 0.000$ & $0.954 \pm 0.002$ & $0.997 \pm 0.000$ & $0.894 \pm 0.009 \, $ & $0.998 \pm 0.000$ & $0.919 \pm 0.006$ \\
&30\%+\method{}+3 & $0.998 \pm 0.000$ & $0.957 \pm 0.000$ & $0.997 \pm 0.000$ & $0.892 \pm 0.007 \, $ & $0.998 \pm 0.000$ & $0.919 \pm 0.004$ \\
&30\%+\method{}+11 & $0.998 \pm 0.000$ & $0.956 \pm 0.002$ & $0.997 \pm 0.000$ & $0.880 \pm 0.011$ & $0.998 \pm 0.000$ & $0.912 \pm 0.006$ \\

\cmidrule(lr){2-8}

&40\%+\method{} & $0.997 \pm 0.000$ & $0.873 \pm 0.006$ & $0.987 \pm 0.003$ & $0.462 \pm 0.044$ & $0.991 \pm 0.002$ & $0.633 \pm 0.027$ \\
&40\%+\method{}+1 & $0.998 \pm 0.000$ & $0.954 \pm 0.004$ & $0.997 \pm 0.001$ & $0.891 \pm 0.010 \, $ & $0.998 \pm 0.000$ & $0.917 \pm 0.007$ \\
&40\%+\method{}+3 & $0.998 \pm 0.000$ & $0.956 \pm 0.002$ & $0.997 \pm 0.000$ & $0.885 \pm 0.010 \, $ & $0.997 \pm 0.000$ & $0.915 \pm 0.005$ \\
&40\%+\method{}+11 & $0.998 \pm 0.000$ & $0.952 \pm 0.006$ & $0.996 \pm 0.000$ & $0.878 \pm 0.013$ & $0.997 \pm 0.000$ & $0.909 \pm 0.006$ \\

\cmidrule(lr){2-8}

&50\%+\method{} & $0.996 \pm 0.000$ & $0.832 \pm 0.005$ & $0.975 \pm 0.007$ & $0.260 \pm 0.013$ & $0.984 \pm 0.007$ & $0.498 \pm 0.009$ \\
&50\%+\method{}+1 & $0.998 \pm 0.000$ & $0.954 \pm 0.006$ & $0.997 \pm 0.000$ & $0.883 \pm 0.008 \, $ & $0.998 \pm 0.000$ & $0.913 \pm 0.006$ \\
&50\%+\method{}+3 & $0.998 \pm 0.000$ & $0.954 \pm 0.003$ & $0.997 \pm 0.000$ & $0.879 \pm 0.004 \, $ & $0.997 \pm 0.000$ & $0.910 \pm 0.001$ \\
&50\%+\method{}+11 & $0.998 \pm 0.000$ & $0.952 \pm 0.006$ & $0.996 \pm 0.001$ & $0.855 \pm 0.004$ & $0.997 \pm 0.000$ & $0.896 \pm 0.005$ \\

\cmidrule(lr){2-8}

&60\%+\method{} & $0.996 \pm 0.000$ & $0.828 \pm 0.010$ & $0.968 \pm 0.013$ & $0.190 \pm 0.019$ & $0.980 \pm 0.008$ & $0.455 \pm 0.015$ \\
&60\%+\method{}+1 & $0.998 \pm 0.000$ & $0.954 \pm 0.002$ & $0.997 \pm 0.000$ & $0.872 \pm 0.015 \, $ & $0.997 \pm 0.000$ & $0.906 \pm 0.009$ \\
&60\%+\method{}+3 & $0.998 \pm 0.000$ & $0.956 \pm 0.001$ & $0.997 \pm 0.000$ & $0.877 \pm 0.008 \, $ & $0.997 \pm 0.000$ & $0.910 \pm 0.004$ \\
&60\%+\method{}+11 & $0.998 \pm 0.000$ & $0.954 \pm 0.004$ & $0.996 \pm 0.000$ & $0.855 \pm 0.009$ & $0.997 \pm 0.000$ & $0.897 \pm 0.006$ \\

\cmidrule(lr){2-8}

&70\%+\method{} & $0.996 \pm 0.000$ & $0.792 \pm 0.001$ & $0.963 \pm 0.002$ & $0.136 \pm 0.044$ & $0.977 \pm 0.001$ & $0.408 \pm 0.026$ \\
&70\%+\method{}+1 & $0.998 \pm 0.000$ & $0.954 \pm 0.005$ & $0.997 \pm 0.000$ & $0.873 \pm 0.002 \, $ & $0.998 \pm 0.000$ & $0.907 \pm 0.000$ \\
&70\%+\method{}+3 & $0.998 \pm 0.000$ & $0.958 \pm 0.001$ & $0.997 \pm 0.001$ & $0.868 \pm 0.004 \, $ & $0.997 \pm 0.000$ & $0.906 \pm 0.002$ \\
&70\%+\method{}+11 & $0.998 \pm 0.000$ & $0.955 \pm 0.002$ & $0.997 \pm 0.001$ & $0.842 \pm 0.013$ & $0.997 \pm 0.000$ & $0.890 \pm 0.008$ \\

\cmidrule(lr){2-8}

&80\%+\method{} & $0.995 \pm 0.000$ & $0.770 \pm 0.015$ & $0.948 \pm 0.014$ & $0.104 \pm 0.028$ & $0.968 \pm 0.008$ & $0.380 \pm 0.021$ \\
&80\%+\method{}+1 & $0.998 \pm 0.000$ & $0.958 \pm 0.001$ & $0.997 \pm 0.001$ & $0.867 \pm 0.011 \, $ & $0.997 \pm 0.000$ & $0.905 \pm 0.007$ \\
&80\%+\method{}+3 & $0.998 \pm 0.000$ & $0.956 \pm 0.003$ & $0.996 \pm 0.001$ & $0.872 \pm 0.001 \, $ & $0.997 \pm 0.000$ & $0.907 \pm 0.001$ \\
&80\%+\method{}+11 & $0.998 \pm 0.000$ & $0.954 \pm 0.004$ & $0.996 \pm 0.001$ & $0.830 \pm 0.008$ & $0.997 \pm 0.000$ & $0.882 \pm 0.003$ \\

\midrule

&10\%+\method{}  & $0.998 \pm 0.000$ & $0.946 \pm 0.007$ & $0.996 \pm 0.001$ & $0.848 \pm 0.036$ & $0.997 \pm 0.000$ & $0.889 \pm 0.021$ \\ 
&10\%+\method{}+1  & $0.999 \pm 0.000$ & $0.960 \pm 0.001$ & $0.998 \pm 0.000$ & $0.913 \pm 0.003 \, $ & $0.998 \pm 0.000$ & $0.933 \pm 0.002$ \\
&10\%+\method{}+3  & $0.999 \pm 0.000$ & $0.959 \pm 0.005$ & $0.998 \pm 0.000$ & $0.907 \pm 0.007 \, $ & $0.998 \pm 0.000$ & $0.929 \pm 0.006$ \\
&10\%+\method{}+11 & $0.999 \pm 0.000$ & $0.960 \pm 0.004$ & $0.997 \pm 0.000$ & $0.902 \pm 0.012$ & $0.998 \pm 0.000$ & $0.927 \pm 0.008$ \\

\cmidrule(lr){2-8}

20\%&20\%+\method{} & $0.998 \pm 0.000$ & $0.920 \pm 0.010$ & $0.993 \pm 0.002$ & $0.736 \pm 0.028$ & $0.995 \pm 0.001$ & $0.813 \pm 0.020$ \\ 
&20\%+\method{}+1 & $0.999 \pm 0.000$ & $0.961 \pm 0.003$ & $0.998 \pm 0.000$ & $0.905 \pm 0.003 \, $ & $0.998 \pm 0.000$ & $0.928 \pm 0.003$ \\
&20\%+\method{}+3  & $0.999 \pm 0.000$ & $0.961 \pm 0.002$ & $0.997 \pm 0.000$ & $0.907 \pm 0.007 \, $ & $0.998 \pm 0.000$ & $0.930 \pm 0.005$ \\
&20\%+\method{}+11 & $0.999 \pm 0.000$ & $0.963 \pm 0.000$ & $0.997 \pm 0.000$ & $0.901 \pm 0.009$ & $0.998 \pm 0.000$ & $0.927 \pm 0.005$ \\

\cmidrule(lr){2-8}

&30\%+\method{} & $0.997 \pm 0.000$ & $0.8888 \pm 0.010$ & $0.987 \pm 0.005$ & $0.577 \pm 0.040$ & $0.991 \pm 0.003$ & $0.707 \pm 0.026$ \\
&30\%+\method{}+1 & $0.999 \pm 0.000$ & $0.961 \pm 0.003$ & $0.997 \pm 0.000$ & $0.891 \pm 0.004 \, $ & $0.998 \pm 0.000$ & $0.920 \pm 0.003$ \\
&30\%+\method{}+3 & $0.999 \pm 0.000$ & $0.962 \pm 0.004$ & $0.998 \pm 0.000$ & $0.888 \pm 0.013 \, $ & $0.998 \pm 0.000$ & $0.919 \pm 0.009$ \\
&30\%+\method{}+11 & $0.999 \pm 0.000$ & $0.960 \pm 0.000$ & $0.997 \pm 0.001$ & $0.883 \pm 0.015$ & $0.998 \pm 0.000$ & $0.915 \pm 0.009$ \\

\cmidrule(lr){2-8}

&40\%+\method{} & $0.997 \pm 0.000$ & $0.882 \pm 0.008$ & $0.983 \pm 0.004$ & $0.455 \pm 0.017$ & $0.989 \pm 0.002$ & $0.633 \pm 0.012$ \\
&40\%+\method{}+1 & $0.999 \pm 0.000$ & $0.960 \pm 0.003$ & $0.997 \pm 0.001$ & $0.889 \pm 0.010 \, $ & $0.998 \pm 0.000$ & $0.919 \pm 0.007$ \\
&40\%+\method{}+3 & $0.999 \pm 0.000$ & $0.962 \pm 0.002$ & $0.996 \pm 0.001$ & $0.893 \pm 0.014 \, $ & $0.997 \pm 0.000$ & $0.922 \pm 0.009$ \\
&40\%+\method{}+11 & $0.998 \pm 0.000$ & $0.957 \pm 0.003$ & $0.996 \pm 0.001$ & $0.868 \pm 0.010$ & $0.997 \pm 0.001$ & $0.905 \pm 0.007$ \\

\cmidrule(lr){2-8}

&50\%+\method{} & $0.996 \pm 0.000$ & $0.839 \pm 0.005$ & $0.975 \pm 0.005$ & $0.271 \pm 0.013$ & $0.984 \pm 0.002$ & $0.508 \pm 0.009$ \\
&50\%+\method{}+1 & $0.999 \pm 0.000$ & $0.959 \pm 0.001$ & $0.997 \pm 0.001$ & $0.885 \pm 0.008 \, $ & $0.997 \pm 0.000$ & $0.916 \pm 0.005$ \\
&50\%+\method{}+3 & $0.999 \pm 0.000$ & $0.960 \pm 0.002$ & $0.996 \pm 0.001$ & $0.887 \pm 0.014 \, $ & $0.997 \pm 0.000$ & $0.918 \pm 0.009$ \\
&50\%+\method{}+11 & $0.999 \pm 9.761$ & $0.962 \pm 0.000$ & $0.997 \pm 0.001$ & $0.878 \pm 0.011$ & $0.997 \pm 0.000$ & $0.913 \pm 0.006$ \\

\cmidrule(lr){2-8}

&60\%+\method{} & $0.996 \pm 0.001$ & $0.798 \pm 0.022$ & $0.954 \pm 0.010$ & $0.125 \pm 0.019$ & $0.971 \pm 0.006$ & $0.404 \pm 0.021$ \\
&60\%+\method{}+1 & $0.999 \pm 0.000$ & $0.961 \pm 0.003$ & $0.997 \pm 0.000$ & $0.883 \pm 0.013$ & $0.997 \pm 0.000$ & $0.916 \pm 0.009$ \\
&60\%+\method{}+3 & $0.998 \pm 0.000$ & $0.957 \pm 0.003$ & $0.996 \pm 0.001$ & $0.864 \pm 0.019$ & $0.997 \pm 0.000$ & $0.903 \pm 0.012$ \\
&60\%+\method{}+11 & $0.998 \pm 7.137$ & $0.958 \pm 0.002$ & $0.996 \pm 0.001$ & $0.847 \pm 0.009$ & $0.997 \pm 0.001$ & $0.894 \pm 0.006$ \\

\cmidrule(lr){2-8}
&70\%+\method{} & $0.996 \pm 0.000$ & $0.786 \pm 0.016$ & $0.955 \pm 0.014$ & $0.115 \pm 0.027$ & $0.972 \pm 0.009$ & $0.394 \pm 0.023$ \\
&70\%+\method{}+1 & $0.999 \pm 0.000$ & $0.957 \pm 0.002$ & $0.997 \pm 0.000$ & $0.884 \pm 0.009$ & $0.997 \pm 0.000$ & $0.915 \pm 0.006$ \\
&70\%+\method{}+3 & $0.999 \pm 0.000$ & $0.959 \pm 0.004$ & $0.996 \pm 0.001$ & $0.871 \pm 0.021$ & $0.997 \pm 0.000$ & $0.908 \pm 0.014$ \\
&70\%+\method{}+11 & $0.998 \pm 0.000$ & $0.954 \pm 0.005$ & $0.995 \pm 0.001$ & $0.814 \pm 0.035$ & $0.997 \pm 0.000$ & $0.873 \pm 0.022$ \\

\cmidrule(lr){2-8}
&80\%+\method{} & $0.995 \pm 0.000$ & $0.747 \pm 0.010$ & $0.939 \pm 0.010$ & $0.087 \pm 0.012$ & $0.962 \pm 0.006$ & $0.360 \pm 0.011$ \\
&80\%+\method{}+1 & $0.998 \pm 0.000$ & $0.957 \pm 0.003$ & $0.996 \pm 0.000$ & $0.877 \pm 0.008$ & $0.997 \pm 0.000$ & $0.910 \pm 0.006$ \\
&80\%+\method{}+3 & $0.999 \pm 0.000$ & $0.960 \pm 0.004$ & $0.997 \pm 0.000$ & $0.879 \pm 0.003$ & $0.998 \pm 0.000$ & $0.913 \pm 0.002$ \\
&80\%+\method{}+11 & $0.998 \pm 0.000$ & $0.949 \pm 0.000$ & $0.995 \pm 0.000$ & $0.806 \pm 0.003$ & $0.996 \pm 0.000$ & $0.866 \pm 0.002$ \\

\bottomrule
\end{tabular}
\end{adjustbox}
\end{table*}

\begin{table*}
\captionsetup{type=table}
\caption{\textcolor{black}{Detection performance of \method{} under varying label noise ratios and different labeled data ratios on the BODMAS dataset. Each experiment is repeated three times, and the mean along with standard deviation is reported.}}
\label{tab:varying_label_noise_label_ratio_bodmas_}
\centering
\begin{adjustbox}{width=1\textwidth,height=9.5cm}
\begin{tabular}{llllllll}
 & \multicolumn{6}{c}{BODMAS} \\
\cmidrule(lr){1-8}
Label ratio & Label noise ratio & seen-AUT (B) & seen-AUT (A) & unseen-AUT (B) & unseen-AUT (A) & overall-AUT (B) & overall-AUT (A)\\
\midrule

5\%
&10\%+\method{} 
& $0.919 \pm 0.003$
& $0.709 \pm 0.039$
& $0.927 \pm 0.031$
& $0.779 \pm 0.025$
& $0.922 \pm 0.016$
& $0.743 \pm 0.021$ \\

&10\%+\method{}+1
& $0.929 \pm 0.002$
& $0.707 \pm 0.033$
& $0.937 \pm 0.020$
& $0.773 \pm 0.016$
& $0.933 \pm 0.009$
& $0.742 \pm 0.022$ \\

&10\%+\method{}+3
& $0.923 \pm 0.008$
& $0.702 \pm 0.029$
& $0.933 \pm 0.025$
& $0.778 \pm 0.019$
& $0.927 \pm 0.013$
& $0.741 \pm 0.022$ \\
&10\%+\method{}+11 & $0.927 \pm 0.007$ & $0.708 \pm 0.053$ & $0.945 \pm 0.010$ & $0.791 \pm 0.023$ & $0.936 \pm 0.005$ & $0.750 \pm 0.023$ \\

\cmidrule(lr){2-8}

&20\%+\method{}
& $0.918 \pm 0.007$
& $0.667 \pm 0.072$
& $0.935 \pm 0.022$
& $0.771 \pm 0.015$
& $0.927 \pm 0.012$
& $0.724 \pm 0.035$ \\

&20\%+\method{}+1
& $0.929 \pm 0.010$
& $0.659 \pm 0.083$
& $0.961 \pm 0.003$
& $0.792 \pm 0.049$
& $0.947 \pm 0.005$
& $0.733 \pm 0.060$ \\

&20\%+\method{}+3
& $0.925 \pm 0.005$
& $0.726 \pm 0.021$
& $0.929 \pm 0.026$
& $0.763 \pm 0.018$
& $0.926 \pm 0.013$
& $0.742 \pm 0.001$ \\
&20\%+\method{}+11 & $0.922 \pm 0.008$ & $0.669 \pm 0.084$ & $0.936 \pm 0.023$ & $0.747 \pm 0.031$ & $0.929 \pm 0.010$ & $0.712 \pm 0.041$ \\

\cmidrule(lr){2-8}

&30\%+\method{}
& $0.916 \pm 0.005$
& $0.707 \pm 0.018$
& $0.925 \pm 0.009$
& $0.743 \pm 0.029$
& $0.919 \pm 0.006$
& $0.721 \pm 0.022$ \\

&30\%+\method{}+1
& $0.925 \pm 0.002$
& $0.727 \pm 0.027$
& $0.927 \pm 0.021$
& $0.767 \pm 0.019$
& $0.924 \pm 0.011$
& $0.744 \pm 0.010$ \\

&30\%+\method{}+3
& $0.920 \pm 0.010$
& $0.688 \pm 0.033$
& $0.928 \pm 0.021$
& $0.770 \pm 0.014$
& $0.923 \pm 0.007$
& $0.730 \pm 0.014$ \\
&30\%+\method{}+11 & $0.926 \pm 0.002$ & $0.715 \pm 0.023$ & $0.948 \pm 0.013$ & $0.804 \pm 0.026$ & $0.938 \pm 0.007$ & $0.762 \pm 0.023$ \\

\cmidrule(lr){2-8}

&40\%+\method{}
& $0.914 \pm 0.002$
& $0.695 \pm 0.014$
& $0.910 \pm 0.023$
& $0.717 \pm 0.036$
& $0.909 \pm 0.012$
& $0.701 \pm 0.015$ \\

&40\%+\method{}+1
& $0.926 \pm 0.005$
& $0.707 \pm 0.005$
& $0.951 \pm 0.001$
& $0.795 \pm 0.011$
& $0.939 \pm 0.002$
& $0.754 \pm 0.006$ \\

&40\%+\method{}+3
& $0.919 \pm 0.006$
& $0.642 \pm 0.070$
& $0.936 \pm 0.028$
& $0.754 \pm 0.029$
& $0.928 \pm 0.015$
& $0.705 \pm 0.034$ \\
&40\%+\method{}+11 & $0.922 \pm 0.007$ & $0.706 \pm 0.020$ & $0.936 \pm 0.022$ & $0.786 \pm 0.015$ & $0.929 \pm 0.012$ & $0.747 \pm 0.013$ \\

\cmidrule(lr){2-8}

&50\%+\method{}
& $0.916 \pm 0.006$
& $0.684 \pm 0.034$
& $0.916 \pm 0.036$
& $0.732 \pm 0.068$
& $0.914 \pm 0.019$
& $0.706 \pm 0.035$ \\

&50\%+\method{}+1
& $0.927 \pm 0.008$
& $0.716 \pm 0.031$
& $0.946 \pm 0.019$
& $0.797 \pm 0.045$
& $0.936 \pm 0.011$
& $0.759 \pm 0.032$ \\

&50\%+\method{}+3
& $0.922 \pm 0.008$
& $0.722 \pm 0.046$
& $0.924 \pm 0.018$
& $0.768 \pm 0.029$
& $0.922 \pm 0.007$
& $0.743 \pm 0.024$ \\
&50\%+\method{}+11 & $0.925 \pm 0.007$ & $0.695 \pm 0.053$ & $0.950 \pm 0.013$ & $0.777 \pm 0.029$ & $0.937 \pm 0.007$ & $0.736 \pm 0.027$ \\

\cmidrule(lr){2-8}
&60\%+\method{}   & $0.912 \pm 0.000$ & $0.695 \pm 0.018$ & $0.894 \pm 0.034$ & $0.719 \pm 0.032$ & $0.899 \pm 0.019$ & $0.702 \pm 0.011$ \\
&60\%+\method{}+1 & $0.923 \pm 0.003$ & $0.697 \pm 0.020$ & $0.948 \pm 0.021$ & $0.783 \pm 0.034$ & $0.936 \pm 0.013$ & $0.742 \pm 0.012$ \\
&60\%+\method{}+3 & $0.920 \pm 0.006$ & $0.655 \pm 0.071$ & $0.941 \pm 0.016$ & $0.748 \pm 0.060$ & $0.931 \pm 0.010$ & $0.706 \pm 0.057$ \\
&60\%+\method{}+11 & $0.923 \pm 0.008$ & $0.683 \pm 0.080$ & $0.946 \pm 0.008$ & $0.780 \pm 0.029$ & $0.935 \pm 0.003$ & $0.735 \pm 0.048$ \\

\cmidrule(lr){2-8}
&70\%+\method{}   & $0.890 \pm 0.008$ & $0.607 \pm 0.036$ & $0.877 \pm 0.001$ & $0.614 \pm 0.016$ & $0.877 \pm 0.003$ & $0.600 \pm 0.023$ \\
&70\%+\method{}+1 & $0.923 \pm 0.006$ & $0.701 \pm 0.062$ & $0.943 \pm 0.021$ & $0.788 \pm 0.044$ & $0.933 \pm 0.011$ & $0.748 \pm 0.043$ \\
&70\%+\method{}+3 & $0.920 \pm 0.005$ & $0.673 \pm 0.079$ & $0.940 \pm 0.023$ & $0.753 \pm 0.055$ & $0.931 \pm 0.012$ & $0.717 \pm 0.037$ \\
&70\%+\method{}+11 & $0.923 \pm 0.007$ & $0.633 \pm 0.075$ & $0.948 \pm 0.015$ & $0.761 \pm 0.014$ & $0.937 \pm 0.006$ & $0.707 \pm 0.038$ \\

\cmidrule(lr){2-8}
&80\%+\method{}   & $0.894 \pm 0.020$ & $0.633 \pm 0.074$ & $0.875 \pm 0.018$ & $0.647 \pm 0.070$ & $0.878 \pm 0.019$ & $0.631 \pm 0.072$ \\
&80\%+\method{}+1 & $0.914 \pm 0.004$ & $0.674 \pm 0.054$ & $0.921 \pm 0.032$ & $0.723 \pm 0.038$ & $0.917 \pm 0.018$ & $0.697 \pm 0.013$ \\
&80\%+\method{}+3 & $0.923 \pm 0.001$ & $0.676 \pm 0.067$ & $0.940 \pm 0.019$ & $0.754 \pm 0.026$ & $0.931 \pm 0.010$ & $0.718 \pm 0.032$ \\
&80\%+\method{}+11 & $0.918 \pm 0.021$ & $0.655 \pm 0.068$ & $0.948 \pm 0.010$ & $0.772 \pm 0.040$ & $0.934 \pm 0.015$ & $0.717 \pm 0.050$ \\

\midrule

10\%
&10\%+\method{}
& $0.918 \pm 0.003$
& $0.661 \pm 0.066$
& $0.933 \pm 0.020$
& $0.732 \pm 0.015$
& $0.925 \pm 0.012$
& $0.697 \pm 0.025$ \\

&10\%+\method{}+1
& $0.930 \pm 0.006$
& $0.704 \pm 0.035$
& $0.960 \pm 0.002$
& $0.814 \pm 0.033$
& $0.946 \pm 0.003$
& $0.762 \pm 0.032$ \\

&10\%+\method{}+3
& $0.918 \pm 0.005$
& $0.622 \pm 0.074$
& $0.954 \pm 0.005$
& $0.756 \pm 0.040$
& $0.938 \pm 0.005$
& $0.701 \pm 0.050$ \\
&10\%+\method{}+11 & $0.933 \pm 0.001$ & $0.738 \pm 0.018$ & $0.945 \pm 0.007$ & $0.804 \pm 0.018$ & $0.938 \pm 0.004$ & $0.771 \pm 0.015$ \\

\cmidrule(lr){2-8}

&20\%+\method{}
& $0.936 \pm 0.007$
& $0.734 \pm 0.029$
& $0.960 \pm 0.005$
& $0.824 \pm 0.016$
& $0.947 \pm 0.005$
& $0.778 \pm 0.021$ \\

&20\%+\method{}+1
& $0.932 \pm 0.002$
& $0.744 \pm 0.020$
& $0.933 \pm 0.025$
& $0.794 \pm 0.007$
& $0.931 \pm 0.014$
& $0.766 \pm 0.013$ \\

&20\%+\method{}+3
& $0.931 \pm 0.009$
& $0.733 \pm 0.031$
& $0.944 \pm 0.007$
& $0.778 \pm 0.016$
& $0.937 \pm 0.001$
& $0.753 \pm 0.015$ \\
&20\%+\method{}+11 & $0.933 \pm 0.006$ & $0.751 \pm 0.013$ & $0.952 \pm 0.013$ & $0.816 \pm 0.028$ & $0.942 \pm 0.009$ & $0.782 \pm 0.017$ \\

\cmidrule(lr){2-8}

&30\%+\method{}
& $0.923 \pm 0.011$
& $0.724 \pm 0.018$
& $0.934 \pm 0.009$
& $0.758 \pm 0.025$
& $0.927 \pm 0.010$
& $0.734 \pm 0.022$ \\

&30\%+\method{}+1
& $0.926 \pm 0.010$
& $0.677 \pm 0.045$
& $0.959 \pm 0.004$
& $0.808 \pm 0.037$
& $0.944 \pm 0.006$
& $0.750 \pm 0.037$ \\

&30\%+\method{}+3
& $0.921 \pm 0.022$
& $0.709 \pm 0.087$
& $0.935 \pm 0.013$
& $0.782 \pm 0.051$
& $0.927 \pm 0.017$
& $0.745 \pm 0.066$ \\
&30\%+\method{}+11 & $0.906 \pm 0.008$ & $0.621 \pm 0.042$ & $0.927 \pm 0.023$ & $0.697 \pm 0.045$ & $0.917 \pm 0.015$ & $0.664 \pm 0.010$ \\

\cmidrule(lr){2-8}

&40\%+\method{}
& $0.916 \pm 0.007$
& $0.707 \pm 0.020$
& $0.935 \pm 0.014$
& $0.754 \pm 0.040$
& $0.924 \pm 0.012$
& $0.727 \pm 0.032$ \\

&40\%+\method{}+1
& $0.931 \pm 0.000$
& $0.692 \pm 0.033$
& $0.959 \pm 0.006$
& $0.795 \pm 0.011$
& $0.945 \pm 0.004$
& $0.746 \pm 0.005$ \\

&40\%+\method{}+3
& $0.930 \pm 0.000$
& $0.735 \pm 0.019$
& $0.935 \pm 0.009$
& $0.772 \pm 0.029$
& $0.932 \pm 0.006$
& $0.750 \pm 0.021$ \\
&40\%+\method{}+11 & $0.926 \pm 0.008$ & $0.725 \pm 0.024$ & $0.952 \pm 0.001$ & $0.812 \pm 0.005$ & $0.939 \pm 0.004$ & $0.768 \pm 0.007$ \\

\cmidrule(lr){2-8}

&50\%+\method{}
& $0.915 \pm 0.009$
& $0.644 \pm 0.080$
& $0.927 \pm 0.022$
& $0.700 \pm 0.042$
& $0.918 \pm 0.016$
& $0.668 \pm 0.045$ \\

&50\%+\method{}+1
& $0.931 \pm 0.009$
& $0.756 \pm 0.024$
& $0.939 \pm 0.023$
& $0.779 \pm 0.051$
& $0.934 \pm 0.016$
& $0.765 \pm 0.036$ \\

&50\%+\method{}+3
& $0.922 \pm 0.008$
& $0.722 \pm 0.046$
& $0.924 \pm 0.018$
& $0.768 \pm 0.029$
& $0.922 \pm 0.007$
& $0.743 \pm 0.024$ \\

&50\%+\method{}+11 & $0.926 \pm 0.005$ & $0.701 \pm 0.030$ & $0.952 \pm 0.006$ & $0.788 \pm 0.046$ & $0.939 \pm 0.004$ & $0.744 \pm 0.039$ \\

\cmidrule(lr){2-8}
&60\%+\method{}   & $0.882 \pm 0.020$ & $0.550 \pm 0.056$ & $0.879 \pm 0.023$ & $0.571 \pm 0.052$ & $0.876 \pm 0.022$ & $0.555 \pm 0.056$ \\
&60\%+\method{}+1 & $0.913 \pm 0.004$ & $0.623 \pm 0.053$ & $0.941 \pm 0.020$ & $0.744 \pm 0.069$ & $0.929 \pm 0.010$ & $0.693 \pm 0.056$ \\
&60\%+\method{}+3 & $0.931 \pm 0.008$ & $0.694 \pm 0.041$ & $0.956 \pm 0.008$ & $0.774 \pm 0.028$ & $0.944 \pm 0.007$ & $0.736 \pm 0.030$ \\
&60\%+\method{}+11 & $0.931 \pm 0.008$ & $0.723 \pm 0.020$ & $0.955 \pm 0.004$ & $0.808 \pm 0.005$ & $0.942 \pm 0.003$ & $0.765 \pm 0.008$ \\

\cmidrule(lr){2-8}
&70\%+\method{}   & $0.859 \pm 0.037$ & $0.515 \pm 0.104$ & $0.871 \pm 0.026$ & $0.588 \pm 0.069$ & $0.863 \pm 0.027$ & $0.554 \pm 0.074$ \\
&70\%+\method{}+1 & $0.933 \pm 0.005$ & $0.761 \pm 0.026$ & $0.918 \pm 0.034$ & $0.760 \pm 0.037$ & $0.923 \pm 0.021$ & $0.754 \pm 0.015$ \\
&70\%+\method{}+3 & $0.909 \pm 0.017$ & $0.646 \pm 0.090$ & $0.931 \pm 0.014$ & $0.739 \pm 0.058$ & $0.920 \pm 0.014$ & $0.696 \pm 0.066$ \\
&70\%+\method{}+11 & $0.928 \pm 0.007$ & $0.674 \pm 0.038$ & $0.961 \pm 0.002$ & $0.812 \pm 0.009$ & $0.946 \pm 0.003$ & $0.750 \pm 0.018$ \\

\cmidrule(lr){2-8}
&80\%+\method{}   & $0.869 \pm 0.002$ & $0.590 \pm 0.016$ & $0.835 \pm 0.007$ & $0.580 \pm 0.029$ & $0.843 \pm 0.002$ & $0.572 \pm 0.023$ \\
&80\%+\method{}+1 & $0.926 \pm 0.004$ & $0.738 \pm 0.033$ & $0.932 \pm 0.025$ & $0.768 \pm 0.041$ & $0.928 \pm 0.015$ & $0.750 \pm 0.029$ \\
&80\%+\method{}+3 & $0.933 \pm 0.002$ & $0.715 \pm 0.035$ & $0.949 \pm 0.015$ & $0.792 \pm 0.038$ & $0.940 \pm 0.009$ & $0.754 \pm 0.027$ \\
&80\%+\method{}+11 & $0.933 \pm 0.006$ & $0.738 \pm 0.052$ & $0.939 \pm 0.009$ & $0.776 \pm 0.025$ & $0.935 \pm 0.005$ & $0.754 \pm 0.028$ \\

\midrule

20\%
&10\%+\method{}
& $0.928 \pm 0.002$
& $0.740 \pm 0.015$
& $0.937 \pm 0.028$
& $0.807 \pm 0.052$
& $0.932 \pm 0.015$
& $0.773 \pm 0.036$ \\

&10\%+\method{}+1
& $0.932 \pm 0.005$
& $0.751 \pm 0.029$
& $0.946 \pm 0.006$
& $0.792 \pm 0.018$
& $0.939 \pm 0.003$
& $0.769 \pm 0.019$ \\

&10\%+\method{}+3
& $0.922 \pm 0.008$
& $0.671 \pm 0.067$
& $0.952 \pm 0.002$
& $0.774 \pm 0.030$
& $0.938 \pm 0.002$
& $0.726 \pm 0.041$ \\
&10\%+\method{}+11 & $0.933 \pm 0.005$ & $0.743 \pm 0.009$ & $0.945 \pm 0.007$ & $0.800 \pm 0.013$ & $0.939 \pm 0.004$ & $0.771 \pm 0.008$ \\

\cmidrule(lr){2-8}

&20\%+\method{}
& $0.931 \pm 0.003$
& $0.740 \pm 0.008$
& $0.942 \pm 0.019$
& $0.785 \pm 0.048$
& $0.936 \pm 0.012$
& $0.759 \pm 0.031$ \\

&20\%+\method{}+1
& $0.924 \pm 0.009$
& $0.616 \pm 0.069$
& $0.952 \pm 0.005$
& $0.740 \pm 0.053$
& $0.939 \pm 0.006$
& $0.688 \pm 0.055$ \\

&20\%+\method{}+3
& $0.928 \pm 0.005$
& $0.738 \pm 0.012$
& $0.951 \pm 0.008$
& $0.818 \pm 0.019$
& $0.939 \pm 0.005$
& $0.780 \pm 0.008$ \\
&20\%+\method{}+11 & $0.930 \pm 0.006$ & $0.679 \pm 0.074$ & $0.944 \pm 0.011$ & $0.747 \pm 0.070$ & $0.937 \pm 0.008$ & $0.716 \pm 0.066$ \\

\cmidrule(lr){2-8}

&30\%+\method{}
& $0.917 \pm 0.010$
& $0.700 \pm 0.010$
& $0.930 \pm 0.019$
& $0.739 \pm 0.064$
& $0.923 \pm 0.012$
& $0.718 \pm 0.036$ \\

&30\%+\method{}+1
& $0.930 \pm 0.006$
& $0.741 \pm 0.007$
& $0.924 \pm 0.036$
& $0.778 \pm 0.043$
& $0.925 \pm 0.022$
& $0.755 \pm 0.021$ \\

&30\%+\method{}+3
& $0.940 \pm 0.001$
& $0.767 \pm 0.021$
& $0.952 \pm 0.009$
& $0.820 \pm 0.010$
& $0.944 \pm 0.005$
& $0.789 \pm 0.005$ \\
&30\%+\method{}+11 & $0.924 \pm 0.007$ & $0.719 \pm 0.079$ & $0.927 \pm 0.024$ & $0.761 \pm 0.026$ & $0.925 \pm 0.013$ & $0.740 \pm 0.019$ \\

\cmidrule(lr){2-8}

&40\%+\method{}
& $0.930 \pm 0.002$
& $0.723 \pm 0.039$
& $0.946 \pm 0.006$
& $0.792 \pm 0.017$
& $0.938 \pm 0.003$
& $0.758 \pm 0.018$ \\

&40\%+\method{}+1
& $0.929 \pm 0.004$
& $0.722 \pm 0.055$
& $0.951 \pm 0.006$
& $0.799 \pm 0.034$
& $0.940 \pm 0.005$
& $0.761 \pm 0.041$ \\

&40\%+\method{}+3
& $0.916 \pm 0.015$
& $0.656 \pm 0.104$
& $0.951 \pm 0.011$
& $0.778 \pm 0.061$
& $0.935 \pm 0.011$
& $0.723 \pm 0.072$ \\
&40\%+\method{}+11 & $0.913 \pm 0.025$ & $0.639 \pm 0.070$ & $0.941 \pm 0.025$ & $0.732 \pm 0.086$ & $0.927 \pm 0.026$ & $0.686 \pm 0.081$ \\

\cmidrule(lr){2-8}

&50\%+\method{}
& $0.913 \pm 0.004$
& $0.704 \pm 0.011$
& $0.920 \pm 0.006$
& $0.731 \pm 0.024$
& $0.914 \pm 0.004$
& $0.711 \pm 0.021$ \\

&50\%+\method{}+1
& $0.925 \pm 0.006$
& $0.705 \pm 0.045$
& $0.936 \pm 0.021$
& $0.748 \pm 0.072$
& $0.930 \pm 0.015$
& $0.725 \pm 0.056$ \\

&50\%+\method{}+3
& $0.932 \pm 0.005$
& $0.744 \pm 0.032$
& $0.950 \pm 0.010$
& $0.812 \pm 0.031$
& $0.940 \pm 0.007$
& $0.778 \pm 0.026$ \\
&50\%+\method{}+11 & $0.930 \pm 0.003$ & $0.741 \pm 0.026$ & $0.950 \pm 0.011$ & $0.810 \pm 0.030$ & $0.939 \pm 0.006$ & $0.774 \pm 0.018$ \\

\cmidrule(lr){2-8}
&60\%+\method{}   & $0.830 \pm 0.063$ & $0.471 \pm 0.140$ & $0.858 \pm 0.043$ & $0.571 \pm 0.080$ & $0.843 \pm 0.047$ & $0.524 \pm 0.098$ \\
&60\%+\method{}+1 & $0.926 \pm 0.001$ & $0.694 \pm 0.058$ & $0.944 \pm 0.013$ & $0.781 \pm 0.006$ & $0.935 \pm 0.008$ & $0.739 \pm 0.017$ \\
&60\%+\method{}+3 & $0.934 \pm 0.002$ & $0.745 \pm 0.010$ & $0.948 \pm 0.005$ & $0.802 \pm 0.011$ & $0.940 \pm 0.003$ & $0.771 \pm 0.010$ \\
&60\%+\method{}+11 & $0.930 \pm 0.006$ & $0.716 \pm 0.045$ & $0.955 \pm 0.005$ & $0.805 \pm 0.010$ & $0.943 \pm 0.004$ & $0.763 \pm 0.023$ \\

\cmidrule(lr){2-8}
&70\%+\method{}   & $0.880 \pm 0.010$ & $0.586 \pm 0.031$ & $0.857 \pm 0.002$ & $0.588 \pm 0.025$ & $0.862 \pm 0.006$ & $0.577 \pm 0.027$ \\
&70\%+\method{}+1 & $0.902 \pm 0.015$ & $0.594 \pm 0.064$ & $0.937 \pm 0.013$ & $0.725 \pm 0.014$ & $0.921 \pm 0.013$ & $0.668 \pm 0.029$ \\
&70\%+\method{}+3 & $0.910 \pm 0.020$ & $0.617 \pm 0.110$ & $0.950 \pm 0.000$ & $0.764 \pm 0.037$ & $0.933 \pm 0.007$ & $0.700 \pm 0.063$ \\
&70\%+\method{}+11 & $0.927 \pm 0.009$ & $0.750 \pm 0.029$ & $0.941 \pm 0.010$ & $0.794 \pm 0.031$ & $0.933 \pm 0.010$ & $0.768 \pm 0.030$ \\

\cmidrule(lr){2-8}
&80\%+\method{}   & $0.785 \pm 0.090$ & $0.420 \pm 0.135$ & $0.791 \pm 0.071$ & $0.467 \pm 0.102$ & $0.784 \pm 0.078$ & $0.442 \pm 0.114$ \\
&80\%+\method{}+1 & $0.924 \pm 0.006$ & $0.703 \pm 0.053$ & $0.955 \pm 0.005$ & $0.809 \pm 0.022$ & $0.941 \pm 0.002$ & $0.759 \pm 0.034$ \\
&80\%+\method{}+3 & $0.910 \pm 0.019$ & $0.616 \pm 0.094$ & $0.936 \pm 0.009$ & $0.711 \pm 0.052$ & $0.923 \pm 0.013$ & $0.666 \pm 0.066$ \\
&80\%+\method{}+11 & $0.926 \pm 0.005$ & $0.732 \pm 0.022$ & $0.942 \pm 0.015$ & $0.788 \pm 0.036$ & $0.934 \pm 0.008$ & $0.760 \pm 0.026$ \\

\bottomrule
\end{tabular}
\end{adjustbox}
\end{table*}

\begin{table*}
\captionsetup{type=table}
\caption{\textcolor{black}{Detection performance of \method{} under varying label noise ratios and different labeled data ratios on the AndroZoo dataset. Each experiment is repeated three times, and the mean along with standard deviation is reported.}}
\label{tab:varying_label_noise_label_ratio_androzoo_}
\centering  

\begin{adjustbox}{width=1\textwidth,height=9.5cm}
\begin{tabular}{llllllll}
 & \multicolumn{6}{c}{AndroZoo} \\
\cmidrule(lr){1-8}
 Label ratio & Label noise ratio & seen-AUT (B) & seen-AUT (A) & unseen-AUT (B) & unseen-AUT (A) & overall-AUT (B) & overall-AUT (A)\\
\midrule

&10\% +\method{} & $0.996 \pm 0.000$ & $0.854 \pm 0.023$ & $0.994 \pm 0.000$ & $0.831 \pm 0.007$ & $0.995 \pm 0.000$ & $0.839 \pm 0.005$ \\ 
&10\%+\method{}+1  & $0.995 \pm 0.001$ & $0.899 \pm 0.005$ & $0.995 \pm 0.000$ & $0.879 \pm 0.027 \, $& $0.995 \pm 0.000$ & $0.886 \pm 0.016$ \\
&10\%+\method{}+3  & $0.996 \pm 0.001$ & $0.909 \pm 0.004$ & $0.995 \pm 0.001$ & $0.886 \pm 0.037\, $ & $0.995 \pm 0.001$ & $0.894 \pm 0.026$ \\
&10\%+\method{}+11 & $0.995 \pm 0.002$ & $0.921 \pm 0.004$ & $0.996 \pm 0.001$ & $0.887 \pm 0.013$ & $0.995 \pm 0.001$ & $0.899 \pm 0.007$ \\

\cmidrule(lr){2-8}
5\%&20\%+\method{} & $0.996 \pm 0.000$ & $0.853 \pm 0.008$ & $0.993 \pm 0.000$ & $0.808 \pm 0.022$ & $0.994 \pm 0.001$ & $0.824 \pm 0.017$ \\ 
&20\% +\method{}+1 & $0.996 \pm 0.000$ & $0.907 \pm 0.012$ & $0.996 \pm 0.000$ & $0.876 \pm 0.011\, $ & $0.996 \pm 0.001$ & $0.887 \pm 0.007$ \\
&20\% +\method{}+3 & $0.995 \pm 0.001$ & $0.896 \pm 0.020$ & $0.994 \pm 0.001$ & $0.871 \pm 0.009\, $ & $0.995 \pm 0.001$ & $0.880 \pm 0.009$ \\
&20\%+\method{}+11 & $0.996 \pm 0.000$ & $0.900 \pm 0.011$ & $0.995 \pm 0.001$ & $0.859 \pm 0.015$ & $0.996 \pm 0.001$ & $0.874 \pm 0.014$ \\

\cmidrule(lr){2-8}
&30\% +\method{}& $0.995 \pm 0.000$ & $0.812 \pm 0.031$ & $0.993 \pm 0.001$ & $0.758 \pm 0.018$ & $0.993 \pm 0.001$ & $0.778 \pm 0.021$ \\
&30\%+\method{}+1 & $0.996 \pm 0.001$ & $0.904 \pm 0.014$ & $0.993 \pm 0.001$ & $0.854 \pm 0.037\, $ & $0.994 \pm 0.001$ & $0.871 \pm 0.026$ \\
&30\%+\method{}+3 & $0.996 \pm 0.001$ & $0.904 \pm 0.006$ & $0.995 \pm 0.000$ & $0.878 \pm 0.018\, $ & $0.995 \pm 0.000$ & $0.886 \pm 0.012$ \\
&30\%+\method{}+11 & $0.995 \pm 0.002$ & $0.895 \pm 0.016$ & $0.993 \pm 0.000$ & $0.868 \pm 0.010$ & $0.994 \pm 0.001$ & $0.878 \pm 0.013$ \\

\cmidrule(lr){2-8}
 &40\% + \method{}& $0.996 \pm 0.000$ & $0.793 \pm 0.040$ & $0.989 \pm 0.000$ & $0.558 \pm 0.024$ & $0.992 \pm 0.000$ & $0.639 \pm 0.025$ \\
&40\%+\method{}+1 & $0.995 \pm 0.001$ & $0.891 \pm 0.010$ & $0.995 \pm 0.001$ & $0.875 \pm 0.012\, $ & $0.995 \pm 0.001$ & $0.881 \pm 0.011$ \\
&40\%+\method{}+3 & $0.995 \pm 0.002$ & $0.900 \pm 0.005$ & $0.995 \pm 0.000$ & $0.875 \pm 0.007\, $ & $0.995 \pm 0.001$ & $0.884 \pm 0.003$ \\
&40\%+\method{}+11 & $0.996 \pm 0.001$ & $0.905 \pm 0.008$ & $0.994 \pm 0.000$ & $0.872 \pm 0.004$ & $0.995 \pm 0.000$ & $0.884 \pm 0.005$ \\

\cmidrule(lr){2-8}
&50\%+\method{} & $0.996 \pm 0.001$ & $0.795 \pm 0.033$ & $0.974 \pm 0.006$ & $0.411 \pm 0.082$ & $0.982 \pm 0.004$ & $0.541 \pm 0.064$ \\
&50\%+\method{}+1 & $0.996 \pm 0.001$ & $0.900 \pm 0.016$ & $0.994 \pm 0.001$ & $0.859 \pm 0.029\, $ & $0.995 \pm 0.000$ & $0.873 \pm 0.025$ \\
&50\%+\method{}+3 & $0.995 \pm 0.002$ & $0.908 \pm 0.011$ & $0.995 \pm 0.000$ & $0.857 \pm 0.011\, $ & $0.995 \pm 0.002$ & $0.875 \pm 0.009$ \\
&50\%+\method{}+11 & $0.996 \pm 0.001$ & $0.896 \pm 0.015$ & $0.995 \pm 0.000$ & $0.857 \pm 0.022$ & $0.995 \pm 0.001$ & $0.870 \pm 0.019$ \\

\cmidrule(lr){2-8}
&60\%+\method{}   & $0.996 \pm 0.000$ & $0.719 \pm 0.008$ & $0.969 \pm 0.004$ & $0.297 \pm 0.007$ & $0.978 \pm 0.002$ & $0.441 \pm 0.007$ \\
&60\%+\method{}+1 & $0.996 \pm 0.000$ & $0.892 \pm 0.010$ & $0.994 \pm 0.001$ & $0.869 \pm 0.021$ & $0.995 \pm 0.001$ & $0.877 \pm 0.017$ \\
&60\%+\method{}+3 & $0.996 \pm 0.001$ & $0.902 \pm 0.014$ & $0.995 \pm 0.000$ & $0.864 \pm 0.026$ & $0.995 \pm 0.000$ & $0.876 \pm 0.021$ \\
&60\%+\method{}+11 & $0.995 \pm 0.001$ & $0.892 \pm 0.007$ & $0.994 \pm 0.000$ & $0.860 \pm 0.014$ & $0.994 \pm 0.000$ & $0.872 \pm 0.012$ \\

\cmidrule(lr){2-8}
&70\%+\method{}   & $0.994 \pm 0.000$ & $0.601 \pm 0.101$ & $0.953 \pm 0.000$ & $0.186 \pm 0.037$ & $0.967 \pm 0.000$ & $0.327 \pm 0.060$ \\
&70\%+\method{}+1 & $0.924 \pm 0.102$ & $0.901 \pm 0.016$ & $0.965 \pm 0.038$ & $0.836 \pm 0.043$ & $0.951 \pm 0.060$ & $0.858 \pm 0.032$ \\
&70\%+\method{}+3 & $0.995 \pm 0.002$ & $0.906 \pm 0.003$ & $0.994 \pm 0.001$ & $0.851 \pm 0.012$ & $0.994 \pm 0.001$ & $0.871 \pm 0.007$ \\
&70\%+\method{}+11 & $0.995 \pm 0.002$ & $0.896 \pm 0.018$ & $0.993 \pm 0.001$ & $0.864 \pm 0.023$ & $0.994 \pm 0.002$ & $0.875 \pm 0.019$ \\

\cmidrule(lr){2-8}
&80\%+\method{}   & $0.993 \pm 0.000$ & $0.616 \pm 0.085$ & $0.931 \pm 0.014$ & $0.190 \pm 0.065$ & $0.952 \pm 0.009$ & $0.334 \pm 0.061$ \\
&80\%+\method{}+1 & $0.994 \pm 0.002$ & $0.890 \pm 0.011$ & $0.994 \pm 0.001$ & $0.843 \pm 0.012$ & $0.994 \pm 0.001$ & $0.860 \pm 0.008$ \\
&80\%+\method{}+3 & $0.996 \pm 0.001$ & $0.902 \pm 0.016$ & $0.994 \pm 0.001$ & $0.847 \pm 0.023$ & $0.995 \pm 0.001$ & $0.866 \pm 0.015$ \\
&80\%+\method{}+11 & $0.995 \pm 0.002$ & $0.905 \pm 0.006$ & $0.994 \pm 0.000$ & $0.851 \pm 0.007$ & $0.994 \pm 0.001$ & $0.870 \pm 0.003$ \\

\midrule

&10\% + \method{} & $0.995 \pm 0.001$ & $0.881 \pm 0.023$ & $0.994 \pm 0.001$ & $0.824 \pm 0.016$ & $0.994 \pm 0.001$ & $0.843 \pm 0.017$ \\ 
&10\% +\method{}+1 & $0.995 \pm 0.002$ & $0.909 \pm 0.015$ & $0.995 \pm 0.000$ & $0.863 \pm 0.037\, $ & $0.995 \pm 0.000$ & $0.879 \pm 0.020$ \\
&10\% +\method{}+3 & $0.995 \pm 0.002$ & $0.904 \pm 0.020$ & $0.994 \pm 0.001$ & $0.858 \pm 0.022\, $ & $0.994 \pm 0.001$ & $0.873 \pm 0.020$ \\
&10\%+\method{}+11 & $0.997 \pm 0.001$ & $0.909 \pm 0.015$ & $0.996 \pm 0.000$ & $0.894 \pm 0.005$ & $0.996 \pm 0.000$ & $0.899 \pm 0.002$ \\

\cmidrule(lr){2-8}
10\%&20\%+\method{} & $0.996 \pm 0.000$ & $0.850 \pm 0.018$ & $0.992 \pm 0.001$ & $0.746 \pm 0.012$ & $0.994 \pm 0.001$ & $0.782 \pm 0.014$ \\ 
&20\%+\method{}+1  & $0.996 \pm 0.001$ & $0.897 \pm 0.010$ & $0.996 \pm 0.001$ & $0.870 \pm 0.024\, $ & $0.996 \pm 0.001$ & $0.885 \pm 0.014$ \\
&20\%+\method{}+3  & $0.997 \pm 0.000$ & $0.914 \pm 0.006$ & $0.996 \pm 0.000$ & $0.898 \pm 0.008\, $ & $0.997 \pm 0.000$ & $0.903 \pm 0.007$ \\
&20\%+\method{}+11 & $0.996 \pm 0.001$ & $0.916 \pm 0.007$ & $0.996 \pm 0.000$ & $0.883 \pm 0.010$ & $0.996 \pm 0.000$ & $0.893 \pm 0.005$ \\

\cmidrule(lr){2-8}
&30\%+\method{} & $0.996 \pm 0.000$ & $0.867 \pm 0.011$ & $0.993 \pm 0.000$ & $0.718 \pm 0.015$ & $0.994 \pm 0.000$ & $0.768 \pm 0.009$ \\
&30\% +\method{}+1& $0.995 \pm 0.001$ & $0.903 \pm 0.016$ & $0.995 \pm 0.000$ & $0.864 \pm 0.016\, $ & $0.995 \pm 0.000$ & $0.877 \pm 0.011$ \\
&30\% +\method{}+3& $0.998 \pm 0.000$ & $0.912 \pm 0.003$ & $0.996 \pm 0.000$ & $0.876 \pm 0.035\, $ & $0.997 \pm 0.000$ & $0.888 \pm 0.022$ \\
&30\%+\method{}+11 & $0.998 \pm 0.000$ & $0.920 \pm 0.005$ & $0.996 \pm 0.000$ & $0.879 \pm 0.010$ & $0.996 \pm 0.000$ & $0.893 \pm 0.004$ \\

\cmidrule(lr){2-8}
&40\%+\method{} & $0.997 \pm 0.000$ & $0.838 \pm 0.014$ & $0.987 \pm 0.003$ & $0.518 \pm 0.066$ & $0.990 \pm 0.002$ & $0.627 \pm 0.043$ \\
&40\%+\method{}+1 & $0.996 \pm 0.001$ & $0.901 \pm 0.006$ & $0.996 \pm 0.000$ & $0.867 \pm 0.019\, $ & $0.996 \pm 0.001$ & $0.879 \pm 0.011$ \\
&40\%+\method{}+3 & $0.995 \pm 0.001$ & $0.895 \pm 0.016$ & $0.995 \pm 0.000$ & $0.870 \pm 0.014\, $ & $0.995 \pm 0.001$ & $0.879 \pm 0.005$ \\
&40\%+\method{}+11 & $0.995 \pm 0.001$ & $0.910 \pm 0.006$ & $0.995 \pm 0.000$ & $0.874 \pm 0.012$ & $0.995 \pm 0.001$ & $0.887 \pm 0.009$ \\

\cmidrule(lr){2-8}
&50\%+\method{} & $0.996 \pm 0.000$ & $0.814 \pm 0.032$ & $0.977 \pm 0.004$ & $0.390 \pm 0.033$ & $0.984 \pm 0.003$ & $0.534 \pm 0.027$ \\
&50\% +\method{}+1& $0.995 \pm 0.001$ & $0.901 \pm 0.019$ & $0.995 \pm 0.000$ & $0.863 \pm 0.017\, $ & $0.995 \pm 0.000$ & $0.876 \pm 0.013$ \\
&50\% +\method{}+3& $0.995 \pm 0.001$ & $0.895 \pm 0.019$ & $0.995 \pm 0.000$ & $0.840 \pm 0.038\, $ & $0.995 \pm 0.000$ & $0.858 \pm 0.023$ \\
&50\%+\method{}+11 & $0.996 \pm 0.002$ & $0.905 \pm 0.008$ & $0.995 \pm 0.001$ & $0.846 \pm 0.037$ & $0.995 \pm 0.001$ & $0.866 \pm 0.026$ \\

\cmidrule(lr){2-8}
&60\%+\method{}   & $0.996 \pm 0.000$ & $0.756 \pm 0.044$ & $0.971 \pm 0.000$ & $0.344 \pm 0.067$ & $0.979 \pm 0.000$ & $0.484 \pm 0.030$ \\
&60\%+\method{}+1 & $0.996 \pm 0.002$ & $0.905 \pm 0.011$ & $0.994 \pm 0.001$ & $0.815 \pm 0.006$ & $0.994 \pm 0.001$ & $0.846 \pm 0.007$ \\
&60\%+\method{}+3 & $0.996 \pm 0.002$ & $0.903 \pm 0.013$ & $0.995 \pm 0.000$ & $0.853 \pm 0.034$ & $0.995 \pm 0.001$ & $0.871 \pm 0.017$ \\
&60\%+\method{}+11 & $0.998 \pm 0.000$ & $0.918 \pm 0.006$ & $0.995 \pm 0.000$ & $0.848 \pm 0.057$ & $0.996 \pm 0.000$ & $0.872 \pm 0.036$ \\

\cmidrule(lr){2-8}
&70\%+\method{}   & $0.994 \pm 0.001$ & $0.672 \pm 0.031$ & $0.952 \pm 0.012$ & $0.291 \pm 0.050$ & $0.966 \pm 0.008$ & $0.420 \pm 0.028$  \\
&70\%+\method{}+1 & $0.996 \pm 0.001$ & $0.913 \pm 0.009$ & $0.995 \pm 0.000$ & $0.866 \pm 0.019$ & $0.995 \pm 0.000$ & $0.882 \pm 0.013$ \\
&70\%+\method{}+3 & $0.995 \pm 0.001$ & $0.903 \pm 0.005$ & $0.994 \pm 0.000$ & $0.826 \pm 0.023$ & $0.994 \pm 0.000$ & $0.852 \pm 0.013$ \\
&70\%+\method{}+11 & $0.996 \pm 0.002$ & $0.908 \pm 0.008$ & $0.994 \pm 0.000$ & $0.827 \pm 0.032$ & $0.995 \pm 0.001$ & $0.854 \pm 0.020$ \\

\cmidrule(lr){2-8}
&80\%+\method{}   & $0.993 \pm 0.002$ & $0.645 \pm 0.045$ & $0.931 \pm 0.002$ & $0.225 \pm 0.062$ & $0.952 \pm 0.000$ & $0.367 \pm 0.025$ \\
&80\%+\method{}+1 & $0.996 \pm 0.002$ & $0.898 \pm 0.021$ & $0.994 \pm 0.000$ & $0.812 \pm 0.046$ & $0.994 \pm 0.000$ & $0.841 \pm 0.025$ \\
&80\%+\method{}+3 & $0.997 \pm 0.001$ & $0.907 \pm 0.011$ & $0.995 \pm 0.000$ & $0.809 \pm 0.039$ & $0.996 \pm 0.000$ & $0.842 \pm 0.023$ \\
&80\%+\method{}+11 & $0.998 \pm 0.000$ & $0.893 \pm 0.022$ & $0.994 \pm 0.000$ & $0.802 \pm 0.051$ & $0.995 \pm 0.000$ & $0.832 \pm 0.025$ \\

\midrule

&10\% +\method{} & $0.994 \pm 0.000$ & $0.845 \pm 0.012$ & $0.994 \pm 0.000$ & $0.857 \pm 0.026$ & $0.994 \pm 0.000$ & $0.852 \pm 0.018$ \\ 
&10\%+\method{}+1  & $0.994 \pm 0.000$ & $0.899 \pm 0.001$ & $0.994 \pm 0.000$ & $0.887 \pm 0.011\, $ & $0.994 \pm 0.000$ & $0.890 \pm 0.007$ \\
&10\%+\method{}+3  & $0.993 \pm 0.001$ & $0.896 \pm 0.003$ & $0.995 \pm 0.000$ & $0.883 \pm 0.021\, $ & $0.994 \pm 0.000$ & $0.887 \pm 0.013$ \\
&10\%+\method{}+11 & $0.993 \pm 0.002$ & $0.901 \pm 0.003$ & $0.995 \pm 0.000$ & $0.882 \pm 0.024$ & $0.994 \pm 0.000$ & $0.889 \pm 0.015$ \\

\cmidrule(lr){2-8}
20\%&20\% +\method{}& $0.995 \pm 0.000$ & $0.871 \pm 0.006$ & $0.994 \pm 0.000$ & $0.787 \pm 0.009$ & $0.994 \pm 0.000$ & $0.815 \pm 0.006$ \\ 
&20\% +\method{}+1 & $0.994 \pm 0.001$ & $0.892 \pm 0.032$ & $0.995 \pm 0.000$ & $0.885 \pm 0.010\, $ & $0.994 \pm 0.001$ & $0.888 \pm 0.014$ \\
&20\% +\method{}+3 & $0.993 \pm 0.000$ & $0.893 \pm 0.008$ & $0.994 \pm 0.000$ & $0.873 \pm 0.008\, $ & $0.993 \pm 0.000$ & $0.879 \pm 0.007$ \\
&20\%+\method{}+11 & $0.993 \pm 0.000$ & $0.905 \pm 0.006$ & $0.996 \pm 0.000$ & $0.904 \pm 0.013$ & $0.995 \pm 0.000$ & $0.904 \pm 0.010$ \\

\cmidrule(lr){2-8}
&30\%+\method{} & $0.995 \pm 0.001$ & $0.855 \pm 0.032$ & $0.992 \pm 0.000$ & $0.721 \pm 0.006$ & $0.993 \pm 0.000$ & $0.766 \pm 0.012$ \\
&30\%+\method{}+1 & $0.991 \pm 0.001$ & $0.881 \pm 0.006$ & $0.993 \pm 0.000$ & $0.869 \pm 0.012\, $ & $0.993 \pm 0.001$ & $0.873 \pm 0.011$ \\
&30\%+\method{}+3 & $0.994 \pm 0.001$ & $0.862 \pm 0.054$ & $0.995 \pm 0.000$ & $0.898 \pm 0.007\, $ & $0.995 \pm 0.000$ & $0.886 \pm 0.017$ \\
&30\%+\method{}+11 & $0.995 \pm 0.001$ & $0.905 \pm 0.018$ & $0.996 \pm 0.000$ & $0.884 \pm 0.012$ & $0.995 \pm 0.001$ & $0.891 \pm 0.015$ \\

\cmidrule(lr){2-8}
&40\% & $0.995 \pm 0.000$ & $0.828 \pm 0.014$ & $0.989 \pm 0.002$ & $0.590 \pm 0.059$ & $0.991 \pm 0.002$ & $0.671 \pm 0.036$ \\
&40\%+\method{}+1 & $0.994 \pm 0.000$ & $0.900 \pm 0.005$ & $0.995 \pm 0.000$ & $0.871 \pm 0.008\, $ & $0.994 \pm 0.000$ & $0.880 \pm 0.004$ \\
&40\%+\method{}+3 & $0.995 \pm 0.001$ & $0.901 \pm 0.010$ & $0.995 \pm 0.000$ & $0.873 \pm 0.015\, $ & $0.995 \pm 0.000$ & $0.883 \pm 0.007$ \\
&40\%+\method{}+11 & $0.997 \pm 0.000$ & $0.919 \pm 0.003$ & $0.996 \pm 0.000$ & $0.883 \pm 0.013$ & $0.997 \pm 0.000$ & $0.895 \pm 0.008$ \\

\cmidrule(lr){2-8}
&50\%+\method{}& $0.994 \pm 0.001$ & $0.786 \pm 0.032$ & $0.984 \pm 0.001$ & $0.431 \pm 0.028$ & $0.987 \pm 0.001$ & $0.551 \pm 0.029$ \\
&50\%+\method{}+1 & $0.995 \pm 0.000$ & $0.877 \pm 0.025$ & $0.995 \pm 0.000$ & $0.865 \pm 0.023\, $ & $0.995 \pm 0.000$ & $0.869 \pm 0.013$ \\
&50\%+\method{}+3 & $0.995 \pm 0.001$ & $0.899 \pm 0.003$ & $0.994 \pm 0.000$ & $0.869\pm 0.016\, $ & $0.994 \pm 0.001$ & $0.879 \pm 0.009$ \\
&50\%+\method{}+11 & $0.993 \pm 0.002$ & $0.889 \pm 0.006$ & $0.992 \pm 0.003$ & $0.877 \pm 0.021$ & $0.992 \pm 0.002$ & $0.881 \pm 0.015$ \\

\cmidrule(lr){2-8}
&60\%+\method{}   & $0.996 \pm 0.000$ & $0.771 \pm 0.054$ & $0.964 \pm 0.009$ & $0.264 \pm 0.048$ & $0.975 \pm 0.006$ & $0.436 \pm 0.044$ \\
&60\%+\method{}+1 & $0.993 \pm 0.000$ & $0.898 \pm 0.011$ & $0.993 \pm 0.000$ & $0.839 \pm 0.027$ & $0.993 \pm 0.000$ & $0.858 \pm 0.020$ \\
&60\%+\method{}+3 & $0.995 \pm 0.001$ & $0.890 \pm 0.025$ & $0.995 \pm 0.001$ & $0.849 \pm 0.039$ & $0.995 \pm 0.001$ & $0.863 \pm 0.035$ \\
&60\%+\method{}+11 & $0.997 \pm 0.000$ & $0.914 \pm 0.010$ & $0.996 \pm 0.001$ & $0.893 \pm 0.012$ & $0.997 \pm 0.000$ & $0.901 \pm 0.011$ \\

\cmidrule(lr){2-8}
&70\%+\method{}   & $0.995 \pm 0.000$ & $0.650 \pm 0.146$ & $0.953 \pm 0.011$ & $0.198 \pm 0.085$ & $0.967 \pm 0.007$ & $0.349 \pm 0.104$ \\
&70\%+\method{}+1 & $0.995 \pm 0.001$ & $0.886 \pm 0.019$ & $0.995 \pm 0.000$ & $0.837 \pm 0.018$ & $0.995 \pm 0.000$ & $0.854 \pm 0.017$ \\
&70\%+\method{}+3 & $0.995 \pm 0.000$ & $0.891 \pm 0.002$ & $0.994 \pm 0.000$ & $0.834 \pm 0.015$ & $0.995 \pm 0.000$ & $0.853 \pm 0.009$ \\
&70\%+\method{}+11 & $0.997 \pm 0.000$ & $0.882 \pm 0.021$ & $0.995 \pm 0.001$ & $0.859 \pm 0.023$ & $0.996 \pm 0.000$ & $0.867 \pm 0.009$ \\

\cmidrule(lr){2-8}
&80\%+\method{}   & $0.993 \pm 0.000$ & $0.469 \pm 0.101$ & $0.923 \pm 0.005$ & $0.116 \pm 0.038$ & $0.947 \pm 0.003$ & $0.234 \pm 0.058$ \\
&80\%+\method{}+1 & $0.995 \pm 0.000$ & $0.905 \pm 0.003$ & $0.994 \pm 0.000$ & $0.811 \pm 0.017$ & $0.994 \pm 0.000$ & $0.843 \pm 0.009$ \\
&80\%+\method{}+3 & $0.996 \pm 0.002$ & $0.890 \pm 0.013$ & $0.995 \pm 0.001$ & $0.839 \pm 0.021$ & $0.995 \pm 0.001$ & $0.856 \pm 0.017$ \\
&80\%+\method{}+11 & $0.996 \pm 0.000$ & $0.888 \pm 0.037$ & $0.995 \pm 0.000$ & $0.834 \pm 0.042$ & $0.995 \pm 0.000$ & $0.851 \pm 0.035$ \\

\bottomrule
\end{tabular}
\end{adjustbox}
\end{table*}

\begin{figure*}[!ht]

\centering 
\subfloat[AndroZoo\label{fig:androzoo-labelnoise}]{\includegraphics[scale=0.15]{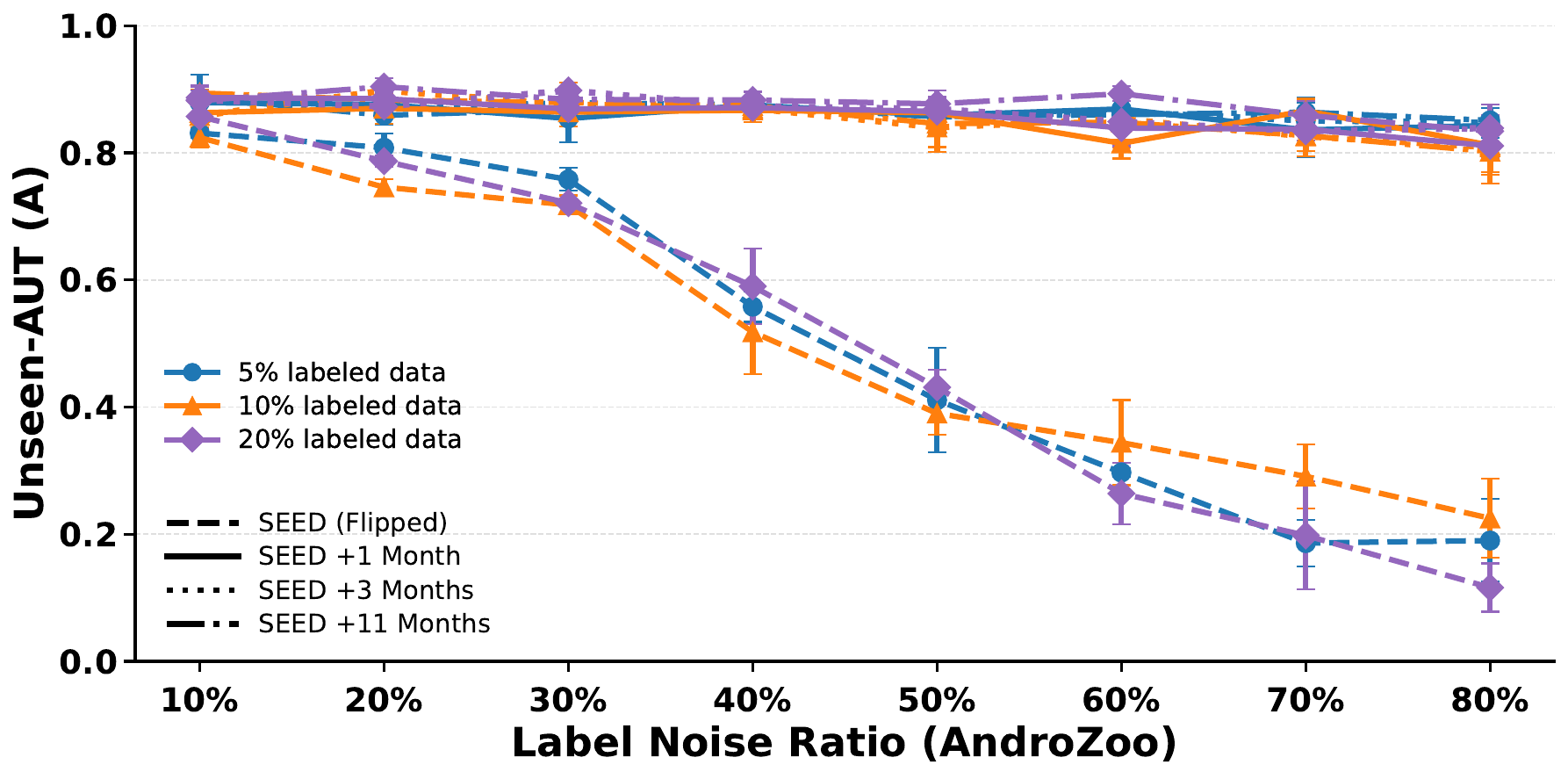}}
\subfloat[APIGraph\label{fig:apigraph-labelnoise}]{\includegraphics[scale=0.15]{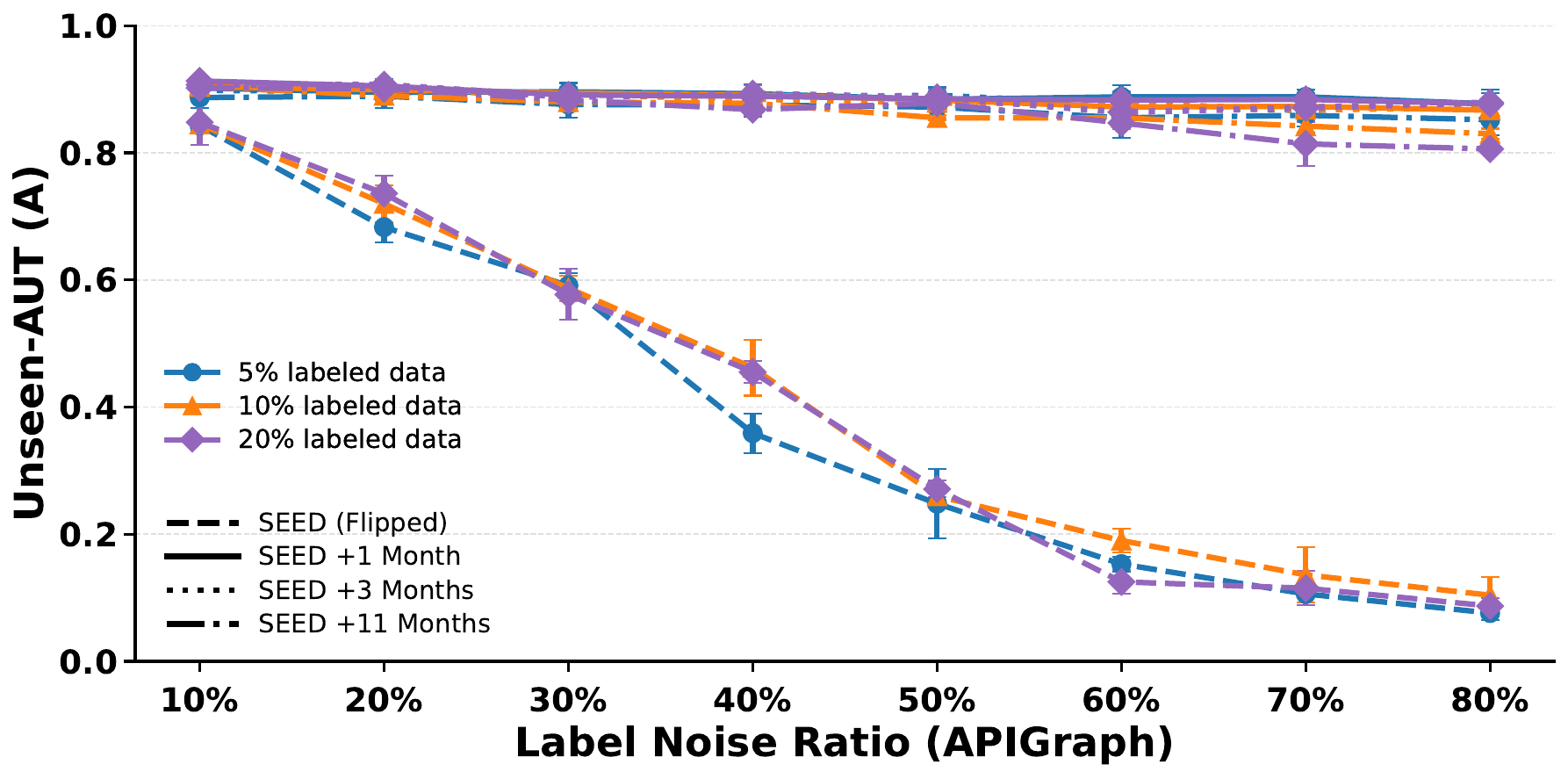}}
\subfloat[BODMAS\label{fig:bodmas-labelnoise}]{\includegraphics[scale=0.15]{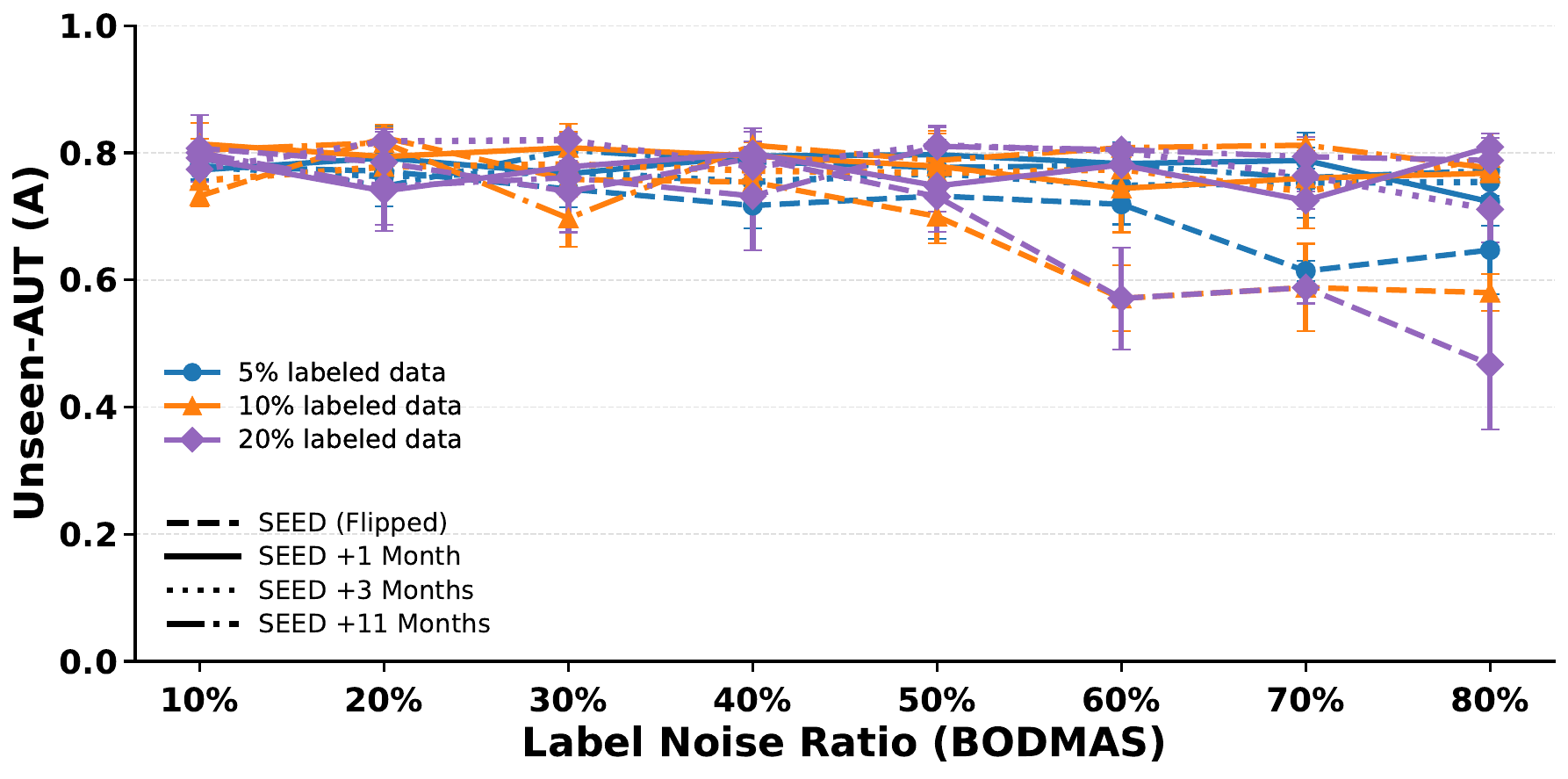}}

 \Description{about the figure}
\caption{\textcolor{black}{Comparing the effect of the label noise with varying label ratio and mitigating its effect using the \method{} method on BODMAS, AndroZoo and APIGraph} }
\label{fig:unseen-tasks-malware-pr-auc}
\end{figure*}

\subsection{Effect of Label Noise on Detection Performance}

\textcolor{black}{In this section, we study how increasing levels of label noise (via label flipping) affect the detection performance on unseen malware. Specifically, we vary the label-noise ratio from $10\%$ to $80\%$ and analyze how the proposed delayed buffer strategy with \method{} mitigates the degradation caused by corrupted labels under different labeled data ratios (5\%, 10\%, 20\%).}

\paragraph{AndroZoo} \textcolor{black}{As shown in Figure~\ref{fig:androzoo-labelnoise}, the detection performance when trained directly with flipped labels degrades sharply as the noise level increases. For instance, at $5\%$ labeled data, unseen-AUT(A) decreases steadily from approximately $0.83$ at $10\%$ noise to around $0.56$ at $40\%$, and further drops to about $0.29$, $0.19$, and $0.19$ at $60\%$, $70\%$, and $80\%$ noise, respectively. Similar degradation patterns are observed for the $10\%$ and $20\%$ labeled-data settings. In contrast, after retraining with noise-free labels obtained after one/three/eleven months, unseen-AUT(A) consistently remains around $0.85$--$0.89$ across the entire noise range, even when the noise ratio reaches $70\%$ or $80\%$. Thus, demonstrating that \method{} effectively mitigates the impact of label corruption and preserves generalization capability on unseen malware families. The complete results of these experiments are available in Table~\ref{tab:varying_label_noise_label_ratio_androzoo_}.}

\paragraph{APIGraph} \textcolor{black}{The severity in the reduction of unseen-AUT(A) on the APIGraph dataset is even greater as the label-noise ratio increases from $10\%$ to $80\%$, as shown in Figure~\ref{fig:apigraph-labelnoise}. For example, at $5\%$ labeled data, unseen-AUT(A) drops dramatically from approximately $0.84$ at $10\%$ noise to $0.08$ at $80\%$ noise. Comparable degradation trends are observed for the $10\%$ and $20\%$ labeled-data settings. In contrast, after incorporating noise-free labels, unseen-AUT(A) remains stable around $0.86$--$0.90$ for all labeled-data ratios, even under extremely noisy scenarios. The complete results of these experiments are available in Table~\ref{tab:varying_label_noise_label_ratio_api_graph_}.}

\paragraph{BODMAS} \textcolor{black}{Figure~\ref{fig:bodmas-labelnoise} shows that the detection performance degradation is less abrupt as the noise ratio increases. For instance, with $5\%$ labeled data, unseen-AUT(A) decreases gradually from approximately $0.78$ at $10\%$ noise to $0.65$ at $80\%$ noise. A similar trend is observed for $10\%$ and $20\%$ labeled data. After retraining with noise-free labels, unseen-AUT(A) values are restored to the range of $0.75$--$0.82$ for most noise ratios, even when the noise reaches $80\%$. Overall, these results demonstrate that \method{} effectively stabilizes detection performance on BODMAS under high label-noise conditions. This comparatively slower degradation may be attributed to the more stable feature distribution in BODMAS. It may also be related to the shorter temporal span of the dataset (approximately one year) compared to AndroZoo and APIGraph, which cover longer time periods and may introduce greater variability and concept drift. The complete results of these experiments are available in Table~\ref{tab:varying_label_noise_label_ratio_bodmas_}.}

\subsection{Operating-Point Evaluation at Fixed False Positive Rates}
\textcolor{black}{In this section, we study the operating-point behavior of SEED and the baseline detectors under fixed false positive rate (FPR) constraints. This evaluation examines how the true positive rate (TPR) evolves across sequential tasks when the detectors are required to operate at specific FPR thresholds.}

\begin{figure*}[!b]

\begin{subfigure}{0.45\textwidth}
\subfloat[FPR = $10^{-1}$]{\label{bodmas} \includegraphics[scale=0.062]{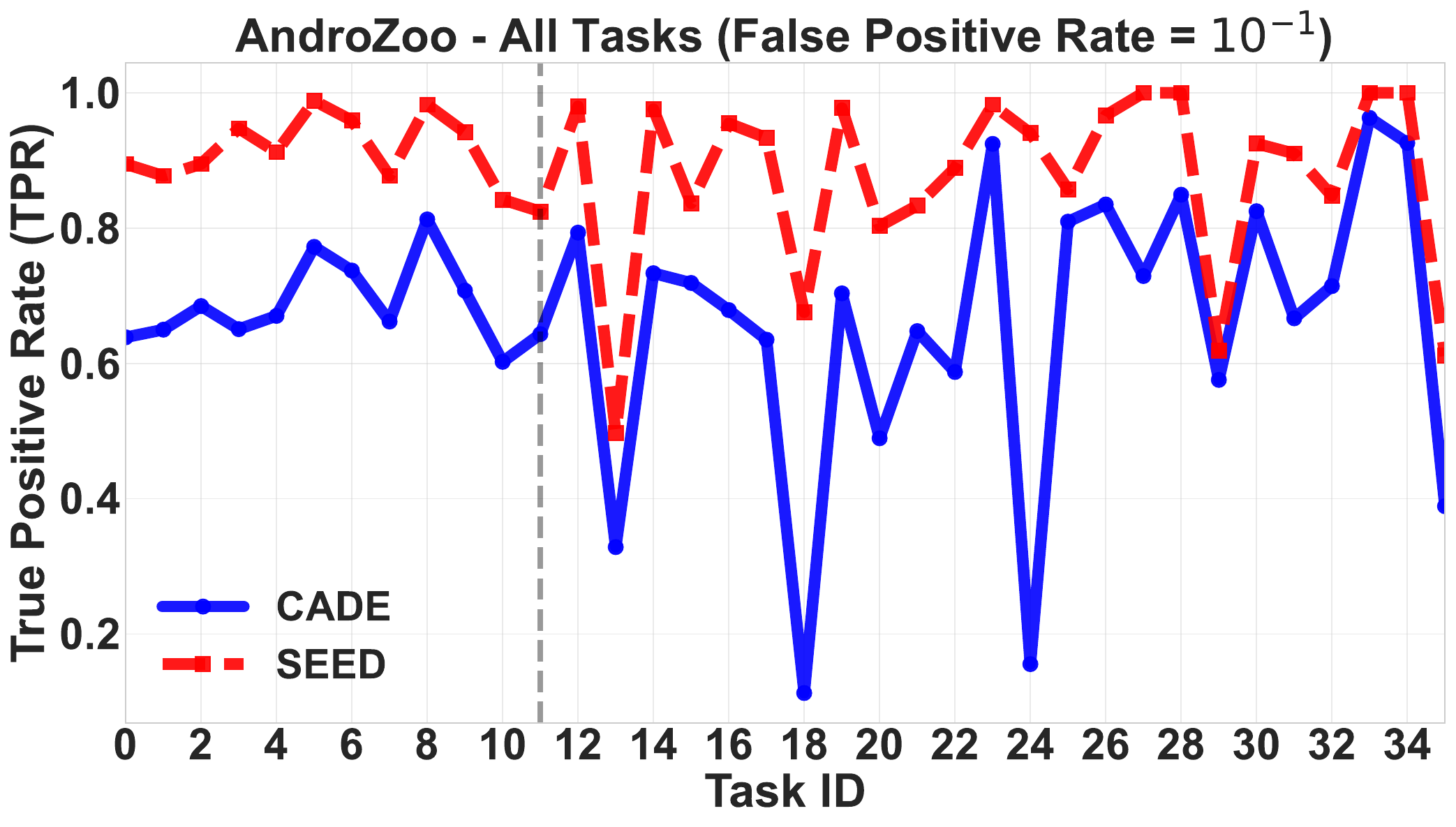}}
\subfloat[FPR = $10^{-2}$]{\label{androzoo} \includegraphics[scale=0.062]{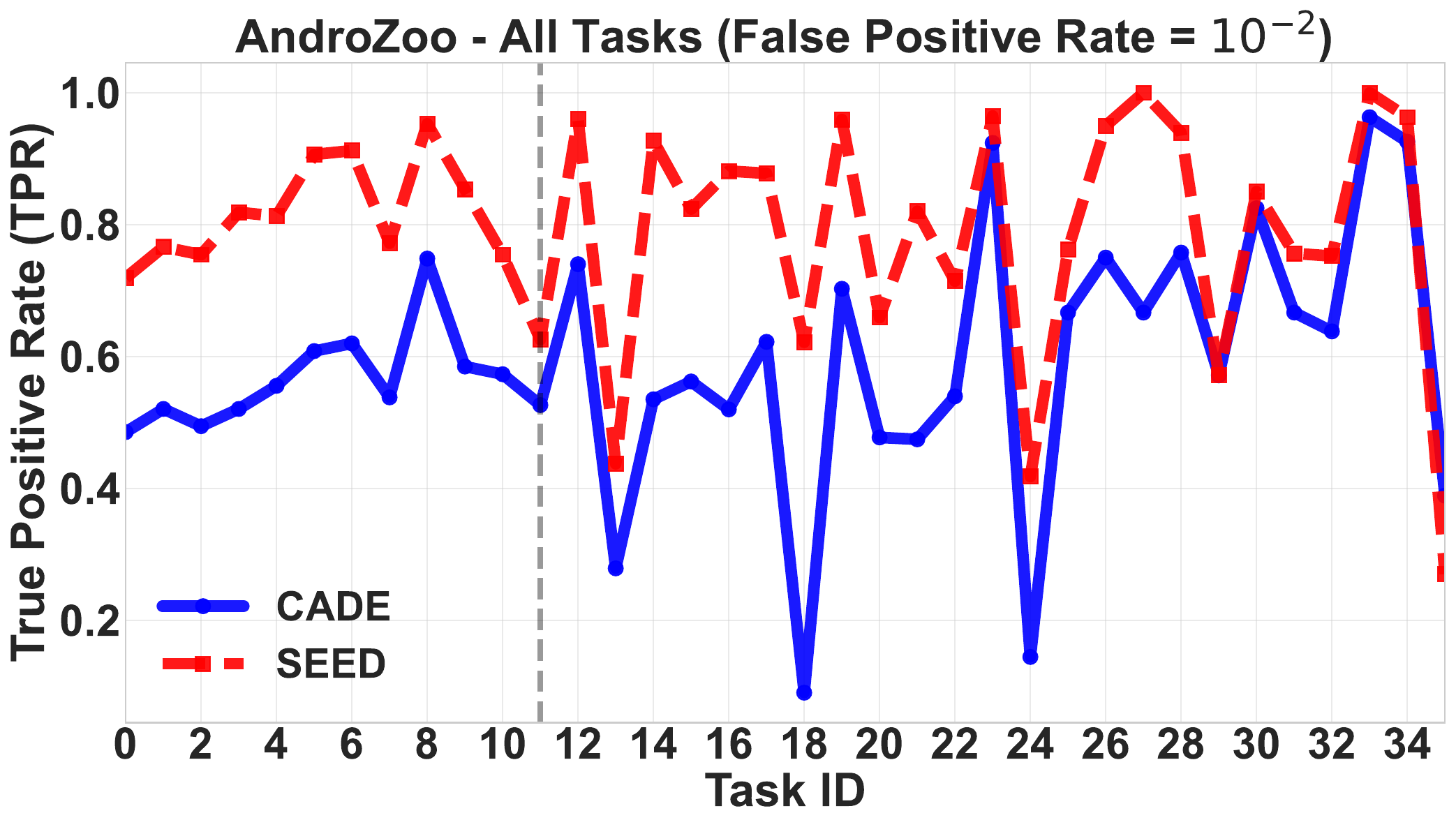}}%
\subfloat[FPR = $10^{-3}$]{\label{apigraph} \includegraphics[scale=0.062]{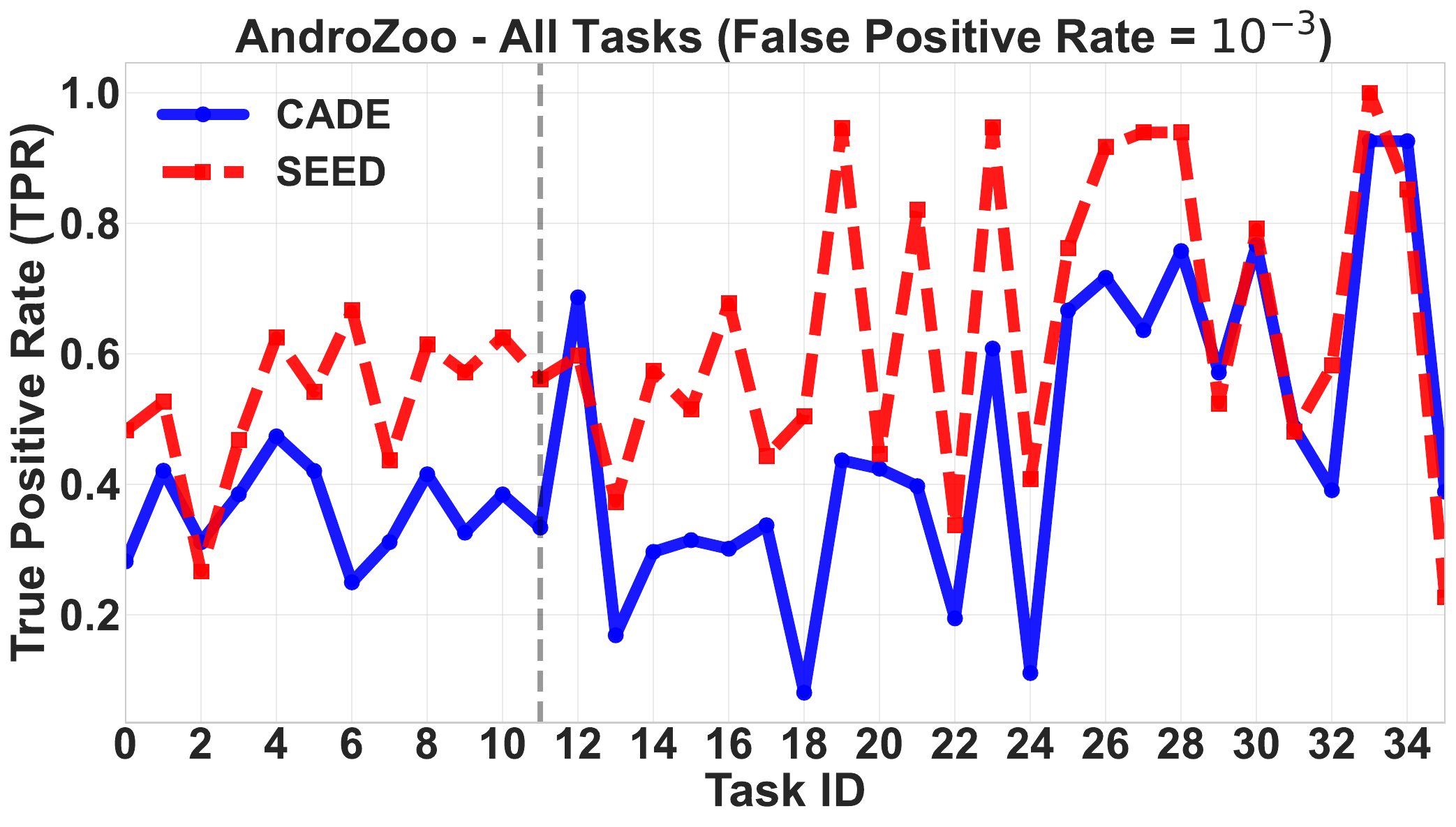}}%
\vspace{2mm}
\makebox[\linewidth][c]{%
\begin{tikzpicture}
\draw[decorate,decoration={brace,mirror,amplitude=6pt}]
(-0.2cm,0) -- (\linewidth+0.2cm,0)
node[midway,yshift=-12pt]{\textbf{CADE Vs SEED}};
\end{tikzpicture}%
}
\end{subfigure}
\hfill
\begin{subfigure}{0.45\textwidth}
\subfloat[FPR = $10^{-1}$]{\label{bodmas} \includegraphics[scale=0.062]{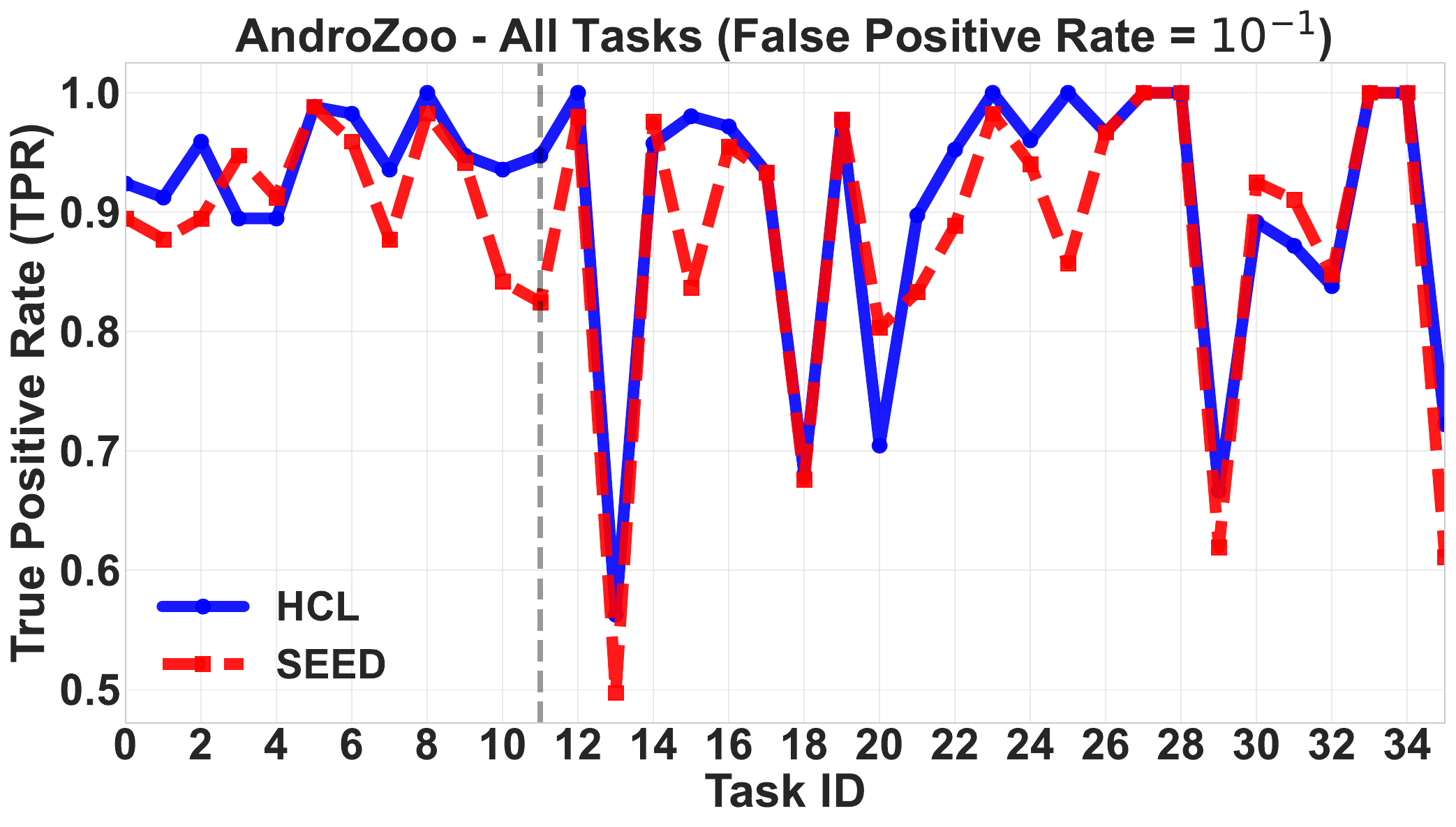}}
\subfloat[FPR = $10^{-2}$]{\label{androzoo} \includegraphics[scale=0.062]{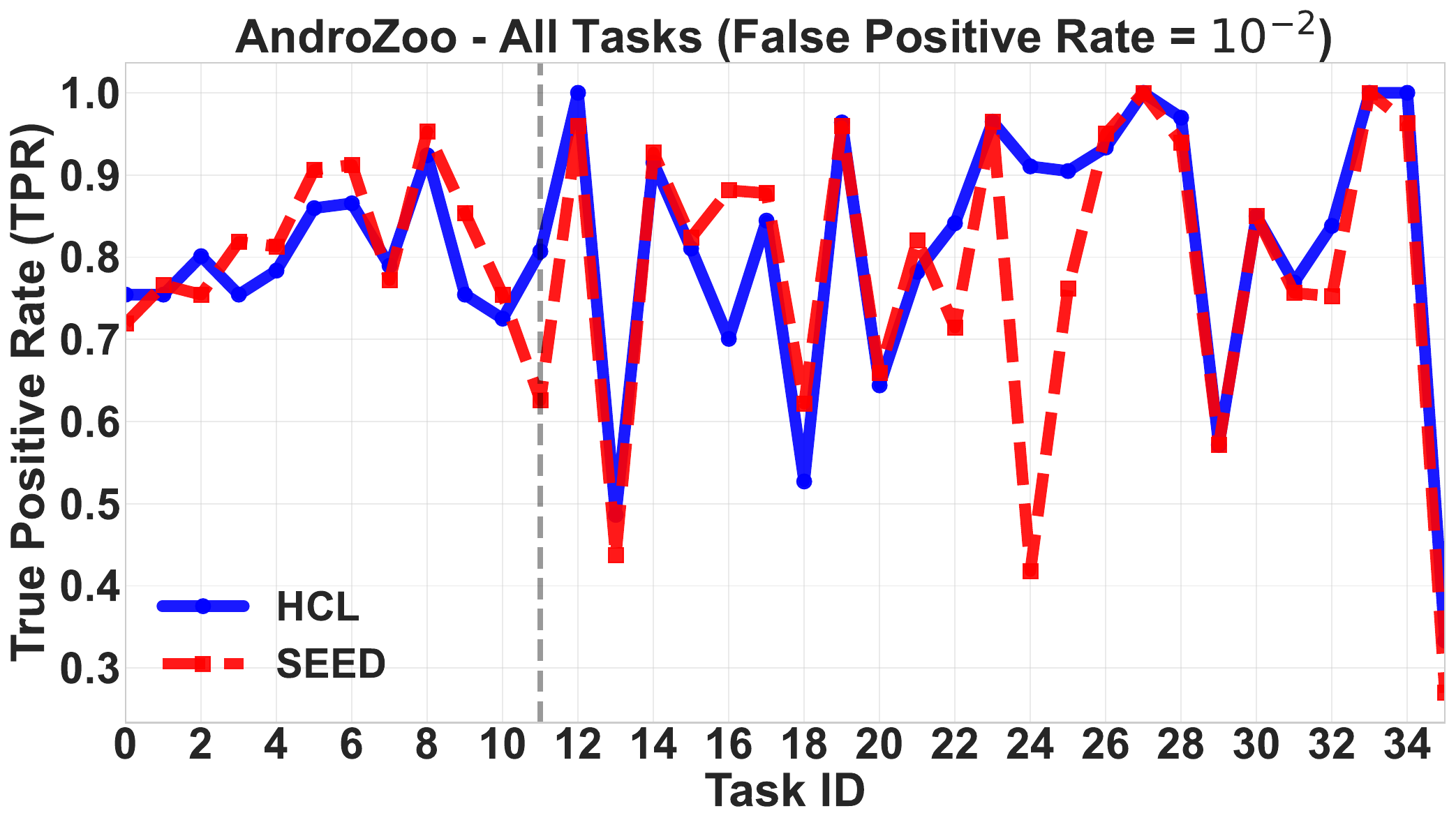}}%
\subfloat[FPR = $10^{-3}$]{\label{apigraph} \includegraphics[scale=0.062]{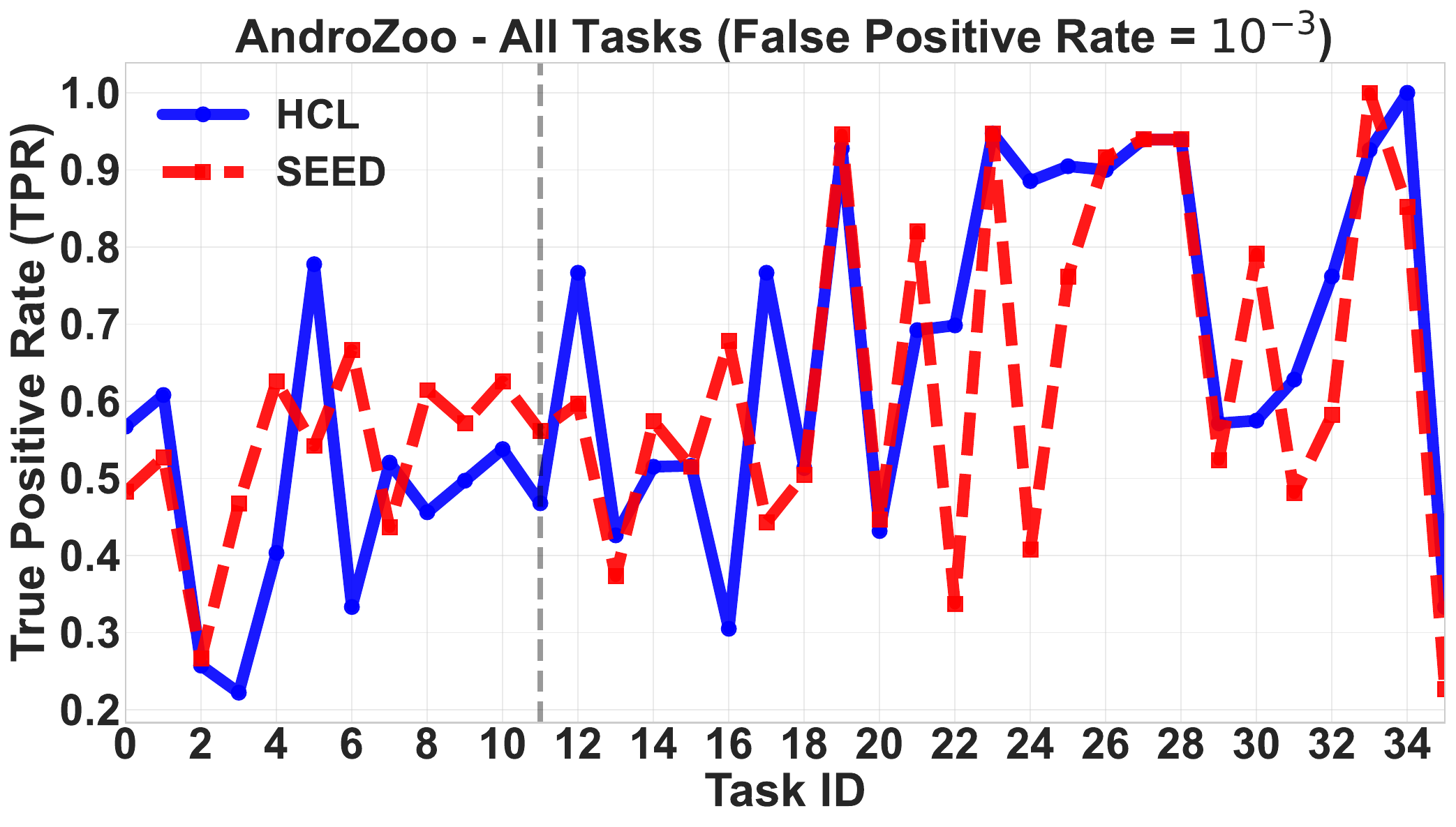}}  
\vspace{2mm}
\makebox[\linewidth][c]{%
\begin{tikzpicture}
\draw[decorate,decoration={brace,mirror,amplitude=6pt}]
(-0.2cm,0) -- (\linewidth+0.2cm,0)
node[midway,yshift=-12pt]{\textbf{HCL Vs SEED}};
\end{tikzpicture}%
}
\end{subfigure}
\hfill
\vspace{0.3cm}
\begin{subfigure}{0.47\textwidth}
\subfloat[FPR = $10^{-1}$]{\label{bodmas} \includegraphics[scale=0.062]{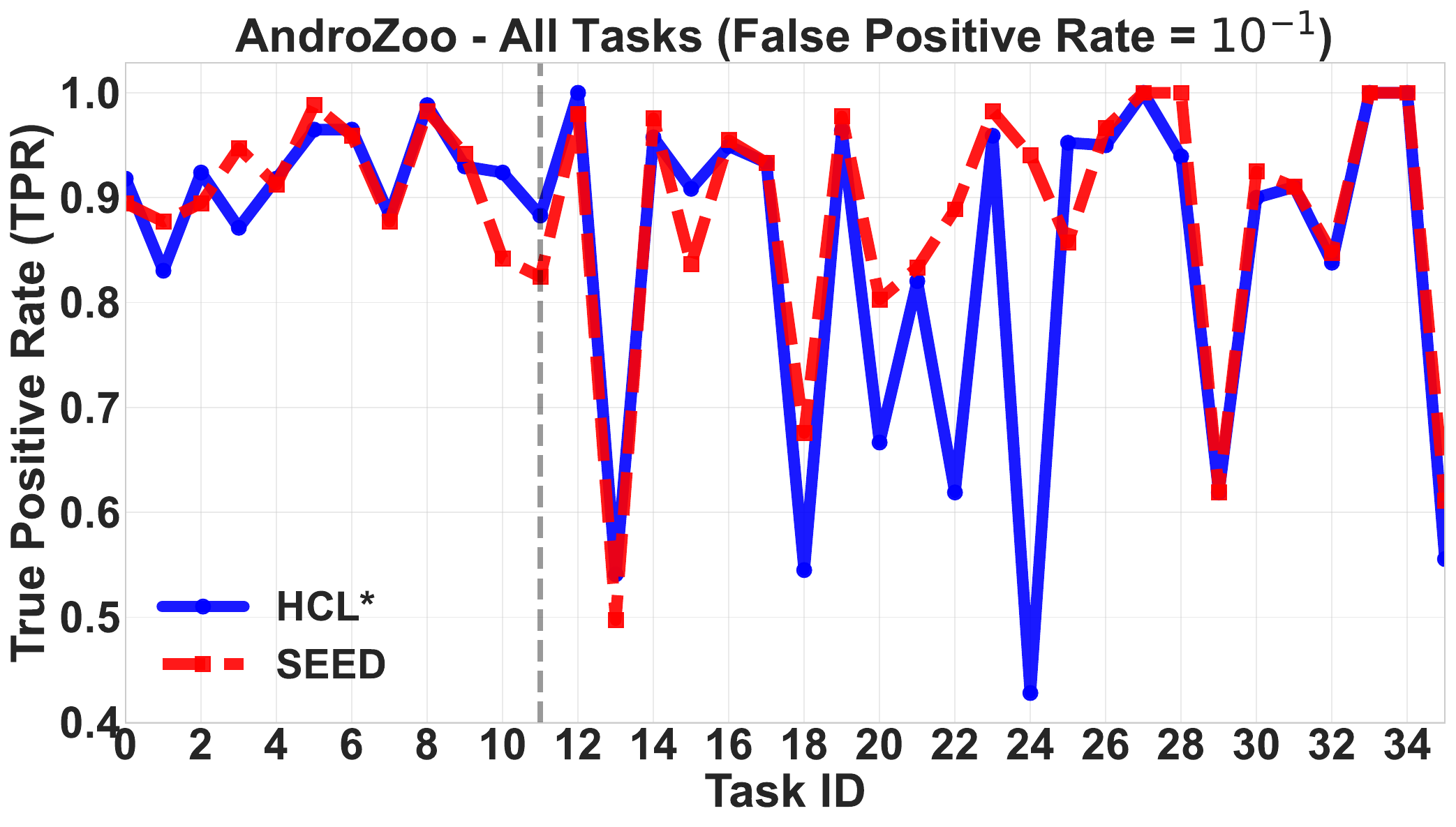}}
\subfloat[FPR = $10^{-2}$]{\label{androzoo} \includegraphics[scale=0.062]{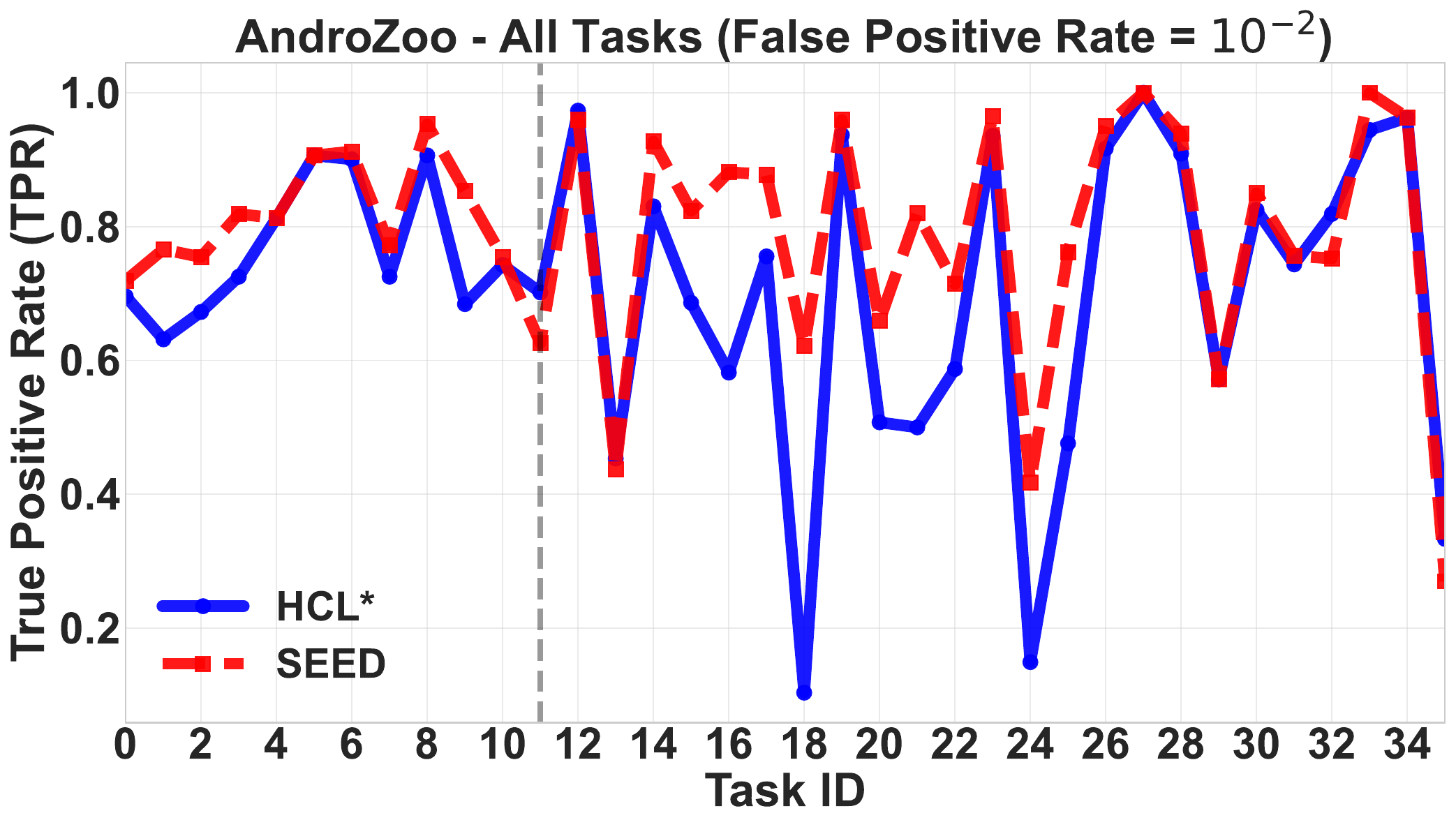}}
\subfloat[FPR = $10^{-3}$]{\label{apigraph} \includegraphics[scale=0.062]{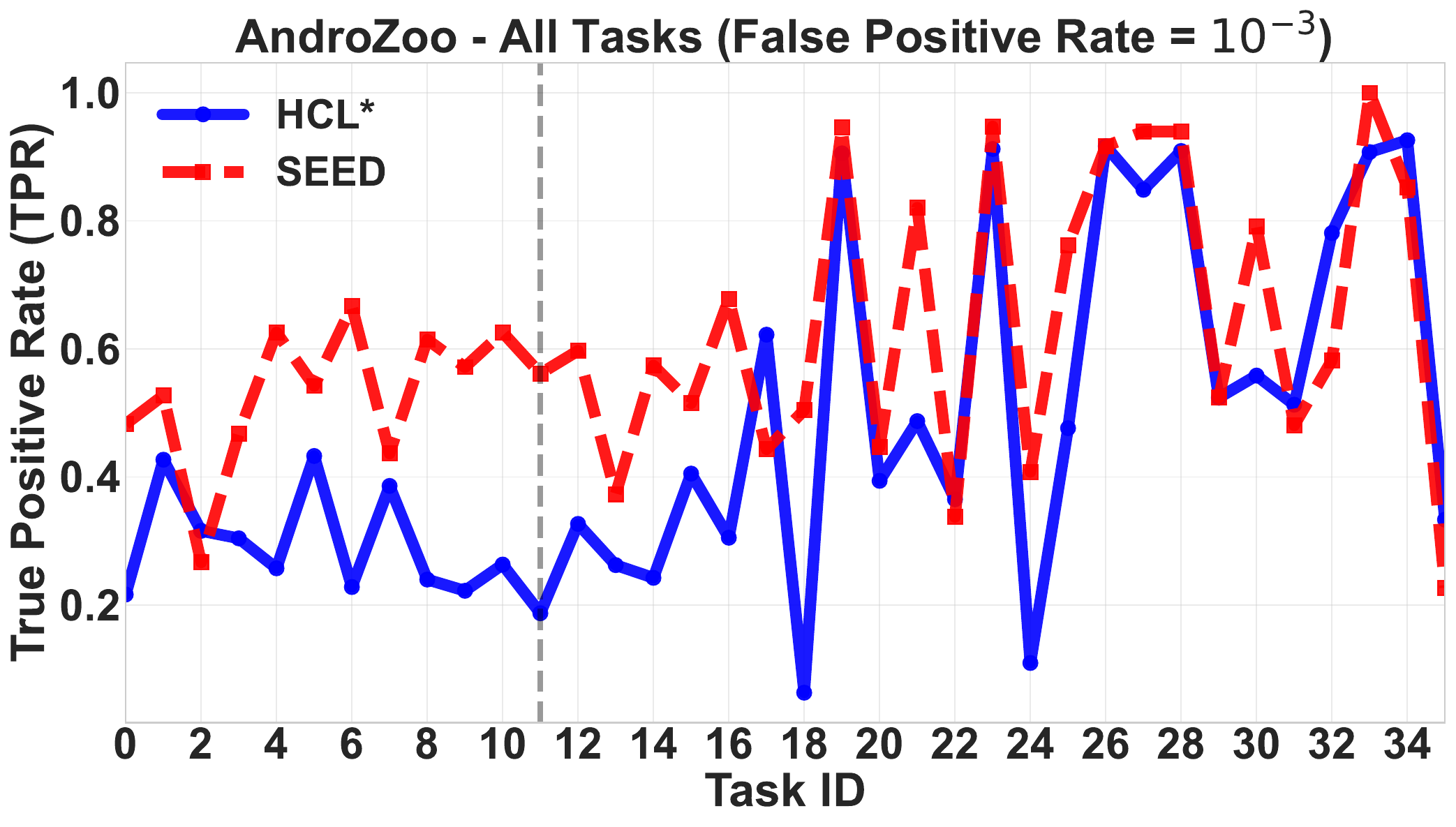}}
 \vspace{2mm}
\makebox[\linewidth][c]{%
\begin{tikzpicture}
\draw[decorate,decoration={brace,mirror,amplitude=6pt}]
(-0.2cm,0) -- (\linewidth+0.2cm,0)
node[midway,yshift=-12pt]{\textbf{HCL$^*$ Vs SEED}};
\end{tikzpicture}%
}
\end{subfigure}

 \Description{about the figure}
\caption{\textcolor{black}{Operating-point comparison of SEED with baseline detectors on the AndroZoo dataset at fixed false positive rate (FPR) thresholds. 
The left column presents CADE vs SEED, the right column shows HCL vs SEED, and the bottom panel shows HCL$^{*}$ vs SEED. 
Within each comparison group, the three plots correspond to FPR levels $10^{-1}$, $10^{-2}$, and $10^{-3}$. 
Across all plots, the red curves denote SEED, whereas the blue curves represent the corresponding baseline methods (CADE, HCL, or HCL$^{*}$).}}
\label{fig:operating-comparison-androzoo}
\end{figure*}
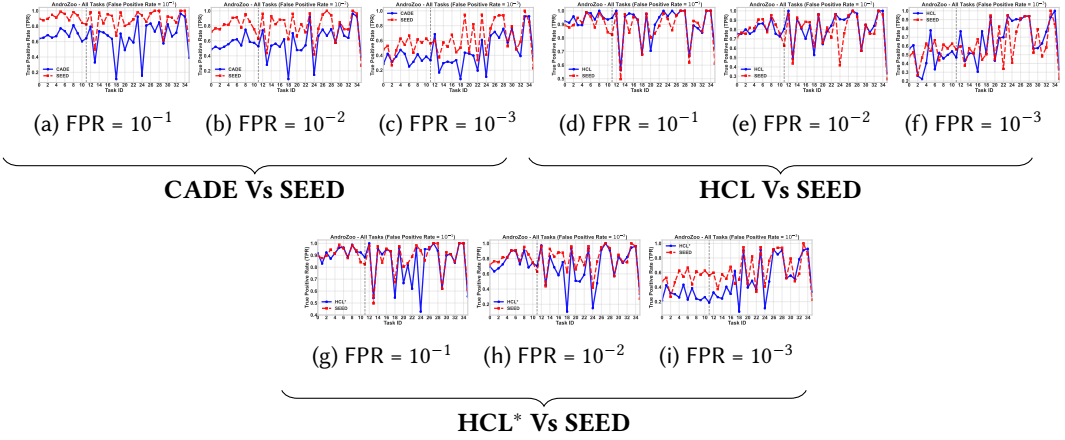
\begin{figure*}[!b]

\centering 
\begin{subfigure}{0.47\textwidth}
  \subfloat[FPR = $10^{-1}$]{\includegraphics[scale=0.062]{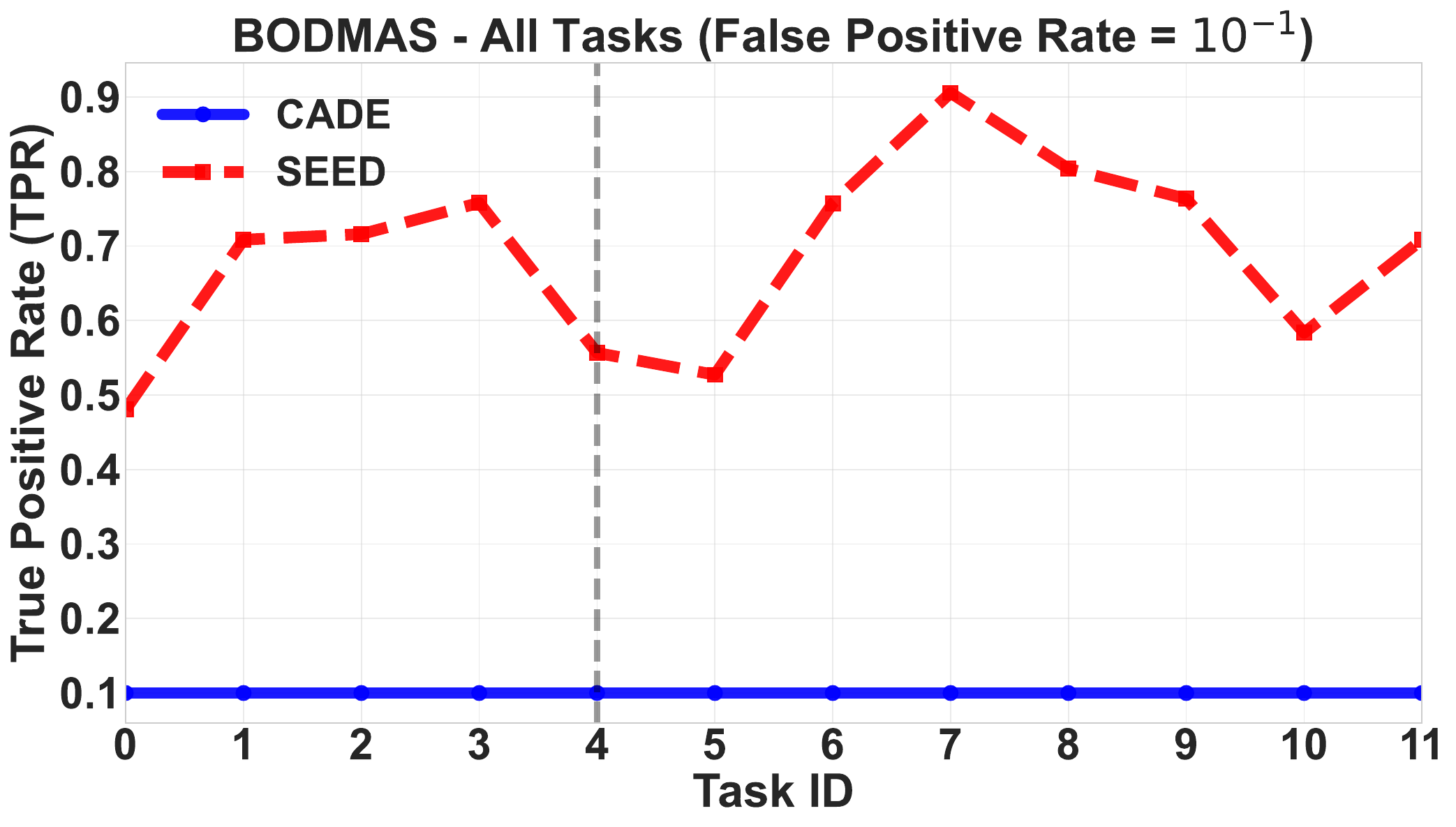}}
\subfloat[FPR = $10^{-2}$]{\includegraphics[scale=0.062]{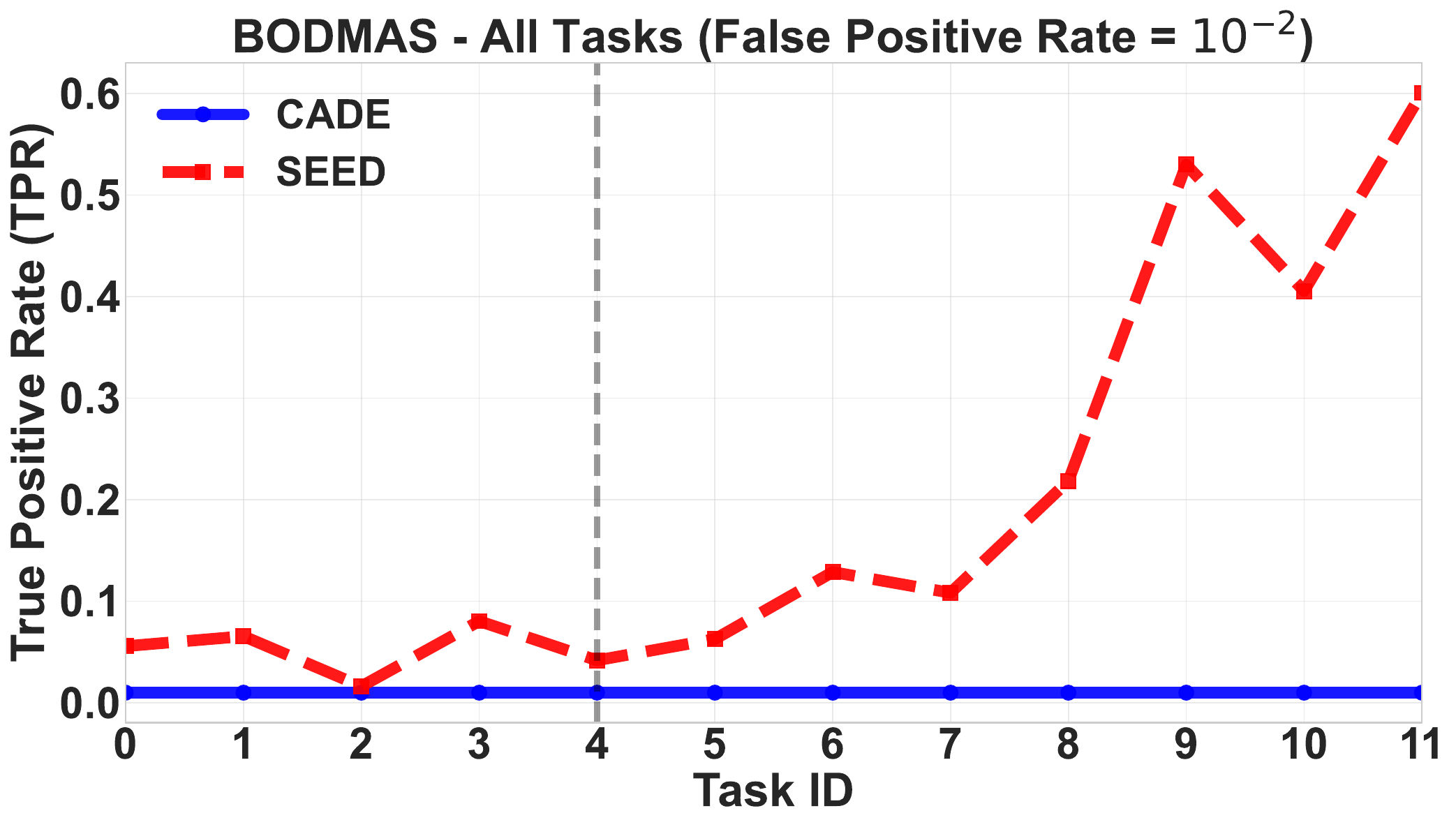}}
\subfloat[FPR = $10^{-3}$]{\includegraphics[scale=0.062]{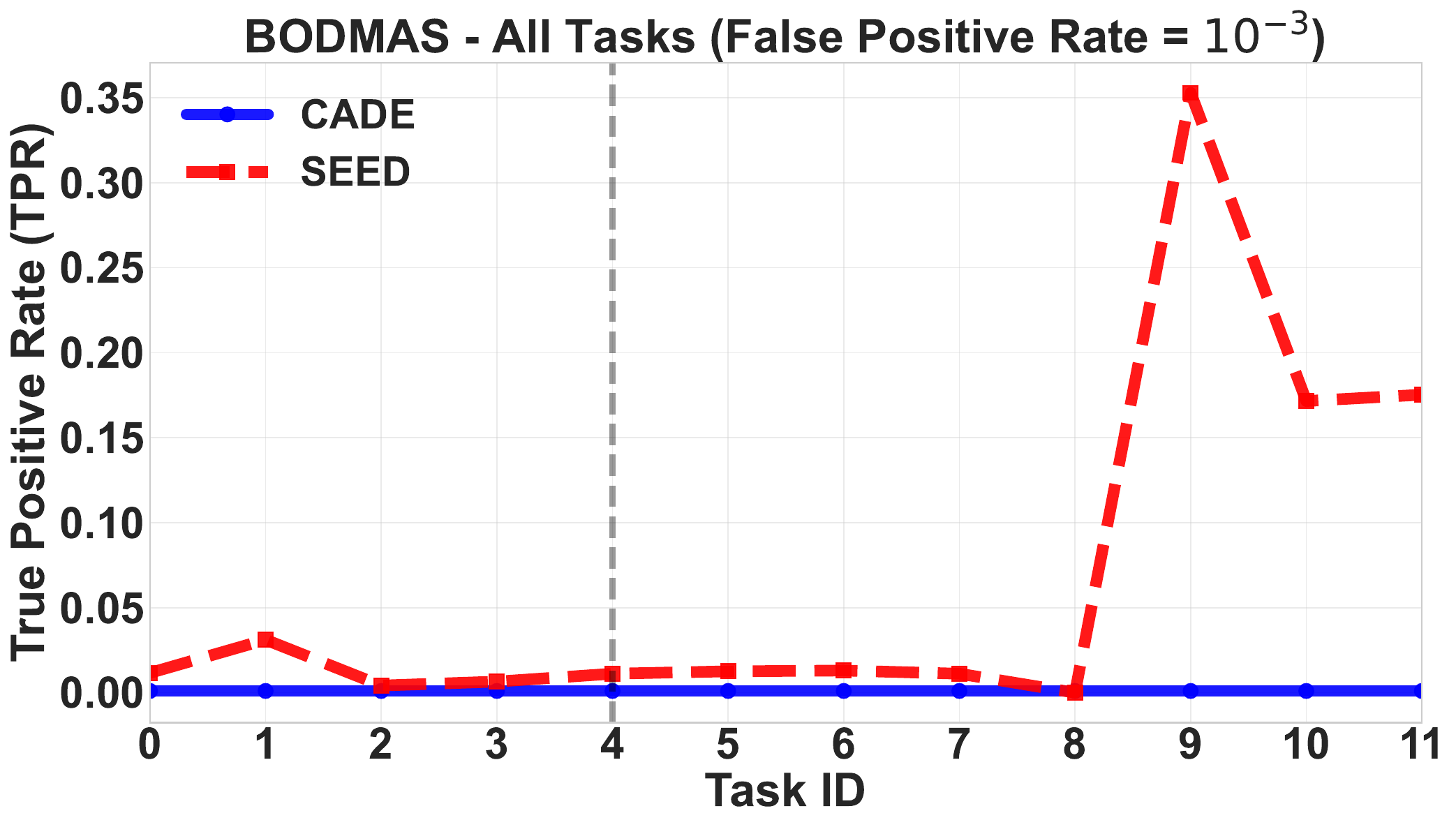}}%
\vspace{2mm}
\makebox[\linewidth][c]{%
\begin{tikzpicture}
\draw[decorate,decoration={brace,mirror,amplitude=6pt}]
(-0.2cm,0) -- (\linewidth+0.2cm,0)
node[midway,yshift=-12pt]{\textbf{CADE Vs SEED}};
\end{tikzpicture}%
}
\end{subfigure}
\hfill
\begin{subfigure}{0.47\textwidth}
\subfloat[FPR = $10^{-1}$]{\includegraphics[scale=0.062]{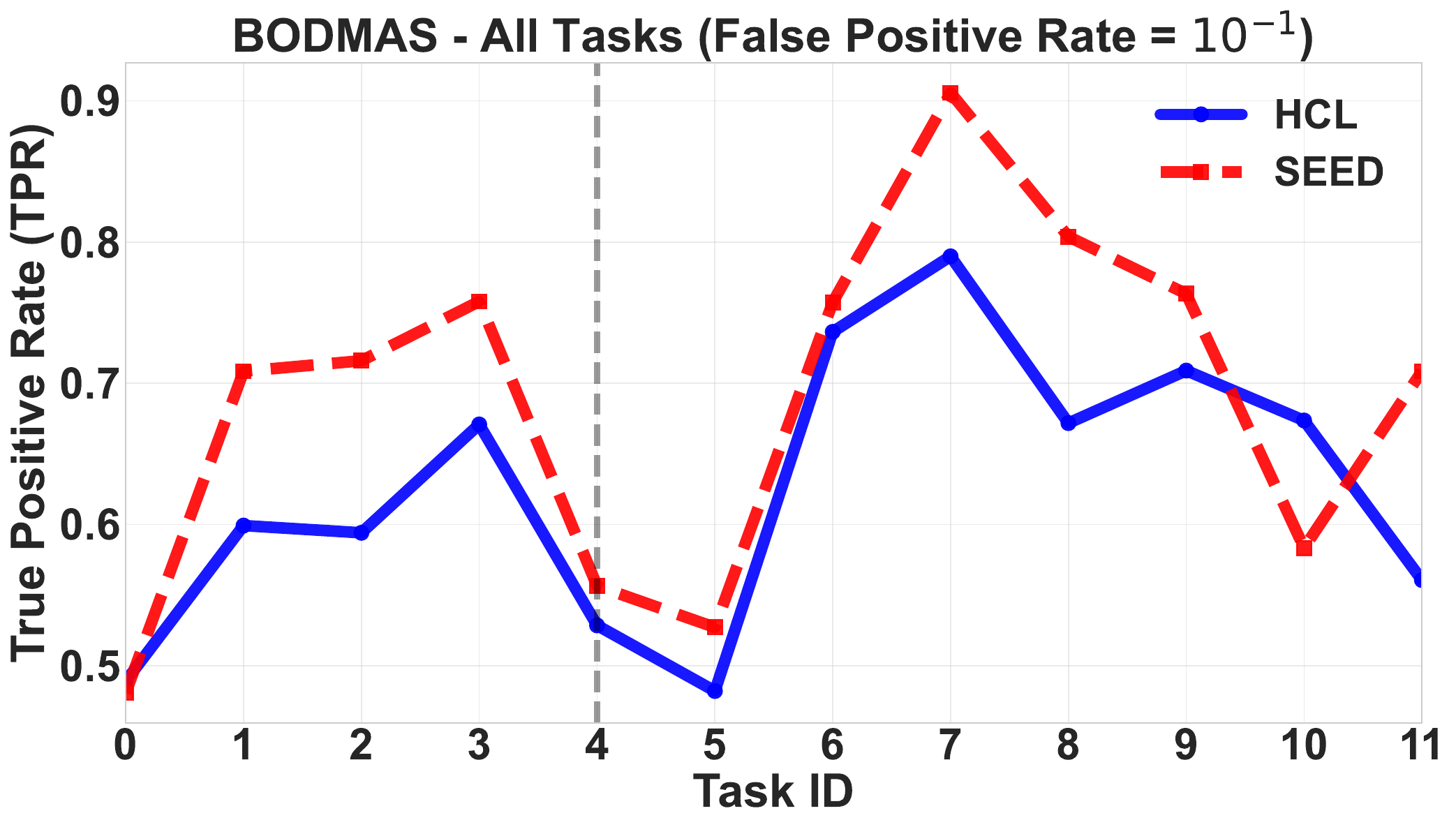}}
\subfloat[FPR = $10^{-2}$]{\includegraphics[scale=0.062]{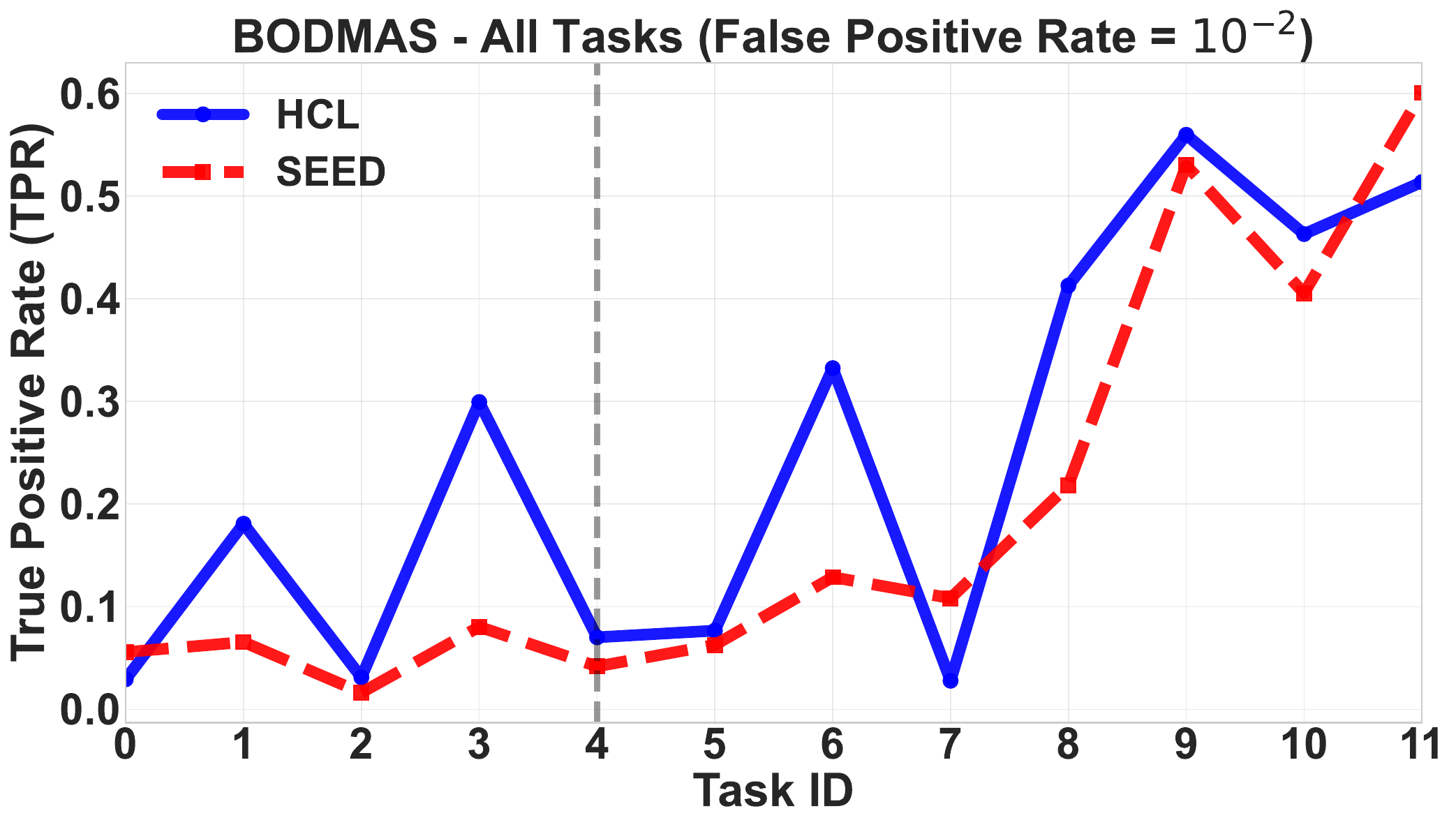}}
\subfloat[FPR = $10^{-3}$]{\includegraphics[scale=0.062]{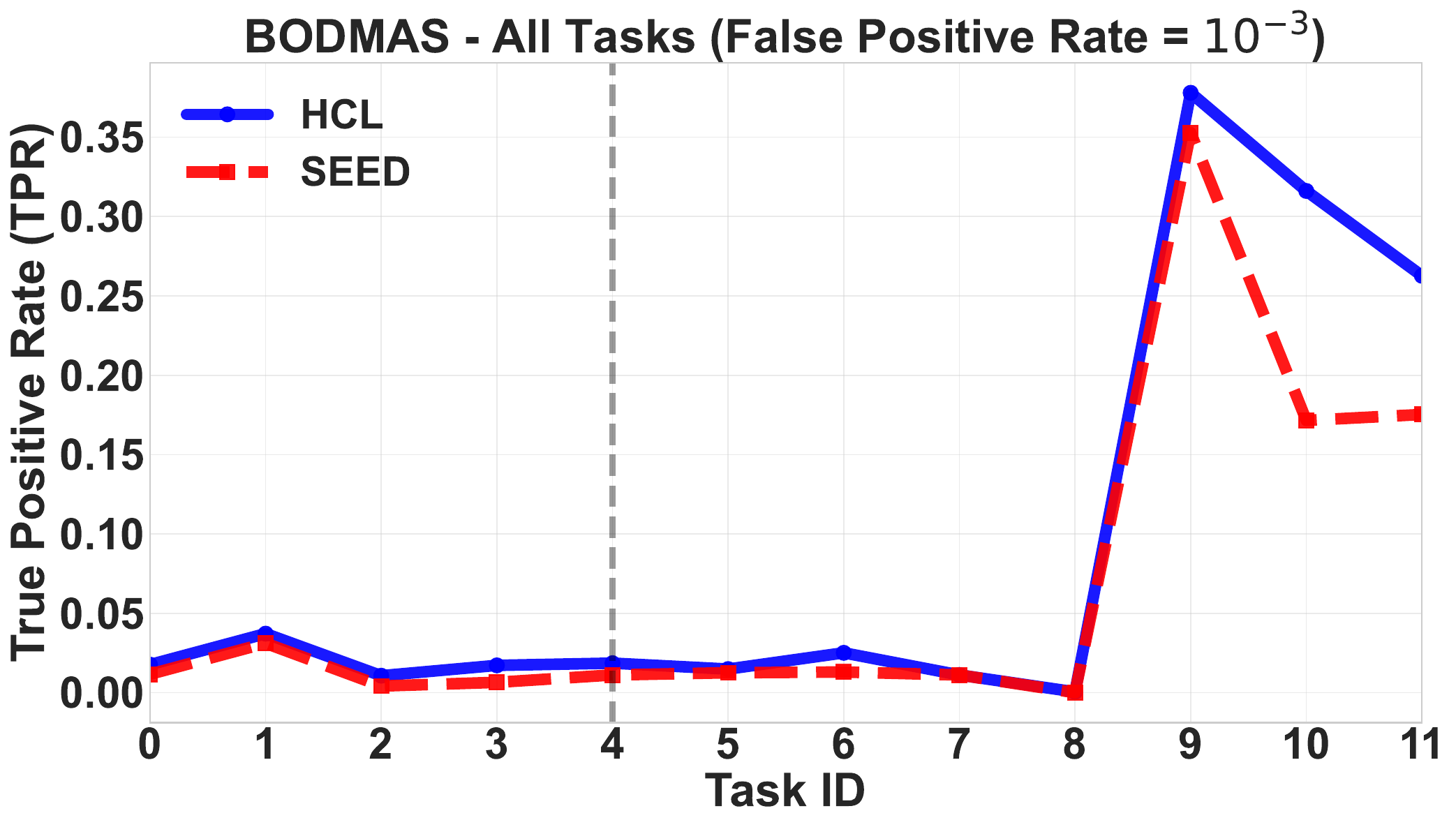}}
\vspace{2mm}
\makebox[\linewidth][c]{%
\begin{tikzpicture}
\draw[decorate,decoration={brace,mirror,amplitude=6pt}]
(-0.2cm,0) -- (\linewidth+0.2cm,0)
node[midway,yshift=-12pt]{\textbf{HCL Vs SEED}};
\end{tikzpicture}%
}
\end{subfigure}
\hfill
\vspace{0.3cm}
\begin{subfigure}{0.48\textwidth}
\subfloat[FPR = $10^{-1}$]{\includegraphics[scale=0.065]{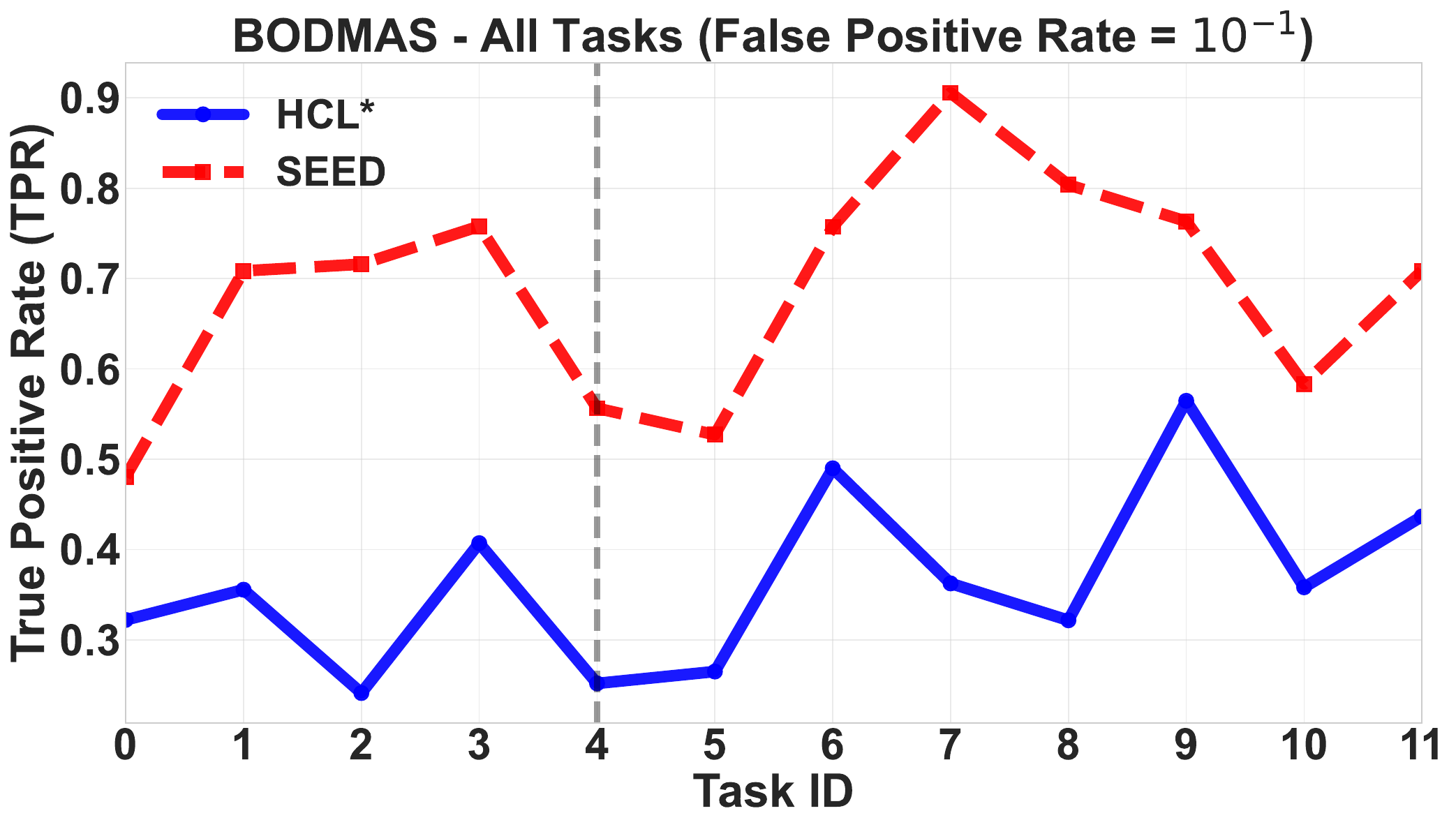}}
\subfloat[FPR = $10^{-2}$]{\includegraphics[scale=0.065]{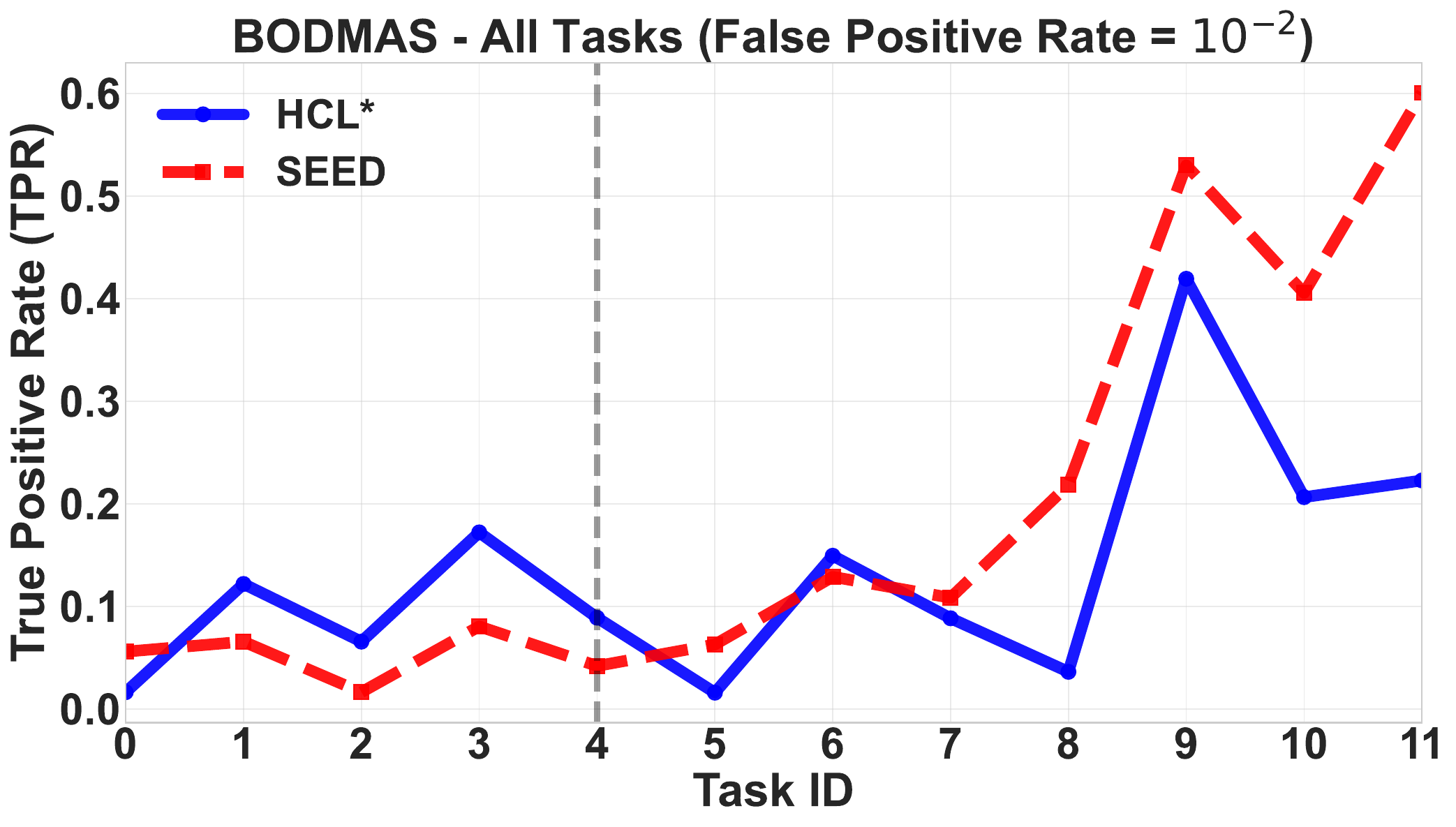}}
\subfloat[FPR = $10^{-3}$]{ \includegraphics[scale=0.065]{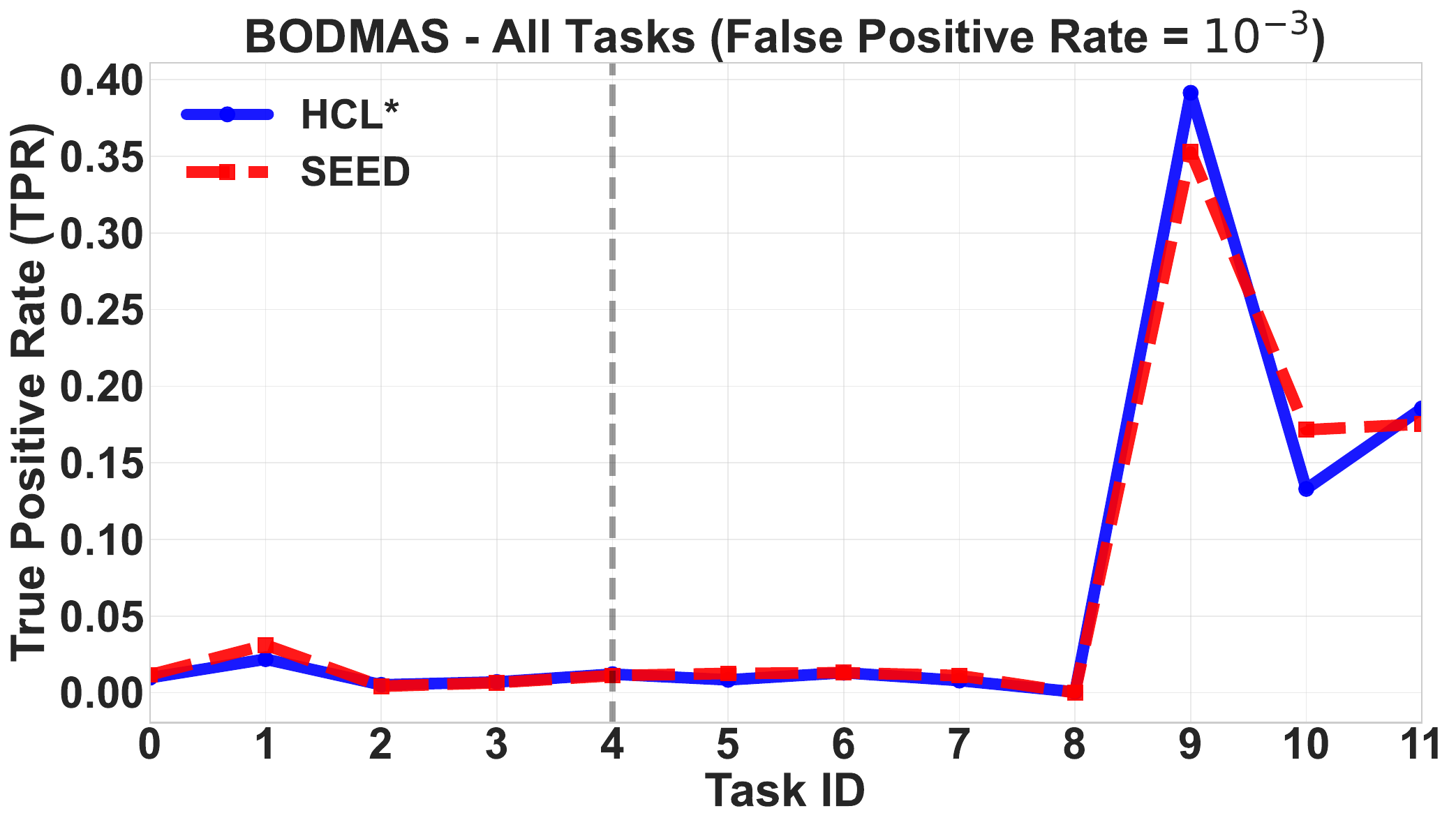}}
\vspace{2mm}
\makebox[\linewidth][c]{%
\begin{tikzpicture}
\draw[decorate,decoration={brace,mirror,amplitude=6pt}]
(-0.2cm,0) -- (\linewidth+0.2cm,0)
node[midway,yshift=-12pt]{\textbf{HCL$^*$ Vs SEED}};
\end{tikzpicture}%
}

\end{subfigure}
 \Description{about the figure}
\caption{\textcolor{black}{Operating-point comparison of SEED with baseline detectors on the BODMAS dataset at fixed false positive rate (FPR) thresholds. 
The left column presents CADE vs SEED, the right column shows HCL vs SEED, and the bottom panel shows HCL$^{*}$ vs SEED. 
Within each comparison group, the three plots correspond to FPR levels $10^{-1}$, $10^{-2}$, and $10^{-3}$. 
Across all plots, the red curves denote SEED, whereas the blue curves represent the corresponding baseline methods (CADE, HCL, or HCL$^{*}$).}}
\label{fig:fig:operating-comparison-bodmas}
\end{figure*}
\begin{figure*}

\centering 
\begin{subfigure}{0.99\textwidth}
\subfloat[FPR = $10^{-1}$]{\includegraphics[scale=0.05]{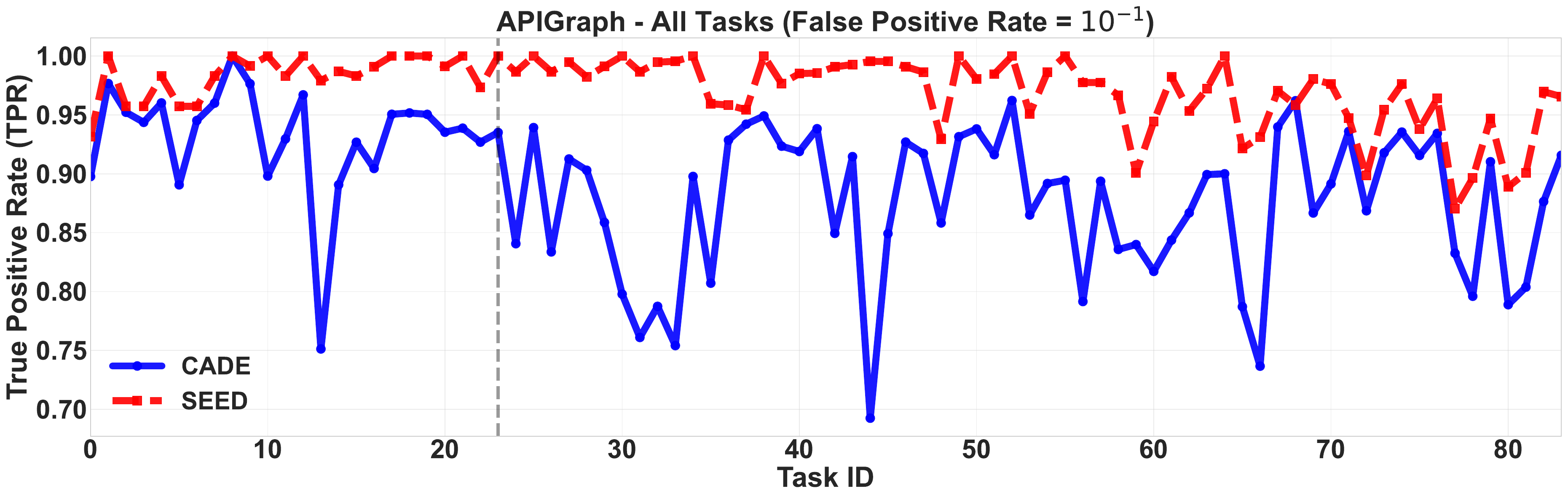}}
\hspace{0.1cm}
\subfloat[FPR = $10^{-2}$]{\includegraphics[scale=0.05]{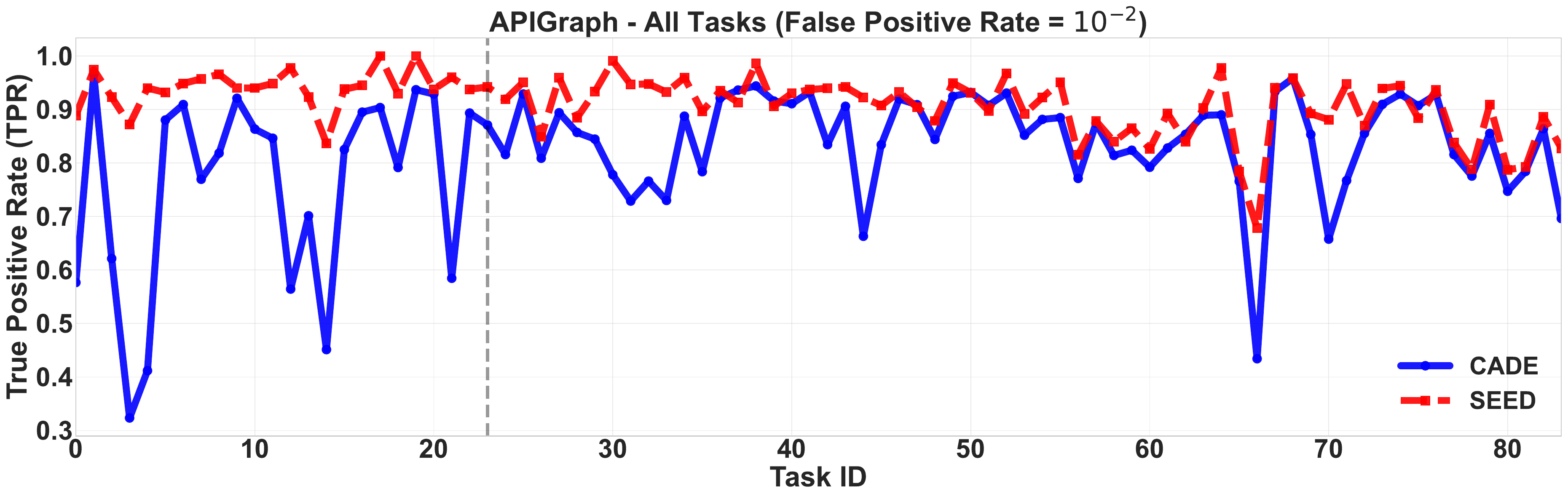}}
\hspace{0.1cm}
\subfloat[FPR = $10^{-3}$]{\includegraphics[scale=0.05]{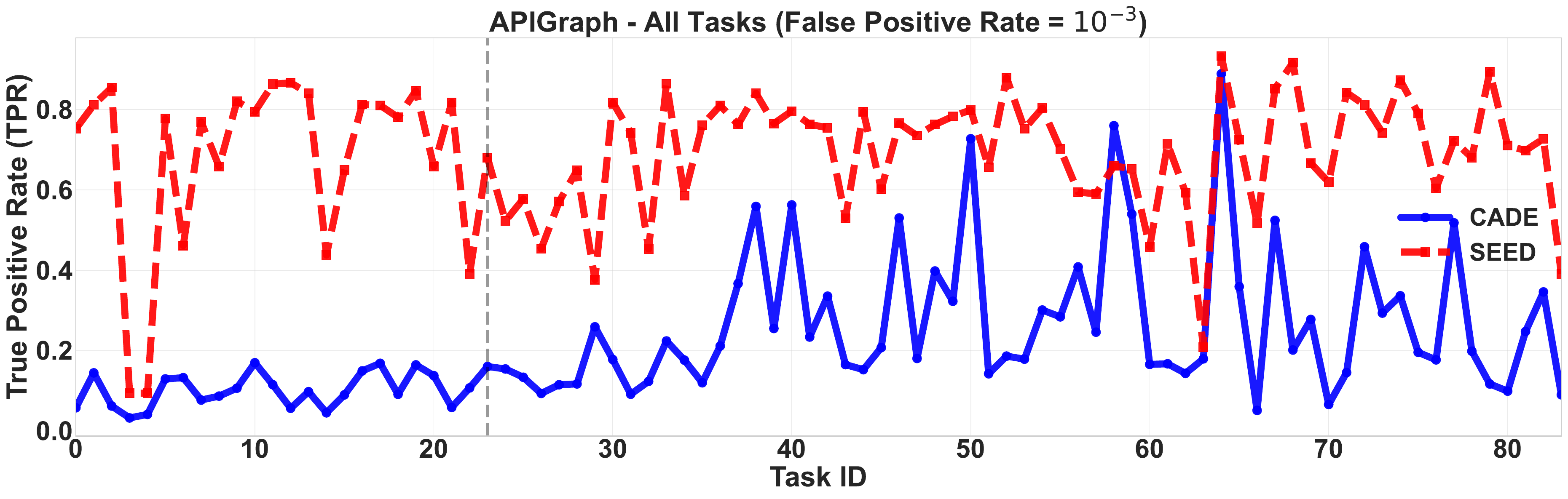}}
\hspace{0.1cm}
\subfloat[FPR = $10^{-4}$]{\includegraphics[scale=0.05]{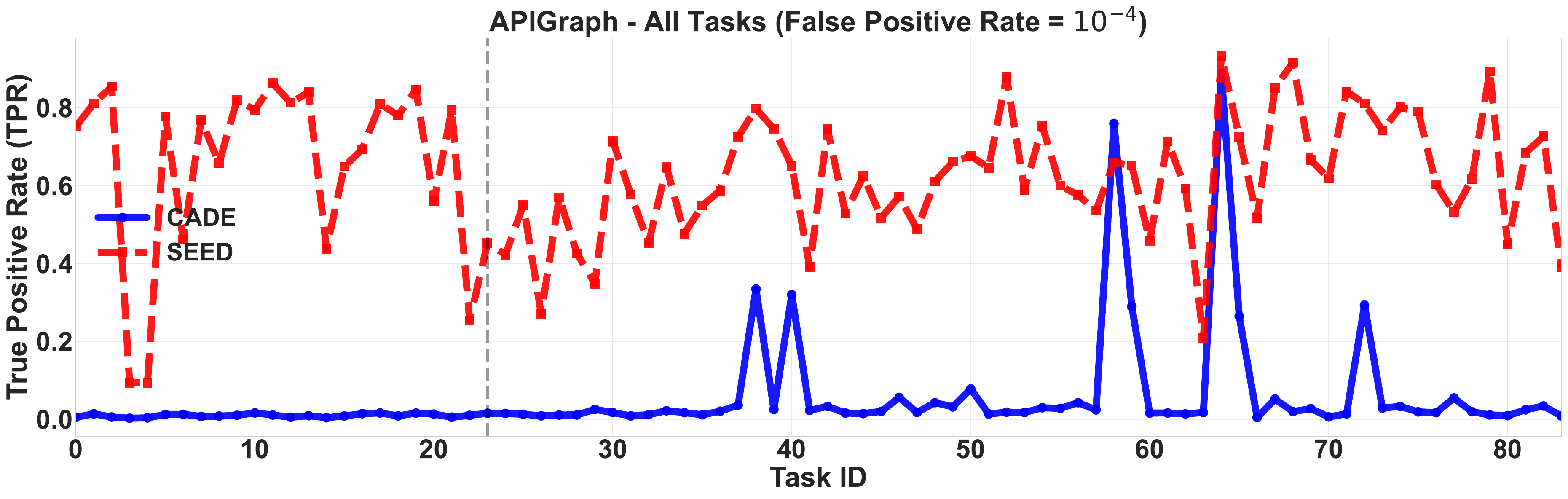}}%
\vspace{2mm}
\makebox[\linewidth][c]{%
\begin{tikzpicture}
\draw[decorate,decoration={brace,mirror,amplitude=6pt}]
(-0.2cm,0) -- (\linewidth+0.2cm,0)
node[midway,yshift=-12pt]{\textbf{CADE Vs SEED}};
\end{tikzpicture}%
}
\end{subfigure}
\hfill
\hspace{0.3cm}

\begin{subfigure}{0.99\textwidth}
\subfloat[FPR = $10^{-1}$]{\includegraphics[scale=0.05]{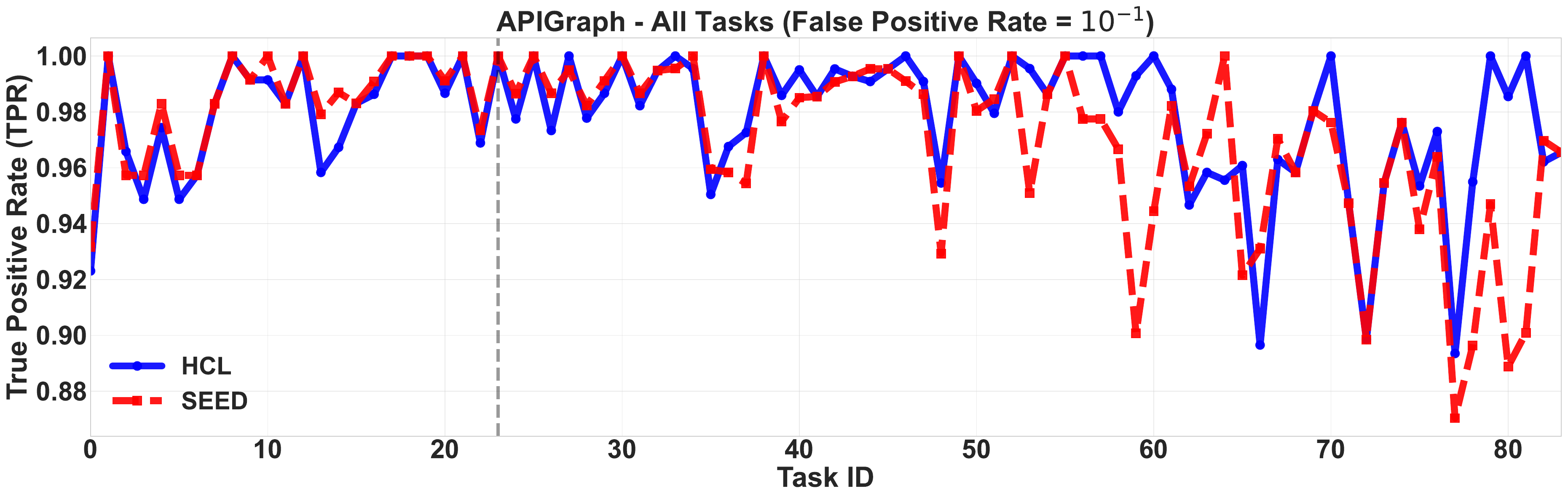}}
\hspace{0.1cm}
\subfloat[FPR = $10^{-2}$]{\includegraphics[scale=0.05]{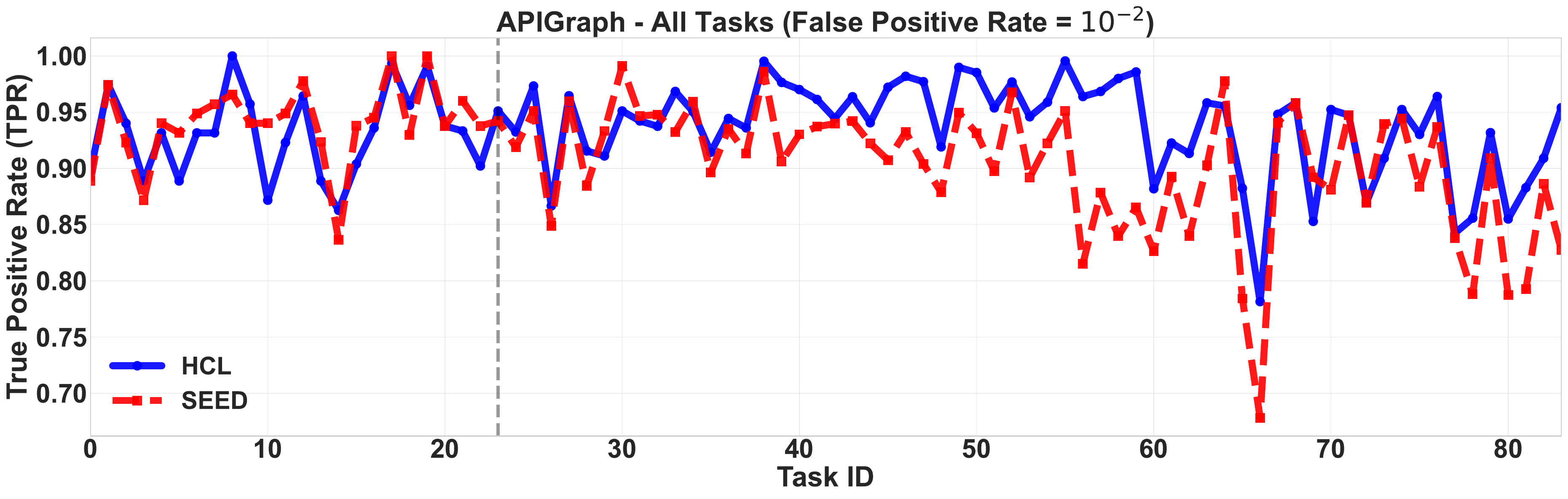}}
\hspace{0.1cm}
\subfloat[FPR = $10^{-3}$]{\includegraphics[scale=0.05]{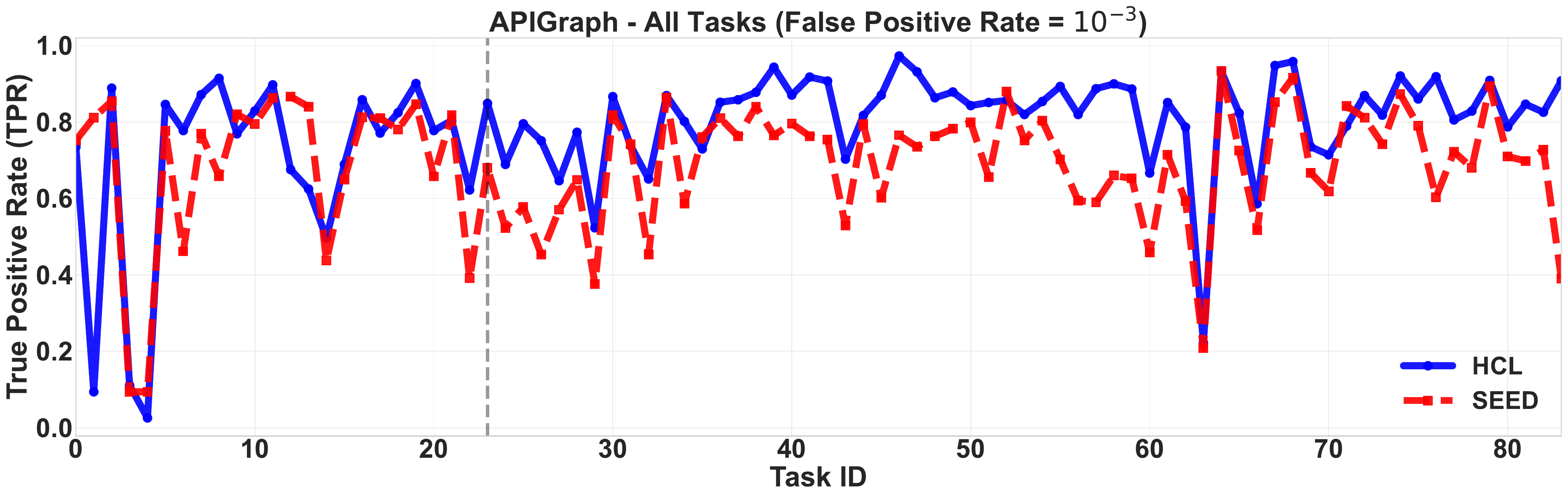}}
\hspace{0.1cm}
\subfloat[FPR = $10^{-4}$]{\includegraphics[scale=0.05]{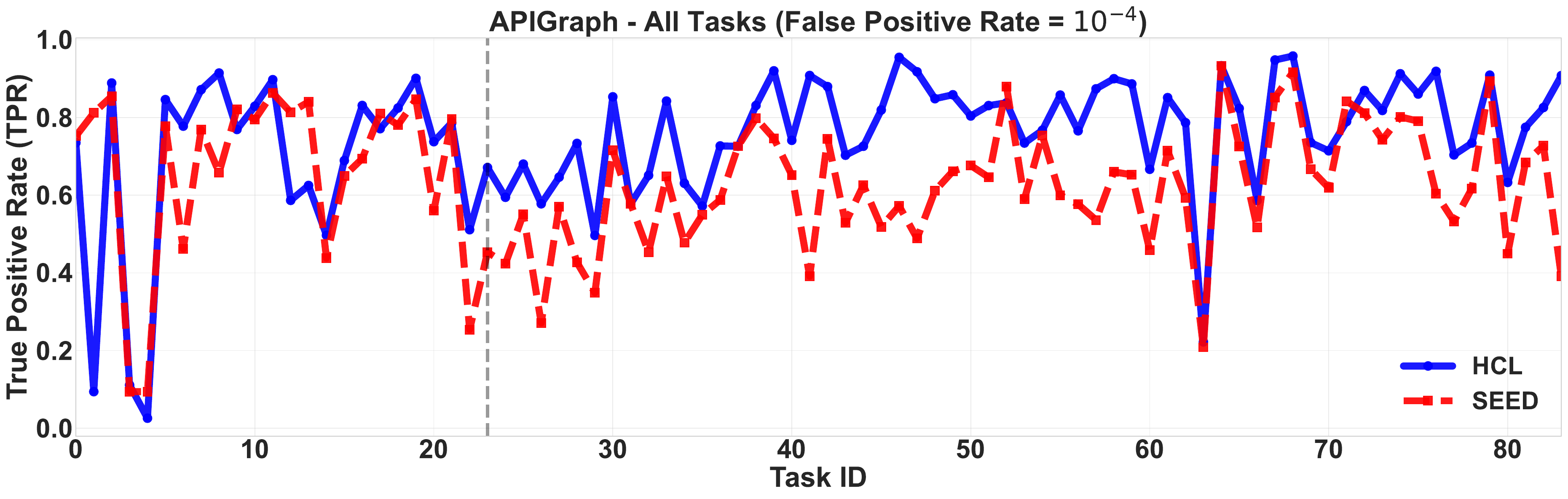}}
\vspace{2mm}
\makebox[\linewidth][c]{%
\begin{tikzpicture}
\draw[decorate,decoration={brace,mirror,amplitude=6pt}]
(-0.2cm,0) -- (\linewidth+0.2cm,0)
node[midway,yshift=-12pt]{\textbf{HCL Vs SEED}};
\end{tikzpicture}%
}
\end{subfigure}
\hfill
\hspace{0.3cm}

\begin{subfigure}{0.99\textwidth}
\subfloat[FPR = $10^{-1}$]{\includegraphics[scale=0.05]{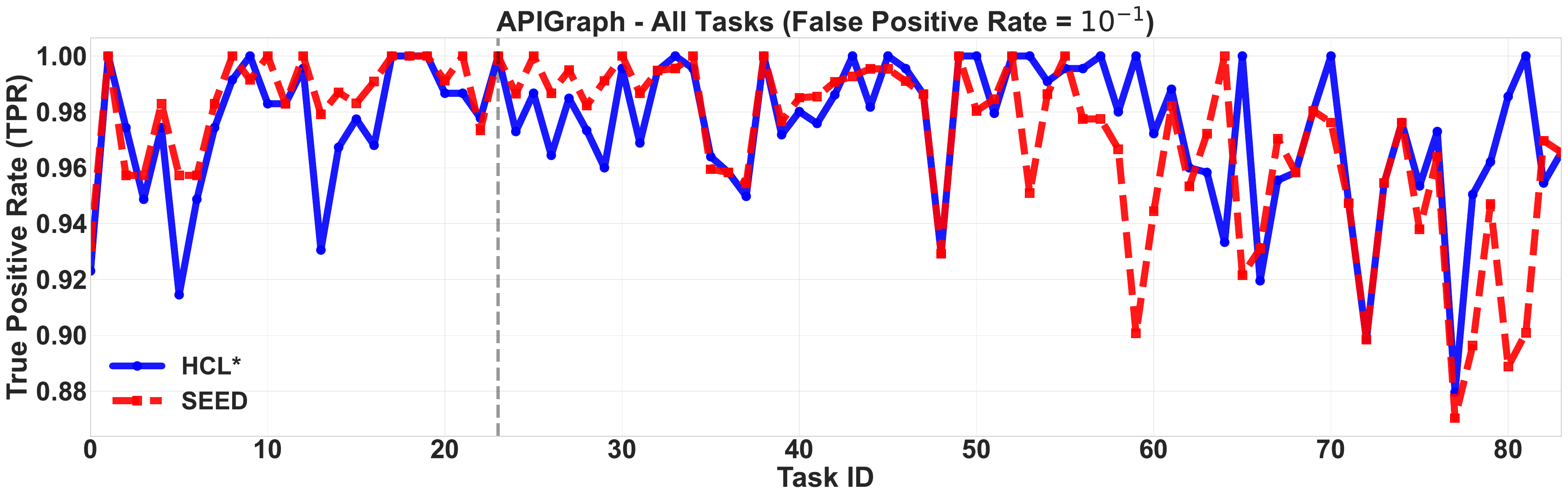}}
\hspace{0.1cm}
\subfloat[FPR = $10^{-2}$]{\includegraphics[scale=0.05]{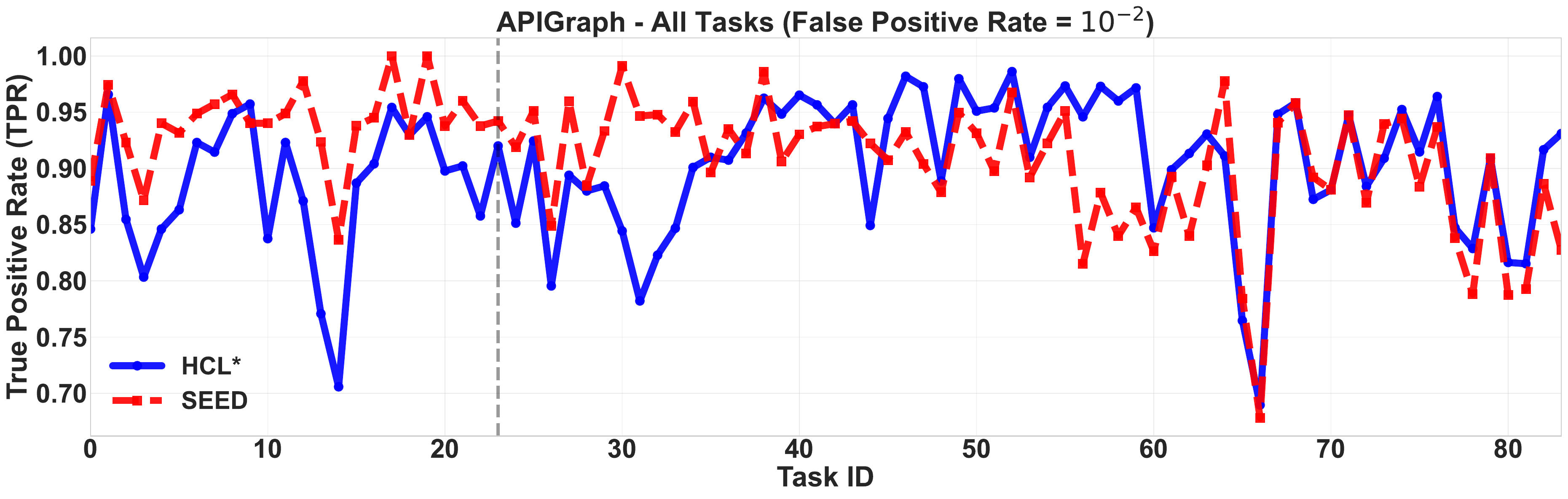}}
\hspace{0.1cm}
\subfloat[FPR = $10^{-3}$]{\includegraphics[scale=0.05]{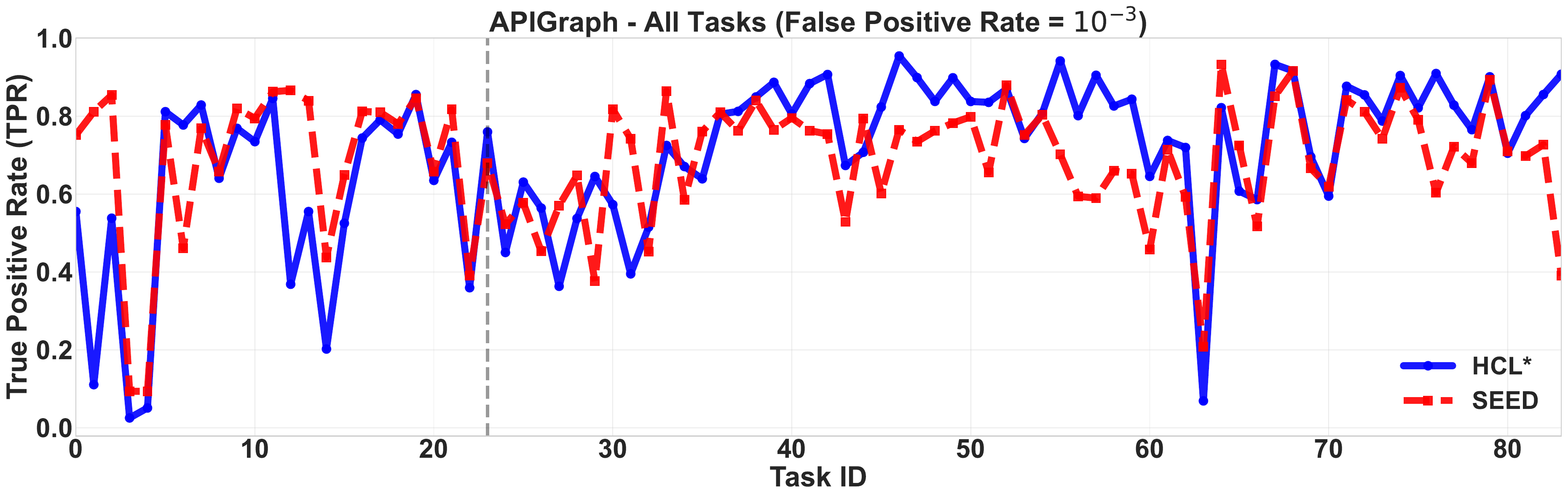}}
\hspace{0.1cm}
\subfloat[FPR = $10^{-4}$]{ \includegraphics[scale=0.05]{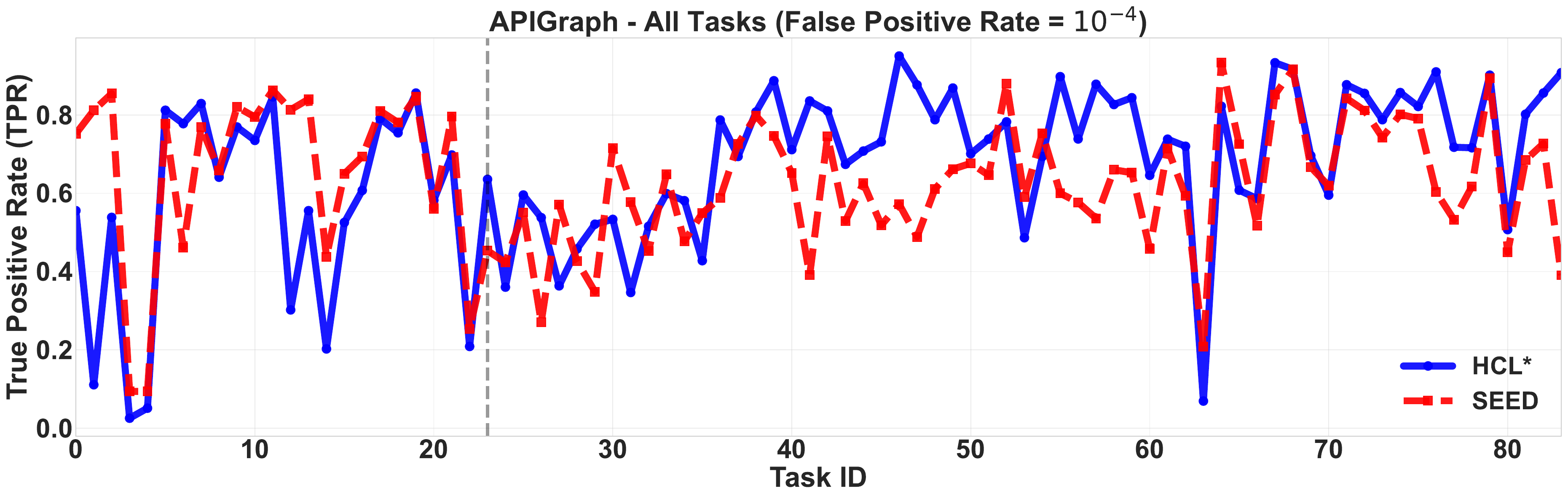}}
\vspace{2mm}
\makebox[\linewidth][c]{%
\begin{tikzpicture}
\draw[decorate,decoration={brace,mirror,amplitude=6pt}]
(-0.2cm,0) -- (\linewidth+0.2cm,0)
node[midway,yshift=-12pt]{\textbf{HCL$^*$ Vs SEED}};
\end{tikzpicture}%
}
\end{subfigure}
 \Description{about the figure}
\caption{\textcolor{black}{Operating-point comparison of SEED with baseline detectors on the APIGraph dataset at fixed false positive rate (FPR) thresholds. 
The top, middle, and bottom rows show CADE vs SEED, HCL vs SEED, and HCL$^{*}$ vs SEED, respectively. 
Each column corresponds to an FPR constraint of $10^{-1}$, $10^{-2}$, $10^{-3}$, and $10^{-4}$. 
Across all plots, red curves denote SEED and blue curves denote the corresponding baseline method.}}
\label{fig:fig:operating-comparison-apigraph}
\end{figure*}


\paragraph{AndroZoo.} \textcolor{black}{Figure~\ref{fig:operating-comparison-androzoo} presents the operating-point analysis of SEED against CADE, HCL, and HCL$^{*}$ on the AndroZoo dataset. In the CADE vs SEED comparison (left block), both methods achieve high detection rates at FPR=$10^{-1}$; however, SEED shows smoother performance evolution across tasks, while CADE exhibits more variability. As the operating constraint becomes stricter (FPR=$10^{-2}$ and $10^{-3}$), SEED preserves higher detection rates across a larger portion of the task sequence. In the HCL vs SEED comparison (right block), SEED maintains a relatively stable TPR across tasks, particularly at stricter operating points. Finally, in the HCL$^{*}$ vs SEED comparison (bottom block), SEED achieves detection performance comparable to the contrastive baseline, with \method{} often maintaining competitive or slightly higher performance in later tasks.}

\paragraph{BODMAS} \textcolor{black}{The operating-point comparison on the BODMAS dataset is illustrated in Figure~\ref{fig:fig:operating-comparison-bodmas}. In the CADE vs SEED comparison (left block of Figure~\ref{fig:fig:operating-comparison-bodmas}), SEED consistently maintains a higher or comparable detection rate across most tasks. In the HCL vs SEED comparison (right block), at FPR=$10^{-1}$, both methods achieve relatively strong performance across the early tasks. However, at the stricter FPRs ($10^{-2}$ and $10^{-3}$), SEED consistently maintains higher TPR values across several tasks, whereas HCL exhibits more pronounced fluctuations. The comparison with HCL$^{*}$ (bottom block) shows that SEED achieves detection performance that is consistently comparable to the contrastive baseline across the task sequence. Further, SEED maintains a stable detection trajectory across tasks.}

\paragraph{APIGraph.} \textcolor{black}{Figure~\ref{fig:fig:operating-comparison-apigraph} presents the operating-point evaluation on the APIGraph dataset. In the CADE vs SEED comparison (top row of Figure~\ref{fig:fig:operating-comparison-apigraph}), SEED maintains detection performance that is consistently comparable to or slightly higher than CADE across most tasks. The HCL vs SEED comparison (middle row) shows that both detectors achieve high detection rates across the task sequence; in particular, SEED maintains a more stable trajectory across tasks at the stricter operating points. The comparison with HCL$^{*}$ (bottom row) reveals that SEED achieves performance that remains highly competitive with the contrastive baseline across all evaluated operating points. Although the performance gap is smaller in this dataset, SEED maintains stable detection performance even at the strictest operating point (FPR=$10^{-4}$).}

\subsection{Effect of Self-labeling on Detection Performance}

\begin{wrapfigure}{l}{0.45\textwidth}
\vspace{-4mm }
\includegraphics[scale=0.30]{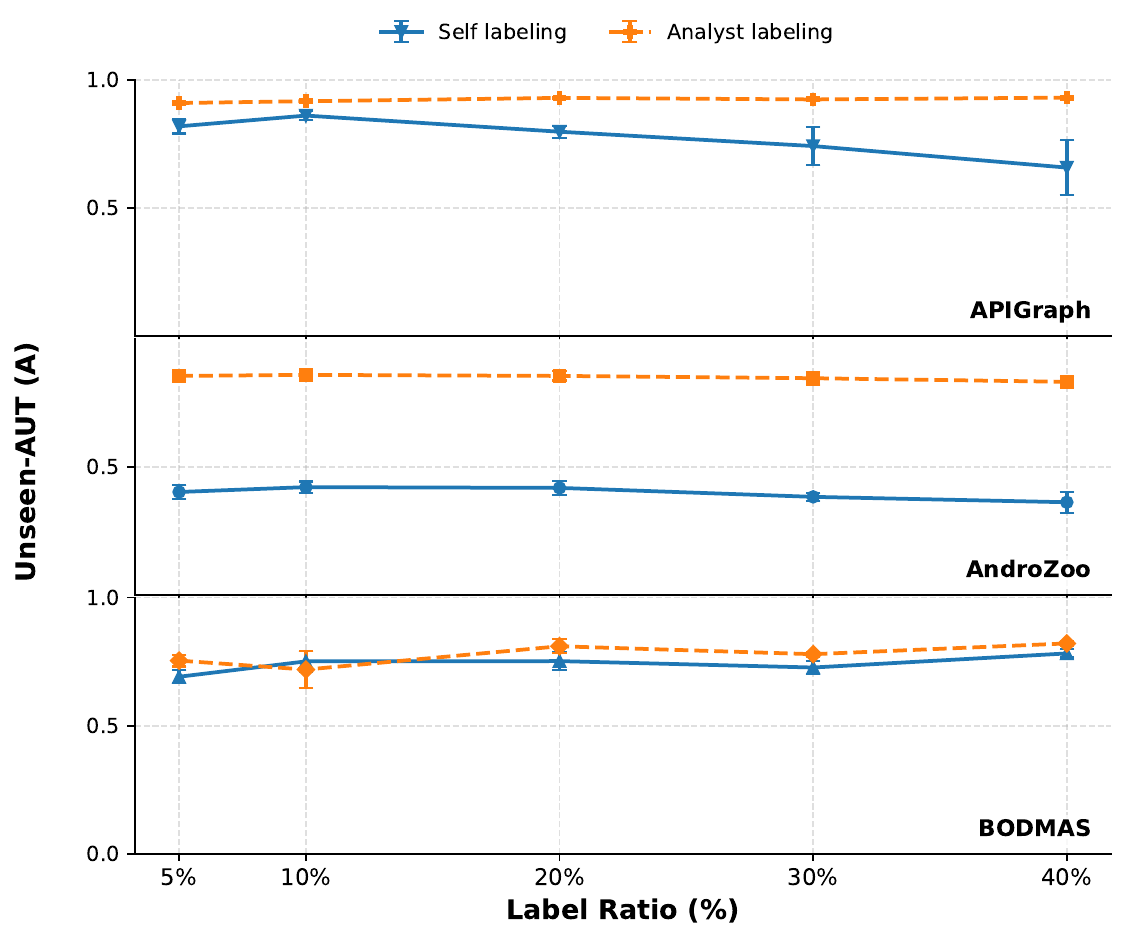}
\caption{\textcolor{black}{Comparing the detection performance (Unseen-AUT(A)) for self-labeling and analyst labeling with varying labeled data ratio. A pronounced gap is observed on AndroZoo, BODMAS remains comparatively close, and APIGraph shows stable analyst performance while self-labeling degrades at higher ratios.}}
\label{fig:seld_labeling}
\end{wrapfigure}

\textcolor{black}{In this subsection, we compare two labeling strategies for samples from unseen tasks: self-labeling, where the detector assigns labels and analyst labeling, where a security expert provides ground-truth annotations (see Figure~\ref{fig:seld_labeling}). We focus on generalization of self-labeling strategy and report the using unseen-AUT(A) values, since this setting best reflects deployment-time robustness. Across datasets, the results show that increasing the amount of labeled data is not always sufficient and the quality of the labels can be the dominant factor.}

\paragraph{APIGraph} \textcolor{black}{Analyst labeling remains both high and stable across all the labeled data ratios, consistently around $0.91$--$0.93$ unseen-AUT(A), with only minor variation. Self-labeling, however, shows a clear degradation as the labeled data ratio increases. The  unseen-AUT(A) value is $0.819$ at $5\%$, becomes $0.861$ at $10\%$, and then declines sharply to about $0.742$ at $30\%$ and roughly $0.658$ at $40\%$, with larger variability at the highest labeled data ratio.} 

\noindent \paragraph{AndroZoo} \textcolor{black}{On AndroZoo, analyst labeling yields consistently strong unseen generalization across all labeled data ratios. In contrast, self-labeling remains low. Specifically, unseen-AUT(A) value is about $0.402$ at $5\%$ and only reaches roughly $0.421$ at $10\%$, before dropping 
to around $0.362$ by $40\%$. This creates a persistent gap of roughly $0.4$--$0.5$ absolute points between the two strategies, indicating that self-labeling disproportionately harms performance on truly unseen tasks.}
\paragraph{BODMAS} \textcolor{black}{The performance of the two strategies are closer, and self-labeling is often competitive at lower labeled-data ratios. Unseen-AUT(A) under self-labeling rises from approximately $0.691$ at $5\%$ to about $0.782$ at $40\%$, while analyst labeling increases from around $0.754$ at $5\%$ to roughly $0.821$ at $40\%$. A small reversal is visible at $10\%$ (self-labeling at $\sim 0.751$ versus analyst labeling at $\sim 0.719$), but from $20\%$ onwards, analyst labeling consistently leads, with a modest but steady advantage. Overall, BODMAS appears relatively tolerant to self-labeling noise compared to the other datasets, although analyst-provided labels still deliver the best unseen-task performance at higher label ratios.}

\subsection{Effect of Distance Computation in Latent Space}

\begin{wrapfigure}{l}{0.45\textwidth}
\centering
\includegraphics[width=0.45\textwidth]{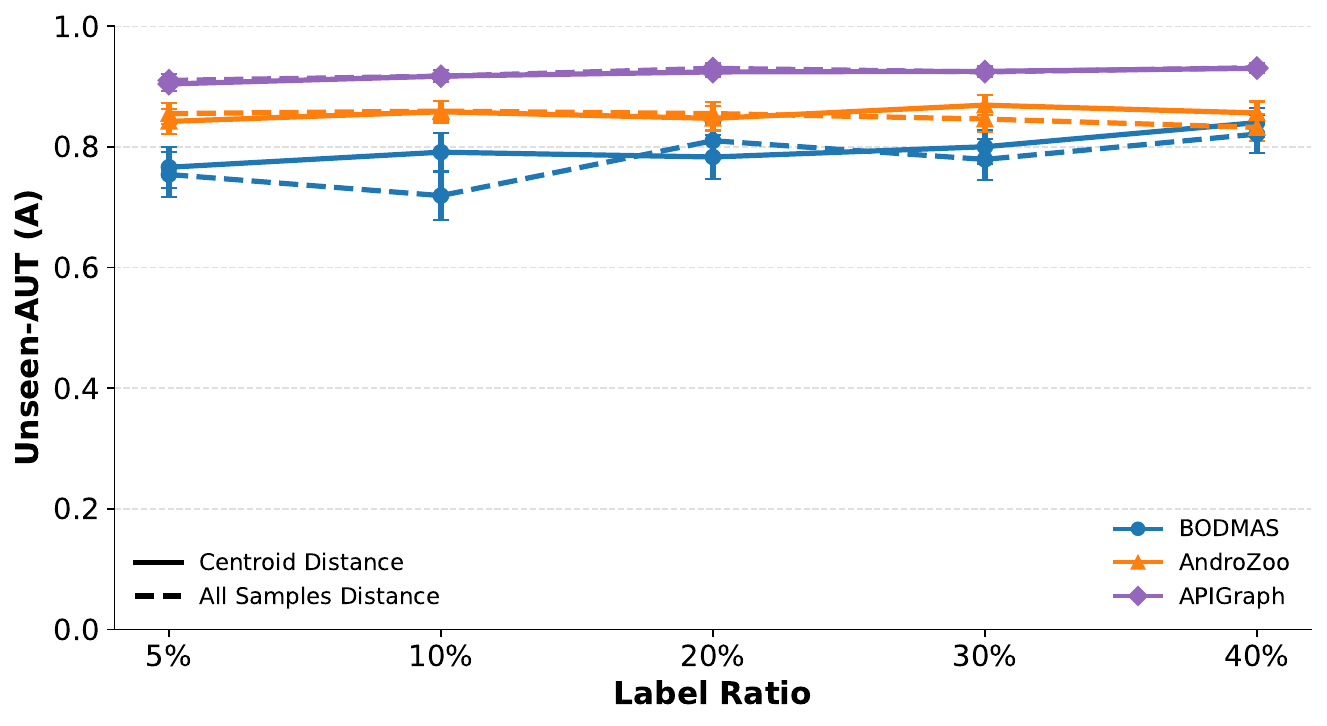}
\caption{\textcolor{black}{Comparing the detection performance (Unseen-AUT(A)) under centroid distance and all-samples distance strategies with varying labeled data ratios on BODMAS, AndroZoo, and APIGraph datasets. APIGraph consistently achieves the highest performance, while BODMAS and AndroZoo show gradual improvements as the labeled data ratio increases.}}
\label{fig:fig:comp_centroid}
\end{wrapfigure}

\textcolor{black}{In this section, we investigate how the choice of distance computation in the latent space affects detection performance under varying labeled data ratios. Specifically, we compare representing each class using a centroid against cosine distance computed against all labeled samples. Across datasets, the relative behavior between centroid and cosine distance varies depending on dataset characteristics and the amount of labeled data} (see Figure~\ref{fig:fig:comp_centroid}).

\paragraph{AndroZoo} \textcolor{black}{Both strategies maintain consistently high performance across all labeled ratios. With 5\% labeled data, the unseen AUT(A) exceeds $0.84$ for both variants, with centroid distance producing $0.842$ and the cosine distance strategy reaching $0.855$. As the labeled ratio increases to 10\%, the performance difference between the two approaches becomes minimal. }

\paragraph{APIGraph} \textcolor{black}{The APIGraph dataset demonstrates strong performance for both distance strategies even under limited labeled data. With only 5\% labeled data, the centroid distance method achieves an unseen AUT(A) of $0.904$, whereas the cosine distance method slightly improves this to $0.910$. Increasing the labeled ratio leads to gradual improvements for both methods, reaching approximately $0.917$ at 10\% labeled data. When the labeled ratio increases further, the gap between the two strategies becomes negligible.}

\paragraph{BODMAS} \textcolor{black}{On BODMAS, with only 5\% labeled data, the centroid distance method achieves an unseen AUT(A) of $0.766$, slightly higher than the $0.754$ obtained using cosine distance. Increasing the labeled ratio to 10\% improves the centroid-based performance to $0.791$, while the cosine-distance-based performance drops to $0.719$. As the labeled data ratio grows, both approaches improve, although their relative ranking fluctuates slightly. For instance, at 20\% labeled data, the cosine distance strategy reaches $0.810$, exceeding the centroid score of $0.783$, while at 30\% and 40\% labeled data the difference becomes subtle. Overall, the results indicate that summarizing each class using a centroid does not significantly improve generalization to unseen tasks.}

\subsection{Effect of SVD Threshold on Detection Performance}

\begin{wrapfigure}{l}{0.45\textwidth}
\includegraphics[scale=0.27]{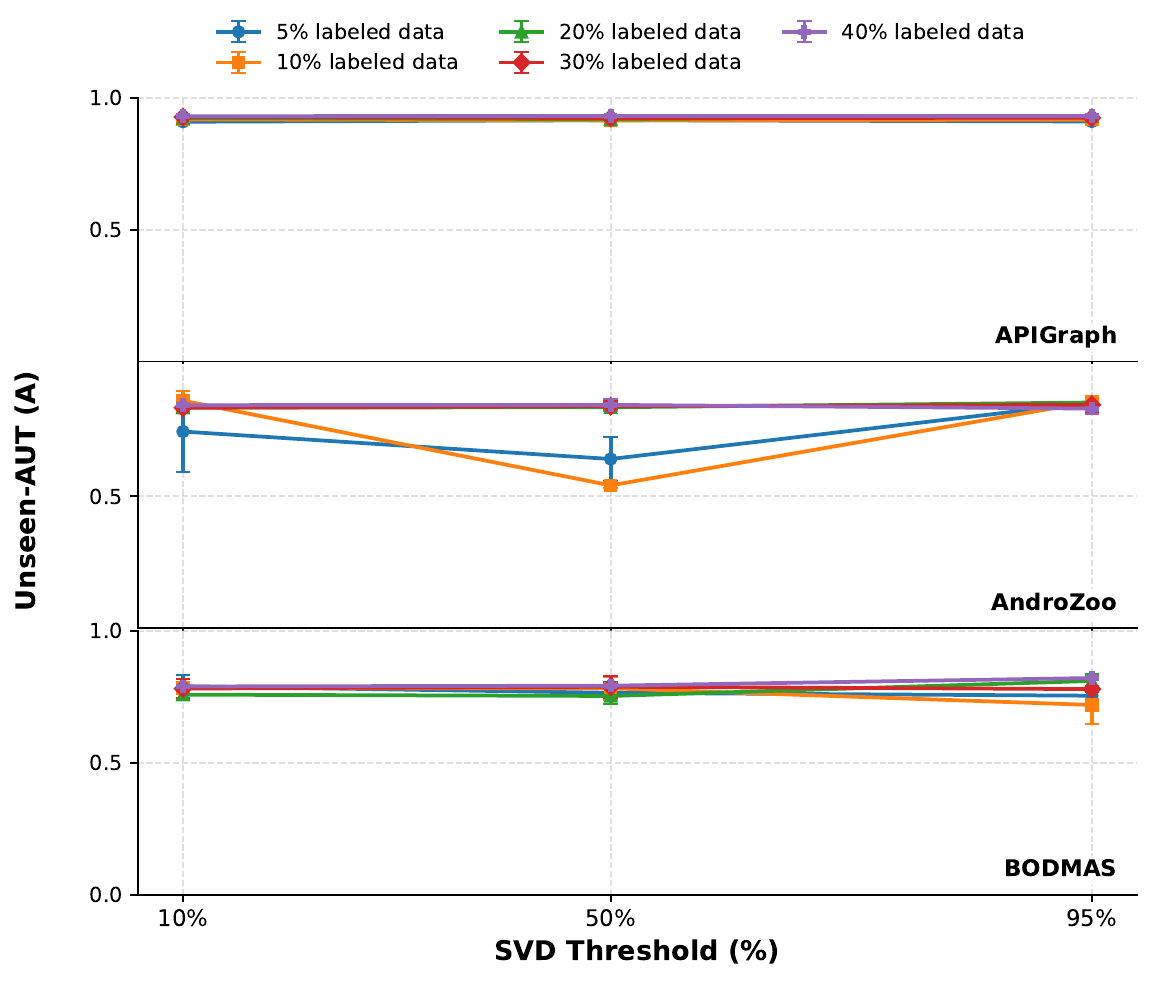}
\caption{\textcolor{black}{Comparing the detection performance (Unseen-AUT(A)) under varying SVD threshold with different labeled data ratios on BODMAS, AndroZoo, and APIGraph datasets. Overall, the $95\%$ SVD threshold consistently achieves competitive or superior performance across datasets, indicating that retaining a larger proportion of latent variance preserves discriminative information for unseen malware detection.}}
\label{fig:svd_threshold2}
\end{wrapfigure}

\textcolor{black}{In this section, we analyze how varying the SVD threshold affects the unseen malware detection performance under different labeled data ratios. We compare SVD thresholds of $10\%$, $50\%$, and $95\%$, and focus on unseen-AUT(A) to understand how retaining principal components influences generalization to unseen tasks. The overall trend in the figure indicates that the $95\%$ threshold consistently remains a strong and reliable choice across datasets (refer Figure~\ref{fig:svd_threshold2}).}

\paragraph{BODMAS} \textcolor{black}{The effect of the SVD threshold is visible but relatively stable compared to the other datasets. At lower labeled-data ratios, the performance under $95\%$ is already competitive, and at higher labeled-data ratios it becomes the strongest setting. For example, unseen-AUT(A) reaches $0.810$ at $20\%$ labeled data and $0.821$ at $40\%$ labeled data under the $95\%$ threshold, which are the best results among the compared thresholds for those settings. Even in cases where $95\%$ is not the absolute best, it remains very close to the top-performing configuration. This suggests that retaining a larger proportion of the latent structure is beneficial and does not introduce instability on BODMAS.}

\paragraph{AndroZoo} \textcolor{black}{On AndroZoo, the benefit of using a higher SVD threshold is even more pronounced. The $50\%$ threshold can lead to clear performance drops in some settings, whereas the $95\%$ threshold consistently restores strong unseen detection performance. For instance, at $5\%$ labeled data, unseen-AUT(A) increases from $0.641$ under $50\%$ to $0.855$ under $95\%$, and at $10\%$ labeled data it rises from $0.541$ to $0.859$. Across the remaining labeled-data ratios, the $95\%$ threshold either achieves the best result or remains very close to the best-performing setting. These observations indicate that a more conservative dimensionality reduction strategy is important for preserving discriminative information on AndroZoo.} 

\paragraph{APIGraph} \textcolor{black}{On APIGraph, all three thresholds yield strong performance, but the $95\%$ threshold remains consistently competitive and often achieves the best or near-best result. For example, unseen-AUT(A) reaches $0.917$ at $10\%$ labeled data, $0.930$ at $20\%$, and $0.931$ at $40\%$ labeled data under the $95\%$ threshold. Although the absolute differences across thresholds are smaller than those observed on AndroZoo, the results show that retaining $95\%$ of the variance does not hurt performance and instead provides a stable operating point across all labeled-data ratios. Taken together with the results on BODMAS and AndroZoo, these trends support the use of a $95\%$ SVD threshold as a generally good default choice for the proposed framework. }

\subsection{Computational Overhead of SVD and GPM}

\textcolor{black}{In this section, we analyze the computational overhead introduced by the SVD and the GPM under different labeled-data ratios. Specifically, we measure the run time required for the SVD computation and the GPM update step across the APIGraph, AndroZoo, and BODMAS datasets. The results are summarized in} Table~\ref{tab:runtime_memory_overhead_all}.

\textcolor{black}{Across all datasets, the time required for SVD computation remains small and stable as the labeled data ratio increases. On APIGraph, the SVD computation takes approximately $4.80$--$5.10$ seconds while the total training time ranges between $1290$ and $1467$ seconds, whereas on AndroZoo it varies from $10.63$ to $13.16$ seconds with total run time between $1189$ and $1453$ seconds. On BODMAS, the SVD overhead is even smaller, around $1.09$--$1.21$ seconds compared to a total run time of $532$--$710$ seconds. The GPM update introduces negligible overhead across all datasets, requiring only $0.17$--$0.20$ seconds on APIGraph, $0.30$--$0.41$ seconds on AndroZoo, and $0.02$--$0.04$ seconds on BODMAS. Overall, these results indicate that incorporating SVD and GPM adds minimal computational cost relative to the total training time. This demonstrates that the proposed framework remains computationally efficient and practical for real-world deployments.}

\subsection{Memory Overhead Analysis}

\textcolor{black}{We analyze the buffer memory reorganization overhead across datasets under varying labeled data ratios. Table~\ref{tab:runtime_memory_overhead_all} summarizes the reorganization time and the resulting buffer size for APIGraph, AndroZoo, and BODMAS. The results show that the memory footprint remains modest even when the memory grows with tasks. For instance, on AndroZoo the buffer memory stabilizes around $118$--$130$ MBytes with reorganization times between $29$ and $38$ seconds. In contrast, the requirements for BODMAS and APIGraph are significantly smaller, remaining around $14.5$ MBytes and $8.5$ MBytes respectively, with reorganization times below $1$ second and around $3$--$4$ seconds. Importantly, the buffer size changes only marginally as the labeled data ratio increases. These results indicate that the memory growth remains controlled in practice and the absolute memory requirement is small relative to modern system resources, demonstrating that the proposed framework remains practical even when the memory expands across tasks.}

\begin{table}[htbp]
\captionsetup{type=table}
\caption{\textcolor{black}{Runtime and memory overhead analysis across APIGraph, AndroZoo, and BODMAS with varying labeled data ratios. Buffer memory reorganization time corresponds to total memory population time, and memory size is reported in Mega Bytes (MBytes).}}
\label{tab:runtime_memory_overhead_all}
\centering
\begin{adjustbox}{width=0.95\textwidth}
\begin{tabular}{lllllll}
\toprule
Dataset & Label ratio & Buffer reorg. time (s) & Buffer size (MBytes) & SVD time (s) & GPM time (s) & Total run time (s) \\
\midrule

\multirow{5}{*}{APIGraph}
& 5\%  & 3.2609 & 8.44 & 4.8047 & 0.1822 & 1454.7591 \\
& 10\% & 3.5565 & 8.56 & 4.8579 & 0.1933 & 1345.0489 \\
& 20\% & 3.6356 & 8.56 & 5.0114 & 0.1992 & 1290.6780 \\
& 30\% & 3.9652 & 8.56 & 5.1027 & 0.1715 & 1455.4202 \\
& 40\% & 3.8255 & 8.50 & 5.0134 & 0.1841 & 1467.6545 \\
\cmidrule(lr){1-7}

\multirow{5}{*}{AndroZoo}
& 5\%  & 29.1482 & 118.33 & 11.3073 & 0.3810 & 1356.7664 \\
& 10\% & 34.0686 & 128.59 & 10.6384 & 0.3636 & 1189.8041 \\
& 20\% & 36.8561 & 128.38 & 13.1620 & 0.3916 & 1283.7342 \\
& 30\% & 37.7255 & 128.45 & 12.7992 & 0.4120 & 1322.1672 \\
& 40\% & 37.1821 & 129.81 & 11.4308 & 0.3030 & 1453.0899 \\
\cmidrule(lr){1-7}

\multirow{5}{*}{BODMAS}
& 5\%  & 0.3379 & 14.48 & 1.2069 & 0.0247 & 665.4773 \\
& 10\% & 0.3509 & 14.60 & 1.1618 & 0.0285 & 532.4300 \\
& 20\% & 0.3665 & 14.46 & 1.1728 & 0.0245 & 599.1412 \\
& 30\% & 0.3338 & 14.60 & 1.0872 & 0.0287 & 568.9392 \\
& 40\% & 0.3384 & 14.60 & 1.1490 & 0.0366 & 709.7778 \\

\bottomrule
\end{tabular}
\end{adjustbox}
\end{table}
\section{Ablation Study}
\label{app:ablation_study}
\par In this section, we study the sensitivity of the proposed approach to various components. Specifically, we focus on three components namely buffer memory, GPM, and gradient direction of GPM. 

\subsection{Sensitivity to the Buffer Memory and GPM}
Buffer memory is used to store and replay the training exemplars to restore the knowledge of previous tasks to improve plasticity. GPM reduces catastrophic forgetting by projecting gradient updates away from subspaces that encode knowledge from earlier tasks, which is particularly important in semi-supervised continual learning.


\begin{table}[!hb]
  \caption{Ablation study demonstrating the sensitivity of the proposed method on various components including buffer memory and GPM. The best values are marked in \textbf{bold}.}
  \label{tab:ablation_study}
  \centering
 

   
    
\footnotesize
 \begin{tabular}{llllllll}
 
  \multicolumn{5}{c}{unseen-AUT (A)}\\
      \cmidrule(lr){3-5} 
   Memory & GPM & BODMAS  &  AndroZoo &  API Graph\\
   \midrule
   
    \nomark & \nomark & 0.514 $\pm$ 0.040 & 0.533 $\pm$ 0.000 & 0.577 $\pm$ 0.065\\
    \yesmark & \nomark & 0.585 $\pm$ 0.081 & 0.533 $\pm$ 0.000 & 0.554 $\pm$ 0.053\\
      \nomark& \yesmark & 0.695 $\pm$ 0.052 & 0.533 $\pm$ 0.000 & 0.854 $\pm$ 0.053\\
      \midrule
      \yesmark & \yesmark & \textbf{0.810 $\pm$ 0.027} & \textbf{0.855 $\pm$ 0.018} & \textbf{0.930 $\pm$ 0.002}\\  
    
    \bottomrule    
  \end{tabular}

   \end{table}
   We compare the impact of each component on unseen-AUT (A) and the results are presented in Table~\ref{tab:ablation_study}. We made the following observations from these results. On all datasets, absence of both the components severely impacts the detection performance of the unseen malware samples. As a result, lowest unseen-AUT (A) is observed. GPM has a greater impact on detection performance than buffer memory on the BODMAS and APIGraph datasets.

\begin{table}[htb]
  \caption{Ablation study demonstrating the sensitivity of the proposed method in detecting unseen malware (unseen-AUT (A)) to the gradient directions in the GPM. The best values are marked in \textbf{bold}.}
  \label{tab:ablation_study_threshold}
  \centering
     \begin{tabular}{llllllll}
  \multicolumn{4}{c}{unseen-AUT (A)}\\
      \cmidrule(lr){2-4} 
     Threshold& BODMAS  &  AndroZoo &  API Graph\\
   \midrule
   
    10\%  & 0.802 $\pm$ 0.020 & \textbf{0.855 $\pm$ 0.018} & \textbf{0.930 $\pm$ 0.002}\\
    50\%  & 0.763 $\pm$ 0.054 & 0.831 $\pm$ 0.023 & 0.926 $\pm$ 0.003\\
      99\% & \textbf{0.810 $\pm$ 0.027} & 0.829 $\pm$ 0.039 & 0.901 $\pm$ 0.010\\

    \bottomrule    
  \end{tabular}

   \end{table}

\subsection{Sensitivity to the Gradient Directions in GPM}
\label{app:svd_threshold}
\par In this section, we study the sensitivity of the proposed method in detecting unseen malware (unseen-AUT (A)) to the gradient directions of past tasks (refer to Table~\ref{tab:ablation_study_threshold}). A higher threshold implies considering all the gradient directions of the past tasks. On the BODMAS dataset, detection performance improves as the threshold increases. Intuitively, this suggests that to enhance detection performance on the current task, the current gradient directions must be orthogonal to those of all past tasks to minimize interference.

In contrast, detection performance decreases with increasing threshold on the APIGraph and AndroZoo datasets. This can be attributed to the data curation process and the inherent semantic structure of the dataset. Intuitively, in the presence of such semantic structure across successive tasks, it is sufficient to project the current gradient directions orthogonal to only a few of the past tasks’ gradient directions. Consequently, increasing the threshold degrades detection performance, as enforcing orthogonality to all gradient directions of all past tasks may disrupt the learning of the dataset’s inherent semantic structure.

\section{Limitations}
\label{app:limitations}
In this section, we describe the limitations of the proposed method from the design perspective and observations made during the empirical studies.

\noindent \textit{Size of buffer memory:} This work assumes growing memory size to store the partial labeled samples until the last seen task. However, there after the size grows steadily based on the labeling budget. 

\noindent \textit{Security of ML model:} It is important to acknowledge that the developed model could potentially be used by adversaries to create more sophisticated malware. 
By publishing our findings, we aim to strike a balance between advancing defensive capabilities and mitigating the risk of misuse. Addressing issues such as adversarial attacks is not considered in this work.

\section{Additional Details About Our Proposed Method}
\label{app:model}
\par The model used in our experiment is composed of two subnetworks: encoder and classifier. The encoder first reduces the dimensionality to 100 and then progressively increases it to 250 and 500 before decreasing it to 150 and 50. Output after each layer is subjected to batchnorm, dropout (with ratio of 0.2) and a ReLU activation. Weights in the layers are initialized using Kaiming uniform method to improve the convergence of our model. The classifier consists of a simple linear layer that outputs 2 neurons with one neuron for each class (benign and malware). The model outputs raw logits from the classifier.

The hyperparameters used in our method are the percentage of batch size that contains the samples from the memory (b\_m), percentage of malware samples (bma) to be present in b\_m, learning rate, weight decay, maximum threshold ($\tau$), and analyst labels. We find the best hyperparameters using grid search method. The search space for b\_m is $0.1$ to $0.7$, for bma $0.1$ to $0.9$, learning rate is $\{10^{-1},10^{-2},10^{-3},10^{-4},10^{-5}\}$, weight decay is $\{10^{-1},10^{-2},10^{-3},10^{-4},10^{-5},10^{-6},10^{-7},10^{-8},10^{-9}\}$, and $\tau_{max}$ is $0.1$ to $0.3$. The best hyperparameters for BODMAS datset are: b\_m is $0.5$, bma is $0.8$, learning rate is $10^{-1}$, weight decay is $10^{-9}$, and $\tau_{max}$ is $0.09$. For AndroZoo dataset, b\_m is $0.3$, bma is $0.4$, learning rate is $10^{-2}$, weight decay is $10^{-1}$, and $\tau_{max}$ is $0.05$. For APIGraph, b\_m is $0.6$, bma is $0.7$, learning rate is $10^{-1}$, weight decay is $10^{-4}$, and $\tau_{max}$ is $0.05$.   

\section{Details About Baseline Methods}
\label{app:baseline}
\par \emph{HCL.} For hierarchical contrastive learning, we use the same model as used by the original paper~\cite{291253}. The model consists of an encoder that  reduces the dimensions from input feature size to 512, 384, 256 and then finally to 128 and the classifier that has two hidden layers of 100 neurons each and finally a 2-neuron output layer. The model outputs normalized softmax probabilities.
Output of each hidden layer is subjected to ReLU. The model is trained for 50 epochs with early stopping of patience 3 based on the PR-AUC value of the validation set. A batch size of 64 is used for all datasets with margin value set to 1 and $\lambda$ to 100. Adam optimizer is used with default values and a grid search is conducted for finding the best learning rate and weight decay. The search space for best hyperparameters is  $\{10^{-1},10^{-2},10^{-3},10^{-4},10^{-5},10^{-6},10^{-7}\}$. 
The best hyperparameters for the datasets found are as follows: For the API Graph, the learning rate is \( 0.0001 \) and the weight decay is \( 10^{-6} \). For BODMAS, the learning rate remains \( 0.0001 \), with the weight decay adjusted to \( 10^{-7} \). Lastly, for AndroZoo, the learning rate is configured as \( 10^{-5} \) and the weight decay is set to \( 10^{-5} \).

\par \emph{CADE.} For CADE, we use the same model architecture as introduced by the original paper which consists of 512-128-32-7 dimensions with ReLU after each layer except the last layers in both the encoder and decoder. We use a MLP with two hidden layers of 100 neurons that finally outputs two neurons as in the case of HCL. The auto encoder is trained using the same setup as previously discussed for hierarchical contrastive learning. The MLP is also trained with Adam Optimizer and has the same learning rate as the ones used in CADE. MLP is trained for 50 epochs with early stopping of patience 7 based on the PR-AUC value of the validation set. The search space for best hyperparameters is  $\{10^{-1},10^{-2},10^{-3},10^{-4},10^{-5},10^{-6},10^{-7}\}$. 
The best hyperparameters for the datasets found are as follows: for the API Graph, the learning rate of \( 0.001 \) and the weight decay \( 10^{-7} \) gave the best results. For BODMAS, the learning rate is \( 0.01 \) with the weight decay adjusted to \( 10^{-7} \). Lastly, for AndroZoo, the learning rate is set to \( 0.001 \) and the weight decay is set to \( 10^{-6} \).

\par \emph{Continual learning baselines.} We use AGEM and EWC continual learning methods provided by the avalanche library. In AGEM, we set the patterns\_per\_exp to 256, sample size of 64 and used a mini-batch size of 64 for training. In case of EWC, we set the regularization strength (ewc\_lambda) to 0.4 and also used the same mini-batch size. While training on these methods we use the same optimizer and model as used in CADE and HCL. We implemented MIR and CBRS methods and for both we use a step scheduler with a gamma of 0.96 and early stopping with patience 3. We use SGD as our optimizer and do a grid search for learning rate and weight decay in the same space as before. The replay size is set to 1500 and memory size of 2000 for all datasets except for APIGraph in CBRS, where we use a replay size of 500 and a memory size of 1000, with minority allocation set to 0.8. The search space for best hyperparameters is  $\{10^{-1},10^{-2},10^{-3},10^{-4},10^{-5},10^{-6},10^{-7}\}$.

For MIR, the best hyperparameters for the APIGraph were when the learning rate was \( 0.01 \) and the weight decay was \( 0.01 \). Similarly, for BODMAS, the learning rate was found to be \( 0.001 \) and the weight decay was \( 10^{-6} \). Lastly, for AndroZoo, the learning rate was \( 0.01 \) and the weight decay was \( 0.001 \).

On CBRS, we found the final best hyperparameters for the APIGraph: a learning rate of \( 0.1 \) and a weight decay of \( 0.001 \). For BODMAS, the learning rate was \( 0.01 \) and the weight decay was \( 10^{-6} \). Finally, for AndroZoo, the learning rate was \( 0.1 \) and the weight decay was \( 0.001 \).

Similarly, for EWC, on the APIGraph dataset, the learning rate was \( 0.01 \) and the weight decay was \( 10^{-5} \). On the BODMAS dataset, the learning rate was \( 0.1 \) with a weight decay of \( 10^{-7} \). For AndroZoo, the learning rate was \( 0.01 \) and the weight decay was \( 10^{-5} \).

Lastly, for AGEM, on the APIGraph dataset, the learning rate was \( 0.01 \) and the weight decay was \( 10^{-6} \). On the BODMAS dataset, the learning rate was \( 0.01 \) and the weight decay was \( 10^{-7} \). Similarly, for AndroZoo, the learning rate was \( 0.01 \) with a weight decay of \( 10^{-7} \).

\section{Hardware and ML Frameworks}
\label{app:hardware}
\par Our experiments were carried out on a server equipped with 376 GB of memory, 104 cores (Intel(R) Xeon(R) Gold 6230R CPU @ 2.10 GHz), and 2 Nvidia Quadro RTX 5000 GPUs. We leverage the open-source continual learning library Avalanche (version 0.2.1) to implement baseline methods like EWC and A-GEM whereas other baselines like MIR, CBRS, and the proposed method are implemented using PyTorch version 1.13 with CUDA version V11.6.124. The Python version is 3.8.13.

\subsection{Ethics Considerations}
\label{app:ethics}
We utilized datasets stored as NPZ files at ~\cite{291253} containing numerical representations of preprocessed data. These files were provided `as-is,' and we did not engage in any data extraction from malware or benign software packages ourselves. The files were devoid of any identifiable information, ensuring no sensitive data was present.

It is important to acknowledge that the developed model could potentially be used by adversaries. 
However, these challenges are not unique to our research; they are common considerations in any new development within the field of malware detection. By publishing our findings, we aim to strike a balance between advancing defensive capabilities and minimizing the risk of misuse. Addressing issues such as adversarial attacks on the proposed method is beyond the scope of this work. We plan to explore adversarial robustness as part of our future work.

\subsection{Reproducibility}
\label{app:reproduce}
The code and results are presented in an accessible way, allowing the research community to build upon them. The code repository is available at \url{https://github.com/amalapuram/SEED}.

For reproducibility of the reported results, first create a virtual environment using the \textcolor{blue}{requirements.txt} file located in the GitHub repository. Ensure that the version numbers of the software packages match exactly. The following software versions were used: Python 3.8.13, PyTorch 1.13 with CUDA version V11.6.124, and the open-source continual learning library \textcolor{blue}{Avalanche} version 0.2.1. After creating the environment, use the commands provided in \textcolor{blue}{configurations.md} to reproduce the results.

\section{Conclusions and Future Work}
In this paper, we introduced a novel semi-supervised continual learning-based method designed to adapt to distribution shifts in both benign and malicious applications under limited labeled data. The proposed method constructs a representation space using basis vectors derived from the limited labeled data via singular value decomposition. This representation space is then used to rank the unlabeled data—which may contain novel malware and benign samples—based on their distance from the labeled data. The most challenging samples are selected for labeling according to the available labeling budget. In this way, we reformulated the problem of unsupervised malware detection as a semi-supervised learning task under a limited labeling budget. Our empirical evaluations on Android and Windows PE malware datasets demonstrated that the proposed approach achieved competitive results using only partially labeled data. We believe our work will inspire and encourage further research in this direction.


As part of future work, we will focus on developing unsupervised anomaly detection methodologies for malware detection that are robust to distribution shifts, as such approaches effectively eliminate the need for human analysts to label novel malware families. Additionally, we believe our research has broader applicability in domains such as network intrusion detection, anomaly detection, fraud detection, among others.

\section*{Acknowledgments}
This work was partially supported by the SERB/ANRF, New Delhi, India, Grant number: JBR/2021/000005 and
the DST-EPSRC project \emph{Synergy: Taking openness to the next level in 6G networks}, Grant number: DST/INT/UK/Telecom/P-181/2024(G).

\printbibliography


\end{document}